\documentclass[a4paper,12pt,openright,oneside]{book}
\usepackage{ucy}
\usepackage[shortlabels]{enumitem}
\usepackage{hyperref}
\usepackage{fancyhdr}
\usepackage[ruled,vlined,linesnumbered]{algorithm2e}
\usepackage{framed}
\usepackage{import}
\usepackage{adjustbox}
\usepackage{multirow}
\newcolumntype{C}[1]{>{\centering\let\newline\\\arraybackslash\hspace{0pt}}m{#1}}
\newcolumntype{L}[1]{>{\let\newline\\\arraybackslash\hspace{0pt}}m{#1}}
\usepackage{pdfpages}
\usepackage{url}

\usepackage[T1]{fontenc}
\usepackage{amssymb}

\usepackage{afterpage}

\usepackage{lscape}
\usepackage{longtable}
\usepackage{times}
\usepackage{breakurl}             
\usepackage{underscore}           
\usepackage{macros}
\usepackage{graphicx}
\usepackage{epstopdf}
\usepackage{float}
\usepackage{bm}
\usepackage{listings}
\newcommand{\remove}[1]{} 
\usepackage{parallel,enumitem}
\usepackage{graphicx}
\newtheorem{definition}{Definition}
\newtheorem{lemma}{Lemma}
\newtheorem{theorem}{Theorem}
\newtheorem{proposition}{Proposition}
\newtheorem{corollary}{Corollary}

\makenomenclature

\fancypagestyle{plain}{%
    \fancyhf{}%
    \fancyfoot[R]{\thepage}%
}

\majoradviser{Associate Professor Anna Philippou}
\committeechair{Associate Professor Chryssis Georgiou}
\associateadviser{Professor   Yannis Dimopoulos}
\associateadviser{Professor Maciej Koutny}
\associateadviser{Associate Professor Ivan Lanese }

\makeatletter
\renewcommand{\paragraph}[1]{%
  \@startsection{paragraph}{4}%
  {\z@}{3.25ex \@plus 1ex \@minus .2ex}{-1em}%
  {\normalfont\normalsize\bfseries }%
	{#1.}
}
\makeatother

\begin{document}
\onehalfspace

\pagenumbering{roman}

\coverpage
\copyrightpage
\clearpage
\approvalpage
\declarationpage
\setcounter{page}{1}


\begin{otherlanguage}{greek}	
\chapter*{Περίληψη}

Ο αναστρέψιμος υπολογισμός είναι μια μη συμβατική μορφή υπολογισμού που επεκτείνει τον τυπικό τρόπο υπολογισμού με τη δυνατότητα αντίστροφης εκτέλεσηs λειτουργιών. Η αναστρεψιμότητα προσέλκυσε πρόσφατα αυξανόμενη προσοχή σε διάφορες ερευνητικές κοινότητες καθώς από τη μία υπόσχεται υπολογισμούς χαμηλής ισχύος και, από την άλλη, είναι εφαρμόσιμη σε μια ποικιλία εφαρμογών.

Η διερεύνηση της αναστρεψιμότητας μέσω τυπικών μοντέλων καθορίζει τα θεωρητικά θεμέλια για το τι είναι η αναστρεψιμότητα, ποιο σκοπό εξυπηρετεί, και πως ωφελεί τα φυσικά και τεχνητά συστήματα.  Ως εκ τούτου, προτείνουμε μια αναστρέψιμη προσέγγιση για τα δίκτυα Πέτρι, εισάγοντας μηχανισμούς και σχετική λειτουργική σημασιολογία για την αντιμετώπιση των προκλήσεων που έχουν οι κύριες μορφές αναστρεψιμότητας. 

Τα δίκτυα Πέτρι είναι μια μαθηματική γλώσσα για μοντελοποίηση και συλλογισμό κατανεμημένων συστημάτων.  Η πρόταση  μας αφορά μία παραλλαγή των δικτύων Πέτρι, που ονομάζεται Αναστρέψιμα Δίκτυα Πέτρι,  όπου τα διακριτικά ενός δικτύου ξεχωρίζουν μεταξύ τους με μοναδικές ταυτότητες. Δείχνουμε τη δυνατότητα εφαρμογής της προσέγγισής μας σε ένα μοντέλο μεταβολικής διαδρομής και ένα σύστημα επεξεργασίας συναλλαγών όπου και τα δύο εκδηλώνουν αναστρέψιμη συμπεριφορά.

Μια άμεση επέκταση του αρχικού μοντέλου συμπεριλαμβάνει την παροχή πολλαπλών διακριτικών που εκπροσωπούν τον ίδιο τύπο. Μία τέτοια επέκταση σε ένα μοντέλο όπως τα δίκτυα Πέτρι, έχει ως αποτέλεσμα αντίστροφες συγκρούσεις όπου ένα διακριτικό μπορεί να έχει τοποθετηθεί σε μία θέση από διαφορετικές μεταβάσεις. Προτείνουμε λοιπόν μια επέκταση των αναστρέψιμων δικτύων Πέτρι που επιτρέπει πολλαπλά διακριτικά του ίδιου τύπου σε ένα μοντέλο, ενώ παράλληλα διασφαλίζεται ο ντετερμινισμός κατά την αναστροφή. Συγκεκριμένα,  στην προσέγγιση την οποία διερευνούμε, διαφορετικά διακριτικά που βρίσκονται στην ίδια θέση μπορούν να διακριθούν με βάση την πορεία που έχουν ακολουθήσει στο δίκτυο. Αποδεικνύουμε ότι η εκφραστική ισχύς των αναστρέψιμων δικτύων Πέτρι με πολλαπλά διακριτικά είναι ισοδύναμη με εκείνη των αναστρέψιμων δικτύων Πέτρι με μοναδικά διακριτικά. Προτείνουμε επίσης την αντίθετη προσέγγιση, η οποία θεωρεί ότι όλα τα διακριτικά ενός συγκεκριμένου τύπου είναι πανομοιότυπα, αγνοώντας την πορεία που ακολούθησαν κατά την εκτέλεση του δικτύου. Δείχνουμε την ευρωστία αυτής της προσέγγισης ως τεχνική μοντελοποίησης συστημάτων που αφορούν πόρους μέσω ενός παραδείγματος από τη βιοχημεία, γνωστό ως αυτοπροτόλυση του νερού.

Και τα δύο προτεινόμενα μοντέλα αναστρέψιμων δικτύων Πέτρι (με μοναδικά ή πολλαπλά διακριτικά) επιτρέπουν την αναστροφή μεταβάσεων χωρίς περιορισμούς ως προς το πότε και αν θα αναστραφεί η εκτέλεση ή όχι.  Με στόχο να περιορίσουμε την αναστρεψιμότητα, επεκτείνουμε τη σημασιολογία μας συσχετίζοντας τις μεταβάσεις με συνθήκες των οποίων η ικανοποίηση επιτρέπει την εκτέλεση μεταβάσεων προς τα εμπρός/πίσω.

Καταλήγοντας,  για να διευκολύνουμε την ανάλυση της συμπεριφοράς μοντέλων αναστρέψιμου υπολογισμού διατυπώνουμε στο πλαίσιο μας βασικές ιδιότητες όπως η ασφάλεια και η προσβασημότητα όταν εφαρμόζονται διαφορετικές στρατηγικές αναστρεψιμότητας. Παρουσιάζουμε το πλαίσιο μαζί με τις σχετικές ιδιότητες με ένα μοντέλο ενός καινοτόμου, κατανεμημένου αλγορίθμου που επιλέγει κεραίες σε κατανεμημένες σειρές κεραιών.

\end{otherlanguage}
\newpage


\chapter*{Abstract}

Reversible computation is an unconventional form of computing that extends the standard forward-only mode of computation with the ability to execute a sequence of operations in reverse at any point during computation. 
Reversibility has recently been attracting increasing attention in various research communities, as on the one hand it promises low-power computation, and on the other hand it is inherent or of interest in a variety of applications.

Exploring reversibility through formal models formulates the theoretical foundations of what reversibility is, what purpose it serves, and how it benefits natural and artificial systems. As such, in this thesis we propose a reversible approach to Petri nets by introducing machinery and associated operational semantics to tackle the challenges of the main forms of reversibility. Petri nets are a mathematical language for modelling and reasoning about distributed systems. Our proposal concerns a variation of cyclic Petri nets, called Reversing Petri Nets (RPNs) where tokens are persistent and distinguished from each other by an identity. We demonstrate the applicability of our approach with a model of the ERK signalling pathway and an example of a transaction-processing system both featuring reversible behaviour.

An immediate extension of the original model includes allowing multiple tokens of the same base/type to occur in a model. The addition of token multiplicity into a model like Petri nets results in various backward conflicts where a token can be generated in a place because of different transition firings. We therefore propose an extension of reversing Petri nets that allows multiple tokens of the same base/type to occur in a model while still ensuring backward determinism. Specifically, we explore the individual token interpretation where one distinguishes different tokens residing in the same place by keeping track of where they come from. We prove that the expressive power of RPNs with multi tokens is equivalent to that of RPNs with single tokens, and we measure the expressiveness in terms of Labelled Transition Systems (LTSs) up to isomorphism of reachable parts that can be denoted by nets of the respective RPN models. We also propose the collective token interpretation, as the opposite approach to token ambiguity, which considers all tokens of a certain type to be identical, disregarding their history during execution. We show the robustness of this approach as a modelling technique for resource-aware systems by modelling an example from biochemistry, known as the autoprotolysis of water.

Both of the proposed models of RPNs (with single or multi tokens) implement the notion of uncontrolled reversibility, meaning that it specifies how to reverse an execution and allows to do so freely, yet it places no restrictions as to when and whether to prefer backward execution over forward execution or vice versa. In this respect, a further aim is to control reversibility  by extending our formal semantics where transitions are associated with conditions whose satisfaction allows the execution of transitions in the forward/reversed direction.

Finally, in order to  facilitate the analysis of the behaviour of reversible  models, we formulate the basic properties of our framework such as safety, reachability, precedence and exception  when different notions and strategies of reversibility are applied. We illustrate the framework along with the associated properties with a model of a novel, distributed algorithm for antenna selection in distributed antenna arrays.
 

\newpage


\chapter*{Acknowledgments}

Undertaking this PhD has been a truly life-changing experience for me and it would not have been possible to do without the support and guidance that I received from many people. I would like to thank the following people, without whom I would not have been able to complete this research.

I would like to start by expressing my sincere gratitude  and my indebtedness to my supervisor Dr. Anna Philippou, for giving me the opportunity to do a PhD thesis under her guidance. Her continuous support, inspiring guidance, invaluable encouragement, and immense knowledge pushed me to sharpen my way of thinking and brought my work to a higher standard.  I would like to thank her for all the practical and financial support, and particularly, for the precious time that she invested in me.  As a token of my gratitude I would like to expose her virtues of patience, kindness and hard work.

I would also like to offer my heartfelt thanks to Kamila  Barylska, Anna Gogolinska, Lukasz Mikulski, and  Marcin Piatkowski  for hosting me at the University of Torun, and whose ideas and guidance helped me to accomplish an important part of this work. I also appreciate the valuable contribution of my co-authors  Bogdan Aman, Gabriel Ciobanu, Yiannis Demopoulos, Eleutheria Kouppari, Stefan Kuhn,  Harun Siljak, and Irek Ulidowski.  Many thanks to the committee members  Chryssis Georgiou, Yiannis Demopoulos,  Maciej Koutny, and Ivan Lanese for accepting to review my thesis, as without them I could not have completed this dissertation. I would also like to thank the EU COST ACTION IC1405 and its participating members that gave me the opportunity to present preliminary results of my thesis in order to receive constructive feedback.

I am very thankful to the team members of the “Foundations of Computing Systems and Theoretical Computer Science Laboratory” at the University of Cyprus for being around at times of very intense effort, expressing their support, and providing useful feedback. I am thankful to the members  of the faculty of the department of Computer Science with whom I have collaborated over these years as part of my teaching assistance duties, and for always displaying a constructive high standard of professionalism in their duties. Similarly, I would also like to thank the Department’s staff who willingly and patiently provided their important services whenever required.

To conclude, I would like to say a heartfelt thank you to my family for all the support they have shown me, for their wise counsel and sympathetic ear, for always believing in me and encouraging me through this research. I cannot forget to thank my friends, for providing stimulating discussions as well as happy distractions to rest my mind outside of my research.


\newpage


\chapter*{Thesis Contributions}
\singlespace
	The following papers were published as a result of the research done for the requirements of this dissertation and are its primary contributing sources. 

\begin{enumerate}

		\item Philippou A. and \textbf{Psara K.}, 2018. Reversible computation in Petri nets. In  Proceedings of the 10th  International Conference on Reversible Computation (pp. 84-101). Lecture Notes in Computer Science volume 11497. Springer.
		
		\item  Barylska  K.,  Gogolinska A.,  Mikulski L.,  Philippou A.,  Piatkowski, M. and  \textbf{Psara. K.}, 2018. Reversing computations modelled by coloured Petri nets. In Proceedings of the International Workshop on Algorithms {\&} Theories
	for the Analysis of Event Data 2018 (pp. 91–111). {CEUR} Workshop Proceedings volume 2115. CEUR-WS.org.

	\item  Philippou A., \textbf{Psara K.} and Siljak H., 2019. Controlling reversibility in reversing Petri nets with application to wireless communications. In Proceedings of the 11th International Conference on Reversible Computation (pp. 238-245). Lecture Notes in Computer Science volume  11497. Springer.
	
	
	
	
	\item Siljak H., \textbf{Psara K.} and Philippou A., 2019. Distributed antenna selection for massive MIMO using reversing Petri nets. IEEE Wireless Communications Letters, volume 8(5), pp.1427-1430.
	
	
	

	\item Dimopoulos Y., Kouppari E., Philippou A. and \textbf{Psara K.}, 2020. Encoding Reversing Petri Nets in Answer Set Programming. In Proceedings of the 12th International Conference on Reversible Computation (pp. 264-271). Lecture Notes in Computer Science volume 12227. Springer.

	\item Kuhn S., Aman B., Ciobanu G., Philippou A., \textbf{Psara K.} and Ulidowski I., 2020. Reversibility in Chemical Reactions. In Reversible Computation: Extending Horizons of Computing - Selected
	Results of the {COST} Action {IC1405} (pp. 151-176), Lecture Notes in Computer Science volume 1270. Springer.

\end{enumerate}

\remove{
This thesis is founded on the knowledge acquired by the author's involvement in the authorship of the following journal articles and conference papers:

\singlespace
\subsection*{Journal Articles}

\begin{enumerate}

\end{enumerate}

\subsection*{Conference and Workshop Proceedings}

\begin{enumerate}
	\setcounter{enumi}{1}

\end{enumerate}
}

\newpage

\onehalfspace

\tableofcontents
\newpage
\vspace{1in}
\printnomenclature




\listoffigures
\newpage



\pagenumbering{arabic}


\chapter{Introduction}\label{sec:Introduction}
\section{Motivation}
\emph{Reversible computation}~bi\cite{RCwille} is an unconventional form of computing where computation can be executed in the backward direction as effortlessly as it can be executed in the standard forward direction. In particular, individual operations can be carried out reversibly and thus, at any point of the execution we are able to uniquely identify the forward or backward state. Hence, every reversible computation process can be traced backward uniquely from end to start whilst exhibiting both forward and backward determinism. Its characteristics make reversibility a very promising paradigm that extends the current irreversible mode of computation by delivering novel computing devices and software.

The study of reversibility originated in the 1960s when scientists and mathematicians started to be concerned with energy efficient computation. In particular, Landauer~\cite{Landauer} answered many questions by proving that logically irreversible operations result in bit erasure that causes heat dissipation and, in general, loss of energy. In particular, a proportion of electrical power consumed by current computers is lost in the form of heat because every time a computer throws away bits of information it generates at least $kTln2$ ($k$ is the Boltzmann constant which is approximately $1.38	\times10^{-23}$ $J/K$, $T$ is the temperature of the heat sink in kelvins, and $ln2$ is the natural logarithm of 2 which is approximately 0.69315) of entropy for each bit of information it erases. 

 Reversibility offers the potential for computationally proceeding in the forward direction as well as in reverse resulting in going back to states visited before or even states that cannot be reached by going forwards alone. 
It is encountered  in a wide range of systems.
 For instance, it is a property of biochemical systems~\cite{ERK}, where reactions can be executed in both the forward and backward direction based on the imposed physical conditions. Another application is quantum computing which in contrast to classical computing is always reversible~\cite{feynman}. At the same time, it is on many occasions a desirable system property. To begin with, reversible computing comes to solve the miniaturisation limitations of current technology that aim to increase the speed and capacity of circuits. recovery from failures such as corrupted data, deadlocked programs and breached security is crucial and could be effortlessly obtained in the presence of reversibility. Further applications are encountered in programming languages and concurrent transactions.


One line of work in the study of reversible computation has been the investigation of its theoretical foundations~\cite{causality}. Understanding the role that reversibility plays in natural systems calls for the development of realistic formal models for concurrent and distributed systems.  
Reversible models can be based on already existing abstract formalisms or specially proposed languages, and can be used not only for modelling reversible systems but also for investigating suitable notions of behavioural equivalences, logics and other analysis techniques. Furthermore, the study of reversible formalisms  may aid  the understanding of the foundation of reversibility and can help towards the understanding, modelling and implementing reversible actions as a feature of computation. 

In the context of the theoretical study of reversible computation, the different strategies of reversing and their relationships are  being investigated and have led to the definition of different forms of reversibility: While in the sequential setting reversibility is generally understood as the ability to execute past actions in the exact inverse order in which they have occurred, a process commonly referred to as \emph{backtracking}, in a concurrent scenario it can be argued that reversal of actions can take place in a more liberal fashion. The main alternatives proposed are those of \emph{causal order reversibility}\cite{causality}, a form of reversing where an action can be undone provided that all of its effects (if any) have been undone beforehand, and \emph{out-of-causal order reversibility}\cite{Bonding}, a form of reversing featured most notably in biochemical systems. 

\section{Previous Work}

Even though reversing computational processes in concurrent and distributed systems has many promising applications, it also has many technical and conceptual challenges. 
The main challenge being the ability to identify the legitimate backward moves by  maintaining the information needed to reverse executed computation, e.g., to keep
track of the history of execution and the choices that have not been made. In contrast to the sequential setting that is well understood, the concurrent setting poses the conceptual question of how do we define a causally-respecting  order of execution. \emph{Causal-consistent reversibility}~\cite{causality} is the most common notion of reversibility in the concurrent and distributed setting. Since its definition, various approaches in formal models and applications of causal-consistent reversibility have been considered. The first works 
handling reversibility in process calculi are
the Chemical Abstract Machine~\cite{abstract}, a calculus inspired by reactions between 
molecules whose operational semantics define both forward and reverse computations, and
RCCS~\cite{RCCS}, an extension of  the Calculus of Communicating Systems (CCS)~\cite{CCS}
equipped with a reversible mechanism that uses memory stacks for concurrent 
threads, further developed in~\cite{TransactionsRCCS,BiologyCCSR}.  This mechanism was 
represented at an abstract level using categories with an application to Petri 
nets~\cite{DBLP:journals/entcs/DanosKS07}. Subsequently,  a general
method for reversing process calculi with CCSK being a special 
instance of the methodology was proposed in~\cite{Algebraic}. This proposal introduced the use
of communication keys to bind together 
communication actions as needed for isolating communicating partners during action reversal. 
Reversible versions of the $\pi$-calculus 
include  $\rho\pi$~\cite{LaneseMS10} and $R\pi$~\cite{DBLP:conf/lics/CristescuKV13}.

 
 

While all the above concentrate on the notion of causal reversibility, approaches considering
other forms of reversibility have also been proposed. 
Consider every state of the execution to be a result of a series of actions that have causally contributed to the existence of the current state. If the actions were to be reversed in a causally-respecting manner then we would only be able to move back and forth through previously visited states. Therefore, one might wish to apply \emph{out-of-causal-order reversibility} in order to create fresh alternatives of current states that were formerly inaccessible by any forward-only execution path. This has been achieved in~\cite{LocalRev} by introducing a new operator for modelling local reversibility,
a form of out-of-causal-order reversibility, whereas a mechanism for controlling out-of-causal reversibility has also been considered in~\cite{ERK}. 
The modelling of bonding within reversible processes and event structures 
was considered in~\cite{Bonding}, whereas a reversible computational calculus for modelling 
chemical systems composed of signals and gates was  proposed in~\cite{CardelliL11}.
The study of reversible process calculi has also triggered research on various
other models of concurrent computation such as reversible event 
structures~\cite{ConRev}. 

A distinguishing feature between the cited approaches
is that of  \emph{controlling reversibility}: while various
frameworks make no restriction as to when an action can be reversed (uncontrolled
reversibility), it can be argued that some means of controlling the conditions
of reversal is often useful in practice. For
instance, when dealing with fault recovery, reversal should only be triggered when a fault is encountered. Based
on this observation, a number of strategies for controlling reversibility have
been proposed: \cite{TransactionsRCCS} introduces the concept of irreversible actions, and \cite{DBLP:conf/rc/LaneseMS12} introduces compensations to deal with these irreversible actions and to avoid repeating past errors. 
Another approach is that of~\cite{ERK} which proposes the usage
of an external entity for capturing the order in
which transitions can be executed in the forward or the backward direction. In another line of work,~\cite{LaneseMSS11} defines a roll-back primitive for reversing computation, and in~\cite{LaneseLMSS13}
roll-back  is extended with the possibility of
specifying the alternatives to be taken on resuming the 
forward execution. 
Finally, in~\cite{statistical} the authors associate
the direction of action reversal with energy parameters
capturing environmental conditions of the modelled systems.

Research on reversible models from process calculi continues in Petri nets, the first approach being that of~\cite{PetriNets,BoundedPNs} which implemented a liberal way of reversing computation in Petri nets by introducing additional reversed transitions. In these works, the authors investigate the effects of adding reversed versions of selected transitions in a Petri net and they explore decidability problems regarding reachability and coverability in the resulting Petri nets. Towards examining causal consistent reversibility in Petri nets, the work in~\cite{Unbounded} investigates whether it is possible to add a complete set of effect-reverses for a given transition without changing the set of reachable markings. The authors show that this problem is in general undecidable however it can be decidable in \emph{cyclic Petri nets} where with the addition of new places these non-reversible Petri nets can become reversible while preserving their behaviour. Recently, an alternative approach~\cite{RPT,RON} on reversing Petri nets introduces reversibility in Petri nets by unfolding the original Petri net into \emph{occurrence nets} and \emph{coloured Petri nets}. The authors encode causal memories while preserving the original computation by adding for each transition its reversible counterpart. In~\cite{ReversingSteps}, the authors examine the possibility of reversing the effect of the execution of groups of various transitions (steps). They then present a number of properties which arise in this context and show that there is a crucial difference between reversing steps which are sets and those which are true multisets. 

\section{Thesis Aims}
 
%

%
In this thesis, we shall consider a particular model of computation,  known as \emph{Petri nets},  that will be extended to a reversible variant. Petri nets are a basic model of parallel and distributed systems, designed by Carl Adam Petri in 1962 in his PhD Thesis: "Kommunikation mit Automaten''~\cite{PNs,reisig2013understanding}. They constitute a graphical mathematical language that can be used for the specification and analysis of discrete event systems and they support both action-based and state-based modelling and reasoning.

In contrast to the extensive research carried out in process calculi and event structures, work done on reversing Petri nets is still at an initial stage. Thus, a first aim of this thesis is  exploring the  several results discussed in process calculi, such as the flexibility of reversible actions allowed in causal reversibility, within Petri nets. This enables us to investigate how and whether these can be embedded within the Petri net model. At the same time, while understanding the theoretical properties of reversibility within Petri Nets, an extension of Petri nets with reversibility offers an added benefit. Petri nets can be applied informally to any area or system that needs some means of representing parallel or concurrent activities as well as systems that can be described graphically like flow charts. The easy applicability of Petri nets is inherent due to their generality and permissiveness~\cite{PNs}.  Since Petri nets are visually comprehensible and simple in their application, they can be used for modelling by both practitioners and theoreticians.

However, classical Petri nets are not reversible by nature, in the sense that every transition cannot be executed in both directions. The reason being the nondeterministic nature of Petri nets.  Specifically, by observing the state of a Petri net with tokens scattered along its places it is not possible to discern the history that led to the specific state and consequently the precise transitions that can be undone. Therefore, an inverse action in classical Petri nets, needs to be added as a supplementary forward transition for achieving the undoing of a previous action. This explicit approach of modelling both forward and reverse transitions can prove cumbersome in systems that express multiple reversible patterns of execution, resulting in larger and more complex systems. Furthermore, it fails to capture reversibility as a mode of computation. Motivated by this, our intention is to study an approach for modelling reversible computation that does not require the addition of new, reversed transitions but instead allows to execute transitions in both the forward as well as the backward direction. This framework should be able to identify at each point in time the history of execution, a necessary aspect for all forms of reversibility. As such, this thesis  aims to propose a reversible approach to Petri nets which introduces machinery and associated operational semantics where executed transitions can be reversed according to three different semantic relations capturing the notions of backtracking, causal reversibility and out-of-causal-order reversibility. 

 Reversible formal models used to model reversible systems need to be able to identify the legitimate backward moves according to forward execution. When it comes to Petri nets, the ability to formally express causal dependencies based on an appropriate causality based concept is one of the most well-known concepts of Petri net theory~\cite{causal}. As such, when proposing a reversible variant of Petri nets the interplay between reversibility and concurrency should be investigated. Specifically, investigating the notion of causal dependence in Petri nets equipped with the ability to reverse is one of the primitive aims of this thesis.
 
  At the same time, out-of-causal reversibility has been observed in many important reversible examples where concurrent systems violate causality. Nonetheless, this body of research is still at a preliminary stage, and while interesting ideas have been discovered, a systematic study of the related problems and of the possible application areas is still missing.  This means that research on how to generalise causality or how it relates to out-of-causal reversibility, in order to deal with such systems, deserves much further investigation. For example, studying the properties of out-of-causal reversibility could potentially prove that both backtracking and causal reversibility are in some essence subsets of out-of-causal-order reversibility, which could potentially yield a universal approach on the strategy used for reversing.  
  Since out-of-order reversibility comes with its own peculiarities that need to be taken into consideration while modelling reversible systems, this thesis aims to understand these peculiarities and obtain an approach that addresses out-of-causal reversibility within Petri nets.  

Understanding the basics of reversibility through reversible models of concurrent computation is useful but it is not directly suitable for most applications, since they do not determine when and whether to prefer a forward over a backward action.  One of the objectives of this thesis is to  consider a strategy for controlling reversibility in Petri nets which along with the wide use of Petri nets can find application in various domains.  The resulting framework will enable us to study and understand reversibility through various case studies thus overcoming some limitations in the understanding, modelling and implementing reversible actions as a feature of computation. 

\section{Obtained Results}

 In Chapter~\ref{sec:cycles} we propose the \emph{first reversible approach to Petri nets}
  which introduces reversing Petri nets (RPNs), a variation of cyclic Petri nets where executed transitions can be reversed according to three different semantic relations capturing the notions of \emph{backtracking}, \emph{causal reversibility} and \emph{out-of-causal-order reversibility}. Furthermore, during a transition firing, tokens can be bonded with each other. The creation of bonds is considered to be the effect of a transition, whereas their destruction is the effect of the transition's reversal.
  

{\bf Causality.} When it comes to causality, cyclic structures make causality quite non-trivial since the presence of cycles exposes the need to define causality of actions within repeated executions of transitions. Indeed there are different ways of introducing reversible behaviour depending on how causality is defined. 
 In our approach, we follow the notion of causality as defined by Carl Adam Petri for one-safe nets that provides the notion of a run of a system where causal dependencies are reflected in terms of a partial order~\cite{PNs}. A causal link is considered to exist between two transitions if one produces tokens that are used to fire the other. In this partial order, a causal dependence relation is explicitly defined as an unfolding of an occurrence net which is an acyclic net that does not have backward conflicts. 
 Based on this notion of causality we handle cyclic structures by adopting ``lists of histories" associated with each transition recording all of its previous occurrences. Also, additional machinery has been necessary
 that captures the causal dependencies in the presence of cycles. We prove that the amount of flexibility allowed in causal reversibility indeed yields causally consistent semantics.

{\bf Out-of-causal order.} Until the proposal of RPNs, it had yet to be proposed a reversible approach to Petri nets which introduces machinery and associated operational semantics to tackle the challenges of all three  forms of reversibility. We  therefore propose the reversible semantics of out-of-causal-order reversibility and demonstrate that this form of reversing  is able to create new states unreachable by forward-only execution. Additionally, we establish the relationship between the three forms of reversing and define a transition relation that can capture each of the three strategies modulo the enabledness condition for each strategy. This allows us to provide a uniform treatment of the basic theoretical results.


{\bf Multiple tokens. } The  proposed model of reversing Petri nets considers tokens to be distinct from each other and assigns unique names to them. Hence, a natural extension is to allow multiple tokens of the same base/type to occur in a model. Allowing multiple instances of identical tokens results in ambiguities when it comes to causal dependencies. Depending on how we treat these ambiguities we define two different approaches when it comes to causal-order reversibility, the first one being the individual token interpretation and the second one the collective token interpretation~\cite{ConfStruct,individual,ZeroSafe}.

In chapter~\ref{sec:multi} we explore the individual token interpretation of reversing Petri nets 
where  tokens of the same type are distinguished as individual. 
 The model keeps track of where the tokens come from and therefore causal dependencies between transitions are reflected in terms of a partial order similar to the partial order of reversing Petri nets with single tokens. Thus, we allow identical tokens to fire the same transition when going forward, however when going backwards tokens are  able to reverse only the transitions that they have fired.  
Additionally we  provide the reversible semantics for out-of-causal-order reversibility in the presence bond destruction. 
 We then proceed to translate reversing Petri nets into labelled transition systems (LTSs) as an event-oriented representation of the operational behaviour of the model. We compare the expressive power offered by multi tokens against that of single tokens, in terms of the associated Labelled Transition Systems, denoted up to isomorphism of reachable parts. As a result, we find that reversing Petri nets with single tokens are equally expressive as reversing Petri nets with multi tokens. As an alternative direction, we then propose the collective token interpretation of reversing Petri nets which is inspired from biochemical reactions and resource allocation systems.  
  In this approach all tokens of a certain type are identical, disregarding their history during execution and therefore assuming the ambiguities between them to be equivalent. 

{\bf Controlled reversibility.} The framework of reversing Petri nets has been extended with a mechanism for \emph{controlling reversibility} in Chapter~\ref{sec:control}. 
%
 In our model control is enforced with the aid of conditions associated with transitions, whose satisfaction acts as a guard for executing the transition in the forward/backward direction. The conditions are enunciated within a simple logical language expressing properties relating to available tokens. The mechanism may capture environmental conditions, e.g., changes in temperature, or the presence of faults. We present a reversible semantics of the resulting framework. The resulting model is general enough to capture a wide range of systems, in this context we give an overview of   several properties  of reversing Petri nets that could be used to analyse the behaviour of these systems. 

  {\bf Case studies.} Our approach is motivated by applications from biochemistry, but it can be applied to a wide range of problems featuring reversibility.  Specifically, we demonstrate the original RPN framework with various examples including a model of the ERK pathway, and a model of a transaction processing system, examples that inherently feature (out-of-causal-order) reversibility. We  then show the same transaction processing system modelled by RPNs with multiple tokens  under the individual token interpretation. Multi tokens are also demonstrated  under the collective token interpretation by modelling a case study from biochemistry, known as the autoprotolysis of water, where instances of the same atom are indistinguishable. Finally, we show the robustness of our control mechanism by modelling an example from telecommunications of a  distributed algorithm for antenna selection illustrating the ability of RPNs to not only formalise complex distributed systems but also naturally capture reversible, controlled execution and conservation of information in a system.

\section{Document Outline}

In Chapter 2 we present the basic background theory for reversible computation, Petri nets, and reversible models of concurrency. This chapter presents an overview of the main approaches, results, potential benefits, and applications of reversible computation. In particular, we focus on reversible formal models of concurrency used for modelling reversible systems or developing techniques to analyse descriptions of reversible protocols. Furthermore, we provide an overview of the traditional model of Petri nets. We present the main extensions of Petri nets and discuss techniques for system validation and verification for the model. Finally, we discuss causality, a concept of high relevance in the context of Petri net semantics.

In Chapter 3 we 
propose the formalism of reversing Petri nets 
by  introducing machinery, associated operational semantics and results for transition enabledness as captured by forward execution, backtracking, causal and out-of-causal-order reversibility. We illustrate the RPN framework with a model of the ERK pathway and a transaction processing system.

In Chapter 4 we extend this formalism by allowing multiple tokens of the same type to occur in a system as well as the ability of forward transitions to break bonds. 
We then compare the two models and show that the expressive power of RPNS with multi tokens is equivalent to the expressive power of RPNs with single tokens. The final contribution of this chapter is a demonstration of a biological case study, namely the autoprotolysis of water reaction. 

In Chapter 5 we introduce another extension of the RPN model with a mechanism for controlling reversibility.  We use the resulting formalism of Controlled RPNs to model a novel distributed algorithm for antenna selection. 

The last part of the thesis, Chapter 6, concludes this work by comparing the results of this proposal with the related literature and proposing the current and future work after the completion of this thesis.

\chapter{Background}\label{sec:Background}
\section{Reversible Computation}\label{sec:Reversible Computation}
Mechanical Computing essentially dates back to the 1800s, followed by Electronic Computing, starting at least six decades ago, and ever since both of them have been supporting and enhancing all aspects of our lives ~\cite{perumalla}. 
  Forward-only computing  has been extensively researched, developed  and analysed in academia, industry and government across the globe.
    Forward direction is the standard kind of execution whereas backward direction is the ability to go back to previous states by undoing previously executed actions. 

Since current technology is not invertible, it leads to  loss of information where previous states cannot be recovered from the current state. It follows an irreversible computation paradigm where ordinary computer chips do not qualify for reversible operations. For example, a simple standard operation, like the logical AND, illustrates that given the output 0 it is impossible to determine the input values as one of combinations of 1 and 0, 0 and 1 or 0 and 0~\cite{RCwille}. Therefore, such logically irreversible operations lack determinism due to the fact that the partial function that maps each machine state onto its successor generates more than one inverse values.  

This entails that, if there is a logic gate that generates an output from a given input, the gate is reversible only when there is an inverse operation that performs a bijective transformation of its local configuration space. Crucially, the ability to have two-way determinism requires an one-to-one mapping, where each input produces a unique output. These bijective reversible operations have at most one previous configuration giving the ability to uniquely identify the forward or backward state at any point of the execution. Such backward deterministic systems are the foundations of an alternative computation paradigm, called \emph{Reversible Computation}~\cite{RCwille}. As such, a computation is logically reversible if it is always possible to efficiently reconstruct the previous state of the computation from its current state.

Reversible computation is an emerging paradigm that extends the standard forward-only mode of computation by allowing one to execute programs in the backward direction as effortlessly as it can be executed in the standard forward direction. In particular, individual operations become time reversible that can easily and exactly be reversed, or undone at any point of the execution. Computer scientists believe that reversible computation is an unconventional but promising form of computing, which is able to deliver novel computing devices and software~\cite{RCwille}. More importantly, reversibility is emerging as one of the most exciting new dimensions in computing for the future, positioned for inevitable progress and expansion in the coming decades. 

\subsection{Motivation}
Reversible computing combines thermodynamics and information theory in order to reflect physical reversibility one of the fundamental microscopic physical properties of Nature. Since all successful fundamental physical theories share the property of reversibility, future computing could also follow rather trivially certain basic facts of fundamental physics. Such properties can be effectively used in computing in order to create an interface between computation and the laws of physics where logical reversibility implements physical reversibility~\cite{RCA-survey}.  

Back to the 1960's, three foundational studies of information lossless computations made their appearance. Having different motivations, Huffman studied finite state machines that do not erase information~\cite{Huffman},  Lecerf studied the theoretical properties of reversible Turing machines~\cite{lecerf} and Landauer studied the thermodynamics of reversible logics. Of these  studies, Landauer's work is the most prominent, but the other two have laid the foundation of the theoretical study of reversible computing.

Physicist Rolf Landauer was the first to argue the relation between thermodynamics and the irreversible character of conventional computers. Specifically, reversible computation originated in the 60's when Landauer published a paper titled ``Irreversibility and Heat Generation in the Computing Process"~\cite{Landauer}, where he attempted to apply the most fundamental, reversible laws  of physics to digital computers.  Landauer's key insight follows directly as an immediate logical consequence of our most thorough, battle-tested understanding of fundamental physics. He noted that, while classical mechanics and quantum mechanics are fundamentally reversible~\cite{Structural} by obeying the laws of motion, their logical state often evolves irreversibly since it is not backward deterministic. This means that since all of the fundamental laws of physical dynamics are reversible, then conceptually any machine should be able to run the laws of physics backwards and thus be able to determine the system's backward states.   
 
Specifically, Landauer observed the direct implications on the thermodynamic behaviour of a device that is carrying out irreversible operations. His reasoning can be understood by realising that reversibility at the lowest level of physics means that we can never truly erase information in a computer. He notes that for the entire history of computers, our computing machines have been erasing bits of information in the process of performing irreversible computations. 

If we return to the example of the logic gate AND, we can observe that given the output 0, the input 1 has been erased after the execution of the operation. Whenever a bit of information gets overwritten by a new value or whenever a logic gate  produces several unused outputs, the previous information might get lost but it will not be physically destroyed. Instead, bit erasure pushes bits out into the computer's thermal environment, where they become entropy causing heat dissipation and, in general, loss of energy. This is known as the von Neumann-Landauer (VNL) principle~\cite{maxwell} where one bit's worth of lost logical information always leads to at least  $kTln2 $ ($k$ is the Boltzmann constant which is approximately $1.38	\times10^{-23}$ $J/K$, $T$ is the temperature of the heat sink in kelvins, and $ln2$ is the natural logarithm of 2 which is approximately 0.69315) amount of physical energy dissipation. 

This result is of great interest because it makes plausible the existence of thermodynamically reversible computers which could perform computations while dissipating considerably less energy per logical step. The energy used in reversible bit operations can be fully recovered and reused for subsequent operations so that every computation can be performed without bit erasures.  This means that a computation is physically reversible when it can be carried out without loss of energy or, more formally, with no increase in physical entropy and it is thus energy efficient.

Landauer's theoretical lower bound has since been experimentally confirmed and it has been argued many times that efficient operations of future computers require them to be reversible~\cite{berut2012experimental}. In the scenario of low-energy computing, the gap between computation and reality needs to be bridged by introducing the reversible or possibly the quantum mechanical  mode of computation. Currently, computers are commonly irreversible with their technology rapidly approaching the  elementary particle level extrapolating towards Landauer's limit. 

In particular, based on Moore's law, computer power roughly doubles every 18 months for the last half century~\cite{TimeSpaceBound}. The miniaturisation of  transistors increases their per-area leakage current and standby power; meanwhile, the reduction of signal energies, causes significant  thermal fluctuations  which eventually prevent any further progress within the traditional computing paradigm~\cite{back}. Efforts are being made within the semiconductor industry in order to try to reduce and forestall these problems, but the solutions are becoming ever more expensive to deploy where eventually no level of spending can ever defeat the laws of physics. Smaller transistors in new conventionally-designed computers would no longer be any cheaper, faster, or more energy-efficient than any predecessors, and at that point, the progress of conventional semiconductor technology will stop being any longer economically justifiable.  Landaure's limit threatens to end improvements in practical computer performance within the next few decades and to avoid this a solution could be to avoid losing track of logical information.
\remove{
\begin{figure}[t]
\centering
\subfigure{\includegraphics[width=13cm]{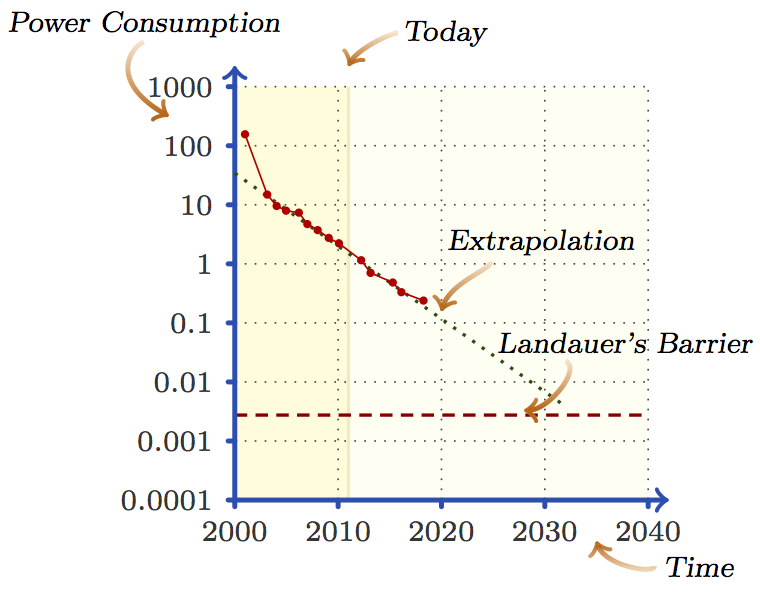}}
\caption{Landauer's Limit}
\label{Landauer}
\end{figure}
}

However, for several decades now, we have known that reversible computing is a theoretically possible alternative paradigm which is in fact the only possible way, within the laws of physics, to have energy and cost efficient computers. However, when reversible computations are implemented on the right hardware, they should be able to circumvent Landauer's limit. The technical motivation  given by Landauer has inspired theoretical work in the area of computational models.  Lecerf~\cite{lecerf} was the  first to describe reversible computations executed on reversible Turing machines, and invented the Lecerf reversal technique to uncompute histories, but he was unaware of Landauer's thermodynamic applications, and  therefore his machines did not save their outputs. Bennett~\cite{Bennett} then reinvented Lecerf's  reversal based on Landauer's point of view that any desired logically irreversible computational operation could be embedded in a reversible one, by simply saving aside any information that it would otherwise erase. For example, the machine might be given an extra tape to record each operation as it was being performed, in order to be able to uniquely determine the preceding state by the present state and the last record on the tape. However, Landauer noted that this method was only going to postpone the inevitable, because the tape would still need to be erased eventually, when the available memory filled up. 

Bennet managed to prove that it is possible to construct fully reversible Turing machines capable of performing any computation whilst erasing garbage information on its tape when it halts and therefore leaving behind only the desired output and the originally furnished input. The trick is to decompute the operations that produced the intermediate results and therefore erasing the temporary data from the memory. This would allow any temporary memory to be reused for subsequent computations without ever having to erase or overwrite it. He also pointed out the possibility of a physically reversible computer where dissipation of energy is arbitrarily small. 

However Bennett's construction only addressed the logical level where any arguments based on thermodynamics needed to be applied on specific hardware technologies.  Toffoli and Fredkin~\cite{FredkinToffoli} were the first to address precisely how to construct a practical physical mechanism for computation that would also be physically reversible. They have reinvented reversible computing in the form of conservative logic circuits, and proved their universality. Toffoli~\cite{Toffoli} invented the Toffoli gate which is perhaps the most convenient universal reversible logic primitive. All these pioneering developments together incrementally set the stage for the field of reversible computing.

As a result, reversible computing has the potential to alleviate the ever-increasing demand for electricity by designing revolutionary reversible logic gates and circuits that lead to low-power computing. Hardware-wise, the potential benefits of reversible computing come to solve the miniaturisation limitations of current technology that aim to increase the speed and capacity of circuits. On the other hand, there already exist various occasions where reversibility is naturally embedded in computation. For example, recovery from failures such as corrupted data, deadlocked programs and breached security is crucial and could be effortlessly obtained in a reversible manner.  Hence, such a mode of computation aligns naturally with many computational tasks such as the treatment of faults and recovery in distributed systems, coding and decoding and many others. 

So far, it is considered to be highly challenging to implement reversibility effectively, because it comes with many underlying problems and the alternative of advancing conventional technology is much easier. Even though it comes with many promising benefits and applications it also comes with its own limitations. Since, the theory of reversible computation is based on the idea of computing and uncomputing operations, it means that arbitrarily large computations executed in reverse would result in almost twice as many steps as an ordinary computation and may require a large amount of temporary storage. This means that there is an underlying trade-off between the efficiency of such recovery and the speed of computation.  

On a practical level achieving efficient reversible computing will likely require new hardware materials, new design tools and device structures, new hardware description languages with the supporting software and overall a thorough remodelling of the entire computer design infrastructure. This also means that a large part of computer scientists and digital engineering workforce will have to be trained to use novel reversible design methodologies.

Nevertheless, the upside potential of reversible computing has attracted many researchers that made significant conceptual progress over the past few decades. This effort of addressing the challenges of reversible computation is highly worthwhile, because with the current rapidly advancing technology, it is now time to focus on reversible computing, and begin collaborative effort to materialise this idea. Committed attention can eventually improve current information technology by making it many orders of magnitude greater than any existing irreversible technology~\cite{DBLP:series/lncs/12070}. 

\subsection{General Overview}
As discussed above, computer scientists, mathematicians and physicians believe that reversible computation will be a key technique in the not so distant future of computer models. 
As such, it attracts much interest for its potential in a growing number of application areas ranging from cellular automata, software architectures, reversible programming languages, digital circuit design to quantum computing. Below we present the main advances in these fields.

 There exist several notions of logical  reversibility  on computing models with a finite number of discrete internal states that evolve in discrete time. Their precise impact on the computational capacities and decidability properties of devices has been considered from different points of view. In the literature there exist various models, including the massively parallel model of cellular automata, the weakly parallel model of multi-head finite automata~\cite{oneWay,RA,multiHead,twoWay} as well as sequential models such as Turing machines~\cite{Bennett,tapeReduction}, pushdown~\cite{RPA,inputDriven} and queue automata~\cite{queue}, and finite state machines~\cite{minimal}. 
 In order to examine whether reversibility increases computational capacities it is useful to study the properties and the impact on suitable models when different notions of reversibility are applied. These models have been equipped with additional resources or structural properties in order to examine whether a computational model can be made reversible and the associated costs.

 A number of interesting reversible programming languages have been developed since 1986.  The first reversible programming language follows the imperative paradigm~\cite{Janus,Flowchart} followed by another simple reversible imperative languages named R~\cite{R}. Other general purpose functional programming languages are RFUN~\cite{RFUN}, muOz~\cite{muoz} the causal-consistent reversible extension of Oz and Theseus~\cite{Theseus}. The family of quantum programming languages consists of languages based on the imperative paradigm such as QCL (Quantum Computation Language)~\cite{QCL}, LanQ~\cite{LanQ} and languages based on the functional paradigm such as cQPL~\cite{cQPL}, and QML~\cite{QML}. Research on compiler technology for reversible languages has also progressed in the last several years~\cite{R,Translation}. The main challenge in this area is that these languages are still prototype languages. Thus, the code base for each of these languages is limited, and the languages do not offer many of the usual programming abstractions. This in turn has hindered the developments of reversible algorithms and useful data structures.
  Persistent (immutable) data structures~\cite{DataStructures} offer more efficient storing of multiple versions of a data structure, sharing structure where possible. 

 The idea of using reversibility for the development of reliable software is quite natural since  backward recovery is an instance of reversible computation in which errors trigger inverse actions. 
  In case of trouble fault treatment seeks to handle certain system errors after their occurrence and therefore stop them from causing a failure. Then the system can go backwards to a past safe state of the system and try to explore new directions, avoiding the troublesome actions and therefore bringing the system to a consistent state.  
 %
 If a fault is detected, checkpointing can be used as a recovery mechanism that restores the system to a previously saved state which is essentially a snapshot of the entire system that is safe from errors~\cite{RevComProc,Transactors,ML}. 
 %
 The past 40 years reversibility has also been naturally applied in the area of  debugging~\cite{AIDS,RevEx} because it gives the ability to explore the computation in both forward and backward, and therefore assisting the programmer in the search of possible misbehaviours. 
 Indeed, many reversible debugging tools exist ~\cite{RevDeb,Igor,RevDebugger,Java} and some reversible debugging features are available in mainstream debuggers. 
There have also been proposed reversible process calculi used to build constructs for reliability, and in particular communicating transactions with compensations~\cite{Kutavas} where interacting transactions with compensations have been mapped into a reversible calculus with alternatives in~\cite{LaneseLMSS13}. Behavioural equivalences for communicating transactions with compensation have been studied in~\cite{Kutavas2,Kutavas3}. 
 Another area that can benefit from reversibility is that of control systems and robotics. Robots are generic mechatronic devices controlled by a computer that can essentially be made reversible. Reversibility plays vital role in different programming paradigms that are used to operate robots. 
%
 Many operations in the field of robotics are naturally reversible both for single robots as well as multi-robot swarms.  These operations assist 
 systems that can autonomously accumulate and revise knowledge from their own experience via self-programming~\cite{SelfReconfig,murRoCE}. Control systems operate concurrently during forward execution in order to predict the behaviour of the environment under constrains of limited time and computational resources. Whereas, during backward execution they retrieve the goals the system was designed to achieve. 
 Another operation is that of reversing the forward execution of an assembly sequence in order to generate the backward disassembly process as well as changing the direction of a mobile robot from forward to backward. 
The increase likelihood of errors in industrial robots 
can be addressed using reverse execution in order to withdraw an erroneous situation and thereafter automatically retry the assembly operation~\cite{ErrorRecover,RASQ}.
 As motivated  by Landauer~\cite{Landauer}, reversible circuits  have several promising applications such as low-power circuit design and quantum computing. The inherent properties of reversibility can be exploited in the design of conventional circuits with many advantages. One of them is the ability to undo operations in case the system reaches an erroneous state as well as full connectivity which detects errors by applying randomly generated stimuli. Conventional computing also benefits from reversibility by achieving perfect observability and controllability which provides easy testability. Another application of reversible circuits is quantum computing~\cite{QuantumComputing} which is inherently reversible and allows reversible computation to be exploited as a subset of quantum operations.
%
 The goal of conventional circuit design is to find a logic circuit that implements the Boolean function and minimises the number of gates or the circuit depth.  Reversible circuit synthesis is a special case of conventional circuit design where all gates are made reversible by disallowing any fanout. In order to avoid any fanout, it must be that the number of input wires of a gate is the same as the number of output wires~\cite{regular}.
%
 The most well known gates in reversible and quantum computing are the Fredkin gate~\cite{fredkin}  together with Toffoli~\cite{Toffoli}  and Feynman~\cite{feynman} gates.
 
\subsection{ Reversible Models of Concurrency}\label{sec:Reversible Models of Concurrency}


There exist many questions that need to be addressed when it comes to reversible computing, including what are the main approaches, results, potential benefits, and applications.
Exploring reversibility through formal models formulates the theoretical foundations of what reversibility is, what purpose it serves, and how it benefits natural and artificial systems. 

In particular, reversible formal models can be used for modelling reversible systems or developing techniques to analyse  them. In order to comprehend the way reversibility works, it is useful to study the properties of these models when different notions and strategies of reversibility are applied. Understanding reversibility through various case studies could potentially propose a unified theory for reversibility in distributed systems, including behavioural and logic semantics, and explore how reversibility can help in specification, verification, and testing. 

Creating expressive reversible formalisms that can be easily understood and simulated, even by scientists with expertise outside Computer Science, can prove very useful to understand, model, and design complex systems. The expressive power and descriptive nature offered by formal models coupled with reversible computation has the potential of providing an attractive setting for studying, analysing, or even imagining alternatives in a wide range of systems. For example, reversibility-inspired theories and formal methods will enable the software industry to deliver safer and more reliable distributed software and systems.

Such reversible formalisms will also assist scientists from other disciplines -for example biochemistry, mathematics and material science (superconductors)- since there exists various systems in the world of artificial and natural sciences where reversibility is inherent or it could be of interest. For instance, biochemical reactions, such as the isomerisation of glucose to fructose, are typically bidirectional, meaning that the direction of the computational system is fixed by an appropriate injection of energy or a change of entropy from environmental conditions like temperature or pressure~\cite{ERK,Bonding} 
 Similarly, quantum computations are also inherently reversible because many of the components in quantum computers, such as databases or modular exponentiation, are reversible~\cite{feynman}.  
%
%
%
Reversibility is also used in software engineering to better explore a computation and analyse different possibilities, as in the exploration of a program state-space toward a solution, or in constructions of mechanisms for system reliability. 
\remove{
 In the setting of fault tolerance schemes, there exists various recovery techniques that imply some form of reversal, including checkpointing and transactions. The main idea is that in case the system encounters an error, the program automatically backtracks to a more stable state where the decisions leading to the error have not been taken yet, and a new forward execution may start avoiding the same error from happening again. Of particular interest is the application of reversibility as a common framework to build reliable distributed systems, including database and workflow management systems. 
%

}
In the same category belong systems of industrial robots often used in production for assembly and disassembly and are normally controlled by a single host computer.
\remove{ Since, various forms of reversibility for debugging and interactive programming have traditionally been a feature found in many major commercial industrial robot systems, then there is also an explicit connection between reversibility in robot programming languages and reversible computing.  Similarly to software debugging, certain classes of errors during assembly operations can be resolved using reversible execution, which allows the robot to temporarily back out of an erroneous situation and then the assembly operation can be automatically retried.
}

 Even though the physical implementations of the computational steps of such systems are naturally reversible, most abstract computation frameworks usually model the progress of computations through a sequence of forward irreversible steps. 
Therefore, the construction of reversible modelling languages can indicate how to capture the behaviour of reversible actions in order to implement or even extend the primitive processes of biological reactions, quantum computation, reliable systems, and  movement in robotics. 

These abstract computation models can be based on existing abstract formalisms and can be used not only for modelling reversible systems but also for investigating suitable notions of behavioural equivalences. The natural and artificial processes in these formalisms can be made reversible in order to facilitate more efficient  model checking of new formulations of useful properties such as reachability, safety,  exception and precedence. We can also explore whether adding the reversibility feature to these abstract models can increase considerably their computational and descriptional complexities. Hence, research on  suitable behavioural semantics and modal logics for reversibility can result in sound foundations to commercial reversible modelling, debugging and testing software tools.  

\subsubsection{Challenges}
Even though reversing computational processes in concurrent and distributed systems has many promising applications it also has many technical and conceptual challenges. In order to create the theoretical foundations of reversible formal models and to discover their purpose and benefits in natural and artificial systems we have to ascertain the costs and limitations that come with reversibility, and to explore the challenges and open problems. A formal model for concurrent systems that embeds reversible computation needs to address two challenges. The first one being the ability to identify the legitimate backward moves at any point during execution and the second one is the ability to compute without forgetting. 

The first challenge depends on the choice of the computation's semantics that determine the order of forward and reverse actions. There are several forms of undoing computation that have been studied in the literature over the past years. In the sequential scenario, the legitimate backtracking moves can be trivially determined based on the order of execution. The computational steps are reversed based on the time of their execution and hence are undone in the exact inverse order of the forward execution.

Although, in the concurrent scenario speaking about backtracking in time is immaterial and the interplay with reversibility is no longer trivial. 
Therefore understanding this interplay is fundamental in many of the areas above, e.g., for biological or reliable distributed systems, which are usually naturally concurrent. In such concurrent systems we do not want to reverse the actions precisely in the opposite order than the one in which they were executed during forward computation, as this order is irrelevant. The concurrency relation between forward actions has to be taken into account and independent threads of execution should be reversed independently, whereas causal dependencies between related threads should remain protected. 

There are however, many important examples, such as mechanisms driving long-running transactions~\cite{TransactionsRCCS,Transactors} and biochemical reactions~\cite{ERK,Bonding}, where concurrent systems violate causality. Causally dependent threads are allowed to freely backtrack in an out-of-causal order which in a way would result in losing the initial computation structure. This means that reversing in out-of-causal order will not return a thread into a previously executed state  but it would give it the ability to reach computation states which were formerly inaccessible.

The second challenge, of forgetting previously executed actions, applies  to both concurrent and sequential systems. Since processes do not remember their past states if we want to reverse a standard process we will generate multiple possibilities. This challenge can be addressed by making the system exactly reversible using memories that remember the position and momentum of each action. When building or extending a reversible variant of a formal models, the syntax can be extended to allow the appropriate syntactic representations for computation memories that allow processes to keep track of everything that has been executed. The resulting mechanism should be light in terms of memory without the need of a global control.

\subsubsection{Forms of Reversibility}

The first challenge of identifying legitimate backward moves during reversal calls to identify possible strategies for going backwards. A large amount of work focused on identifying such strategies within process calculi~\cite{ERK,Algebraic,RCCS,LaneseMSS11}. Behind the insights of the theoretical study of reversible computation lie the challenging quest of understanding the nature of reversibility while formally representing various computational concepts. Understanding the role that reversibility plays in natural systems, helps in the development of realistic formal models for concurrent and distributed systems. Reversibility could initially be divided into two main categories: \emph{Rigid} and \emph{Uncontrolled}~\cite{causality}. 

{\bf\em Rigid} means that the execution of a forward step followed by the corresponding backward step leads back to the starting state, where an identical computation can restart. However rigid reversibility may not always be the best choice especially in the case of reliable systems. If the error that we are trying to recover was a transient fault then going back to the state that the error occurred and retrying the computations might solve the problem. Although, if the failure was permanent going back and, forth by following the same computational steps will infinitely result in the same error. 

{\bf\em Uncontrolled }means that there is no hint as to when to go forward and backward. Uncontrolled reversibility defines how to reverse a process execution by determining the necessary history and the associated causal transformation yet it does not specify when and whether to prefer backward execution over forward execution or vice versa. Uncontrolled reversibility gives good understanding on how reversible computation works, but it does not exploit  it into applications because different application areas need different mechanisms to control reversibility.  

Uncontrolled reversibility can be further subcategorised into several approaches for performing and undoing steps, which differ in the order in which steps are taken backwards and forwards. The most prominent of these are \emph{backtracking}, \emph{causal reversibility} and \emph{out-of-causal-order reversibility}. 

{\bf \em Backtracking} is the process of rewinding one's computation trace, that is, computation steps are undone in the exact inverse order to the one in which they  occurred. It does not allow any thread to freely backtrack because it might result in losing the initial computation structure and reaching computation states which were formerly inaccessible. This form of reversing ensures that at any state in a computation there is at most one predecessor 
state, yielding the property of \emph{backwards determinism}. In the context of concurrent systems, this form of reversibility can be thought of as overly restrictive since, undoing moves only in the order in which they were taken, induces fake causal dependencies on backward sequences of actions: actions, which could have been reversed in any order are forced to be undone in the precise order in which they occurred.

\begin{figure}[h]
	\centering
	\subfloat{\includegraphics[width=7cm]{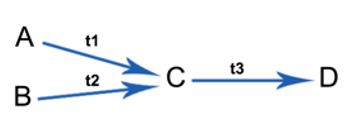}}
	\caption{Causal dependencies}
	\label{dependencies}
\end{figure}

\begin{figure}[h]
	\centering
	\subfloat{\includegraphics[width=4cm]{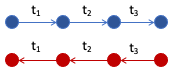}}
	\caption{Backtracking}
	\label{backDep}
\end{figure}

Consider the following example with order of forward execution $t_1,t_2,t_3$. As indicated in Figure~\ref{dependencies} let us assume that the action $t_1$ occurs independently of action $t_2$ and when both $t_1$ and $t_2$ occur they cause the execution of action $t_3$. Figure~\ref{backDep} shows that in backtracking mode there exists only one order of reverse execution which is the exact opposite direction of the forward one $t_3,t_2,t_1$.

Relaxing the rigidity of backtracking,  a second approach to reversibility,  {\bf \em causal reversibility},  allows a more flexible form of reversing by allowing events to reverse in an arbitrary order, assuming that they respect the causal dependencies that hold between them. Thus,  in the context of causal reversibility, reversing does not have to follow the exact inverse order for independent events as long  as caused actions, also known as effects, are undone before the reversal of the actions that have caused  them. This form of reversibility is called causal, meaning that it respects causality  a binary irreflexive relation of events that identifies which events cause others, and therefore need to 
be reversed last. Thus, causally backtracking a trace could be allowed along any path that respects causality also known as a causally equivalent path. A main
feature of causal reversibility is that
reversing an action returns a thread into a previously
executed state, thus, any continuation of 
the computation after the reversal would also be possible in 
a forward-only execution where the specific step was not taken in the
first place. 

\begin{figure}[h]
	\centering
	\subfloat{\includegraphics[width=4cm]{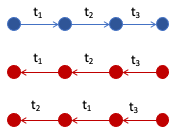}}\\
	\caption{Causal reversing}
	\label{coDep}
\end{figure}

Consider the same example as before with forward execution $t_1,t_2,t_3$. Since $t_1$ occurs independently of action $t_2$ we can now reverse $t_1$ and $t_2$ in any order we want, although we can never reverse them before $t_3$. As indicated in Figure~\ref{coDep}, causal order reversal gives an additional reverse path which is the execution of $t_3,t_1,t_2$, as well as, the backtracking path $t_3,t_2,t_1$.

Both backtracking and causal reversing are cause-respecting. There are however, many important examples where undoing events in an {\bf \em out-of-causal} order is either inherent or could be beneficial. In fact, this form of undoing plays vital role on mechanisms driving long-running transactions and biochemical reactions~\cite{Bonding}. This flexible notion of reversibility cancels out soundness since some backtracking computations could give access to formerly unreachable states.  Consider every state of the execution to be a result of a series of actions that have causally contributed to the existence of the current state. If the actions were to be reversed in a causally-respecting manner 
then we would only be able to move back and forth through previously visited states. Therefore, one might wish to apply out-of-order reversibility in order to create fresh alternatives of current states that 
were formerly inaccessible by any forward-only execution path. 

\begin{figure}[h]
	\centering
	\subfloat{\includegraphics[width=4cm]{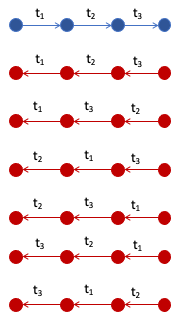}}\\
	\caption{Out-of-causal reversing}
	\label{ocoDep}
\end{figure}

Again, consider the above example where now actions can be reversed in any possible execution path. Given the forward execution of $t_1,t_2,t_3$, six alternative reversing paths are produced as potential reverse executions based on out-of-causal reversibility. In Figure~\ref{ocoDep} can be observed that these paths include paths produced by backtracking execution as well as, paths produced during causal reversal.

\remove{
Since out-of-order reversibility contradicts program order by violating the laws of causality, it comes with its own peculiarities that need to be taken into consideration while designing reversible systems. To appreciate these peculiarities and obtain insights towards our approach on addressing reversibility in such systems, we will use a standard example from the literature, namely the process of catalysis from biochemistry. For instance consider the process of catalysis from biochemistry where a catalyst $c$ helps the otherwise inactive molecules $a$ and $b$ to bond. The process is as follows: initially element $c$ bonds with $a$ which then enables the bonding between $a$ and $b$. Next, the catalyst is no longer needed and its bond to the other two molecules is released.

\begin{figure}[h]
	\centering
	\subfloat{\includegraphics[width=14cm]{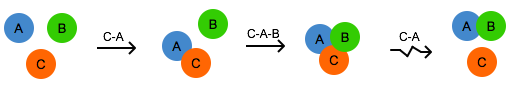}}\\
	\caption{Catalytic reaction}
\end{figure} 

}

\subsubsection{Reversible Formalisms} 

Both challenges have been addressed within various computational models ranging from process calculi~\cite{RCCS,Algebraic} to event structures~\cite{CardelliL11,ConfStruct}. The second challenge, of forgetting previously executed actions, has been addressed using external mechanisms such as memories or identifiers that remember the position and momentum of each action. Since processes do not remember their past states, various reversible formalisms make a system exactly reversible by extending the syntax  to include mechanisms that serve as histories of executed actions of the corresponding  processes. 

Research on reversible formal languages can be  traced back to a publication from Berry and Boudol titled ``Chemical Abstract Machine"~\cite{abstract}. The authors propose a calculus, inspired by chemical reactions, whose operational semantics define forward and the corresponding reverse reduction relation. They have introduced the notion of a chemical abstract machine called ``cham" which is based on the chemical metaphor used in the $\Gamma$ language. They have illustrated the descriptive powers of the chemical abstract machine by showing that it is suited to model concurrent systems and reversible computations. 

The first attempt to reverse classical process calculi was explored by Vincent Danos and Jean Krivine~\cite{BiologyCCSR} who built a notion of distributed backtracking on top of Milner's CCS~\cite{CCS}. The proposed process calculus was named CCS-R which is essentially a reversible extension of CCS motivated by the desire to represent reversible biological systems as the evolution of biological processes. Reversibility is embedded in the syntax of CCS as a distributed monitoring system and meshes well with the forward only syntax of the host calculus. 

Some of the limitations of the model of CCS-R have been later addressed on a newer version named RCCS~\cite{RCCS} where CCS-R is extended to deal with recursion, and uses unique names to identify threads. RCCS is again a process algebra in the style of CCS where processes have the ability to backtrack. Their calculus is essentially Milner's CCS with the added bonus that some observable actions in the standard labelled transition system semantics can be understood to be reversible. Their seminal paper was the first to discuss the notion of causality as a suitable requirement for reversibility in a concurrent scenario and paved the way to the definition of causal-consistent reversibility. RCCS is a causal-consistent reversible extension of CCS that  uses memory stacks in order to keep track of past communications, further  developed  in~\cite{TransactionsRCCS}. 

 RCCS~\cite{RCCS}  and the work of ~\cite{Structural} on mapping functional programs into reversible automata inspired I. Phillips and I. Ulidowski to proposed another approach on reversing CCS, named CCSK~\cite{Algebraic}. CSSK is a reversible version of CCS based on the use of communication keys. It can be used to model and analyse the bidirectional behaviour of systems that are able to choose the direction of execution spontaneously, for example the binding and unbinding of molecules in biochemical reactions. Given a forward transition relation their reversible algebraic process calculi is able to obtain the inverse because it has the ability to remember previously executed actions. To achieve this they have introduced a method for converting the standard irreversible operators of CCS into reversible operators, while preserving their operational semantics. Similarly to RCCS, they use a memory mechanism which contains a history of past communication keys and it can be used to reverse computation in a causally preserving manner. Contrary to the global control and extensive record keeping of RCCS, their motivation was to produce a reversible process calculus that does not rely heavily on external devices such as memories. The most crucial component of their procedure is the notion of communication keys which are a more expressive form of past actions. These are unique identifiers that are used to ``mark" a previously executed action $a$ by a fresh identifier $k$ and record it in the syntax of the action's occurrence as $a[k]$. 

The creators of RCCS  continued their work by employing their reversible mechanism to $\pi$-calculus by proposing a reversible labelled transition semantics called $R\pi$~\cite{DBLP:conf/lics/CristescuKV13}. They introduce the syntax and semantics for the reversible $\pi$-calculus and 
they prove similar results to the ones proved for RCCS, such as equivalence between any backtracking path and forward computation as well as causal equivalence up-to permutation which means that computations are maximally liberal with respect to the structural causality of the reduction semantics.

  This work continues in~\cite{LaneseMS16} where Lanese et al. presented  a reversible asynchronous higher-order $\pi$-calculus, called $\rho \pi$, which has been shown to be causally consistent and gives two original contributions. The first one is a novel reversible machinery which, in the contrary of the previously proposed machineries in CCS, preserves the classical structural congruence laws of the $\pi$-calculus, and  relies on simple name tags for identifying threads and explicit memory processes.  The second contribution is a faithful encoding on $\rho \pi$ calculus into a variant of HO$\pi$, showing that adding reversibility does not change substantially the expressive power of HO$\pi$. The work on reversible $\pi$-calculus continues from $\rho\pi$ to roll-$\pi$~\cite{LaneseMSS11}, which is a fine-grained rollback primitive for higher-order $\pi$-calculus, that builds on the reversibility apparatus of  $\rho\pi$ in order to adopt the ability to undo every single step in a concurrent execution. In~\cite{LaneseLMSS13} Lanese et al. continued their work on reversible $\pi$-calculus by proposing a new concurrent process calculus, named croll-$\pi$, as a framework for flexible reversibility and compensating roll-$\pi$. Croll-$\pi$ features flexible reversibility, where it is possible to specify alternatives to a computation, that can be used upon explicit rollback. 
  
  On a more general note, the work in~\cite{General} proposes a general and automatic technique which defines a causal-consistent reversible extension for forward models. These models include a variety of formalisms studied in the literature on causal-consistent reversibility such as  Higher-Order $\pi$-calculus and Core Erlang. Another work aiming to generalise  causal reversibility in formalisms, is that of~\cite{Axiomatic}. This work examines the various properties that a reversible system should enjoy and shows how they relate to the already suggested properties such as the parabolic lemma and the causal consistency property. Specifically, a generic labelled transition system has been used to capture these properties as a set of axioms which can then be used by reversible formalisms in order to verify their properties.  Additionally, two new notions of causal consistent reversibility are derived from these axioms, namely safety and causal liveness. 

Most recently, the study of out-of-causal-order  reversibility continued with the introduction of a new operator  for modelling local reversibility in~\cite{LocalRev}. The authors here also consider controlled reversibility in CCS, in the form of a reversible process calculus called Calculus of Covalent Bonding (CCB). Their reversible calculus has a novel and purely local in character prefixing operator. This operator has been inspired by the mechanism of covalent bonding, which is the most common type of chemical bonds between atoms,  that allows modelling of locally controlled reversibility. In their proposal actions can be undone spontaneously or as pairs of concerted actions, where performing a weak action forces reversing of another action. The new operator in a restricted version of their calculus preserves causal consistency, however in its full generality it also allows modelling in out-of-causal order, where effects are undone before their causes.  

Reversibility has also been extended to quantum process calculi, that are used to describe and model the behaviour of systems that combine classical and quantum communication and computation the most prominent being qCCS ~\cite{qccs} and CQP~\cite{cqp}.  qCCS is a natural quantum extension of CCS which can deal with input and output of quantum states, and unitary transformations and measurements on quantum systems. The operational semantics of qCCS is given in terms of probabilistic labeled transition system. CQP (Communicating Quantum Processes) has been defined for modelling systems which combine quantum and classical communication and computation. CQP combines the communication primitives of the pi-calculus with primitives for measurement and transformation of quantum state; in particular, it has a static type system which classifies channels, distinguishes between quantum and classical data, and controls the use of quantum state.   

The study of reversible process calculi triggered also research on more abstract models for describing concurrent systems such as event structures~\cite{modelsConcurr,Bonding}. In particular, the work on CCSK has also continued in a paper titled ``Reversibility and models of Concurrency"~\cite{modelsConcurr} where the authors studied the impact of allowing events to be undone in prime event structures. They proposed prime graphs to prove that  transition systems associated with CCSK and other reversible process algebras are equivalent as models to labelled prime event structures.  This study continued in~\cite{Bonding} where they proposed how to model reaction systems that consist of objects that are combined together by the means of bonds or dissolved via reduction-style semantics. Motivated from the initial study of~\cite{Bonding} research on reversible event structures continued with introducing reversible forms of prime event structures and asymmetric event structures~\cite{Conflict}. In order to control the manner in which events are reversed, the authors focused on analysing asymmetric conflict and causation of events in the reversible, and not necessarily causal, setting. Ulidowski et al. continued their research on reversible event structures in a publication titled ``Concurrency and reversibility"~\cite{ConRev}, where they have shown how to model reversibility in concurrent computation as realised abstractly in terms of event structures. The authors have introduced two different forms of events structures: event structures defined in terms of the causation and precedence relations, and event structures defined by the enabling relation. The proposed forms of event structures have been illustrated in various examples that demonstrate how to model causally consistent reversibility as well as out-of-causal-order reversibility. 

In the literature there exists another line of research concerning reversible process calculi that focuses on dealing with reduction semantics describing the evolution of processes in isolation. This approach is usually simpler and hence more easily applicable to expressive calculi such as CCS and $\pi$-calculus. In this line of research we are able to find Reversible structures  a reversible computational calculus for modelling chemical systems, composed of signals and gates~\cite{CardelliL11}. Reversible structures are computational units that may progress in forward and backward direction. They are amenable to biological implementations in terms of DNA circuits and are expressive enough to encode a reversible process calculus such as asynchronous RCCS.

The first study of reversible computation within Petri nets was proposed  in~\cite{PetriNets,BoundedPNs}. In these works, the authors investigate the effects of adding \emph{reversed} versions of selected transitions in a Petri net, where these transitions are obtained by reversing the directions of a transition's arcs. They then explore decidability problems regarding reachability and coverability in the resulting Petri nets. However, non-deterministically deciding to reverse any of the transitions causes the reversal of the ``wrong" transition which might lead to new states that have not been reached through forward execution only. The reason behind this is that the addition of reversibility into a model like Petri nets results in various backward conflicts where a token can be generated in a place because of different transition firings. The marking of that particular place is not enough to deduce whether the token has been produced because of a particular transition. This approach on reversible computation violates causality which is more challenging than randomly selecting reversed transitions since in a concurrent setting there is no natural way for totally ordering events. 

Towards examining causal consistent reversibility in Petri nets, the work in~\cite{Unbounded} investigates whether it is possible to add a complete set of effect-reverses for a given transition without changing the set of reachable markings. The authors show that this problem is in general undecidable however it can be decidable in cyclic Petri nets where with the addition of new places these non-reversible Petri nets can become reversible while preserving their behaviour.  Moreover, the works of~\cite{RPT,RPlaceTrans} propose a causal semantics for P/T nets by identifying the causalities and conflicts of a P/T net through unfolding it into an equivalent occurrence net and subsequently introducing appropriate reverse transitions to create a coloured Petri net that captures a causal-consistent reversible semantics. The colours in this coloured Petri net capture causal histories. On a similar note, \cite{RON}   introduces the notion of reversible occurrence nets and associate a reversible occurrence net to a causal reversible prime event structure, and vice versa. 
 In~\cite{ReversingSteps} the authors examine the possibility of reversing the effect of the execution of groups of various transitions (steps). They then present a number of properties which arise in this context and show that there is a crucial difference between reversing steps which are sets and those which are true multisets.



On another note, having models that express controlled reversibility is more useful in real life applications. For instance, various biological phenomena control the direction of the computation based on physical conditions such as temperature, pressure and reaction rates.  Therefore, a distinguishing feature of reversible computation
is that of {\emph controlling} reversibility: while various
frameworks make no restriction as to when a transition can be reversed (uncontrolled reversibility), it can be argued that some means of controlling the conditions
of transition reversal is often useful in practice. For
instance, when dealing with fault recovery, 
reversal should only be triggered when a fault is encountered. Based
on this observation, a number of strategies for controlling reversibility have
been proposed: \cite{TransactionsRCCS} introduces the concept of irreversible actions, and \cite{DBLP:conf/rc/LaneseMS12} introduces compensations to deal with  irreversible actions in the context of programming abstractions for distributed systems.  
Another approach for controlling reversibility is proposed in~\cite{ERK} where an external entity is employed for capturing the order in
which transitions can be executed in the forward or the backward direction. In another line of work,~\cite{LaneseMSS11} defines a roll-back primitive for reversing computation, and in~\cite{LaneseLMSS13}
roll-back  is extended with the possibility of
specifying the alternatives to be taken on resuming the 
forward execution. 
Finally, in~\cite{statistical} the authors associate
the direction of action reversal with energy parameters
capturing environmental conditions of the modelled systems. 


Reversible calculi were born  with mainly biological motivation. Since many biological phenomena are naturally reversible, a reversible formalism seems to be suitable to model such systems. Indeed, efforts have been made to model biological systems~\cite{CardelliL11,BiologyCCSR,ERK,Bonding} as well as chemical reactions~\cite{LocalRev} using reversible process calculi. We highlight~\cite{CardelliL11} which illustrates a compilation from asynchronous CCS to DNA circuits.
Given their formal definition, process calculi are suitable to formally verify properties of systems. There is a strong line of work concerned with applications of reversible process calculi such as session types, contracts, biological phenomena, and constructs for reliability.  ~\cite{sessionBasedPi} shows how the session type discipline of $\pi$-calculus extends to its reversible variants. In~\cite{singleSessions}, (binary and multiparty) session type systems are used to restrict the study of reversibility in $\pi$-calculus to single sessions. Instead, ~\cite{ServerClient,retractable} study the compliance of a client and a server when both of them have the ability to backtrack to past states.
In a different setting, reversible process calculi have also been used to build constructs for reliability, and in particular communicating transactions with compensations~\cite{Kutavas}. Transactions with compensations are computations that either succeed, or their effects are compensated by a dedicated ad-hoc piece of code. In~\cite{Kutavas}, the effect of the transaction is first undone, and then a compensation is executed. Behavioural equivalences for communicating transactions with compensation have also been studied  in~\cite{Kutavas2,Kutavas3}. In~\cite{LaneseLMSS13}, interacting transactions with compensations are mapped into a reversible calculus with alternatives.

\section{Petri Nets}\label{sec:Petri Nets}

In this work, we shall consider a particular model of concurrency,  known as Petri nets,  that in this thesis will be extended to its reversible variant. It is a basic model of parallel and distributed systems, designed by Carl Adam Petri in 1962 in his PhD Thesis:``Kommunikation mit Automaten"~\cite{PNs,reisig2013understanding}. Petri Nets are a graphical mathematical language that can be used for the specification and analysis of discrete event systems. Petri nets are a formal model of concurrent systems which supports both action-based and state-based modelling and reasoning where the basic idea behind it is to describe state changes in a system with transitions.

\begin{figure}[h]
	\centering
	\subfloat{\includegraphics[width=9cm]{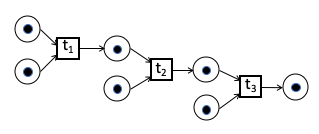}}
	\caption{Petri net}
	\label{PN}
\end{figure}

Petri nets are extensions of directed, finite, bipartite graphs, typically without isolated nodes as seen in Figure \ref{PN}. They are also known as place/transition (PT) nets based on their four main components: places, transitions, arcs and tokens. Formally:

\begin{definition}{\rm$\!\!$
		A \emph{Petri Net}
		is a tuple $\mathit{N} = (P,T,F, W,M_0)$ where:
		\begin{enumerate}
			\item $P$ is a finite set of \emph{places}.
			\item $T$ is a finite set of \emph{transitions} such that
			$P\cap T = \emptyset$.
			\item $F$ is a set of arcs ( or flow relations)  $F \subset (P \times T) \cup (T \times P) $
			\item $W:((P\times T)\cup(T\times P))\rightarrow \mathbb{N}$ is the arc weight mapping where $W(f)=0$ for all $f \not\in F$, and $W(f)>0$ for all $f \in F$, and
			\item $M_0 : P \rightarrow \mathbb{N}$ is the initial marking representing the initial distribution of tokens
		\end{enumerate}
}\end{definition}

\begin{figure}[h]
	\centering
	\subfloat{\includegraphics[width=1.5cm]{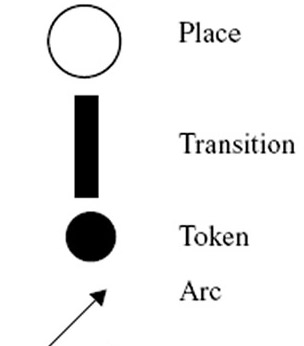}}
	\caption{Petri net components}
	\label{components}
\end{figure}

{\bf Main Components.} As seen in Figure \ref{components} the first component is places which are illustrated  by circles  and refer to as conditions or states. They are passive nodes used to store, accumulate or show tokens to indicate the current state of the execution. Places are used to define conditions that need to be satisfied in order to execute a specific action. Therefore, if a place has an incoming directed arrow then it is called a post-place and if it has outgoing directed arrows then it is called a pre-place. Places filled by tokens indicate the current state of the execution and the overall distribution of tokens across places is known as the marking $M$. The input state is indicated by the initial marking $M_0$ and every other consecutive state can be reached by relocating the required tokens.  

The second component is transitions which are indicated by bars or boxes containing the respective label  and model activities which can occur when a transition fires. They are active nodes  that when fired they can produce, consume transport or change tokens, indicating the execution of the corresponding event. A transition is enabled to be fired only when the number of tokens in each of its input places is at least equal to the number of arcs going from the place to the transition meaning that all of its pre places need to be filled with tokens. Firing a transition means consuming tokens from its pre-places and then distributing them to each output place. Thus, after the firing of a transition  the marking of the net is changed to a new reachable marking, where some transitions are no longer enabled while others become enabled.

Places  and transitions are connected to each other by directed arcs. Graphically an arc is represented by an arrow indicating the relation between  the components such as logical connections, access rights, spatial proximities or immediate linkings.  An arc never connects two places or two transitions. It  rather runs from a place to a transition or from a transition to a place.  The order in which transitions and places appear amongst the directed arcs defines which places are pre-places and therefore are required in order to fire the transition and which are post-places. Labelled incoming arcs may be drawn from a place into a transition labelled with a natural number indicating the finite number of tokens in the associated place that need to be consumed by the occurrence of the transition. Labelled outgoing arcs may be drawn from a transition to a place labelled with a natural number indicating the number of tokens to be deposited in the place by the occurrence of an event.

A Petri net is a particular kind of directed graph, together with an initial state called the initial marking. A marking assigns to each place a non-negative integer indicating the is the configuration of tokens distributed over an entire Petri net diagram. Pictorially, we place black bullets representing tokens in various places of a net where semantically a marking is denoted by $M$, an $m$-vector, where $m$ is the total number of places.  If a marking assigns to place $p$ a non-negative integer $k$, we say that $p$ is marked with $k$ tokens.   A marking is depicted by placing a dot (token) in each of its places. The dynamic behaviour of the represented system, in terms of system state and its evolution, is defined by describing the possible moves between markings based on the marking evolution rule. The marking of the net changes through the occurrence of transitions according to what is commonly called the token game for nets.

As in many formal models the concept of conditional events can be used in order to represent the dynamics of a system. In Petri net modelling places represent conditions, and transitions represent events. A transition has a certain number of input and output places representing the pre-conditions and post-conditions of the event, respectively. The presence of a required number of tokens in a place is interpreted as holding the truth of the condition associated with the incoming arc. A transition is said to be enabled if all input places have sufficient number of tokens for the firing consumptions to be possible. Meaning that the number of tokens in an input place should be  at least equal to the weight of the arc joining such a place with the considered transition. Formally, a transition $t$ is said to be enabled if each input place $p$ of $t$ is marked with at least $w(p,t)$ tokens, where $w(p, t)$ is the label of the arc from $p$ to $t$.

Once a transition is enabled it may or may not fire depending on whether or not the event actually takes place. The firing of an enabled transition is an instantaneous operation which evolves a marking into a new marking. In that case, a number of tokens is consumed by removing from each input place of the transition a number of tokens equal to the weight of the incoming arc leading to a transition. Tokens are consumed by the firing, but also new tokens are produced in its outgoing places, namely a number of tokens are created to each output place equal to the weight of the arc joining the considered transition with such a place. 


{\bf Extensions.} There are many extensions of Petri nets. Some of them are completely backwards-compatible with the original Petri nets, while some add properties that cannot be modelled in the original Petri net formalism. Although backwards-compatible models do not extend the computational power of Petri nets, they may have more succinct representations and may be more convenient for modelling. Extensions that cannot be transformed into Petri nets are sometimes very powerful, but usually lack the full range of mathematical tools available to analyse ordinary Petri nets. An extension of Petri nets is the addition of new types of arcs; such as  inhibitor arcs which impose the precondition that the transition may only fire when the place is empty and reset arcs~\cite{reset} which do not impose a precondition on firing, and empties the place when the transition fires. Reset arcs make reachability undecidable, while some other properties, such as termination, remain decidable~\cite{resetarcs}; whereas inhibitor arcs allow arbitrary computations on numbers of tokens to be expressed, which makes the formalism Turing complete and implies existence of a universal net~\cite{inhibitor}. 

The term \emph{high-level Petri net} is used for many extensions of Petri nets although, the term is mostly used for the type of Coloured Petri nets~\cite{CPN} supported by CPN Tools. In a standard Petri net, tokens are indistinguishable whereas in a coloured Petri nets, every token has a colour. This allows tokens to have a data value attached to them which can be of arbitrarily complex type where places in CPNs usually contain tokens of one type, which is called a coloured set. In popular tools for coloured Petri nets such as CPN Tools, the values of tokens are typed, and can be tested using guard expressions and manipulated with a functional programming language. Coloured Petri nets preserve useful properties of Petri nets and at the same time extend the initial formalism to allow the distinction between tokens. 

Another extension is that of  \emph{timed Petri nets}~\cite{TPN} used to model the timing of a model, where time constraints restrict the causal behaviour of the system and limit its state space by forcing events to occur and keep others from happening following the constraints. In time Petri nets, there is an upper and a lower bound for the time an event can remain enabled without occurring after its preconditions are met. The upper time bound of one potential event can limit the time when another conflicting event can occur, creating dependencies not seen in the simple causal view of the system. 

The qualitative notion of time is implicitly represented in Petri nets in the sense that each firing of a transition is associated with a timestamp or clock cycle. In such a representation, the firing of transitions depends not only on the marking, but also on the elapsed time since the occurrence of some other events. The elapsed time is not represented by the number of internal ticks since the start of the clock but is represented based on the configuration of a marking at the given clock cycle.  

There are several variations of timed Petri nets incorporating the notion of time to virtually every component of the Petri nets framework, namely transition, tokens, arcs, and, places. A subsidiary of timed Petri nets are the \emph{stochastic Petri nets}~\cite{SPN} that add nondeterministic time through adjustable randomness of the transitions. The exponential random distribution is usually used to time these nets. In this case, the nets' reachability graph can be used as a continuous time Markov chain (CTMC).

{\bf Modelling and Analysis.} Various kinds of Petri nets are applied in different disciplines, including Computer Science (formal languages~\cite{rozenberg1983subset}, logic programs~\cite{thieler1977petri}), business process modelling (decision models~\cite{tabak1985petri}), information management (distributed-database systems~\cite{voss1980using}), software engineering (concurrent and parallel programs~\cite{murata1980relevance}), and systems engineering (multiprocessor memory systems~\cite{marsan1986performance}, asynchronous circuits and structures~\cite{dennis1971speed,jump1972equivalence}, compiler and operating systems~\cite{baer1977model}). 
The reason is that Petri nets are a promising tool for describing and studying information processing systems, that are characterised as being concurrent, asynchronous, distributed, parallel, nondeterministic, and/or stochastic. Specifically, executable modelling languages, applied with proper tool support, are expected to automate system validation and verification, simulation and code generation from the modelling language representation. Another advantage, is that formal modelling is more rigorous by nature because it explores every possibility to ensure correctness and completeness. Formal techniques mainly include process algebras, temporal logic, automata theoretic techniques, Petri nets and partial order models. 

 As a mathematical tool, it is possible to set up state equations, algebraic equations, and other mathematical models governing the behaviour of systems. The simplicity of the basic user interface of Petri nets has easily enabled extensive tool support over the years, particularly in the areas of model checking, graphically oriented simulation, and software verification.  As a graphical tool, Petri nets can be used as a visual-communication aid similar to flow charts, block diagrams, and networks. The use of computer-aided tools is a necessity for practical applications of Petri nets and thus most Petri-net research groups have their own software packages and tools to assist the drawing, analysis, and simulation of various applications.

A major strength of Petri nets is their support for analysis of many properties and problems associated with concurrent systems~\cite{murata}. They can be used to study the reachability and coverability  problems as well as study properties such as liveness, boundedness, invariance and conservativeness. Reachability is a fundamental basis for studying the dynamic properties of any system and is essentially the ability to identify whether a given state is reachable from the initial state. However, coverability represents precisely the coverable sets rather than the reachable sets and  is closely related to liveness which is the possibility to ultimately fire any transition of the net by progressing through some further firing sequence. A Petri net is also said to be $k$-bounded or simply bounded if the number of tokens in each place does not exceed a finite number $k$ for any marking reachable from the initial marking. It is also said to be conservative if there exists a positive integer for every place such that the weighted sum of tokens is the same for every marking and for any fixed initial marking. Another important feature of Petri nets is that their structural properties can be obtained by linear algebraic techniques. Such properties are called invariants because they are the properties that depend on only the topological structure of a Petri net and are independent of the initial marking. One of the properties studied in the context of Petri nets is that of Petri net reversibility which describes the ability of  a system to return to the initial state from any reachable state. This, however, is in contrast to the notion of reversible computation as discussed in this work where the intention is not to return to a state via arbitrary execution but to reverse the effect of already executed transitions.

{\bf Petri Net Causality.} \emph{Causality} is one of the most interesting notions in Petri net theory since it allows the explicit representation 
of causal dependencies between action occurrences when modelling reactive systems.
In fact,  how to formalise causal dependencies based on an appropriate causality based concept is a 
well-known topic in Petri net theory~\cite{causal}. The investigation of Petri nets has given rise into 
two different approaches when it comes to causality, one of them being disjunctive causality implemented by the collective token interpretation and the other 
one being partial order causality implemented by the individual token interpretation~\cite{ConfStruct,individual,ZeroSafe}.  Many different semantics have been proposed in the literature for both views, all of them aiming to remain abstract enough while doing justice to the truly concurrent nature of Petri nets. Each philosophy can be justified either by the theoretical properties of the modelled systems, or by the implementation of possible applications.

A common concern between most of the theoretical models of computation is expressing causality in concurrent systems. In contrast to the sequential setting that is well understood, the concurrent setting poses the conceptual question of how do we define the causal order of execution.  When it comes to Petri nets the ability to formally express causal dependencies based on an appropriate causality based concept is one of the most well-known problems of Petri nets but also one of the  most interesting properties~\cite{causal}. 
%

Most of the behavioural models for Petri nets have firing rules that embody the collective token philosophy rather than the individual token philosophy. The collective token philosophy has been investigated in~\cite{Collective} and is considered to be the standard firing rule of Petri nets were tokens residing in the same place are indistinguishable. In the collective token interpretation when multiple tokens of the same type reside in the same place then these  tokens are not distinguished. This means that all that is known by the model is the amount of token occurrences of a specific type and their location in the marking. In the collective token philosophy were we assume unambiguous tokens to be equivalent because when focussing on the net behaviour these tokens are operationally equivalent. The collective approach fits well with resource allocation systems where tokens represent resources and their identity is indistinguishable since their behavioural capabilities are identical.

The computational interpretation  of the collective token philosophy has been extended to the individual token approach, where tokens residing in the same place are distinguished based on their causal path~\cite{NonSeq,IndividualTokens}. 
As such, the individual token interpretation distinguishes tokens of the same type as individual and it has been formally described by the notion of a process in~\cite{NonSeq,IndividualTokens}.  In this approach the model keeps track of where the tokens come from and therefore identifies the causal links between transitions as means of partial order. The semantics of the individual token interpretation are more complicated since this approach requires precise correspondence between the token instances and their past. This approach solves the ambiguity between tokens of the same type by allowing tokens to carry information about their mappings.  This distinction between tokens allows us to give a precise account of the causal and distributed nature of the net as a partial order. The causal relations between the transitions in a distributed
run of a net can also be described by means of causal net~\cite{causal}.
In the standard approach to causality~\cite{PNs} a causal
link is considered to exist between two transitions if one produces
tokens that are used to fire the other.  This relation 
is used to define ``causal order" which is transitive so that if a transition $t_1$ causally precedes $t_2$ and $t_2$ causally 
precedes $t_3$ then  $t_1$ also causally precedes $t_3$. Furthermore, it is an irreflexive relation, i.e., no transition
causally precedes itself.

\chapter{Reversible Computation in Petri Nets} \label{sec:cycles}
During the last few years a number of formal models have been
developed aiming to provide understanding of the basic principles of reversibility along with its
costs and limitations, and to explore
how it can be used to support the solution of complex problems. In this chapter, we set out to study reversible computation in the context of Petri Nets and to explore the modelling of the main strategies for reversing computation. We aim to address the challenges of capturing the notions of backtracking, causal reversibility and out-of-causal-order reversibility within the Petri Net framework, thus proposing a novel, graphical methodology for studying reversibility in a model where transitions can be taken in either direction.
Our proposal is motivated by applications from biochemistry where out-of- causal-order reversibility is inherent, and it supports all the forms of reversibility that have been discussed above.  
%

Adding reversibility to Petri nets turns out to be quite nontrivial since the presence of cycles
exposes the need to define causality of actions within a cyclic structure. Indeed, there are different
ways of introducing reversible behaviour depending on how causality is defined. 
%
In our approach, we follow the 
notion of causality as defined by Carl Adam Petri  for one-safe nets that  provides the notion of a run 
of a system where causal dependencies are reflected in terms of a partial order~\cite{PNs}. A causal
link is considered to exist between two transitions if one produces tokens that are used to fire the other. In this partial order, causal dependencies  are explicitly defined as an unfolding of an occurrence net which is 
 an acyclic net that does not have backward conflicts. 
 We prove 
 that the amount of flexibility allowed in causal reversibility indeed yields  causally consistent semantics. 
 We also demonstrate 
that out-of-causal-order  reversibility is  able to create new states unreachable by forward-only execution.
Additionally, we establish the relationship between the three forms of reversing and define
a transition relation that can capture each of the three strategies modulo the enabledness condition for
each strategy. This allows us to provide a uniform treatment of the basic theoretical results.
We demonstrate the framework with a model of the Ras-Raf-MEK-ERK pathway~\cite{ERK}, and a transaction processing system, examples that inherently feature (out-of-causal-order) reversibility.

\section{Forms of Reversibility and Petri Nets}

Reversing computational processes in concurrent and distributed systems has many
promising applications but also presents some  technical and conceptual challenges. 
In particular, a formal model for concurrent systems that embeds reversible computation needs to 
be able to compute without forgetting and 
to identify the legitimate backward moves at any point during computation.

The first challenge applies to both concurrent and sequential systems. Since
processes typically do not remember their past states, reversing their execution is not directly supported.
This challenge can be addressed with the use of memories. When building a reversible variant
of a formal language, its syntax can be extended to include appropriate representations for computation 
memories to allow processes to keep track of past execution.

The second challenge regards the strategy to be applied when going backwards.
As already mentioned, the most prominent approaches for performing and undoing steps are \emph{backtracking}, 
\emph{causal reversibility}, and \emph{out-of-causal-order reversibility}.
{\em Backtracking} is well understood as the process of rewinding one's
computation trace,
whereas in causal reversibility, reversing does not have to follow the 
exact inverse order for events as long as caused actions, also
known as effects, are undone before 
the reversal of the actions that have caused them. 

\begin{figure}
	\centering
	\subfloat{\includegraphics[width=7.5cm]{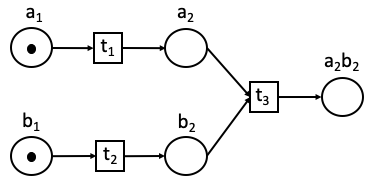}}
	\caption{ Causal reversibility}
	\label{causal}
\end{figure}

For example, consider the Petri net in Figure \ref{causal}.  We may observe
that transitions $t_1$ and $t_2$ are independent from each other as they may be taken in any
order, and they are both prerequisites for transition $t_3$. 
Backtracking the sequence of transitions  $\langle t_1, t_2,t_3\rangle$ would require that
the three transitions should be reversed in exactly the reverse order, i.e. 
$\langle t_3, t_2,t_1\rangle$. Instead, causal flexibility 
allows the inverse computation to rewind $t_3$ and then $t_1$ and $t_2$ in any order, but never
$t_1$ or $t_2$  before $t_3$.

Both backtracking and causal reversing are cause-respecting. There are, however, many important examples 
where concurrent systems 
execute in out-of-causal-order reversibility in order to
allow a system to discover states that are inaccessible in any forward-only execution. This
can be achieved since, reversing in out-of-causal order allows reversing an action before its effects are undone,
and subsequently exploring new computations while the effects of the reversed action are still present.
As such, out-of-order reversibility can  create new alternatives of current states that 
were formerly inaccessible by any forward-only execution path.

\begin{figure}
	\centering
	\subfloat{\includegraphics[width=9cm]{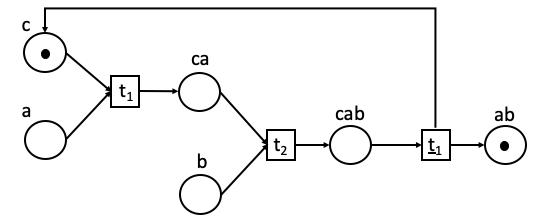}}
	\caption{ Catalysis in classic Petri nets}
	\label{catalyst2}
\end{figure}

Since out-of-order reversibility  contradicts program order, it comes with its 
own peculiarities that need to be taken into consideration while designing reversible systems. To appreciate
these peculiarities and obtain insights towards our approach on addressing reversibility within Petri nets, consider the process of catalysis from biochemistry, 
whereby a substance called \emph{catalyst} enables a chemical reaction between a set of other elements.  
Specifically consider a catalyst $c$ that helps the otherwise inactive molecules $a$  and $b$ to bond. This is achieved
as follows: catalyst $c$ initially bonds with $a$ which then enables the bonding 
between $a$ and $b$. Finally, the catalyst is no longer needed and its bond to
the other two molecules is released. A Petri net model of this process is illustrated in Figure~\ref{catalyst2}.
The Petri net executes transition $t_1$ via which the bond $ca$ is created, followed by
action $t_2$ to produce $cab$. Finally, action
$\underline{t_1}$ ``reverses'' the bond between $a$ and $c$, yielding $ab$ and releasing $c$. (The figure portrays the final state of the execution assuming that initially exactly one token existed in places $a$, $b$, and $c$.)

This example illustrates that Petri nets are not reversible by 
nature, in the sense that every transition cannot be executed
in both directions. Therefore an inverse action (e.g., transition $\underline{t_1}$  for
undoing the effect of transition $t_1$), needs to be added as 
a supplementary forward transition for achieving the undoing of a previous action. 
This explicit approach of modelling reversibility can prove 
cumbersome in systems that express multiple reversible
patterns of execution, resulting in larger and more intricate systems. Furthermore, it
fails to capture reversibility as a mode of computation.
The intention of our work is to study an approach for modelling 
reversible computation that does not require the addition of new, 
reversed transitions but
instead offers as a basic building block transitions that can 
be taken in both the forward as well as the backward direction, and, 
thereby, explore the theory
of reversible computation within Petri nets.

However, when attempting to model the catalysis example
while executing transitions in both
the forward and the backward directions, we may observe a 
number of obstacles. At an abstract level, the behaviour of 
the system should exhibit a sequence of three transitions: 
execution of $t_1$ and $t_2$, followed by the reversal of transition $t_1$. 
The reversal of transition $t_1$ should implement the release of $c$ from the bond $cab$ and make
it available for further instantiations of transitions, 
if needed, while the bond $ab$ should remain in place.
This implies that a reversing Petri net model should provide resources
$a$, $b$ and $c$ as well as $ca$, $cab$ and $ab$ and 
implement the reversal of action $t_1$ as the
transformation of resource $cab$ into $c$ and $ab$. Note that resource $ab$ is inaccessible during the forward execution 
of transitions $t_1$ and $t_2$ and only materialises after the 
reversal of transition $t_1$,
i.e., only once the bond between $a$ and $c$ is broken. Given 
the static nature of a Petri net, this suggests that
resources such as $ab$ should be  represented at the token
level (as opposed to the place level).    
As a result, the concept of token individuality is of particular relevance to
reversible computation in Petri nets while other constructs/functions 
at token level are needed to capture the effect and reversal
of a transition.

Indeed, reversing a transition in an out-of-causal order may imply 
that while some of the effects of the transition
can be reversed (e.g., the release of the catalyst back to the 
initial state), others must be retained due to computation that 
succeeded the forward execution of the next transition (e.g., 
token $a$ cannot be released during the reversal of $t_1$ since
it has bonded with $b$ in transition $t_2$). This latter point 
is especially challenging since it requires to specify a model 
in a precise manner so as to identify which effects are allowed 
to be ``undone'' when reversing a transition. 
Thus, as highlighted by the catalysis example, reversing transitions 
in a Petri net model requires close monitoring of
token manipulation within a net and clear enunciation of the effects of a transition.

\remove{
	\begin{figure}
		\centering
		\subfloat{\includegraphics[width=8cm]{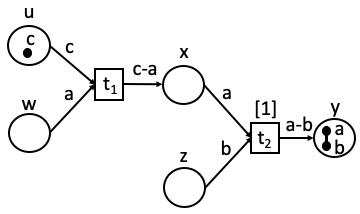}}
		\caption{ Catalysis in reversing Petri nets}
		\label{catalyst}
	\end{figure}
}

As already mentioned, the concept of token individuality can 
prove useful to handle these challenges.
This concept has also been handled in various works, 
e.g.,~\cite{causal,individual,Glabbeek},
where each token is associated with information regarding its 
causal path, i.e., the places and transitions
it has traversed before reaching its current state.  
In our approach, we also implement  the notion of token individuality 
where instead of maintaining extensive histories for recording
the precise evolution of each token through transitions and places, 
we  employ a novel approach inspired
by out-of-causal reversibility in biochemistry as well as approaches 
from related literature~\cite{Bonding}. The resulting 
framework is light in
the sense that no memory needs to be stored per token to retrieve its causal path
while enabling reversible semantics for the three main types of reversibility. 
Specifically, 
we introduce two notions that intuitively capture tokens and their history: 
the notion of a \emph{base} and a new type of tokens called \emph{bonds}. 
A base is a persistent type of token which cannot be
consumed and therefore preserves its individuality through various transitions. 
For a transition to fire, the incoming arcs identify the required tokens/bonds 
and the outgoing arcs may create new bonds or transfer already existing 
tokens/bonds along the places of a Petri Net.  
Therefore, the effect of a transition is the creation of new \emph{bonds} 
between the tokens it takes as input and the reversal of such a transition 
involves undoing the respective bonds. 
In other words, a token can be a base or a  coalition of bases 
connected via bonds into a structure. 

\begin{figure}
	\centering
	\subfloat{\includegraphics[width=8cm]{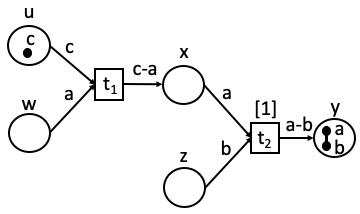}}
	\caption{ Catalysis in reversing Petri nets}
	\label{catalyst}
\end{figure}

Based on these ideas, we may describe the catalysis example in 
our proposed framework as shown in Figure~\ref{catalyst}. 
In this setting $a$ and $c$ are bases which  are connected via a 
bond into place $x$ during transition $t_1$, while transition 
$t_2$ brings into place a new bond between $a$ and $b$.  In Figure~\ref{catalyst} 
we see the state that arises after the execution of $t_1$ and $t_2$ and the reversal 
of transition $t_1$. In this state, base $c$ has returned to its initial 
place $u$ whereas bond $a-b$ has remained in place $y$. A thorough 
explanation of the notation is given in the next section.

Finally, in order to identify at each point in time the
history of execution, thus to discern the transitions that can be reversed
given the presence of backward nondeterminism of Petri nets, we associate 
transitions with  history storing keys in increasing order each time an 
instance of the transition is executed. This allows to backtrack computation 
as well as to extract the causes of bonds as
needed in causal and out-of-causal-order reversibility. 

\section{ Reversing Petri Nets}

We define reversing Petri nets as follows:

\begin{definition}{\rm
		A \emph{\PN}(RPN) is a tuple $(A, P,B,T,F)$ where:
		\begin{enumerate}
			\item $A$ is a finite set of \emph{bases} or \emph{tokens} ranged over by $a$, $b$, $\ldots$  $\overline{A} = \{\overline{a}\mid a\in A\}$
			contains a \emph{negative instance} for each token and we write ${\cal{A}}=A \cup \overline{A}$.
			\item $P$ is a finite set of \emph{places}.
			\item $B\subseteq A\times A$ is a set of undirected \emph{bonds} ranged over by $\beta$, $\gamma$, $\ldots$
			We use the notation $a \bond b$ for a bond $(a,b)\in B$. $\overline{B} = \{\overline{\beta}\mid \beta\in B\}$
			contains a \emph{negative instance} for each bond and we write ${\cal{B}}=B \cup \overline{B}$.
			\item $T$ is a finite set of \emph{transitions}.
			\item $F : (P\times T  \cup T \times P)\rightarrow 2^{{\cal{A}}\cup {\cal{B}}}$ defines a set of directed \emph{arcs} each associated with a subset of ${\cal{A}}\cup {\cal{B}}$.
		\end{enumerate}
}\end{definition}

A \PN is built on the basis of a set of \emph{bases} or \emph{tokens}. We 
consider each token to have a unique name. In this way, tokens may be distinguished from
each other, their persistence can be guaranteed and their history inferred from the structure
of a Petri net (as implemented by function $F$, discussed below). Tokens correspond to
the basic entities that occur in a system. 
They may occur as stand-alone elements but they may also merge together to form \emph{bonds}. \emph{Places}
and \emph{transitions} have the standard meaning. 

Directed arcs connect places to transitions and vice
versa and are labelled by a subset of ${\cal{A}}\cup {\cal{B}}$ where  $\overline{A} = \{\overline{a}\mid a\in A\}$
is the set of \emph{negative} tokens expressing token absence, and  $\overline{B} = \{\overline{\beta}\mid \beta\in B\}$ is the set of \emph{negative} bonds expressing bond absence.
For a label $\ell= F(x,t)$ or $\ell = F(t,x)$, we assume that each token $a$ 
can appear in $\ell$ at most once, either as $a$ or as $\overline{a}$,
and that if a bond $(a,b)\in\ell$ then $a,b\in\ell$. Furthermore, for $\ell = F(t,x)$,
it must be that $\ell\cap (\overline{A}\cup \overline{B}) = \emptyset$, that is, negative tokens/bonds may only occur
on arcs incoming to a transition. 
Intuitively, these labels express the requirements for a transition
to fire when placed on arcs incoming the transition, and the effects of the transition when placed on the
outgoing arcs. Thus, if $a\in F(x,t)$ this implies that token $a$ is required for the transition $t$ to
fire, and similarly for a bond $\beta\in F(x,t)$. On the other hand, $\overline{a}\in F(x,t)$ expresses that
token $a$ should not be present in the incoming place $x$ of $t$ for the transition to fire and similarly for a bond $\beta$, 
$\overline\beta \in F(x,t)$. Note
that negative tokens/bonds are close in spirit to the inhibitor arcs of extended Petri nets. Finally, note that 
$F(x,t)= \emptyset$ implies that there is no arc from place $x$ to transition $t$ and similarly for
$F(t,x) = \emptyset$.

We introduce the following notations. We write 
$\circ t =   \{x\in P\mid  F(x,t)\neq \emptyset\}$ and  
$ t\circ = \{x\in P\mid F(t,x)\neq \emptyset\}$
for the incoming and outgoing places of transition
$t$, respectively. Furthermore, we write
$\guard{t}  =   \bigcup_{x\in P} F(x,t)$ for the union of all labels on the incoming arcs of  transition $t$, and
$\effects{t}  =   \bigcup_{x\in P} F(t,x)$ for the union of all labels on the outgoing arcs of transition $t$.
\begin{definition}\label{well-formed}{\rm 
		A \PN $(A,P,B,T,F)$ is \emph{well-formed} if it satisfies the following conditions for all $t\in T$:
		\begin{enumerate}
			\item $A\cap \guard{t} = A\cap \effects{t}$,
			\item If $ a \bond b \in \guard{t}$ then $ a \bond b \in \effects{t}$,
			\item $ F(t,x)\cap F(t,y)=\emptyset$ for all $x,y\in P$, $x\neq y $. 
		\end{enumerate}
}\end{definition}

According to the above we have that: (1) transitions do not
erase tokens or create new ones, (2) transitions do not destroy bonds, that is, if a bond $a\bond b$ exists in an input place of a transition, then it is
maintained in some output place, and 
(3) tokens/bonds cannot be cloned into more than one outgoing place.

As with standard Petri nets, we employ the notion of a \emph{marking}. A marking is a distribution
of tokens and bonds across places,  $M: P\rightarrow 2^{A\cup B}$ where $a \bond b \in M(x)$, for some $x\in P$, implies
$a,b\in M(x)$. 
In addition, we employ the notion of a \emph{history}, which assigns a memory to each transition of a reversing Petri net as
$H : T\rightarrow 2^\mathbb{N}$. Intuitively, a history of $H(t) = \emptyset$ for some $t \in T$ captures that the transition
has not taken place, and a history of 
$H(t) = \{k_1,\ldots,k_n\}$ captures that the transition was executed
and not reversed $n$ times where $k_i$, $1\leq i \leq n$, indicates the order of execution of the $i^{th}$ instance
amongst non-reversed actions. Note that this machinery, 
is needed to accommodate the presence of cycles, which yield the
possibility of repeatedly executing the same transitions.
$H_0$ denotes the initial history where $H_0(t) = \emptyset$ for all $t\in T$. 
A pair of a marking and a history describes a \emph{state} of a reversing Petri net based on which execution is determined. 
We use the notation $\state{M}{H}$ to denote states. 

In a graphical representation, tokens are indicated by $\bullet$, places by circles, transitions by boxes, and bonds by lines 
between tokens. Furthermore,
histories are presented over the respective transitions as the list  $[k_1,...,k_n]$ when
$H(t) = \{k_1,\ldots,k_n\}$, $n>0$, and omitted when $H(t)=\emptyset$. 

As the last piece of our machinery, we define a notion that identifies connected
components of tokens and their associated bonds within a place. 
Note that more than one connected component may arise
in a place due to the fact that various unconnected tokens may be moved to a place 
simultaneously by a transition, while the reversal of transitions, which results
in the destruction of bonds, may break down a connected component into various
subcomponents. We define $\connected(a,C)$, where $a$ is a base and $C\subseteq A\cup B$ to be the tokens connected
to $a$ via sequences of bonds as well as the bonds creating these connections according to 
set $C$.  
\[\connected(a,C)=(\{a\} \cap C)\cup\{\beta, b, c \mid \exists  w \mbox{ s.t. }  \paths(a,w,C), \beta\in w \mbox{, and } \beta=(b,c) \}\]
where $\paths(a,w,C)$ if $w=\langle\beta_1,\ldots,\beta_n\rangle$, and for all $1\leq i \leq n$, $\beta_i=(a_{i-1},a_i)\in C\cap B$, $a_i\in C\cap A$, and $a_0 = a$.

Returning to the example of Figure~\ref{catalyst}, we may see
a reversing net with three tokens $a$, $b$, and $c$, transition $t_1$, which bonds tokens $a$ and $c$ within place $x$,
and  transition $t_2$, which bonds the $a$ of bond $c \bond a$ with token $b$ into place $y$.
{Note that to avoid overloading figures, we omit writing the bases of
	bonds on the arcs of RPNs, so, e.g., on the arc between $t_1$ and $x$, we write $a \bond b$ as opposed to $\{a \bond b, a, b\}$.}
(The marking depicted in the figure is the one arising after the execution of transitions $t_1$ and $t_2$ and subsequently 
the reversal of transition $t_1$ by the semantic relations to be defined in the next section.) 

We may now define the various types of execution for  reversing Petri nets. In what follows we
restrict our attention to  well-formed RPNs $(A,P,B,T,F)$ with initial
marking $M_0$ such that for all $a\in A$, $|\{ x \mid a\in M_0(x)\} | = 1$.

\subsection{Forward Execution}
In this section we consider the standard, forward execution of RPNs.
\begin{definition}\label{forward}{\rm
		Consider a reversing Petri net $(A, P,B,T,F)$, a transition $t\in T$, 
		and a state $\state{M}{H}$. We say that
		$t$ is \emph{forward-enabled} in $\state{M}{H}$  if the following hold:
		\begin{enumerate}
			\item  if $a\in F(x,t)$,  for some $x\in\circ t$, then $a\in M(x)$, and if   
			$\overline{a}\in F(x,t)$
			for some $x\in\circ t$, then $a\not\in M(x)$, 
			\item  if $\beta\in F(x,t)$,  for some $x\in\circ t$, then 
			$\beta\in M(x)$, and if $\overline{\beta}\in F(x,t)$
			for some $x\in\circ t$, then $\beta\not\in M(x)$, 
			\item if $a\in F(t,y_1)$,  $b\in F(t,y_2)$, $y_1 \neq y_2$, then $b \not\in \connected(a,M(x))$ for all
			$x \in \circ t$, and  
			\item if $\beta\in F(t,x)$ for some $x\in t\circ$ and $\beta\in M(y)$ for some $y\in \circ t$, then $\beta\in F(y,t)$. 
		\end{enumerate}
}\end{definition}

Thus, $t$ is enabled in state $\state{M}{H}$ if (1), (2), all 
tokens and bonds required for the transition to take place
are available in the incoming places of $t$ and  none of the 
tokens/bonds whose absence is
required exists in an incoming place of the transition, (3) if 
a transition forks into outgoing places $y_1$ and
$y_2$ then the tokens transferred to these places are not connected to each other in the
incoming places of the transition,  and (4) if a pre-existing bond 
appears in an outgoing arc of a transition, then it is also a precondition 
of the transition to fire. 
Contrariwise, if the bond appears in an outgoing arc of a 
transition ($\beta\in F(t,x)$ for some $x\in t\circ$)
but is not a requirement for the transition to fire ($\beta\not\in F(y,t)$ for all $y\in \circ t$),
then  the bond should not be present in an incoming place of the transition ($\beta\not\in M(y)$ for all $y\in \circ t$).

We observe that the new bonds created by a transition are
exactly those that occur in the outgoing edges of a transition but not in the incoming edges. Thus, we define the effect of a transition as 
\[\effect{t} = \effects{t} - \guard{t}\]
This will subsequently enable the enunciation of transition reversal by the destruction of exactly
the bonds in $\effect{t}$.

\begin{definition}{\rm \label{forw}
		Given a reversing Petri net $(A, P,B,T,F)$, a state $\langle M, H\rangle$, and a transition $t$ enabled in 
		$\state{M}{H}$, we write $\state{M}{H}
		\trans{t} \state{M'}{H'}$
		where:
		\[
		\begin{array}{rcl}
		M'(x) & = & \left\{
		\begin{array}{ll}
		M(x)-\bigcup_{a\in F(x,t)}\connected(a,M(x))  & \textrm{if } x\in \circ{t} \\
		M(x)\cup F(t,x)\cup \bigcup_{ a\in F(t,x)\cap F(y,t)}\connected(a,M(y)) 
		& \textrm{if }  x\in t\circ\\
		M(x), &\textrm{otherwise}
		\end{array}
		\right.
		\end{array}
		\]
		and
		\[
		\begin{array}{rcl}
		H'(t') & = & \left\{
		\begin{array}{lll}
		H(t')\cup \{ \max( \{0\} \cup\{k \mid k\in H(t''), t''\in T\}) +1\},\hspace{0.1in}
		\textrm{if } t' = t\\
		H(t'),\hspace{3in}\textrm{ otherwise}
		\end{array}
		\right.
		\end{array}
		\]
}\end{definition} 

\begin{figure}[t]
	\centering
\hspace{.9cm}	
\subfloat{\includegraphics[width=7cm]{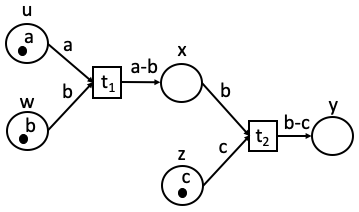}} \\
	\subfloat{\includegraphics[width=.9cm]{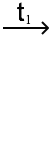}} 
		\subfloat{\includegraphics[width=7cm]{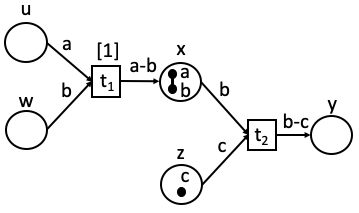}} \\
	\subfloat{\includegraphics[width=.9cm]{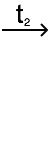}}
	\subfloat{\includegraphics[width=7cm]{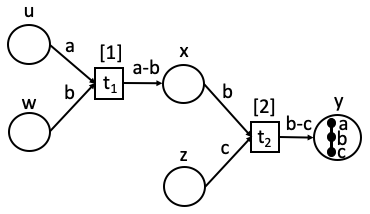}}
	\caption{Forward execution}
	\label{f-example}
\end{figure}

Thus, when a transition $t$ is executed in {the} forward direction, all tokens and bonds
occurring in its incoming arcs are relocated from the input 
places to the output places along with their connected 
components. An example of forward transitions can be seen in 
Figure~\ref{f-example} where transitions $t_1$ and $t_2$ take 
place with the histories of the two transitions 
becoming {\bf[1]} and {\bf[2]}, respectively. 

We may prove the following result, which verifies that bases are preserved during forward execution in the sense that transitions neither erase nor clone them. As far as bonds are concerned, the proposition states that forward execution may
create but not destroy bonds.
\begin{proposition} [Token and bond preservation]  \label{prop1}    {\rm Consider a reversing Petri net $(A, P,B,T, F)$, a state $\langle M, H\rangle$ such that
		for all $a\in A$,  $|\{x\in P\mid a\in M(x)\}| = 1$, and a transition 
		$\state{M}{H} \trans{t} \state{M'}{H'}$. Then:
		\begin{enumerate}
			\item for all $a\in A$,  $|\{x\in P \mid a\in M'(x)\}|=1$,
			and 
			\item for all $\beta\in B$, $|\{x\in P \mid \beta\in M(x)\}| \leq |\{x\in P \mid \beta\in M'(x)\}|\leq 1$.
		\end{enumerate}
}\end{proposition}

\paragraph{Proof}
The proof of the result follows the definition of forward execution and relies on the well-formedness of
RPNs. Consider a reversing Petri net $(A, P,B,T,F)$, a state $\langle M, H\rangle$  such that $|\{x\in P\mid a\in M(x)\}| = 1$ for
all $a\in A$, and suppose 
$\state{M}{H} \trans{t} \state{M'}{H'}$.  

For the proof of clause (1) let $a\in A$. Two cases exist:
\begin{enumerate}
	\item $a\in \connected(b,M(x))$ for some $b\in F(x,t)$. Note that $x$ is unique by the
	assumption that $|\{x\in P\mid a\in M(x)\}| = 1$. Furthermore, according to Definition~\ref{forw}, 
	we have that $M'(x) = M(x) - \{\connected(b,M(x)) \mid b\in F(x,t)\}$, which implies that $a\not\in M'(x)$.
	On the other hand,  by  Definition~\ref{well-formed}(1),
	$b\in \effects{t}$. Thus, there exists $y\in t\circ$, such that $b\in F(t,y)$. Note that this $y$
	is unique by Definition~\ref{well-formed}(3).  As a result, by Definition~\ref{forw}, 
	$M'(y) = M(y)\cup F(t,y)\cup \{\connected(b,M(x)) \mid b\in F(t,y), x\in\circ{t}\}$. Since
	$b\in F(x,t)\cap F(t,y)$, $a\in \connected(b,M(x))$, this implies that $a\in M'(y)$. 
	
	Now suppose that $a\in \connected(c,M(x))$ for some  $c\neq b$, $c\in F(t,y')$. Then, by Definition~\ref{forward}(3), 
	it must be that $y = y'$. As a result, we have that $\{z\in P\mid a\in M'(z)\} = \{y\}$ and 
	the result follows.
	\item $a\not\in \connected(b,M(x))$ for all $b\in F(x,t)$, $x\in P$. This implies that 
	$\{x\in P\mid a\in M'(x)\} = \{x\in P\mid a\in M(x)\}$ and the result follows.
\end{enumerate}

To prove clause (2) of the proposition, consider a bond
$\beta\in B$, $\beta=(a,b)$. We observe that, since $|\{x\in P\mid a\in M(x)\}| = 1$ for
all $a\in A$, $|\{x\in P\mid \beta\in M(x)\}| \leq 1$. The proof follows by case analysis as
follows:
\begin{enumerate}
	\item Suppose $|\{x\in P\mid \beta\in M(x)\}| =0$. Two cases exist:
	\begin{itemize}
		\item Suppose $\beta\not\in F(t,x)$ for all $x\in P$. Then, by 
		Definition~\ref{forw}, $\beta\not\in M'(x)$ 
		for all $x\in P$. Consequently, $|\{x\in P\mid \beta\in M'(x)\}| =0$ and the result follows.
		\item Suppose $\beta\in F(t,x)$ for some $x\in P$. Then, by  Definition~\ref{well-formed}(3),
		$x$ is unique, and by Definition~\ref{forw}, $\beta\in M'(x)$.
		Consequently, $|\{x\in P\mid \beta\in M'(x)\}| =1$ and the result follows.
	\end{itemize}
	\item Suppose $|\{x\in P\mid \beta\in M(x)\}| =1$. Two cases exist:
	\begin{itemize}
		\item $\beta\not\in \connected(c,M(x))$ for all $c\in F(x,t)$. 
		This implies that $\{x\in P\mid \beta\in M'(x)\} = 
		\{x\in P\mid \beta\in M(x)\}$ and the result follows.
		\item $\beta\in \connected(c,M(x))$ for some $c\in F(x,t)$. Then, according to Definition~\ref{forw}, 
		we have that $M'(x) = M(x) - \{\connected(c,M(x)) \mid c\in F(x,t)\}$, which implies that $\beta\not\in M'(x)$.
		On the other hand, by the definition of well-formedness, Definition~\ref{well-formed}(1),
		$c\in \effects{t}$. Thus, there exists $y\in t\circ$, such that $c\in F(t,y)$. Note that this $y$
		is unique by Definition~\ref{well-formed}(3).  As a result, by Definition~\ref{forw}, 
		$M'(y) = M(y)\cup F(t,y)\cup \{\connected(c,M(x)) \mid c\in F(t,y), x\in\circ{t}\}$. Since
		$c\in F(x,t)\cap F(t,y)$, $\beta\in \connected(c,M(x))$, this implies that $\beta\in M'(y)$. 
		
		Now suppose that $\beta\in \connected(d,M(x))$ for some $d\neq c$, $c\in F(d,y')$. Then, by Definition~\ref{forward}, 
		and since $\connected(c,M(x)) = \connected(d,M(x))$,
		it must be that $y = y'$. As a result, we have that $\{z\in P\mid \beta\in M'(z)\} =  \{y\}$ and 
		the result follows.
	\end{itemize}
	\proofend
\end{enumerate}

\subsection{Backtracking}

Let us now proceed to the simplest form of reversibility, namely, backtracking. 
We define a transition to
be \emph{bt}-enabled (backtracking-enabled) if it was the last executed transition:

\begin{definition}\label{bt-enabled}{\rm
		Consider a state $\state{M}{H}$ and a transition $t\in T$. We say that $t$ is \emph{$bt$-enabled} in
		$\state{M}{H}$ if
		$k\in H(t)$ with $k\geq k'$ for all $k' \in H(t')$, $t'\in T$.
}\end{definition}

Thus, a transition $t$ is $bt$-enabled if its history contains the highest value among all transitions. 
The effect of backtracking a transition in a reversing Petri net is as follows:

\begin{definition}\label{br-def}{\rm
		Given a reversing Petri net $(A, P,B,T,F)$, a state $\langle M, H\rangle$, and a transition $t$ that is $bt$-enabled in $\state{M}{H}$, we write $ \state{M}{H}
		\btrans{t} \state{M'}{H'}$
		where:
		\[
		\begin{array}{rcl}
		M'(x) & = & \left\{
		\begin{array}{ll}
		M(x)\cup\bigcup_{ y \in t\circ, a\in F(x,t)\cap F(t,y)}\connected(a,M(y)-\effect{t}),  & \textrm{if } x\in \circ{t} \\
		M(x)- \bigcup_{a\in F(t,x)}\connected(a,M(x)) , & \textrm{if }  x\in t\circ\\
		M(x) &\textrm{otherwise}
		\end{array}
		\right.
		\end{array}
		\]
		and 
		\[
		\begin{array}{rcl}
		H'(t') & = & \left\{
		\begin{array}{lll}
		H(t')- \{ k \},\hspace{0.8in} & \textrm{if } t' = t, k = \max(H(t))\\
		H(t')  & \textrm{otherwise}
		\end{array}
		\right.
		\end{array}
		\]
		
}\end{definition}

\begin{figure}[t]
	\centering
	\hspace{.9cm}
	\subfloat{\includegraphics[width=7cm]{figures/backtracking.png}}
	\subfloat{\includegraphics[width=.9cm]{figures/arrow1.png}}
	\subfloat{\includegraphics[width=7cm]{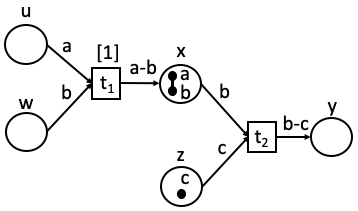}}\\
	\subfloat{\includegraphics[width=.9cm]{figures/arrow2.png}}
	\subfloat{\includegraphics[width=7cm]{figures/backtracking2.png}}
	\subfloat{\includegraphics[width=.9cm]{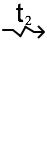}}
	\subfloat{\includegraphics[width=7cm]{figures/backtracking3.png}}\\
	\subfloat{\includegraphics[width=.9cm]{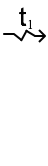}}
	\subfloat{\includegraphics[width=7cm]{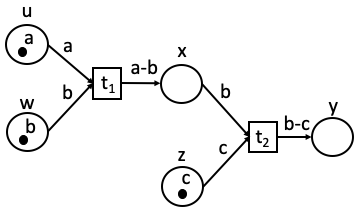}}
	\caption{Backtracking execution}
	\label{b-example}
\end{figure}

When a transition $t$ is reversed in a backtracking fashion  all tokens and bonds in the
postcondition of the transition, as well as their connected components, 
are transferred to the incoming places of the transition and any newly-created bonds are broken. 
Furthermore, the largest key in the history of the transition is removed.

\begin{figure}[t]
	\centering
	\subfloat{\includegraphics[width=7cm]{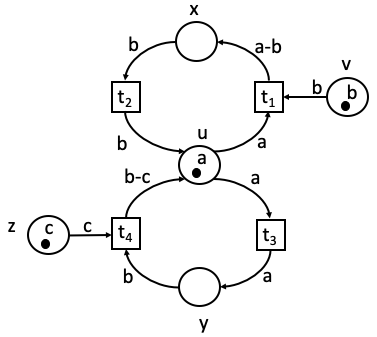}}
	\subfloat{\includegraphics[width=.9cm]{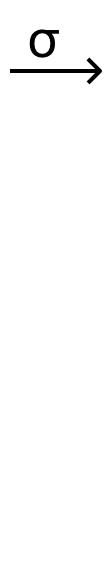}}
	\subfloat{\includegraphics[width=7cm]{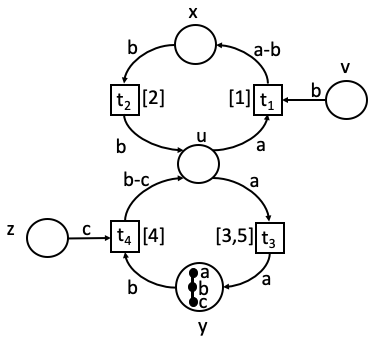}}\\
	\subfloat{\includegraphics[width=.9cm]{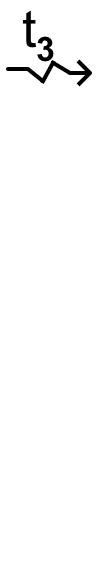}}
	\subfloat{\includegraphics[width=7cm]{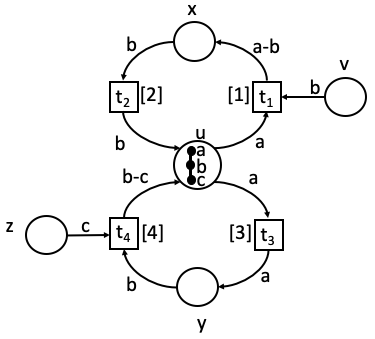}}
	\caption{Backtracking execution where $\sigma =\langle t_1,t_2,t_3,t_4,t_3\rangle$}
	\label{b-cycles}
\end{figure}

An example of backtracking extending the example of Figure~\ref{f-example} can be seen in  Figure~\ref{b-example} where we observe transitions $t_2$ and $t_1$ being reversed  with the histories of the two transitions being eliminated.
A further example can be seen in Figure~\ref{b-cycles} where after the execution of transition sequence $\langle t_1, t_2,t_3,t_4,t_3\rangle$, only transition $t_3$ is $bt$-enabled since it was the last transition to be executed. During its reversal, the component $a \bond b\bond c$ is returned to  place $u$. Furthermore, the largest key of the history of $t_3$ becomes empty.

We may prove the following result, which verifies that bases are preserved during backtracking execution in the sense that
there exists exactly one instance of each base and backtracking transitions neither erase nor clone them. 
As far as bonds are concerned, the proposition states that at any time there may exist at most one 
instance of a bond and that backtracking transitions may only destroy bonds.
\begin{proposition}  [Token preservation and bond destruction] \label{prop2}
	{\rm Consider a reversing Petri net $(A, P,B,T,F)$, a state $\langle M, H\rangle$
		such that for all $a\in A$, $|\{x\in P \mid a\in M(x)\}| = 1$, and a transition 
		$\state{M}{H} \btrans{t} \state{M'}{H'}$. Then: 
		\begin{enumerate} 
			\item for all $a\in A$,  
			$|\{x\in P \mid a\in M'(x)\}| = 1$, and 
			\item  for all $\beta\in B$, $1 \geq |\{x \in P \mid \beta\in M(x)\}| \geq |\{x \in P \mid \beta\in M'(x)\}|$.
		\end{enumerate}
}\end{proposition}
\paragraph{Proof}
The proof of the result follows the definition of backward 
execution and relies on the well-formedness of
reversing Petri nets. 
Consider \RPN $(A, P,B,T,F)$, a state $\langle M, H\rangle$ such that $|\{x\in P \mid a\in M(x)\}| =1$
for all $a\in A$, and suppose  $\state{M}{H} \btrans{t} \state{M'}{H'}$.  

We begin with the proof of clause (1) and let $a\in A$. Two cases exist:
\begin{enumerate}
	
	\item $a\in \connected(b,M(x))$ for some $b\in F(t,x)$. Note that by the assumption 
	of $|\{x\in P \mid a\in M(x)\}| =1$, $x$ must be unique. 
	Let us choose $b$ such that, additionally, $a\in \connected(b,M(x) - \effect{t})$. Note
	that such a $b$ must exist, otherwise 
	the forward execution of $t$ would not have transferred $a$ along with $b$ to place $x$.
	
	According to Definition~\ref{br-def}, 
	we have that $M'(x) = M(x) - \{\connected(b,M(x)) \mid b\in F(t,x)\}$, 
	which implies that $a\not\in M'(x)$.
	On the other hand, note that by the definition of well-formedness, Definition~\ref{well-formed}(1),
	$b\in \guard{t}$. Thus, there exists $y\in \circ t$, such that $b\in F(y,t)$. Note that this $y$
	is unique. If not, then there exist
	$y$ and $y'$ such that $y\neq y'$ with $b\in F(y,t)$ and $b\in F(y',t)$.  By the assumption, however,
	that there exists at most one token of each base, and Proposition~\ref{prop1}, $t$ would never be enabled, 
	which leads to a contradiction.   As a result, by Definition~\ref{br-def}, 
	$M'(y) = M(y)\cup\{\connected(b,M(x)-\effect{t}) \mid b\in F(y,t)\cap F(t,x)\}$. Since
	$b\in F(y,t)\cap F(t,x)$, $a\in \connected(b,M(x)-\effect{t})$, this implies that $a\in M'(y)$. 
	
	Now suppose that $a\in \connected(c,M(x)-\effect{t})$, $c\neq b$, and $c\in F(y',t)$. Since
	$a\in \connected(b,M(x) - \effect{t})$, it must be that
	$\connected(b,M(x) - \effect{t})=\connected(c,M(x)-\effect{t})$. Since $b$ and $c$ are
	connected to each other but the connection was not created by transition $t$ (the connection is
	present in $M(x)-\effect{t}$), it must be that the connection was already present before
	the forward execution of $t$ and, by token uniqueness, we conclude that  $y=y'$.
	\item $a\not\in \connected(b,M(x))$ for all $b\in F(t,x)$, $x\in P$. This implies that 
	$\{x\in P\mid a\in M'(x)\} = \{x\in P\mid a\in M(x)\}$ and the result follows.
\end{enumerate}

Let us now prove clause (2) of the proposition. Consider a bond
$\beta\in B$, $\beta=(a,b)$. We observe that, since $|\{x\in P\mid a\in M(x)\}| = 1$ for
all $a\in A$, $|\{x\in P\mid \beta\in M(x)\}| \leq 1$. The proof follows by case analysis as
follows:
\begin{enumerate}
	\item $\beta\in \connected(c,M(x))$ for some $c\in F(t,x)$, $x\in P$. By the assumption 
	of $|\{x\in P \mid \beta\in M(x)\}| =1$, $x$ must be unique. 
	Then, according to Definition~\ref{br-def}, 
	we have that $M'(x) = M(x) - \{\connected(c,M(x)) \mid c\in F(x,t)\}$, 
	which implies that $\beta\not\in M'(x)$.
	Two cases exist: 
	\begin{itemize}
		\item If $\beta\in \effect{t}$, then $\beta\not\in M'(y)$ for all places $y\in P$.
		\item If $\beta\not\in \effect{t}$ then let us choose $c$ such that 
		$\beta\in \connected(c,M(x) - \effect{t})$. Note
		that such a $c$ must exist, otherwise 
		the forward execution of $t$ would not have connected $\beta$ with $c$. 
		By the definition of well-formedness, Definition~\ref{well-formed}(1),
		$c\in \guard{t}$. Thus, there exists $y\in \circ t$, such that $c\in F(y,t)$. Note that this $y$
		is unique (if not, $t$ would not have been enabled).  As a result, by Definition~\ref{br-def}, 
		$\beta \in M'(y)$.
		
		Now suppose that $\beta\in \connected(d,M(x)-\effect{t})$, $d\neq c$, and $d\in M'(y')$. Since
		$\beta\in \connected(c,M(x) - \effect{t})$, it must be that
		$\connected(c,M(x) - \effect{t})=\connected(d,M(x)-\effect{t})$.  Since $c$ and $d$ are
		connected to each other but the connection was not created by transition $t$ (the connection is
		present in $M(x)-\effect{t}$), it must be that the connection was already present before
		the forward execution of $t$ and, by token uniqueness, we conclude that  $y=y'$.
		This implies that $\{z\in P\mid \beta\in M'(z)\} = \{y\}$. 
	\end{itemize}
	The above imply that $\{z\in P\mid \beta\in M(z)\} = \{x\}$ and  $\{z\in P\mid \beta\in M'(z)\} \subseteq \{y\}$ and 
	the result follows.
	\item $\beta\not\in \connected(c,M(x))$ for all $c\in F(t,x)$, $x\in P$. This implies that 
	$\{x\in P\mid \beta\in M'(x)\} = \{x\in P\mid \beta\in M(x)\}$ and the result follows.
	\proofend
\end{enumerate}

Let us now consider the combination of forward and backward moves in executions. 
We write $\fbtrans{}$ for $\trans{}\cup\btrans{}$.
The following result establishes that in an execution beginning in the initial state of a reversing Petri net, bases are 
preserved, bonds can have at most one instance at any time and a new occurrence of  a bond may be created 
during a forward  transition that features the bond as its effect whereas a bond can be destroyed 
during the backtracking of a transition that features the bond as its effect. This last point clarifies
that the effect of a transition characterises the bonds that are newly-created during the transition's forward execution
and the ones that are destroyed during its reversal.

\begin{proposition}\label{Prop}{\rm Given a reversing Petri net $(A, P,B,T,F)$, an initial state 
		$\langle M_0, H_0\rangle$ and an execution
		$\state{M_0}{H_0} \fbtrans{t_1}\state{M_1}{H_1} \fbtrans{t_2}\ldots \fbtrans{t_n}\state{M_n}{H_n}$, the following hold:
		\begin{enumerate}
			\item For all $a\in A$  and $i$, $0\leq i \leq n$,   $|\{x\in P \mid a\in M_i (x)\}| = 1$.
			\item For all $\beta \in B$ and $i$, $0\leq i \leq n$, 
			\begin{enumerate}
				\item $0 \leq |\{x \in P \mid \beta\in M_i(x)\}| \leq 1$,
				\item if $t_i$ is executed in the forward direction and $\beta\in \effect{t_i}$, then 
				$\beta\in M_{i}(x)$ for some $x\in P$ where $\beta\in F(t_i,x)$, and  $\beta\not\in M_{i-1}(y)$ for all $y\in P$,
				\item if $t_i$ is executed in the forward direction, $\beta\in M_{i-1}(x)$ for some $x\in P$, and $\beta\not\in \effect{t_i}$ then, if $\beta \in \connected(a,M_{i-1}(x))$ and $a\in F(t_i,y)$, then $\beta\in M_{i}(y)$, otherwise $\beta\in M_i(x)$,
				\item if $t_i$ is executed in the reverse direction and $\beta\in \effect{t_i}$ then 
				$\beta\in M_{i-1}(x)$ for some $x\in P$ where $\beta\in F(t_i,x)$, and  $\beta\not\in M_{i}(y)$ for all $y\in P$, and
				\item if $t_i$ is executed in the reverse direction, $\beta\in M_{i-1}(x)$ for some $x\in P$, and $\beta\not\in \effect{t_i}$ then, if $\beta \in \connected(a,M_{i-1}(x))$ and $a\in F(y,t_i)$, then $\beta\in M_{i}(y)$, otherwise $\beta\in M_i(x)$.
			\end{enumerate}
	\end{enumerate}}
\end{proposition}
\paragraph{Proof}
To begin with, we observe that 
the proofs of clauses (1) and (2)(a) follow directly from clauses (1) and (2) of Propositions~\ref{prop1} 
and~\ref{prop2}. Clause (2)(b) follows from Definition~\ref{forward}(4) and Definition~\ref{forw}. Clause (2)(c) follows from Definition~\ref{forw} and the condition refers to whether
the bond is part of a component manipulated by the forward execution of $t_i$. Similarly,to (2)(a)
clause (2)(d) 
stems from Definition~\ref{br-def}. Finally, Clause (2)(e) follows from Definition~\ref{br-def} and the condition refers to whether
the bond is part of a component manipulated by the reverse execution of $t_i$.
\proofend

In this setting we may establish a loop lemma:
\begin{lemma}[Loop]\label{loopb}{\rm 
		For any forward transition $\state{M}{H}\trans{t}\state{M'}{H'}$ there exists a backward transition
		$\state{M'}{H'} \btrans{t} \state{M}{H}$ and vice versa. 
}\end{lemma}
\paragraph{Proof}
{Suppose $\state{M}{H}\trans{t}\state{M'}{H'}$. Then $t$ is clearly $bt$-enabled in $H'$. Furthermore,
	$\state{M'}{H'} \btrans{t} \state{M''}{H''}$ where $H'' = H$. In addition, all tokens and bonds
	involved in transition $t$ (except those in $\effect{t}$) will be returned from the outgoing places
	of transition $t$ back to its incoming places. Specifically, for all $a\in A$, it
	is easy to see by the definition of $\btrans{}$  that $a\in M''(x)$ if and only if $a\in M(x)$.
	Similarly,  for all $\beta\in B$,  $\beta\in M''(x)$ if and only if
	$\beta\in M(x)$. The opposite direction can be argued similarly.
}
\proofend

\subsection{Causal-Order Reversibility}
We now move on to consider causal-order reversibility in RPNs. To define
such as reversible semantics in the presence of cycles, a number of 
issues need to be resolved. To begin with, consider a sequence of transitions
pertaining to the repeated execution of a cycle. Adopting the view that
reversible computation has the ability to rewind \emph{every} executed action of a 
system, we require that each of these transitions is executed in reverse
as many times as it was executed in the forward direction.
Furthermore, the presence of cycles raises questions about the causal relationship 
between transitions of a cycle
as well as of overlapping or even structurally distinct cycles.
In the next subsection we discuss our
adopted notion of transition causality. Subsequently, we develop a theory
for causal-order reversibility in RPNs.

\subsubsection{Causality in cyclic reversing Petri nets}
A cycle in a reversing Petri net is associated with a cyclic path in the net's graph structure. It contains a sequence of transitions
where an outgoing place of the last transition coincides with an incoming place of the first transition.
Note that a cycle in the graph of a reversing Petri net
does not necessarily imply the repeated execution of its transitions since, for instance, entrance
to the cycle may require a token or a bond that has been directed into a different part of the net during execution
of the cycle.

\begin{figure}[t]
	\centering
	\hspace{.9cm}
	\subfloat{\includegraphics[width=8cm]{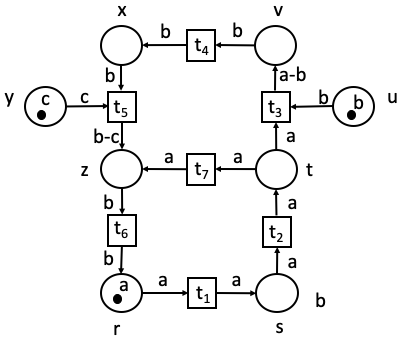}}\\
	\subfloat{	\includegraphics[width=.9cm]{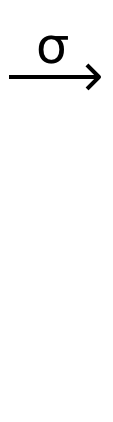}}
	\subfloat{\includegraphics[width=8cm]{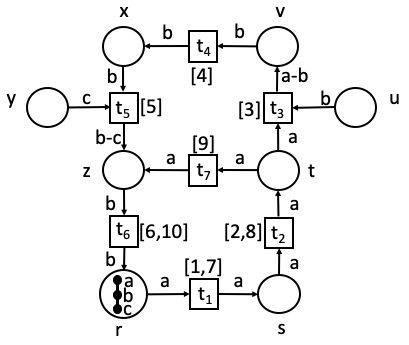}}
	\caption{
		RPN with overlapping cycles $\sigma_1=\langle t_1, t_2, t_3, t_4, t_5, t_6\rangle$ and 
		$\sigma_2 = \langle t_1, t_2, t_7, t_6\rangle$, and the state arising after the forward execution of $\sigma=\sigma_1\sigma_2$}
	\label{cycles1}
\end{figure}

In the standard approach to causality in classical Petri nets~\cite{PNs}, a causal
link is considered to exist between two transitions if one produces
tokens that are used to fire the other.  This relation 
is used to define a ``causal order", $\prec$, which is transitive so that if a transition $t_1$ causally precedes $t_2$ and $t_2$ causally 
precedes $t_3$, then $t_1$ also causally precedes $t_3$.

Adapting this notion in the context of cycle execution, consider  a cycle with
transitions $t_1$ and $t_2$, executed twice yielding the transition instances $t_1^1$, $t_2^1$, $t_1^2$, $t_2^2$, where $t_i^j$
denotes the $j$-th execution of transition $t_i$. Furthermore, suppose that $t_1$ produces tokens that are consumed by $t_2$ and vice versa. This implies the causal order relation
$\prec$, such that $t_1^1\prec t_2^1\prec t_1^2\prec t_2^2$, allowing us to conclude that each execution of the cycle causally
precedes any subsequent executions. This is a natural conclusion in the case of the consecutive execution of cycles,
since a second execution of a cycle cannot be initiated before the first one is completed. This is because the tokens manipulated by the first transition of the cycle need to return to its input places before the transition can be repeated.

\begin{figure}[h]
	\centering
	\hspace{.7cm}
	\subfloat{\includegraphics[width=8cm]{figures/cycle20.png}}\\
	\subfloat{\includegraphics[width=.9cm]{figures/sigma.png}}
	\subfloat{\includegraphics[width=8cm]{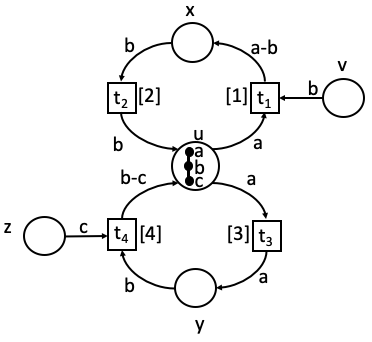}}
	\caption{Causally dependent cycles, where $\sigma=\langle t_1, t_2, t_3,t_4\rangle$ }
	\label{cycles2}
\end{figure}

Let us now move on to determining when a token produced by a transition is consumed by
another. In RPNs this concept acquires an additional complexity due to the fact 
that tokens are distinguished by names and the fact that the creation of bonds
between tokens may disguise the causal relation between transitions.
For instance, consider the example of 
Figure~\ref{cycles1}. This RPN features two overlapping cycles, which can be executed sequentially.
Suppose we execute the outer cycle (transition sequence $\langle t_1, t_2, t_3, t_4, t_5, t_6\rangle$) followed
by the inner cycle (transition sequence $\langle t_1, t_2, t_7,t_6\rangle$).

Observing the token manipulation of the transition
instances as captured by the arcs of the transition, we obtain the order $t_1^1\prec t_2^1\prec t_3^1\prec t_4^1\prec t_5^1\prec t_6^1$ and
$t_1^2\prec t_2^2\prec t_7^1\prec t_6^2$. However by
simply observing the structure of the RPN there is no evidence that $t_1$ consumes tokens
produced by $t_6$. Nonetheless, in this scenario transition instance $t_3^1$ has bonded tokens
$a$ and $b$ and, thus, transition instance $t_1^2$ requires bond $a-b$ to be produced and placed at $r$ by $t_6$ before transition $t_1$ can be executed for the second time. 
Thus, $t_6^1 \prec t_1^2$ also holds.

Note that, if the two cycles were not considered to be causally dependent and were allowed to reverse in any order, then, 
reversal of the first  before the second one would disable the reversal of the second cycle. This is because
reversing transition $t_3$ would return token $b$ to place $u$, thus disabling a second reversal of
transition $t_6$ (and consequently the reversal of the inner cycle).

\comment{This observation becomes more important when one considers the consecutive
	execution of different cycles, be they structurally independent or partially overlapping. Consider the example of 
	Figure~\ref{cycles1}. This RPN features two overlapping cycles, which can be executed sequentially.
	Suppose we first execute the outer cycle (transition sequence $\langle t_1, t_2, t_3, t_4, t_5, t_6\rangle$) followed
	by the inner cycle (transitions $\langle t_1,t_2,t_7,t_6\rangle$). Note that the possibility of executing the inner cycle
	in this execution is lost after execution of transition $t_3$ and regained when the outer cycle is concluded
	via transition $t_6$, and token $a$ is returned to place $r$. The reason behind this is the transitive nature of causal order where the execution of transition $t_3$ causally precedes $t_4$ and also $t_3$ transitively precedes $t_7$ since the tokens produced by $t_3$ are also consumed by $t_7$. As such, we consider the second cycle to be
	causally dependent on the first: had the first cycle not been concluded the second cycle would not have
	been feasible. This approach is also compatible with the expectation for consistency in causal-order reversibility:
	If the cycle execution were not considered to be causally dependent and were allowed to reverse in any order, then, 
	reversal of the first cycle before the second one would disable the reversal of the second cycle. This is because
	reversing transition $t_3$ would return token $b$ to place $u$, thus disabling a second reversal of
	transition $t_6$ (and thus the reversal of the inner cycle). The history function adopted in the RPN definition, allows 
	us to distinguish 
	between instances of transition executions and to determine the causal relationship between nested (and other) cycles.
}

Similarly, in the example of Figure~\ref{cycles2} we observe two cycles that are structurally independent but
where the presence of common tokens between the two cycles creates a dependence between their executions.
For instance, suppose that the upper cycle is initially selected via execution of transition $t_1$.
This choice disables the lower cycle, which is only re-enabled once the upper cycle is completed and token $a$
is returned to place $u$. As a result, the execution of $t_3$, and thus the lower cycle, following an execution of the upper cycle, is
considered to be causally dependent on the execution of $t_2$. 

The above examples highlight that syntactic token independence between two transitions
or cycles does not preclude their causal dependence. Instead, causal dependence is 
determined by the path that tokens follow: two transition occurrences are causally 
dependent, if a token produced by the one occurrence was subsequently used to
fire the other occurrence.
To capture this type of dependencies, we adopt the following 
definitions.
\begin{definition}\label{transOccurr}{\rm
		Consider a state $\state{M}{H}$ and a transition $t$. We refer to $(t,k)$  as a \emph{transition occurrence} in $\state{M}{H}$ if $k \in H(t)$.
}\end{definition}

\begin{definition}\label{co-dep}{\rm
		Consider  
		a state  $\state{M}{H}$ and suppose $\state{M}{H}\trans{t} \state{M'}{H'}$
		with $(t,k)$, $(t',k')$ transition occurrences in $\state{M'}{H'}$, $k=\max(H(t))$.
		We say that $(t,k)$ \emph{causally depends} on $(t',k')$ denoted by $(t',k')\prec 
		(t,k)$, if $k'<k$ and  there exists $a\in F(x,t)$
		where $\connected(a,M(x))\cap \effects{t'}\neq \emptyset$. 
}\end{definition}

Thus, a transition occurrence $(t,k)$ causally depends on a preceding 
transition occurrence $(t',k')$ if one or more tokens used during the 
firing of $(t,k)$ was produced by $(t',k')$. Note that the tokens employed 
during a transition in a specific marking are determined by the connected 
components of $F(x,t)$ in the marking. 
For example,
in Figure~\ref{cycles1} we have $(t_5,5)\prec (t_7,9)$  and in Figure~\ref{cycles2}  
$(t_1,1)\prec(t_4,4)$, where in each case token $a$ has been transferred from its initial place through $(t_5,5)$ to $(t_7,9)$ and through $(t_1,1)$ to $(t_4,4)$.

\subsubsection{Causal reversing}
Following this approach to causality, we now move on to define causal-order
reversibility in reversing Petri nets. As expected,
we consider a transition $t$ to be enabled for causal-order reversal only if all
transitions that are causally dependent on it have either been reversed or not 
executed. To this respect, relation $\prec$ becomes an important piece of machinery
and we extend the notion of a \emph{state} for the purposes of causal
dependence to a triple $\cstate{M}{H}{\prec}$ where $\prec$ captures
the causal dependencies that have formed up to the creation of the state.
We assume that in the initial state $\prec = \emptyset$ and
we extend the definition of forward execution as follows:

\begin{definition}{\rm \label{cforw}
		Given a reversing Petri net $(A,P,B,T,F)$, a state $\langle M,H,\prec\rangle$, 
		and a transition $t$ forward-enabled in 
		$\state{M}{H}$, we write $\cstate{M}{H}{\prec}
		\trans{t} \cstate{M'}{H'}{\prec'}$
		where $M'$ and $H'$ are defined as in Definition~\ref{forw}, and
		\[
		\prec' \;= \;\prec\cup \{((t',k'),\!(t,k))\mid 
		k \!=\!\max(H'(t)), (t,k) \mbox{ causally depends on } (t',k') \}
		\]
}\end{definition} 
We may now define that a transition is enabled for causal-order reversal as follows:
\begin{definition}\label{co-enabled}{\rm
		Consider a state $\cstate{M}{H}{\prec}$ 
		and a transition $t\in T$. Then $t$, $H(t)\neq \emptyset$, is $c$-enabled (causal-order reversal
		enabled) in  $\langle M, H,\prec\rangle$ if
		\begin{enumerate}
			\item for all  $x\in t\circ$, if $a\in F(t,x)$ then $a\in M(x)$ and if $\beta\in F(t,x)$ then $\beta\in M(x)$,
			and
			\item there is no transition occurrence $(t',k') \in \langle M, H,\prec\rangle$
			with $(t,k)\prec (t',k')$, for $k=max(H(t))$. 
		\end{enumerate}
}\end{definition}

According to the definition, an executed transition is 
$c$-enabled if all tokens and bonds required for its reversal
(i.e., in $\effects{t}$) are available in its outgoing places
and there are no transitions which depend on it causally. 
Note that the second condition becomes relevant in the presence of cycles since it is possible that, while
more than one transitions simultaneously have available the tokens required for their
reversal, only one of them is $c$-enabled. Such an example can be seen in the final
state of  Figure~\ref{cycles2} and transitions $t_2$ and $t_4$.

Reversing a transition in a causally-respecting manner is implemented similarly to
backtracking, i.e. the tokens are moved from the outgoing places to the incoming places of the transition and all bonds
created by the transition are broken. In addition, the history function is updated in the same manner as in backtracking, where we remove the key of the reversed transition. 
Finally, the 
causal dependence relation removes all references to the reversed
transition occurrence.

\begin{figure}[t]
	\centering
	\hspace{.9cm}
	\subfloat{\includegraphics[width=7cm]{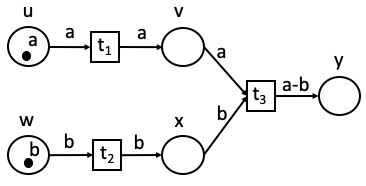}}
	\subfloat{\includegraphics[width=.9cm]{figures/arrow1.png}}
		\subfloat{\includegraphics[width=7cm]{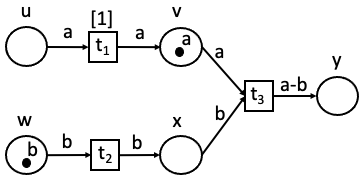}}\\
	\subfloat{\includegraphics[width=.9cm]{figures/arrow2.png}}
		\subfloat{\includegraphics[width=7cm]{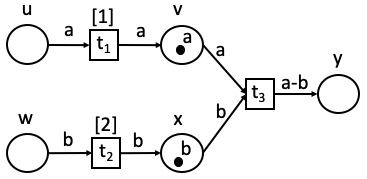}}
	\subfloat{\includegraphics[width=.9cm]{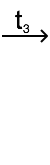}}
	\subfloat{\includegraphics[width=7cm]{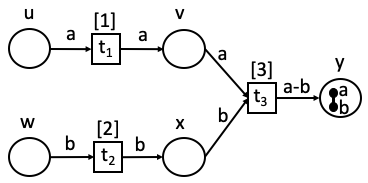}}\\
	\subfloat{\includegraphics[width=.9cm]{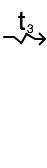}}
	\subfloat{\includegraphics[width=7cm]{figures/causal4.png}}
	\subfloat{\includegraphics[width=.9cm]{figures/arrow1r.png}}
	\subfloat{\includegraphics[width=7cm]{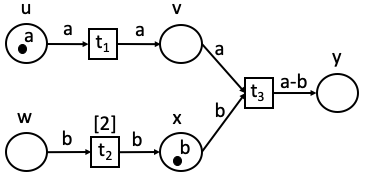}}\\
	\subfloat{\includegraphics[width=.9cm]{figures/arrow2r.png}}
	\subfloat{\includegraphics[width=7cm]{figures/causal6.png}}
	\caption{Causal-order example }\label{co-example}
\end{figure}

\begin{definition}\label{co-def}{\rm
		Given  a state $\langle M, H,\prec\rangle$ and a transition $t$ $c$-enabled in $\langle M, H,\prec\rangle$, we write $\cstate{M}{H}{\prec}
		\ctrans{t} \cstate{M'}{H'}{\prec'}$ for $M'$ and $H'$ as in Definition~\ref{br-def}, and $\prec'$  such that 
		\begin{eqnarray*}
			\prec' &=& \{((t_1,k_1), (t_2,k_2)) \in \prec\; \mid\; k_2\neq k, k=\max(H(t))\}
		\end{eqnarray*}
}\end{definition}

An example of causal-order reversibility can be seen in Figure~\ref{co-example}. Here
we have two independent transitions, $t_1$ and $t_2$ causally preceding transition $t_3$. 
Once the transitions are executed in the order $t_1$, $t_2$, $t_3$, the example 
demonstrates a causally-ordered reversal where $t_3$ is (the only transition that can be) reversed, 
followed by the reversal of its two causes $t_1$ and $t_2$.  In general $t_1$ and $t_2$
can be reversed 
in any order although in the example $t_1$ is reversed before $t_2$. Whenever a transition 
occurrence is reversed its key is eliminated from the history of the transition.

\begin{figure}[t]
	\centering
	\subfloat{\includegraphics[width=7cm]{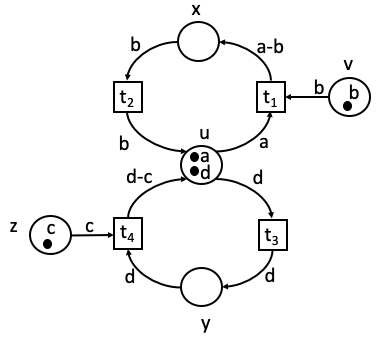}}
	\subfloat{\includegraphics[width=.9cm]{figures/sigmacycle.png}}
	\subfloat{\includegraphics[width=7cm]{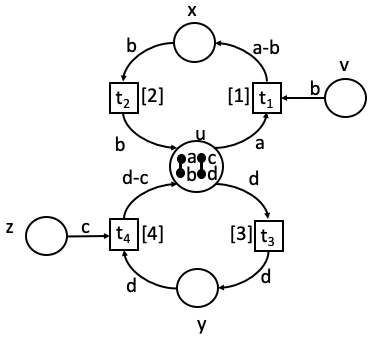}}\\
	
	\subfloat{\includegraphics[width=.9cm]{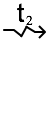}}
	\subfloat{\includegraphics[width=7cm]{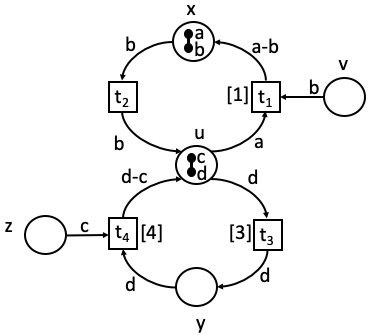}}
	\caption{Causal execution where $\sigma=\langle t_1,t_2,t_3,t_4\rangle$}
	\label{c-cycles}
\end{figure}
As a further example consider the example in Figure~\ref{c-cycles} demonstrating a 
cyclic RPN. Assume that $\sigma=\langle t_1,t_2,t_3,t_4\rangle$, i.e. from the 
initial state of the RPN  the upper cycle is executed followed by the lower 
cycle. The transitions of the two cycles are causally independent since they 
manipulate different sets of tokens and therefore they can be reversed in any 
order. The figure illustrates the reversal of transition $t_2$ before transition $t_4$, which returns
the bond between $a\bond b$ to place $x$.

In what follows we write $\fctrans{}$ for $\trans{}\cup\ctrans{}$.  
The following result, similarly to  Proposition~\ref{Prop}, establishes that under the causal-order reversibility semantics, tokens are unique and preserved, bonds are unique, and they can only be created during forward execution and destroyed during reversal.
Note that in what follows we will often omit the causal dependence
relation and simply write $\state{M}{H}$ for states when it is not relevant to
the discussion.
\begin{proposition}\label{Prop4}{\rm Given a reversing Petri net $(A, P,B,T,F)$, an initial state 
		$\langle M_0, H_0\rangle$ and an execution
		$\state{M_0}{H_0} \fctrans{t_1}\state{M_1}{H_1} \fctrans{t_2}\ldots \fctrans{t_n}\state{M_n}{H_n}$, the following hold:
		\begin{enumerate}
			\item For all $a\in A$  and $i$, $0\leq i \leq n$,   $|\{x\in P \mid a\in M_i (x)\}| = 1$.
			\item For all $\beta \in B$ and $i$,  $0\leq i \leq n$, 
			\begin{enumerate}
				\item $0 \leq |\{x \in P \mid \beta\in M_i(x)\}| \leq 1$,
				\item if $t_i$ is executed in the forward direction and $\beta\in \effect{t_i}$, then 
				$\beta\in M_{i}(x)$ for some $x\in P$ where $\beta\in F(t_i,x)$, and  $\beta\not\in M_{i-1}(y)$ for all $y\in P$,
				\item if $t_i$ is executed in the forward direction, $\beta\in M_{i-1}(x)$ for some $x\in P$, and $\beta\not\in \effect{t_i}$, then, if $\beta \in \connected(a,M_{i-1}(x))$ and $a\in F(t_i,y)$ then $\beta\in M_{i}(y)$, otherwise $\beta\in M_i(x)$
				\item if $t_i$ is executed in the reverse direction and $\beta\in \effect{t_i}$, then 
				$\beta\in M_{i-1}(x)$ for some $x\in P$ where $\beta\in F(t_i,x)$, and  $\beta\not\in M_{i}(y)$ for all $y\in P$, and
				\item if $t_i$ is executed in the reverse direction, $\beta\in M_{i-1}(x)$ for some $x\in P$, and $\beta\not\in \effect{t_i}$, then, if $\beta \in \connected(a,M_{i-1}(x))$ and $a\in F(y,t_i)$ then $\beta\in M_{i}(y)$, otherwise $\beta\in M_i(x)$.
			\end{enumerate}
	\end{enumerate}}
\end{proposition}

\paragraph{Proof} The proof follows along the same lines as that of Proposition~\ref{Prop} with $\btrans{}$ replaced
by $\ctrans{}$.
\proofend

We may now establish the causal consistency of our semantics. 
First we define 
some auxiliary notions. Given a transition $\state{M}{H}\fctrans{t}\state{M'}{H'}$,
we say that the \emph{action} of the transition is
$t$ if $\state{M}{H}\trans{t}\state{M'}{H'}$ and $\underline{t}$ 
if $\state{M}{H}\ctrans{t}\state{M'}{H'}$
and we may write $\state{M}{H}\fctrans{\underline{t}}\state{M'}{H'}$. 
We use $\alpha$ to
range over $\{t,\underline{t} \mid t\in T\}$ and write 
$\underline{\underline{\alpha}} = \alpha$. We extend this
notion to sequences of transitions and, given an execution 
$\state{M_0}{H_0}\fctrans{t_1}\ldots 
\fctrans{t_n}\state{M_n}{H_n}$, we say that the \emph{trace} 
of the execution is
$\sigma=\langle \alpha_1,\alpha_2,\ldots,\alpha_n\rangle$, 
where $\alpha_i$ is the action of transition 
$\state{M_{i-1}}{H_{i-1}}\fctrans{t_i}\state{M_i}{H_i}$,  and write
$\state{M}{H}\fctrans{\sigma}\state{M_n}{H_n}$. Given $\sigma_1 = 
\langle \alpha_1,\ldots,\alpha_k\rangle$, $\sigma_2 = 
\langle \alpha_{k+1},\ldots,\alpha_n\rangle$,
we write $\sigma_1;\sigma_2$ for $\langle \alpha_1,\ldots,\alpha_n\rangle$. 
We may also use the notation $\sigma_1;\sigma_2$ when 
$\sigma_1$ or $\sigma_2$ is a single transition.

An execution of a Petri net can be partitioned
as a set of independent flows of execution
running through the net. We capture these flows by the notion of
causal paths:
\begin{definition}\label{co-path}{\rm
		Given a state $\cstate{M}{H}{\prec}$ and transition occurrences
		$(t_i,k_i)$ in $\cstate{M}{H}{\prec}$, $1\leq i \leq n$, we say that
		$(t_1,k_1),\ldots,(t_n,k_n)$ is a \emph{causal path} in $\cstate{M}{H}{\prec}$,
		if $(t_i,k_i)\prec (t_{i+1}, k_{i+1})$, for all $0\leq i < n$.
}\end{definition}
\begin{figure}[t]
	\centering
	\subfloat{\includegraphics[width=7cm]{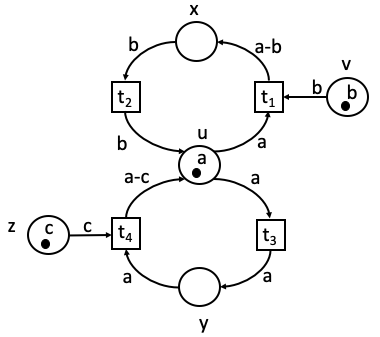}}
	\subfloat{\includegraphics[width=1cm]{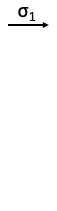}}
	\subfloat{\includegraphics[width=7cm]{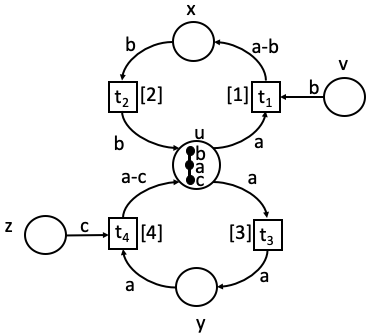}}\\
	\subfloat{\includegraphics[width=7cm]{figures/path2.png}}
	\subfloat{\includegraphics[width=1cm]{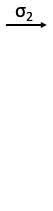}}
	\subfloat{\includegraphics[width=7cm]{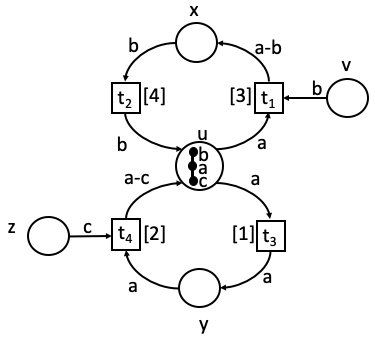}}
	\caption{Causal paths in the context of dependent cycles, where $\sigma_1=\langle t_1,t_2,t_3,t_4\rangle$ and $\sigma_2=\langle t_3,t_4,t_1,t_2\rangle$}
	\label{causalPaths1}
\end{figure}

As an example, consider the reversing Petri net in Figure~\ref{causalPaths1}
where we denote the first execution by $\cstate{M_0}{H_0}{\emptyset} \trans{\sigma_1}\cstate{M_4}{H_4}{\prec}$ 
for  $\sigma_1=\langle t_1,t_2,t_3,t_4\rangle$, 
and the second execution by $\cstate{M_0}{H_0}{\emptyset}
\trans{\sigma_2}\cstate{M_4'}{H_4'}{\prec'}$
for  $\sigma_2=\langle t_3,t_4,t_1,t_2\rangle$. 
In the case of
$\sigma_1$ we have  $\prec$ to be the transitive
closure of $\{((t_1,1), (t_2,2)),$ $  ((t_2,2),(t_3,3)),((t_3,3),(t_4,4))\}$, 
which results in the causal path  
$(t_1,1),(t_2,2),$ $(t_3,3),(t_4,4)$. In the case of
$\sigma_2$ where the cycles are executed in the opposite order,
$\prec'$ is the transitive
closure of $\{((t_3,1),(t_4,2)),  ((t_4,2),(t_1,3)), ((t_1,3),$ $(t_2,4))\}$, 
and the corresponding causal path is $(t_3,1),(t_4,2),(t_1,3),(t_2,4)$.

\begin{figure}[t]
	\centering
	\subfloat{\includegraphics[width=7cm]{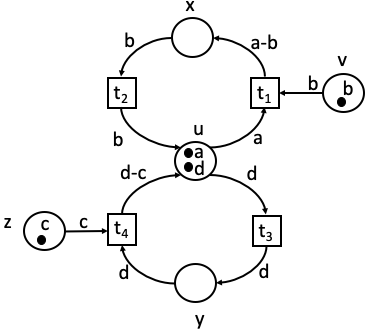}}
	\subfloat{\includegraphics[width=1cm]{figures/sigma1.png}}
	\subfloat{\includegraphics[width=7cm]{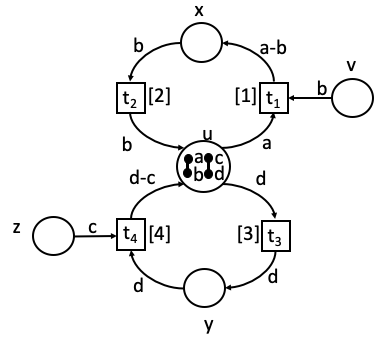}}\\
	\subfloat{\includegraphics[width=7cm]{figures/path5.png}}
	\subfloat{\includegraphics[width=1cm]{figures/sigma2.png}}
	\subfloat{\includegraphics[width=7cm]{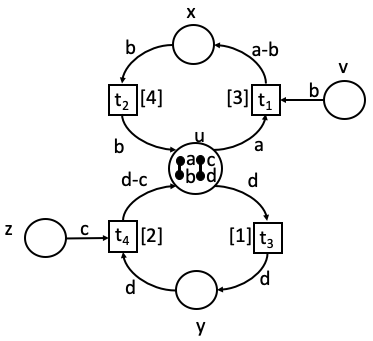}}
	\caption{Causal paths in the context of independent cycles, where $\sigma_1=\langle \pi_1,\pi_2\rangle$ such that $\pi_1=t_1,t_2$,  $\pi_2=t_3,t_4$ and $\sigma_2=\langle \pi_1,\pi_2\rangle$ such that $\pi_1=t_3,t_4$,  $\pi_2=t_1,t_2$}
	\label{causalPaths2}
\end{figure}

This comes in contrast to the RPN of Figure~\ref{causalPaths2}, which 
contains two independent cycles. 
Here, the  causal dependencies of the first execution (trace $\sigma_1$)
are constructed as 
$(t_1,1)\prec (t_2,2)$ and $(t_3,3) \prec (t_4,4)$, which results in the two 
independent causal paths $\langle (t_1,1),(t_2,2)\rangle$ and 
$\langle (t_3,3),(t_4,4)\rangle$.
Similarly, after execution of $\sigma_2$, 
the causal dependencies are $(t_1,3)\prec (t_2,4)$ and 
$(t_3,1) \prec (t_4,2)$, which results in the causal paths 
$\langle (t_1,3),(t_2,4)\rangle$ and $\langle (t_3,1),(t_4,2)\rangle$. 

As seen from the examples in Figures~\ref{causalPaths1} and~\ref{causalPaths2}, 
the causal paths of an execution capture its causal behaviour. 
Based on this concept, we define the notion of causal equivalence for histories by 
requiring that two histories $H$ and $H'$ are causally equivalent if and only
if they contain the same causal paths: 

\begin{definition}\label{eq}{\rm Consider a reversing Petri net $(A,P,B,T,F)$ and two executions
		$\cstate{M}{H}{\prec} \fctrans{\sigma} \cstate{M'}{H'}{\prec'}$ and
		$\cstate{M}{H}{\prec}\fctrans{\sigma'} \cstate{M''}{H''}{\prec''}$. Then the histories
		$H'$ and $H''$ are \emph{causally equivalent}, denoted by $H'\asymp H''$, if
		for each causal path   $(t_1,k_1),\ldots,(t_n,k_n)$ 
		in $\cstate{M'}{H'}{\prec'}$, there is a causal path $(t_1,k_1'),\ldots,
		(t_n,k_n')$ in $\cstate{M''}{H''}{\prec''}$, and vice versa.

		We extend this notion and write $\cstate{M}{H}{\prec}\asymp\cstate{M'}{H'}{\prec'}$ 
		if and only if $M=M'$ and $H\asymp H'$.
}\end{definition}
Returning to the example in Figure~\ref{causalPaths1} we observe that while the two executions result in the same marking,
the resulting states do not have the same causal paths and, as such,
they are not considered as causally equivalent.

We may now establish the Loop lemma. 
\begin{lemma}[Loop]\label{loopc}{\rm 
		For any forward transition $\state{M}{H}\trans{t}\state{M'}{H'}$ there exists a backward transition
		$\state{M'}{H'} \ctrans{t} \state{M}{H}$ and for any backward transition $\state{M}{H} \ctrans{t} \state{M'}{H'}$ there exists a forward transition $\state{M'}{H'}\trans{t}\state{M}{H''}$ where $H\asymp H''$.
}\end{lemma}

\paragraph{Proof} The proof of the first direction follows 
along the same lines as that of Lemma~\ref{loopb} with $\btrans{}$
replaced by $\ctrans{}$. For the other direction suppose $\state{M}{H}\ctrans{t}\state{M'}{H'}\trans{t}\state{M}{H''}$. 
To begin with, we may observe that, as with Lemma~\ref{loopb}, $M=M''$.
To show that $H\asymp H''$, we observe that $H=H''$ with the exception of
$t$, where, if $k=\max(H(t))$,
and $k'=\max(\{0\}\cup\{k''|(t',k'')\in H'(t'), t'\in T\})+1$, then
$H''(t) = (H(t)-\{k\})\cup\{k'\}) $.
Furthermore, since $t$ is $c$-enabled in $\state{M}{H}$, $(t,k)$ must
be the last transition occurrence in all the causal paths it occurs in,
and we may observe that $H''$ contains the same causal paths with 
$(t,k)$ replaced by $(t,k')$.
As a result it must be that $H\asymp H''$ and the result follows.
\proofend

We now proceed to define causal equivalence on traces, a notion that employs the concept of concurrent transitions:
\begin{definition}{\rm
		Actions  $\alpha_1$ and $\alpha_2$ are \emph{concurrent} in state $\cstate{M}{H}{\prec}$, if whenever
		$\cstate{M}{H}{\prec}\fctrans{\alpha_1}\cstate{M_1}{H_1}{\prec_1}$ and $\cstate{M}{H}{\prec}\fctrans{\alpha_2}\cstate{M_2}{H_2}{\prec_2}$
		then $\cstate{M_1}{H_1}{\prec_1}\fctrans{\alpha_2}\cstate{M'}{H'}{\prec'}$ and $\cstate{M_2}{H_2}{\prec_2}\fctrans{\alpha_1}$ $\cstate{M''}{H''}{\prec''}$,
		where $\cstate{M'}{H'}{\prec'} \asymp \cstate{M''}{H''}{\prec''}$.
}\end{definition}

Thus, two actions are concurrent if the execution of the one does not preclude the other and the two execution 
orderings lead to 
causally equivalent states. The condition on final states being
equivalent is required to rule out transitions constituting self-loops
to/from the same place that are causally dependent on each other.
\begin{definition}\label{co-executions}{\rm  \emph{Causal equivalence on traces}, denoted by $\asymp$, is the least
		equivalence relation closed under composition of traces such that 
		(i) if $\alpha_1$ and $\alpha_2$ are concurrent actions then $\alpha_1;\alpha_2\asymp \alpha_2;\alpha_1$ and
		(ii) $\alpha ; \underline{\alpha} \asymp \epsilon$.
}\end{definition}

The first clause states that in two causally-equivalent traces concurrent actions
may occur in any order and the second clause states that it is possible to ignore transitions that have occurred in both
the forward and the reverse direction.

The following proposition establishes that two transition instances belonging
to distinct causal paths are in fact concurrent transitions and thus can be executed in any order. 

\begin{proposition}\label{conc-transitions}{\rm
		Consider a reversing Petri net $(A,P,B,T,F)$ and suppose $\langle M, H,{\prec}\rangle
		\trans{t_1} \cstate{M_1}{H_1}{\prec_1}\trans{t_2} \cstate{M_2}{H_2}{\prec_2}$,
		where the executions of $t_1$ and $t_2$ correspond to transition instances $(t_1,k_1)$ and  $(t_2,k_2)$ in $\cstate{M_2}{H_2}{\prec_2}$. If
		there is no causal path $\pi$ in $\cstate{M_2}{H_2}{\prec_2}$ with $(t_1,k_1)\in\pi$ and $(t_2,k_2)\in\pi$,
		then
		$(t_1,k_1)$ and $(t_2,k_2)$ are concurrent transition occurrences in $\cstate{M}{H}{\prec}$. 
}\end{proposition}

\paragraph{Proof}
Since there is no causal path containing both $(t_1,k_1)$ and $(t_2,k_2)$ in $\cstate{M_2}{H_2}{\prec_2}$,
we conclude that $(t_1,k_1)\not\prec_2 (t_2,k_2)$.
This implies that the two transition occurrences do not handle any common tokens and they can
be executed in any order leading to the same marking. Thus, they are 
concurrent in $\cstate{M}{H}{\prec}$.
\proofend

We note that causally-equivalent states can execute the same transitions.
\begin{proposition}{\rm
		Consider a reversing Petri net $(A,P,B,T,F)$ with causally-equivalent states $\langle M, H_1,{\prec_1}\rangle\asymp\langle M, H_2,{\prec_2}\rangle$. 
		Then $\cstate{M}{H_1}{\prec_1}\fctrans{\alpha}\cstate{M_1}{H_1'}{\prec_1'}$ if and only if  
		$\cstate{M}{H_2}{\prec_2}\fctrans{\alpha}\cstate{M_2}{H_2'}{\prec_2'}$, where $\langle M_1, H_1',{\prec_1'}\rangle\asymp\langle M_2, H_2',{\prec_2'}\rangle$. 
		\label{extendStates}
}\end{proposition}
\paragraph{Proof}
It is easy to see that if a transition $\alpha$ is enabled in $\cstate{M}{H_1}{\prec_1}$
it is also enabled in $\cstate{M}{H_2}{\prec_2}$. Therefore, if  $\cstate{M}{H_1}{\prec_1}\fctrans{\alpha}\cstate{M_1}{H_1'}{\prec_1'}$
then $\cstate{M}{H_2}{\prec_2}\fctrans{\alpha}\cstate{M_2}{H_2'}{\prec_2'}$ where
$M_1=M_2$, and vice versa. In order
to show that $H_1'\asymp H_2'$ two cases exist:
\begin{itemize}
	\item Suppose $\alpha$ is a forward transition corresponding to transition
	occurrence $(t,k_1)$ in $\cstate{M_1}{H_1'}{\prec_1'}$ and transition occurrence $(t,k_2)$ in $\langle M_2, H_2',$ $\prec_2'\rangle$.
	Suppose that $(t',k_1')\prec_1' (t,k_1)$.
	Then, $\effects{t'}\cap \connected(a,M(x))\neq \emptyset$
	for some $a\in F(x,t)$. Since $H_1\asymp H_2$, this implies that $(t',k_2')\prec_2' (t,k_2)$
	where $k_2' = \max(H_2(t'))$. 
	Therefore, for all causal paths $\pi$ in $\cstate{M}{H_1}{\prec_1}$,
	if the last transition occurrence of $\pi$ causes $(t,k_1)$ then
	$\pi;(t,k_1)$ is a causal path of $\cstate{M_1}{H_1'}{\prec_1'}$ and, if not, then $\pi$ is
	a causal path in $\cstate{M_1}{H_1'}{\prec_1'}$. The same holds for causal paths
	in $\cstate{M_2}{H_2'}{\prec_2'}$ and $(t,k_2)$. Consequently, we deduce that $H_1'\asymp H_2'$, as required.
	\item Suppose that $\alpha$ is a reverse transition, i.e. $\alpha = \underline{t}$ 
	for some $t$, and consider the causal paths of
	$H_1'$ and $H_2'$. Since $\alpha$ is a reverse transition, there exists no transition occurrence
	caused by $(t,\max(H_1(t)))$ in $\cstate{M}{H_1}{\prec_1}$ 
	and no transition occurrence caused by $(t,\max(H_2(t)))$ in $\cstate{M}{H_2}{\prec_2}$. 
	As such, $(t,\max(H_1(t)))$ and  $(t,\max(H_2(t)))$ are the last
	transition occurrences in all paths in $\cstate{M}{H_1}{\prec_1}$ and $\cstate{M}{H_2}{\prec_2}$,
	respectively, in which they belong.
	Reversing the transition occurrences results in their elimination from these causal paths. 
	Therefore, we observe that for each causal path in $\cstate{M_1}{H_1'}{\prec_1'}$ there is
	an equivalent causal path in $\cstate{M_2}{H_2'}{\prec_2'}$, and vice versa. 
	Thus $H_1'\asymp H_2'$ as required.
	\proofend
\end{itemize}

Note that the above result can be extended to sequences of transitions:
\begin{corollary}\label{conc-paths}{\rm
		Consider a reversing Petri Net $(A,P,B,T,F)$ with causally-equivalent  states $\langle M, $ $H_1,{\prec_1}\rangle\asymp\langle M, H_2,{\prec_2}\rangle$. 
		Then $\cstate{M}{H_1}{\prec_1}\fctrans{\sigma}\cstate{M_1}{H_1'}{\prec_1'}$ if and only if  
		$\cstate{M}{H_2}{\prec_2}\fctrans{\sigma}\cstate{M_2}{H_2'}{\prec_2'}$, where $\langle M_1, H_1',{\prec_1'}\rangle\asymp\langle M_2, H_2',{\prec_2'}\rangle$.  
}\end{corollary}

\remove{We may also prove that any two simultaneously enabled reverse transitions are in
	fact concurrent.
	\begin{proposition}\label{rev-conc-transitions}{\rm
			Suppose $\langle M, H,{\prec}\rangle \ctrans{t_1} \cstate{M_1}{H_1}{\prec_1}$ 
			and $\langle M, H,{\prec}\rangle \ctrans{t_2} \cstate{M_2}{H_2}{\prec_2}$.
			Then $\underline{t_1}$ and $\underline{t_2}$ are concurrent transitions in 
			$\cstate{M}{H}{\prec}$. 
	}\end{proposition}
	
	\paragraph{Proof}
	Suppose that $\langle M, H,{\prec}\rangle \ctrans{t_1} \cstate{M_1}{H_1}{\prec_1}$ 
	and $\langle M, H,{\prec}\rangle \ctrans{t_2} \cstate{M_2}{H_2}{\prec_2}$.
	...
	\proofend
}

The main result, Theorem~\ref{main} below, states  that  two computations beginning in the same initial state lead to equivalent
states if and only if the sequences of executed transitions of the two computations are causally equivalent. 
This guarantees the consistency of the approach since reversing transitions in causal order is in a sense equivalent
to not executing the transitions in the first place. Reversal does not give rise
to previously unreachable states, on the contrary, it gives rise to exactly the same markings and causally-equivalent
histories due to the different keys being possibly assigned because of the different ordering of transitions. The proof structure along with the intermediate results follow those initially presented in~\cite{RCCS}.

\begin{theorem}\label{main}~\cite{RCCS}{\rm Consider executions $\state{M}{H} \fctrans{\sigma_1} \state{M_1}{H_1}$ and $\state{M}{H}\fctrans{\sigma_2} \state{M_2}{H_2}$. Then, $\sigma_1\asymp\sigma_2$ if and only if   $\state{M_1}{H_1}\asymp\state{M_2}{H_2}$.
	}
\end{theorem}

For the proof of Theorem~\ref{main} we employ some intermediate results. To begin with, the lemma below
states that causal equivalence allows the permutation of reverse and forward transitions that have no causal relations between them. 
Therefore, computations are allowed to reach for the maximum freedom of choice going backward and then continue forward.

\begin{lemma}\label{perm}~\cite{RCCS}{\rm \
		Let $\sigma$ be a trace. Then there exist traces $r,r'$ both forward such that  
		$\sigma\asymp\underline{r};r'$ and if $\state{M}{H} \frtrans{\sigma}\state{M'}{H'}$ then $\state{M}{H} \frtrans{\underline{r};r'}\state{M'}{H''}$, where $H'\asymp H''$.
}\end{lemma}

\paragraph{Proof}
We prove this by induction on the length of  $\sigma$ and the
distance from the beginning of $\sigma$ to the earliest pair of transitions
that contradicts the property $\underline{r};r'$. If there is no such
contradicting pair then the property is trivially satisfied.  If not, we distinguish the following cases:
\begin{enumerate}
	\item If the first contradicting pair is of the form $t;\underline{t}$
	then we have $\state{M}{H}\fctrans{\sigma_1}\state{M_1}{H_1}\fctrans{t}\state{M_2}{H_2}\fctrans{\underline{t}}\state{M_3}{H_3}\fctrans{\sigma_2}\state{M'}{H'}$ where $\sigma=\sigma_1;t;\underline{t};\sigma_2$. By the Loop Lemma $\state{M_1}{H_1}=\state{M_3}{H_3}$, which yields $\state{M}{H} \fctrans{\sigma_1} \state{M_1}{H_1} \fctrans{\sigma_2} \state{M'}{H'}$. Thus we may remove the two transitions from
	the sequence, the length of $\sigma$ decreases, and the proof follows
	by induction.
	\item If the first contradicting pair is of the form $t;\underline{t}'$ then
	we observe that the specific occurrences of $t$ and $\underline{t}'$ must be
	concurrent. Specifically we have $ \state{M}{H} 
	\fctrans{\sigma_1}\state{M_1}{H_1}\fctrans{t}\state{M_2}{H_2}\fctrans{\underline{t'}}
	\state{M_3}{H_3}\fctrans{\sigma_2}\state{M'}{H'}$ where $\sigma=\sigma_1 ; t
	;\underline{t'}; \sigma_2$. Since
	action $t'$ is being reversed, all 
	transition occurrences that are causally dependent on it have either
	not been executed up to this point or they have already been reversed. This implies
	that in $\state{M_2}{H_2}$ it was not the case that $(t,max(H_2(t))$ was causally dependent on $(t',max(H_2(t'))$.
	As such, by Proposition~\ref{conc-transitions}, $\underline{t'}$ and $t$ are concurrent transitions
	and $t'$ can be reversed before
	the execution of $t$
	to yield $ \state{M}{H} \fctrans{\sigma_1}\state{M_1}{H_1}\fctrans{\underline{t'}}\state{M_2'}{H_2'}
	\fctrans{t}\state{M_3}{H_3'}\fctrans{\sigma_2}\state{M'}{H''}$, where $H_3' \asymp H_3$ and $H'\asymp H''$. This results in a
	later earliest contradicting pair and by induction the result follows.
	\proofend
\end{enumerate}

From the above lemma we conclude the following corollary establishing that causal-order
reversibility is consistent with standard forward execution in the sense that causal execution will not generate
states that are unreachable in forward execution:

\begin{corollary}\label{equivalent-executions}~\cite{RCCS}{\rm\ \
		Suppose that  $H_0$ is the initial history. If $\state{M_0}{H_0} \fctrans{\sigma} \state{M}{H}$, and $\sigma$ is a
		trace with both forward and backward transitions then
		there exists a transition $\state{M_0}{H_0}\fctrans{\sigma'}\state{M}{H'}$, where $H\asymp H'$ 
		and $\sigma'$ a trace of forward transitions.
}\end{corollary}
%
\paragraph{Proof} According to
Lemma~\ref{perm}, $\sigma\asymp \underline{r};r'$ where both $r$ and $r'$ are forward
traces. Since, however, $H_0$ is the initial history it must be that $r$ is empty. This
implies that $\state{M_0}{H_0}\fctrans{r'}\state{M}{H'}$, $H\asymp H'$
and $r'$ is a 
forward trace. Consequently, writing $\sigma'$ for $r'$, the result follows.
\proofend

\begin{lemma}\label{short}~\cite{RCCS}{\rm\ \
		Suppose $\state{M}{H}\fctrans{\sigma_1}\state{M'}{H_1}$ and
		$\state{M}{H}\fctrans{\sigma_2}\state{M'}{H_2}$, where $H_1\asymp H_2$ and
		$\sigma_2$ is a forward trace. Then, there exists a forward trace
		$\sigma_1'$ such that $\sigma_1 \asymp \sigma_1'$.
}\end{lemma}

\paragraph{Proof}
If  $\sigma_1$ is forward then $\sigma_1 = \sigma_1'$ and the result follows
trivially. Otherwise, we may prove the lemma by induction on the length of
$\sigma_1$.
We begin by noting that, by Lemma~\ref{perm},
$\sigma_1\asymp\underline{r};r'$ and $\state{M}{H}\fctrans{\underline{r};r'}
\state{M'}{H_1}$. 
Let  $\underline{t}$ be
the last action in $\underline{r}$. 
Given that $\sigma_2$ is a forward execution that 
simulates $\sigma_1$, it must be that $r'$ contains a forward execution 
of transition $t$ so that $\state{M'}{H_1}$ and $\state{M'}{H_2}$ contain the same causal 
paths involving transition $t$ (if not we would have 
$|H_1(t)|<|H_2(t)|$ leading to a contradiction). 
Consider the earliest  occurrence of $t$ in $r'$. If $t$ is the first
transition in $r'$, by the Loop Lemma we may remove the pair of opposite transitions
and the result follows by induction. Otherwise, suppose 
$\state{M}{H}\fctrans{\underline{r_1}}\fctrans{\underline{t}}
\fctrans{{r_1'}} \state{M_1}{H_3} \fctrans{t*}\fctrans{t}\state{M_1'}{H_4}
\fctrans{r_2'}\state{M'}{H_1}$, where $r = r_1;t$ and $r'=r_1';t*;t;r_2$.
Two cases exist:
\begin{enumerate}
	\item Suppose $t*\in\sigma_2$. Let us denote by $num(t,\sigma)$, 
	the number of executions of transition $t$ in a sequence of
	transitions $\sigma$. We observe that since $\sigma_2$ contains no
	reverse executions of $t$, it must be that $num(t,r') = num(t,\sigma_2) + num(t,r)$.
	Suppose that the transition occurrences of $t*$ and $t$ as shown in the execution
	belong to a common causal path. We may extend this path with the succeeding 
	occurrences of $t$ and obtain a causal path such that $t*$ is succeeded by
	$num(t,\sigma_2) + num(t,r)$ occurrences of $t$. We observe that it is impossible to
	obtain such a causal path in $\state{M'}{H_2}$, since $t*$ is followed by fewer occurrences of $t$ in $\sigma_2$.
	This contradicts the assumption that $H_1\asymp H_2$. We conclude that the transition
	occurrences of $t$ and $t*$ above do not belong to any common causal path and
	therefore, by Proposition~\ref{conc-transitions}, the two transition occurrences are
	concurrent in $\state{M_1}{H_3}$.
	\item
	Now suppose that $t*\not\in \sigma_2$. Since $H_1(t*)\neq \emptyset$ it must
	be that $H_2(t*)\neq \emptyset$ and $|H(t*)| = |H_1(t*)|= |H_2(t*)|$. As such, it
	must be that $t*\in r$ and that its reversal has preceded the reversal of $t$.
	Let us suppose that the transition occurrences of $t*$ and $t$ as shown in the execution
	belong to a common causal path. This implies that a causal path
	with $t*$ preceding $t$ also occurs in $H_2$ as well as in $H$. If we
	observe that $t*$ has reversed before $t$ we conclude that  $t*$ does not cause the preceding occurrence of $t$. As such
	there is no causal path within $\state{M}{H}$ or $\state{M'}{H_2}$ containing both $t$ and $t*$, which results in a contradiction. We
	conclude that the forward occurrences of $t$ and $t*$ are,  by Proposition~\ref{conc-transitions}, concurrent in $\state{M_1}{H_3}$.
\end{enumerate}
Given the above, since the occurrences of $t$ and $t*$ are concurrent the two occurrences may be swapped to yield
$\state{M}{H}\fctrans{\underline{r_1}}\fctrans{\underline{t}}
\fctrans{{r_1'}}  
\state{M_1}{H_3} \fctrans{t}\fctrans{t*}\state{M_1'}{H_4'}\fctrans{r_2'}\state{M'}{H_1'}$ where $H_4\asymp H_4'$ and, by Corollary~\ref{conc-paths},  $H_1\asymp H_1'$.
By repeating the process for the remaining transition occurrences in $r_1'$, this implies
that we may permute $t$ with transitions in $r_1'$ to yield the sequence $\underline{t};t$. By the Loop Lemma we may remove the pair of opposite
transitions and obtain a shorter equivalent trace, also
equivalent to $\sigma_2$ and conclude by induction.
\proofend

We now proceed with the proof of Theorem~\ref{main}:

\paragraph{Proof of Theorem~\ref{main}}
Suppose $\state{M}{H} \fctrans{\sigma_1} \state{M_1}{H_1}$, $\state{M}{H} \fctrans{\sigma_2} \state{M_2}{H_2}$
with  $\state{M_1}{H_1}$\\$\asymp\state{M_2}{H_2}$. 
We prove that $\sigma_1\asymp \sigma_2$ by using a lexicographic induction on the pair consisting 
of the sum of the lengths of $\sigma_1$ and $\sigma_2$ and the depth of the earliest disagreement
between them. By Lemma~\ref{perm} we may suppose that $\sigma_1$ and $\sigma_2$
satisfy the property
$\underline{r};r'$. Call $t_1$ and $t_2$ the earliest actions where they disagree. There are three
cases in the argument depending on whether these are forward or backward.
\begin{enumerate}
	
	\item If $t_1$ is backward and $t_2$ is  forward, we have $\sigma_1=\underline{r};\underline{t_1};u$ 
	and $\sigma_2=\underline{r};t_2;v$ for some $r,u,v$. Lemma~\ref{short} applies to $t_2;v$, 
	which is forward, and $\underline{t_1};u$, which contains both forward and backward actions and thus,
	by the lemma, it  has a shorter forward equivalent. Thus, $\sigma_1$ has a shorter forward 
	equivalent and the result follows by induction.
	
	\item If $t_1$ and $t_2$ are both forward  then it must be the
	case that $\sigma_1 = {\underline{r};r'};t_1; u$
	and $\sigma_2 =  {\underline{r};r'}; t_2; v$, for some $r$, $u$, $v$. Note that
	it must be that $t_1\in v$ and $t_2\in u$. 
	If not, we would have $|H_1(t_1)|\neq |H_2(t_1)|$, and similarly for
	$t_2$, which contradicts the assumption that ${H_1}\asymp {H_2}$.
	As such, we may write $\sigma_1 =  {\underline{r};r'};t_1;u_1;t_2;u_2$, 
	where $u=u_1;t_2;u_2$
	and $t_2$ is the first occurrence of $t_2$ in $u$. Consider $t*$ the 
	action immediately preceding $t_2$. We observe that $t*$ and $t_2$ 
	cannot belong to
	a common causal path in $\state{M_1}{H_1}$, since an equivalent causal path is 
	impossible to
	exist in $\state{M_2}{H_2}$. This is due to the assumption that $\sigma_1$ 
	and $\sigma_2$ coincide up to transition sequence  {$\underline{r};r'$}. 
	Thus, we conclude by  Proposition~\ref{conc-transitions} that $t*$ and 
	$t_2$ are in fact concurrent and can be swapped.
	The same reasoning may be used
	for all transitions preceding $t_2$ up to and including
	$t_1$, which leads to the conclusion that
	$\sigma_1\asymp  {\underline{r};r'};t_2;t_1; u_1;u_2$. This results in an 
	equivalent execution of the same length with a later earliest divergence with 
	$\sigma_2$ 
	and the result follows by the induction hypothesis.
	
	\item If $t_1$ and $t_2$ are both backward, we have $\sigma_1=\underline{r};\underline{t_1};u$ 
	and $\sigma_2=\underline{r};\underline{t_2};v$ for some $r,u,v$. Two cases exist:
	\begin{enumerate}
		\item If $\underline{t_1}$ occurs in $v$, then we have that $\sigma_2=\underline{r};\underline{t_2};\underline{v_1};\underline{t_1};v_2$.
		Given that $t_1$ reverses right after $\underline{r}$ in
		$\sigma_1$, we may conclude that there is no transition occurrence
		at this point that causally depends on $t_1$. As such it
		cannot have caused the transition occurrences of $t_2$ and
		${v_1}$ whose reversal precedes it in $\sigma_2$. 
		This implies that the reversal of $t_1$
		may be swapped in $\sigma_2$ with each of the preceding
		transitions, to give
		$\sigma_2\asymp\underline{r};\underline{t_1};\underline{t_2};\underline{v_1};v_2$.
		This results in an equivalent execution of the same length with a later earliest divergence 
		with $\sigma_1$ and the result follows by the induction hypothesis.
		\item If $\underline{t_1}$ does not occur in $v$, this implies that $t_1$ occurs
		in the forward direction in $u$, i.e. $\sigma_1=\underline{r};\underline{t_1};u_1;t_1;u_2$, where $u = u_1;t_1;u_2$ with
		the specific occurrence of $t_1$ being the first such occurrence in $u$. 
		Using similar arguments as those in Lemma~\ref{short},
		we conclude that $\sigma_1\asymp\underline{r};\underline{t_1};t_1;u_1;u_2
		\asymp \underline{r};u_1;u_2$, an equivalent execution of shorter length for $\sigma_1$ and the result follows by the induction hypothesis.
	\end{enumerate}

	We may now prove the opposite direction. Suppose that $\sigma_1 \asymp \sigma_2$
	and $\state{M}{H}\fctrans{\sigma_1}\state{M_1}{H_1}$ and $\state{M}{H}\fctrans{\sigma_2}\state{M_2}{H_2}$. 
	We will show that $\state{M_1}{H_1}\asymp \state{M_2}{H_2}$.
	The proof is by induction on the number of rules, $k$, applied to establish 
	the equivalence $\sigma_1 \asymp \sigma_2$.
	For the base case we have $k=0$, which implies that $\sigma_1=\sigma_2$
	and the result trivially follows. For the inductive step, let us assume that
	$\sigma_1\asymp \sigma_1'\asymp \sigma_2$, where $\sigma_1$ can be transformed
	to $\sigma_1'$ with the use of $k=n-1$ rules and $\sigma_1'$ can be transformed
	to $\sigma_2$ with the use of a single rule. By the induction hypothesis,
	we conclude that $\state{M}{H}\fctrans{\sigma_1'}\state{M_1}{H_1'}$, where 
	$H_1\asymp H_1'$. We need to show that $\state{M_1}{H_1'}\asymp 
	\state{M_2}{H_2}$. Let us write
	$\sigma_1' = u;w;v$ and $\sigma_2 = u;w';v$, where $w$, $w'$ refer
	to the parts of the two executions where the equivalence rule has been applied.
	Furthermore, suppose that
	$\state{M}{H}\fctrans{u}\state{M_u}{H_u}\fctrans{w}\state{M_w}{H_w}\fctrans{v}\state{M_1}{H_1'}$ and
	$\state{M}{H}\fctrans{u}\state{M_u}{H_u}\fctrans{w'}\state{M_w'}{H_w'}\fctrans{v}\state{M_2}{H_2}$.
	Three cases exist:
	\begin{enumerate}
		\item $w= t_1;t_2$ and $w'=t_2;t_1$ with $t_1$ and $t_2$ concurrent
		\item $w=t;\underline{t}$ and $w'=\epsilon$
		\item $w=\underline{t};t$ and $w'=\epsilon$
	\end{enumerate}
	In all the cases above, we have that $\state{M_w}{H_w}\asymp \state{M_w'}{H_w'}$:
	for (a) this follows by the definition of concurrent transitions, whereas
	for (b) and (c) by the Loop Lemma. Given the equivalence of these two
	states, by Corollary~\ref{equivalent-executions}, we have that $\state{M_w}{H_w}\fctrans{v}\state{M_1}{H_1'}$  and
	$\state{M_w'}{H_w'}\fctrans{v}\state{M_2}{H_2}$, where $\state{M_1}{H_1'}
	\asymp \state{M_2}{H_2}$, as required. This completes the proof.
	\proofend
\end{enumerate}

We note that the causal-consistency theorem has been proved using the standard approach of~\cite{RCCS}. An alternative approach, stemming from he recent work of~\cite{Axiomatic} could also be possible, whereby the study of various properties within a general framework for reversible systems is established. More precisely, causal consistency can be guaranteed by proving a set of axioms relating to the parabolic Lemma and the Square property.

\subsection{Out-of-Causal-Order Reversibility}

While in backtracking and causal-order reversibility reversing is cause respecting, there are many examples of systems where undoing
actions in an out-of-causal order is either inherent or
desirable. In this section we consider this type of 
reversibility in the context of RPNs. We begin by specifying that in out-of-causal-order reversibility any executed transition can be reversed at any time.
\begin{definition}\label{o-enabled}{\rm
		Consider a reversing Petri net $(A,P,B,T,F)$, a state $\state{M}{H}$, and a transition $t\in T$. We say that $t$ is \emph{$o$-enabled} in 
		$\state{M}{H}$, if $H(t)\neq \es$.
}\end{definition}

\begin{figure}[t]
	\centering
	\hspace{.6cm}
	\subfloat{\includegraphics[width=9cm]{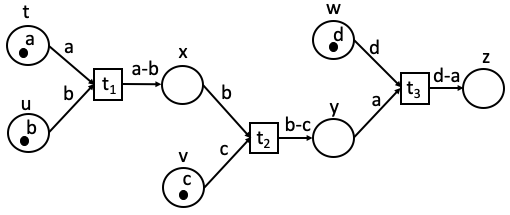}}\\
	\subfloat{\includegraphics[width=.8cm]{figures/arrow1.png}}
	\subfloat{\includegraphics[width=9cm]{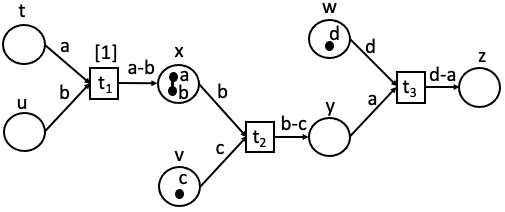}} \\
	\subfloat{\includegraphics[width=.8cm]{figures/arrow2.png}}
	\subfloat{\includegraphics[width=9cm]{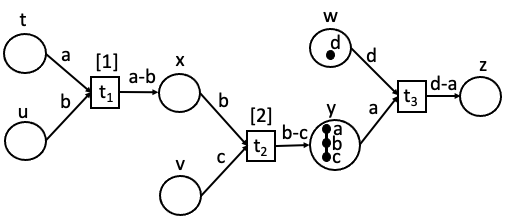}}\\ 
	\subfloat{\includegraphics[width=.8cm]{figures/arrow3.png}}
	\subfloat{\includegraphics[width=9cm]{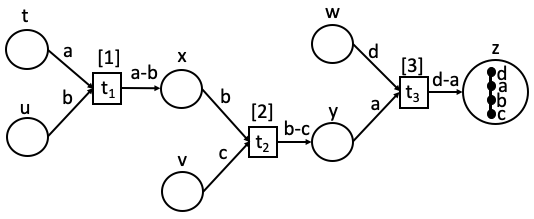}} 
	\caption{Forward execution of out-of-causal-order example}\label{fo-example}
\end{figure}

\begin{figure}[t]
	\centering
	\hspace{.6cm}
	\subfloat{\includegraphics[width=9cm]{figures/outoforder3.png}}\\ 
	\subfloat{\includegraphics[width=.8cm]{figures/arrow1r.png}}
	\subfloat{\includegraphics[width=9cm]{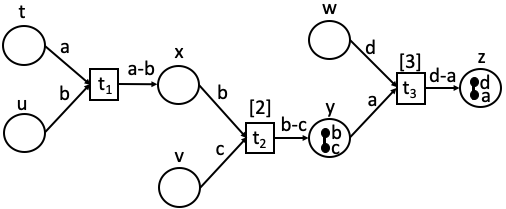}}\\
	\subfloat{\includegraphics[width=.8cm]{figures/arrow2r.png}}
	\subfloat{\includegraphics[width=9cm]{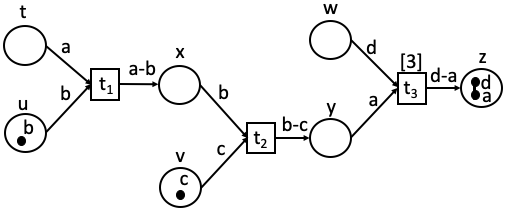}} \\
	\subfloat{\includegraphics[width=.8cm]{figures/arrow3r.png}}
	\subfloat{\includegraphics[width=9cm]{figures/outoforder6.png}}
	\caption{Out-of-causal-order example}\label{o-example}
\end{figure}

Let us begin to consider out-of-causal-order reversibility
via the example
of Figures~\ref{fo-example} and ~\ref{o-example}. The first Figure~\ref{fo-example} presents the forward execution of the transition sequence $\langle t_1,t_2,t_3\rangle$. 
The second Figure~\ref{o-example} represents the out-of-causal-order reversal of transition sequence $\langle t_1,t_2,t_3\rangle$. Suppose that transition $t_1$ is to be reversed out of order. 
The effect of
this reversal should be the destruction of the bond between $a$ and $b$. This means that the component
$d\bond a\bond b\bond c$ is broken into the bonds $d\bond a$ and $b\bond c$, which should
backtrack within the net to capture the reversal of the transition. Nonetheless, the tokens
of $d\bond a$ must remain at place $z$. This is because a bond exists between them that has not been reversed
and was the effect of the immediately preceding transition $t_3$.
However, in the case of $b\bond c$, the bond can be returned to place $y$, which is the place
where the two tokens were connected and from where they could continue to participate in any further
computation requiring their coalition. Once transition $t_2$ is subsequently reversed,
the bond between $b$ and $c$ is destroyed and thus the two tokens are able to return to their
initial places as shown in the third net in the figure. Finally, when subsequently transition $t_3$ is reversed, the bond
between $d$ and $a$ breaks and, given that neither $d$ nor $a$ are connected to other elements,
the tokens return to their initial places. 
As with the other types of reversibility, when reversing a transition histories are updated by removing the greatest key identifier of the executed transition. 

Summing up, the effect of reversing a transition in out-of-causal order is that all bonds created by the transition
are undone. This may result in tokens backtracking in the net. Further,
if the reversal of a transition causes a coalition of bonds 
to be broken down into
a set of subcomponents due to the destruction of bonds, then each of these coalitions should
flow back, as far back as possible, after the last transition in which this sub-coalition participated. 
To capture this notion of ``as far backwards as possible''
we introduce the following: 

\begin{definition}\label{last}{\rm
		{	Given a reversing Petri net $(A,P,B,T,F)$, an initial marking $M_0$,  
			a history $H$, and 
			a set of bases and bonds $C\subseteq A\cup B$ we write:
			\[
			\begin{array}{rcl}
			\lastt{C,H} &=& \left\{
			\begin{array}{ll}
			t , \hspace{2cm}\;\;\textrm{ if }\exists t, \; \effects{t}\cap C\neq \emptyset, \; H(t)\neq \emptyset, \mbox{ and }\\
			\hspace{3.5cm} \not\exists t', \;\effects{t'} \cap C\neq \emptyset, \; H(t') \neq \emptyset, 	\\ 	
			\hspace{5cm} max(H(t'))\geq max (H(t)) \\
			\bot,  \hspace{6.8cm}\;\textrm{ otherwise }
			\end{array}
			\right.
			\end{array}
			\]
			\[
			\begin{array}{rcl}
			\lastp{C,H} &=& \left\{
			\begin{array}{ll}
			x , \;\;\;\textrm{ if }t=\lastt{C,H}, \{x\} = \{y\in t\circ 
			\mid F(t,y)\cap C \neq \emptyset\}\\
			\hspace{0.25in} \;\textrm{or, if } \bot=\lastt{C,H}, C\subseteq M_0(x)\\
			\bot,  \;\;\textrm{ otherwise }
			\end{array}
			\right.
			\end{array}
			\]}
}\end{definition}
Thus, if component $C$ has been manipulated by some previously-executed
transition,  then $\lastt{C,H}$ is the last executed such transition.
Otherwise, if no such transition exists (e.g., because all transitions
involving $C$ have been reversed), then $\lastt{C,H}$ is undefined 
($\bot$). Similarly, $\lastp{C,H}$ is the outgoing place connected to
$\lastt{C,H}\neq \bot$ having common tokens with $C$, assuming that such
a place is unique, or the place in the initial marking in which $C$ 
existed if $\lastt{C,H}= \bot$, and undefined otherwise.

Transition reversal in an out-of-causal order can thus be defined as follows: 
\begin{definition}\label{oco-def}{\rm
		Given a reversing Petri net $(A, P,B,T,F)$, an initial marking $M_0$, a state $\langle M, H\rangle$ and a transition $t$ that is $o$-enabled in $\state{M}{H}$, we write 
		$\state{M}{H}
		\otrans{t} \state{M'}{H'}$
		where $H'$ is defined as in Definition~\ref{br-def} and  we have:
		\begin{eqnarray*}
			M'(x) & = & \Big(M(x)\cup \bigcup_{a\in M(y){\cap \effects{t}}, \lastp{C_{a,y},H'} =x}C_{a,y}\Big ) \\ 	 
			&& -\Big(\effect{t}  \cup \bigcup_{a\in M(x){\cap \effects{t}}, \lastp{C_{a,x},H'}\neq x}C_{a,x}\Big) 
		\end{eqnarray*}
		where we use the shorthand $C_{b,z} = \connected(b,M(z)-\effect{t})$ for $b\in A$, $z\in P$. 
}\end{definition}

Thus, when a transition $t$ is reversed in an out-of-causal-order 
fashion all bonds that were created by the transition in $\effect{t}$ 
are undone. Furthermore, tokens and bonds involved in the transition are
relocated back to the place where they would have existed if 
transition $t$ never took place, as defined by $\lastp{C,H'}$. Note
that if the destruction of a bond divides a component 
into smaller connected sub-components then each of these
sub-components is relocated separately. 
Specifically, the definition states that: if a token $a$ and 
its connected components involved in transition $t$,
last participated in some transition 
with outgoing place $y$ other than $x$,  then the sub-component
is removed from place $x$ and returned to place $y$, otherwise
it is returned to the place where it occurred in the initial marking. 

\begin{figure}[t]
	\centering
	\subfloat{\includegraphics[width=7cm]{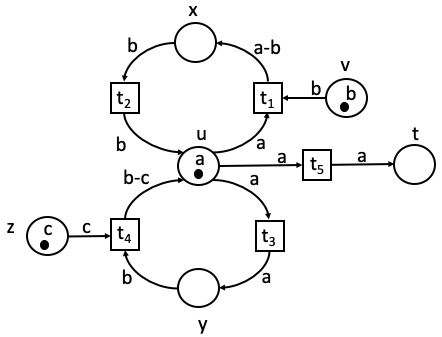}}
	\subfloat{\includegraphics[width=.8cm]{figures/sigmacycle.png}}
	\subfloat{\includegraphics[width=7cm]{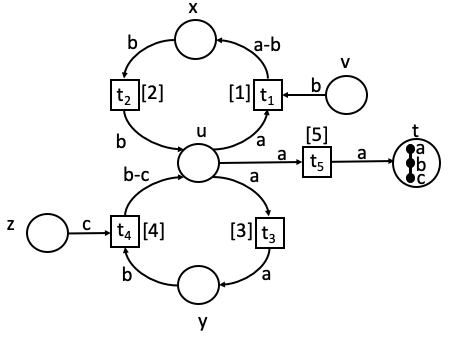}}\\
	\subfloat{\includegraphics[width=.8cm]{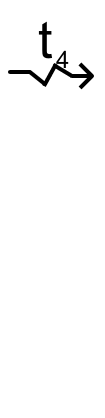}}
	\subfloat{\includegraphics[width=7cm]{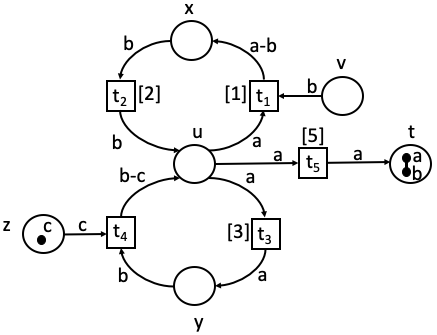}}
	\caption{Out-of-causal-order reversing where $\sigma=\langle t_1,t_2,t_3,t_4,t_5\rangle$}
	\label{o-cycles}
\end{figure}

An example of out-of-causal-order reversibility in a cyclic RPN can be seen in Figure~\ref{o-cycles}. Here the cycles $\langle t_1,t_2\rangle$ and  $\langle t_3,t_4\rangle$ are executed in this order followed by transition $t_5$. We reverse in out-of-causal order transition $t_4$, which breaks the bond between $b \bond c$ and returns token $c$ back to its original place $z$. Moreover, the bond between $a\bond b$ remains in place $t$, which is the outgoing place of the last transition of token $a$. Note that this state did not occur during the forward execution of the RPN.

The following results describe how tokens and bonds are manipulated
during out-of-causal-order reversibility, where we write $\fotrans{}$ for $\trans{}\cup\otrans{}$.
\begin{proposition}\label{markings}{\rm
		Suppose $\state{M}{H} \fotrans{t}\state{M'}{H'}$ and let $a\in A$ where
		$a\in M(x)$ and $a\in M'(y)$.
		Then, 
		$\connected(a,M'(y))=\connected(a,M(x)\cup C)$, where $C=\effect{t} \cup\{ \connected(b,M(u))\mid a-b\in \effect{t}, b\in M(u)\})$,
		if $t$ is a forward transition, and $\connected(a,M'(y))=\connected(a,M(x)-\effect{t})$, if $t$ is a
		reverse transition.
	}
\end{proposition}
\paragraph{Proof}
The proof is straightforward by the definition of the firing rules.
\proofend

\begin{proposition}\label{prop5}
	\rm Given a reversing Petri net $(A, P,B,T,F)$, an initial state 
	$\langle M_0, H_0\rangle$, and an execution
	$\state{M_0}{H_0} \fotrans{t_1}\state{M_1}{H_1} \fotrans{t_2}\ldots 
	\fotrans{t_n}\state{M_n}{H_n}$ the following hold for all $0\leq i \leq n$:
	\begin{enumerate}
		\item For all $a\in A$, $|\{x\in P \mid a\in M_i (x)\}| = 1$, 
		and  $a\in M_i(x)$ where 
		$x=\lastp{\connected(a,M_{i}(x)),H_i}$. 
		\item For all $\beta \in B$,  
		\begin {enumerate}
		\item $0 \leq |\{x \in P \mid \beta\in M_i(x)\}| \leq 1$.
		\item if $|\{x \in P \mid \beta\in M_{i-1}(x)\}|=0$ and $|\{x \in P \mid \beta\in M_i(x)\}|=1$,
		then $t_i$ is a forward transition and  $\beta\in \effect{t_i}$,
		\item if $|\{x \in P \mid \beta\in M_{i-1}(x)\}|=1$ and $|\{x \in P \mid \beta\in M_i(x)\}|=0$,
		then $t_i$ is a reverse transition and  $\beta\in \effect{t_i}$,
		\item if $|\{x \in P \mid \beta\in M_{i-1}(x)\}|=|\{x \in P \mid \beta\in M_i(x)\}|$,
		then $\beta\not\in \effect{t_i}$.
	\end{enumerate}
\end{enumerate}
\end{proposition}
\paragraph{Proof}
Consider a reversing Petri net $(A, P,B,T,F)$, an initial state 
$\langle M_0, H_0\rangle$, and an execution
$\state{M_0}{H_0} \fotrans{t_1}\state{M_1}{H_1} \fotrans{t_2}\ldots \fotrans{t_n}\state{M_n}{H_n}$. The 
proof is given by induction on~$n$.
\paragraph{Base Case} For  $n=0$, by our assumption of token 
uniqueness and the definitions of $\mathsf{last}_P$ and
$\mathsf{last}_T$ the claim follows trivially.
\paragraph{Induction Step} Suppose the claim holds for all but the last transition
and consider
transition $t_n$. Two cases exist, depending on whether $t_n$ is a forward or a reverse transition:
\begin{itemize}
\item Suppose that $t_n$ is a forward transition. Then by Proposition \ref{prop1},  for all $a\in A$,
$|\{x\in P \mid a\in M_n (x)\}| = 1$. 
Additionally, we may see that if $a\in M_n(x)$ two cases exists.
If $a\in \connected(b,M_{n-1}(y))$, for some $b\in F(t_n,z)$
then $x=z=\lastp{\connected(a,M_n(x)), H_n}$. 
Otherwise, it must be that $a\in M_{n-1}(x)$
where, by the induction hypothesis, $x = \lastp{\connected(a,M_{n-1}(x)), H_{n-1}}$. Since $a\not \in \effect{t_n}$, by clause 2(b)  we may deduce that    $\connected(a,M_{n-1}(x))=\connected(a,M_{n}(x))$, which 
leads to 
$x = \lastp{(\connected(a,M_{n-1}(x)), H_{n-1}}=\lastp{\connected(a,M_n(x)), H_n}$. Thus, the result follows.

Now let $\beta \in B$.
To begin with, clause (2)(a) follows by Proposition \ref{prop1}. 
Furthermore, we may see that the forward transition $t_n$ may only create 
exactly the bonds in $\effect{t_n}$ and it maintains all remaining bonds. Thus,
clauses 2(b) and 2(d) follow. 
\item {Suppose that $t_n$ is a reverse transition. 
	Consider $a\in A$ with $a\in M_{n-1}(x)$ for some $x\in P$. Two cases exist:
	\begin{itemize}
		\item Suppose $\lastt{\connected(a,M_{n-1}(x)-\effect{t_n}),H_n} = \bot$. It  must be that $\connected(a,M_{n-1}(x)-\effect{t_n})
		\subseteq M_0(y)$ for some $y$ such that $a\in M_0(y)$. Suppose
		that this is not the case. Then there must exist some $\beta\in\connected(a,M_{n-1}(x)-\effect{t_n})$ with $\beta\not\in M_0(y)$. By the induction hypothesis, there exists some $t_i$
		in the execution such that $\beta\in \effect{t_i}$ which was not
		reversed, i.e. $H_n(t_i)\neq \emptyset$. This however implies
		that $t_i$ is a transition that has manipulated the connected
		component $\connected(a,M_{n-1}(x)-\effect{t_n})$, which contradicts
		our assumption of $\lastt{\connected(a,M_{n-1}(x)-\effect{t_n}),H_n} = \bot$. 
		Therefore, $a\in M_n(y)$, where $a\in M_0(y)$ and by Proposition~\ref{markings} $\connected(a,M_{n-1}(x)-\effect{t_n})=\connected(a,M_{n}(y))$ which gives $y = \lastp{\connected(a,M_n(y)), H_n}$
		and the result follows.
		\item Suppose $\lastt{\connected(a,M_{n-1}(x)-\effect{t_n}),H_n} = t_k$. 
		Then, it  must be that there exists a unique $y\in t_k
		\circ$ 
		such that $\connected(a,M_{n-1}(x)-\effect{t_n})
		\cap F(t_k,z)\neq \emptyset$. Suppose
		that this is not the case. Then there must exist some $\beta=(a,b)\in \connected(a,M_{n-1}(x)-\effect{t_n})$ with $a\in F(t_k,y_1)$, $b\in F(t_k,y_2)$, and $y_1\neq y_2$. Since $\beta\in M_n(y)$,
		by the induction hypothesis, there exists some $t_i$
		in the execution such that $\beta\in \effect{t_i}$, $i>k$ which was not
		reversed, i.e. $H_n(t_i)\neq \emptyset$. This however implies
		that $t_i$ is a transition that has manipulated the connected
		component $\connected(a,M_{n-1}(x)-\effect{t_n})$ later than 
		$t_k$, which contradicts
		our assumption of $\lastt{\connected(a,M_{n-1}(x)-\effect{t_n}),H_n} = t_k$. Therefore, there exists a unique $y\in t_k
		\circ$ 
		such that $\connected(a,M_{n-1}(x)-\effect{t_n})
		\cap F(t_k,z)\neq \emptyset$, $a\in M_n(y)$. Furthermore, by Proposition~\ref{markings} $\connected(a,M_{n-1}(x)-\effect{t_n})=\connected(a,M_{n}(y))$ which gives
		$y = \lastp{\connected(a,M_n(y)), H_n}$
		and the result follows.
	\end{itemize}
	
	Now consider $\beta \in B$. By clause 1, we may deduce clause 2(a). 
	Finally, we may observe that the reverse transition $t_n$ may only remove 
	exactly the bonds in $\effect{t_n}$ and it maintains all remaining bonds, thus,
	clauses 2(b)-2(d) follow. 
}
\proofend
\end{itemize}
As we have already discussed (e.g., see Figures~\ref{catalyst2} and~\ref{o-cycles}), unlike causal-order 
reversibility, out-of-causal-order reversibility may give rise
to states that cannot be reached by forward-only execution.
Nonetheless, note that the proposition establishes
that during out-of-causal-order reversing
it is not the case that tokens and bonds may reach places they
have not previously occurred in. On the contrary,
%
a component will always return to the place following the last
transition that has manipulated it. This observation also gives rise to the following corollary, which 
characterises the marking of a state during computation.
\begin{corollary}\label{oco-tokenplace}{\rm Given a reversing Petri net $(A, P,B,T,F)$, an initial state 
	$\langle M_0, H_0\rangle$, and an execution
	$\state{M_0}{H_0} \fotrans{t_1}\state{M_1}{H_1} \fotrans{t_2}\ldots 
	\fotrans{t_n}\state{M_n}{H_n}$, then for all $x\in P$ we have 
	\[ M_n(x) =\bigcup_{a\in M_n(y),\lastp{C_{a,y},H_n} =x} C_{a,y}
	\]
	where $C_{a,y} = \connected(a,M_n(y))$}.
\end{corollary}
\paragraph{Proof}
According to Proposition~\ref{prop5} clauses (1) and 2(a) the result follows.
\proofend

The dependence of the position of a connected component and a transition
sequence can be exemplified by the following proposition.
\begin{proposition}\label{second}{\rm Consider executions $\state{M_0}{H_0} 
	\fotrans{\sigma_1} \state{M_1}{H_1}$, $\state{M_0}{H_0} 
	\fotrans{\sigma_2} \state{M_2}{H_2}$,
	and a 
	token $a$ such that $a\in M_1(x)$, $a\in M_2(y)$, for some $x$, $y\in P$, and $\connected(a, M_1(x)) = \connected(a,M_2(x))$. Then,  
	$\lastt{\connected(a,M_1(x)),H_1}=\lastt{\connected(a,M_2(y)),H_2}$ implies
	$x=y$.
}\end{proposition}

\paragraph{Proof} 
Consider executions $\state{M_0}{H_0} \fotrans{\sigma_1} 
\state{M_1}{H_1}$, $\state{M_0}{H_0} \fotrans{\sigma_2} 
\state{M_2}{H_2}$  and a token $a$ such that $a\in M_1(x)$,
$a\in M_2(x)$.
Further, let us assume that $\lastt{\connected(a,M_1(x)),H_1} = \lastt{\connected(a,M_2(y)),H_2}$. Two cases exist:
\begin{itemize}
\item $\lastt{\connected(a,M_1(x)),H_1} = \lastt{\connected(a,M_2(y)),H_2}=\bot$.
This implies that no transition has manipulated any of the tokens
and bonds of the two connected components. As such, by Proposition~\ref{prop5}, $\connected(a,M_1(x))\subseteq M_0(x)$ and $\connected(a,M_2(y))\subseteq M_0(y)$,
and by the uniqueness of tokens we conclude that $x=y$ as required.
\item   $\lastt{\connected(a,M_1(x)),H_1} = \lastt{\connected(a,M_2(y)),H_2} = t$. 
This implies that there is $b\in \connected(a,M_1(x))=\connected(a,M_2(y))$ such that $b\in F(t,z)$ for some place $z$.
By definition, we deduce that $ \lastp{\connected(a,M_1(x)),H_1}$ $=z=\lastp{\connected(a,M_2(y)),H_2}$, thus, $x=y$ as required.
\proofend
\end{itemize}

From the above result we may prove the following proposition establishing that 
executing two causally equivalent sequences of transitions
in the out-of-causal-order setting will give rise to causally equivalent states.

\begin{proposition}\label{corTheorem1}{\rm\ \  
	Suppose 
	$\state{M_0}{H_0} \fotrans{\sigma_1} \state{M_1}{H_1}$ and $\state{M_0}{H_0} \fotrans{\sigma_2} \state{M_2}{H_2}$. 
	If $\sigma_1\asymp\sigma_2$ then  $\state{M_1}{H_1}\asymp\state{M_2}{H_2}$.
}
\end{proposition} 
\paragraph{Proof} Suppose $\state{M_0}{H_0} \fotrans{\sigma_1} \state{M_1}{H_1}$,
$\state{M_0}{H_0} \fotrans{\sigma_2} \state{M_2}{H_2}$ and $\sigma_1\asymp\sigma_2$. 
Since $\sigma_1\asymp \sigma_2$ it must be that the two executions
contain the same causal paths, therefore,
$H_1\asymp H_2$. To show that $M_1 = M_2$ consider
token $a$ such that $a\in M_1(x)\cap M_2(y)$. 
Since $\sigma_1\asymp \sigma_2$, we may conclude that the two executions contain the same set of executed and not reversed transitions. Thus, by Proposition~\ref{prop5}(2), 
we have $\connected(a, M_1(x)) = \connected(a, M_2(y))$. Furthermore,
it must be that $t_1=\lastt{\connected(a, M_1(x)), H_1}
=\lastt{\connected(a, M_2(y)), H_2}= t_2$. 
If not, since $\sigma_1 \asymp \sigma_2$, we would have that $t_1$
and $t_2$ are concurrent, which is not possible since they
manipulate the same connected component and thus a
causal relation exists between them. Therefore, by Proposition~\ref{second},
$x=y$. This implies by Corollary~\ref{oco-tokenplace}
that $M_1(x) = M_2(x)$, for all places $x$, which completes the proof.
\proofend

We finally establish a Loop Lemma for out-of-causal reversibility.
\begin{lemma}[Loop]\label{loopo}{\rm 
	For any forward transition $\state{M}{H}\trans{t}\state{M'}{H'}$ there exists a reverse
	transition $\state{M'}{H'} \otrans{t} \state{M}{H}$. 
}\end{lemma}
\paragraph{Proof}
Suppose $\state{M}{H}\trans{t}\state{M'}{H'}$. Then $t$ is clearly $o$-enabled 
in $H'$. Furthermore, $\state{M'}{H'} \otrans{t} \state{M''}{H''}$ where $H''=H$
by the definition of $\otrans{}$. 
In addition, for all $a\in A$, we may prove that $a\in M''(x)$ if and only if $a\in M(x)$. Suppose $a\in M(y)$, we distinguish two cases. 
If $\connected(a,M(y))\cap \guard{t}= \es$, then we may see that $a\in M'(y)$ and 
$a\in M''(y)$, and the result follows. Otherwise, if $\connected(a,M(y))\cap \guard{t}\neq \es$, then 
$a\in M'(z)$, where $F(t,z)\cap \connected(a,M(y))\neq \es$.
Furthermore, suppose that $a\in M''(w)$. By
Proposition~\ref{markings} we have that $\connected(a,M'(z))=\connected(a,M(y)\cup C)$, $C=\effect{t}\cup\{ \connected(b,M(u))\mid a-b\in \effect{t}, b\in M(u)\}$, and 
$\connected(a,M''(w))= \connected(a,M'(z)- \effect{t})
=\connected(a,(M(y)\cup C)-\effect{t}) 
=\connected(a,M(y))$. 
Furthermore, $y = \lastp{\connected(a,M(y)), H}$, by Corollary~\ref{oco-tokenplace}. Since $H= H''$, we
have $w=\lastp{\connected(a,M''(w)), H'')}= \lastp{\connected(a,M(y)), H)} = y$, and the result follows.
\proofend

Note that in the case of out-of-causal-order reversibility, the opposite direction of the 
lemma does not hold. This is because reversing a transition in an out-of-causal-order
fashion may bring a system to a state not reachable by forward-only transitions, and
where the transition is not enabled in the forward direction. As an example, 
consider the RPN of Figure~\ref{o-example} and after the reversal of transition 
$t_2$. In this state, transition $t_2$ is not forward enabled since token $b$ is
not available in place $x$, as required for the transition to fire.

\subsection{Relationship Between Reversibility Notions}
We continue to study the relationship between the three forms of reversibility. Our first result confirms
the relationship between the enabledness conditions for each of backtracking, causal-order, and out-of-causal-order reversibility.
\begin{proposition}\label{enable}{\rm\ \
	Consider  a state $\langle M, H\rangle$, and a transition $t$. Then, if $t$ is $bt$-enabled in  $\langle M, H\rangle$
	it is also $c$-enabled. Furthermore, if  $t$ is $c$-enabled in  $\langle M, H\rangle$
	then it is also $o$-enabled.
}
\end{proposition}
\paragraph{Proof}
The proof is immediate by the respective definitions.
\proofend

We next demonstrate a ``universality'' result of the $\otrans{}$ transition relation by showing that it manipulates the
state of a reversing Petri net in an identical way to $\ctrans{}$, in the case of $c$-enabled transitions,
and to $\btrans{}$, in the case of $bt$-enabled transitions.
Central to the proof is the following result establishing
that during causal-order reversibility a component is
returned to the place following the last transition
that has manipulated it or, if no such transition exists,
in the place where it occurred in the initial marking. 
\begin{proposition}\label{last-as-co}
\rm Given a reversing Petri net $(A, P,B,T,F)$, an initial state 
$\langle M_0, H_0\rangle$, and an execution
$\state{M_0}{H_0} \fctrans{t_1}\state{M_1}{H_1} \fctrans{t_2}\ldots 
\fctrans{t_n}\state{M_n}{H_n}$. Then for all $a\in A$, 
$a\in M_n(x)$ where $x=\lastp{\connected(a,M_{n}(x)),H_n}$.
\end{proposition}
\paragraph{Proof}
The 
proof is by induction on $n$ and it follows along similar lines to the proof of Proposition~\ref{prop5}(1).
\proofend

Propositions~\ref{prop5} and~\ref{last-as-co} yield the following
corollary for forward-only execution.
\begin{corollary}\label{last-as-f}
\rm Given a reversing Petri net $(A, P,B,T,F)$, an initial state 
$\langle M_0, H_0\rangle$, and an execution
$\state{M_0}{H_0} \trans{t_1}\state{M_1}{H_1} \trans{t_2}\ldots 
\trans{t_n}\state{M_n}{H_n}$, for all $a\in A$, 
$a\in M_n(x)$ where $x=\lastp{\connected(a,M_{n}(x)),H_n}$.
\end{corollary}
We may now verify that the causal-order and out-of-causal-order reversibility have the same effect when reversing a $c$-enabled transition.
\begin{proposition}\label{c-to-o}{\rm\ \
	Consider a state $\langle M, H\rangle$ and a transition $t$ $c$-enabled in  $\langle M, H\rangle$.
	Then, $\langle M, H\rangle\ctrans{t}\langle M', H'\rangle$ if and only if $\langle M, H\rangle\otrans{t}\langle M', H'\rangle$.
}
\end{proposition}
\paragraph{Proof}
Let us suppose that transition $t$ is $c$-enabled and $\langle M, H\rangle\ctrans{t}
\langle M_1, H_1\rangle$. 
By Proposition~\ref{enable}, $t$ is also $o$-enabled. Suppose $\langle M, H\rangle\otrans{t}\langle M_2, H_2\rangle$.
It is easy to see that in fact $H_1 = H_2$ (the two histories are 
as $H$ with the exception that $H_1(t) = H_2(t) = H(t)-\{\max(H(t))\}$). 

To show that $M_1=M_2$ first we observe that for all
$a\in A$, by Proposition~\ref{last-as-co} we have $a\in M_1(x)$ where
$x=\lastp{\connected(a,M_1(x)),H_1}$ and by
Proposition~\ref{prop5} we have $a\in M_1(y)$ where
$y=\lastp{\connected(a,M_2(y)),H_2}$. We may also see
that $\connected(a,M_1(x)) = \connected(a,M(z)-\effect{t})= \connected(a,M_2(y))$, where $a\in M(z)$. Since in addition we have $H_1=H_2$
the result follows.

Now let $\beta\in B$. We must show that
$\beta\in M_1(x)$ if and only if $\beta\in M_2(x)$. Two cases exist:
\begin{itemize}
\item If $\beta \in \effect{t}$ then by Propositions~\ref{Prop4} and~\ref{prop5}, $\beta\not\in M_1(x)$ and $\beta\not \in M_2(x)$ for all $x\in P$.
\item if $\beta \not\in \effect{t}$ then by Propositions~\ref{Prop4} and~\ref{prop5},
$|\{x\in P \mid \beta\in M_1(x))\}| = |\{x\in P \mid \beta\in M_2(x))\}|= 1$
and by the analysis on
tokens  $\beta\in M_1(x)$ if and only if $\beta\in M_2(x)$ and
the result follows.
\end{itemize}
This completes the proof.
\proofend

An equivalent result can be obtained for backtracking.

\begin{proposition}\label{b-to-c}{\rm\ \
	Consider  a state $\langle M, H\rangle$, and a transition $t$, $bt$-enabled in  $\langle M, H\rangle$.
	Then, $\langle M, H\rangle\btrans{t}\langle M', H'\rangle$ if and only if $\langle M, H\rangle\otrans{t}\langle M', H'\rangle$.
}
\end{proposition}
\paragraph{Proof}
Consider a state $\langle M, H\rangle$ and suppose
that transition $t$ is $bt$-enabled and $\langle M, H\rangle\btrans{t} \langle M', H'\rangle$. Then,
by Proposition~\ref{enable}{, there exists $k\in H(t)$, such that {for all $t'\in T$, $k'\in H(t')$, it holds that
	$k\geq k'$}. This implies
that} $t$ is also $c$-enabled, and by the definition of $\ctrans{}$, we conclude that 
$\langle M, H\rangle{\ctrans{t}} \langle M', H'\rangle$. Furthermore, by Proposition~\ref{c-to-o}
$\langle M, H\rangle{\otrans{t}} \langle M', H'\rangle$, and the result follows.
\proofend

We obtain the following corollary confirming
the expectation that backtracking is an instance of causal reversing, which in turn is an instance of
out-of-causal-order reversing.  It is easy to see that both inclusions are strict,
as for example illustrated in Figures~\ref{b-example},~\ref{co-example}, and~\ref{o-example}.
\begin{corollary}\label{connect}{\rm\ \
	$\btrans{} \subset\ctrans{}\subset \otrans{}$. 
}
\end{corollary}

\paragraph{Proof}
The proof follows from Propositions~\ref{c-to-o} and~\ref{b-to-c}.
\proofend

We note that in addition to establishing the relationship between the three notions of reversibility,
the above results provide a unification of the different reversal strategies, in the sense that
a single firing rule, $\otrans{}$, may be paired with the three notions of transition enabledness
to provide the three different notions of reversibility.
This fact may be exploited in the proofs of results that span the three notions of reversibility. Such a proof follows in
the following proposition that
establishes a reverse diamond property
for RPNs. According to this property, the execution of a reverse
transition does not preclude the execution of another
reverse transition and their execution leads to the same
state.  In what follows we write $\frtrans{}$ 
for $\trans{}\cup\rtrans{}$ where $\rtrans{}$ could be an instance of one of $\btrans{}$, $\ctrans{}$, and $\otrans{}$.  

\begin{proposition}[Reverse Diamond]\label{reverseDiamond}{\rm 
	Consider a state $\state{M}{H}$, and reverse transitions $\state{ M}{H}$$ \rtrans{t_1}\state{ M_1}{H_1}$ and $ \state{ M}{H} \rtrans{t_2}\state{ M_2}{H_2}$, $t_1\neq t_2$. Then $\state{ M_1}{H_1} \rtrans{t_2}\state{ M'}{H'}$ and $\state{ M_2}{H_2} \rtrans{t_1}\state{ M'}{H'}.$
}\end{proposition}

\paragraph{Proof}
Let us suppose that
$\state{ M}{H} \rtrans{t_1}\state{ M_1}{H_1}$ and $ \state{ M}{H} \rtrans{t_2}\state{ M_2}{H_2}$, $t_1\neq t_2$.
First we note that $\rtrans{}$ may be an instance of $\ctrans{}$ or $\otrans{}$ but not $\btrans{}$, since
in the case of $\btrans{}$ the backward transition is uniquely determined
as the transition with the maximum key. 
Furthermore, we observe that 
$t_2$ remains backward-enabled in $\state{ M_1}{H_1}$ and likewise $t_1$ in
$\state{ M_2}{H_2}$. Specifically, if $\rtrans{}=\ctrans{}$,
since $t_1$ and $t_2$  are $c-$enabled in $\state{M}{H}$, by Definition~\ref{co-enabled} we conclude that
$(t_2,max(H(t_2)))$ is not causally dependent on $(t_1,max(H(t_1)))$ and vice versa, which continues
to hold after the reversal of each of these transitions.
In the case of $\rtrans{}=\otrans{}$ this is straightforward from the definition of $o$-enabledness.

So, let us suppose that $\state{M_1}{H_1} \ctrans{t_2}\state{ M_1'}{H_1'}$ and $\state{M_2}{H_2} \ctrans{t_1}\state{ M_2'}{H_2'}$. It is easy to see that $H_1' = H_2'$ since both of 
these histories are identical with $H$ with the maximum keys
of $t_1$ and $t_2$ removed. 

To show that $M_1'=M_2'$ first we observe that for all
$a\in A$, by Propositions~\ref{prop5} and~\ref{last-as-co}
we have $a\in M_1'(x)$, $a\in M_2'(y)$ where
$x=\lastp{\connected(a,M_1'(x)),$ $H_1'}$, $y=\lastp{\connected(a,M_2'(y)),H_2'}$.
We may see
that $\connected(a,M_1'(x)) = \connected(a,M(z)-(\effect{t_1}\cup \effect{t_2})
= \connected(a,M_2'(y))$, where $a\in M(z)$. Since in addition we have $H_1'=H_2'$
the result follows.

Now let $\beta\in B$. We must show that
$\beta\in M_1'(x)$ if and only if $\beta\in M_2'(x)$. Two cases exist:
\begin{itemize}
\item If $\beta \in \effect{t_1}\cup\effect{t_2}$ then by Propositions~\ref{Prop4} and~\ref{prop5}, $\beta\not\in M_1'(x)$ and $\beta\not \in M_2'(x)$ for all $x\in P$.
\item if $\beta \not\in \effect{t_1}\cup \effect{t_2}$ then by Propositions~\ref{Prop4} and~\ref{prop5},
$|\{x\in P \mid \beta\in M_1'(x))\}| = |\{x\in P \mid \beta\in M_2'(x))\}|$
and, by the analysis on
tokens, $\beta\in M_1'(x)$ if and only if $\beta\in M_2'(x)$.
\end{itemize}
This completes the proof.
\proofend

\begin{corollary}\label{diamondCorollary}{\rm \
	Consider a state  $\state{ M}{H}$, and traces $\sigma_1,\sigma_2$ permutations of the same reverse transitions where $\state{M}{H} \frtrans{\sigma_1}\state{ M'}{H'}$ and  $\state{M}{H} \frtrans{\sigma_2}\state{ M''}{H''}$. Then  $\state{M'}{H'}=\state{M''}{H''}$.
}\end{corollary}

\paragraph{Proof}
The proof follows by induction on the sum of the length $|\sigma_1|= |\sigma_2|$ and the depth of the earliest 
disagreement between the two traces, and uses similar arguments to those
found in the proof of Proposition~\ref{reverseDiamond}.
\proofend

We note that the analogue of Proposition~\ref{reverseDiamond} for forward
transitions, i.e. the Forward Diamond property, does not hold for RPNs.
To begin with $t_1$ and $t_2$ may be in conflict. 
The proposition fails to hold even in
the case of joinable transitions (i.e. transitions that may yield the same
marking after a sequence of forward moves) due to the case of
co-initial, independent cycles: Even though such cycles can be executed in any order, it is
impossible to complete the square for their initial transitions.

\section{Case Studies}

 The framework of reversing Petri nets could be applied in fields outside Computer Science, since the expressive power and visual nature offered by Petri nets coupled with reversible computation has the potential of providing an attractive setting for analysing systems, for instance in biology, chemistry or hardware engineering. The construction of reversible modelling languages can indicate how to capture the behaviour of reversible actions in order to implement or even extend the primitive processes of biological reactions, movement in robotics, quantum computation and reliable systems. Implementing several applications ranging from biochemistry to long-running transactions would give a better understanding on reversible computation especially when it comes to out-of-causal modelling which is still not very well understood in the area of Computer Science. 

\subsection{ ERK Signalling Pathway }

Biochemical systems, such as covalent bonds, constitute the ideal setting to study reversible computation especially in its out-of-causal-order form. In particular, the  MAPK/ERK pathway (also known as the Ras-Raf-MEK-ERK pathway)is one 
of many real-life examples that naturally feature reversibility that violates the causal 
ordering established by forward execution. This pathway has been 
modelled in various formalisms including  \emph{CCSK}~\cite{ERK},  PEPA~\cite{PEPA}, 
BioNetGen~\cite{BioNetGen}, and Kappa~\cite{Kappa}.

 In this section we illustrate how reversing 
Petri nets allow us to capture naturally this form of out-of-causal-order reversible system.  Specifically, our configuration follows that of CCSK in~\cite{ERK} where out-of-causal reversibility is triggered to release tokens form connected component so that the tokens can proceed to participate in other transitions. Additionally, in~\cite{ERK} an execution control operator is used to enforce a particular order of execution between forward and reverse actions. In RPNs this is achieved by using negative tokens that require the reversal of specific transitions in order to reverse negative tokens and therefore allow the forward execution of following transitions. However, in RPNs the execution of concurrent forward transitions can be executed in any order unlike the control operator of CCSK which is able to require a specific order among forward transitions.

\begin{figure}
	\centering
	\subfloat{\includegraphics[width=2.5cm]{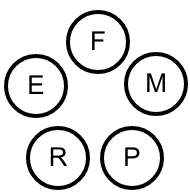}}
	\subfloat{\includegraphics[width=.55cm]{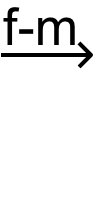}}
	\subfloat{\includegraphics[width=2.5cm]{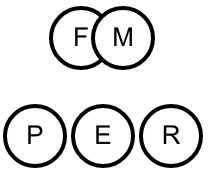}}
	\subfloat{\includegraphics[width=.55cm]{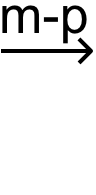}}
	\subfloat{\includegraphics[width=2.5cm]{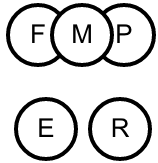}}
	\subfloat{\includegraphics[width=.6cm]{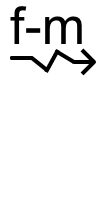}}
	\subfloat{\includegraphics[width=2.5cm]{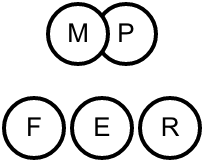}}
	\subfloat{\includegraphics[width=.6cm]{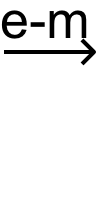}}
	\subfloat{\includegraphics[width=2.5cm]{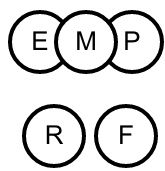}}
	\subfloat{\includegraphics[width=.6cm]{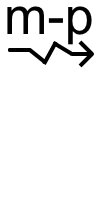}}\\
	\vspace{.5cm}
	\subfloat{\includegraphics[width=2.5cm]{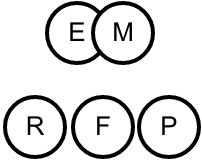}}
	\subfloat{\includegraphics[width=.6cm]{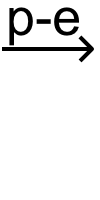}}
	\subfloat{\includegraphics[width=2.5cm]{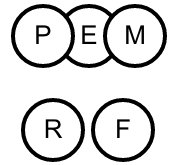}}
	\subfloat{\includegraphics[width=.6cm]{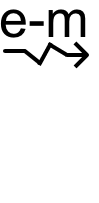}}
	\subfloat{\includegraphics[width=2.5cm]{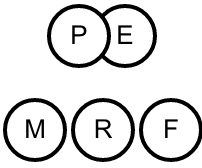}}
	\subfloat{\includegraphics[width=.6cm]{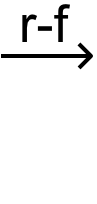}}
	\subfloat{\includegraphics[width=2.5cm]{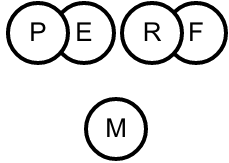}}
	\subfloat{\includegraphics[width=.6cm]{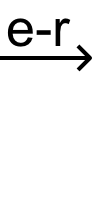}}
	\subfloat{\includegraphics[width=2.5cm]{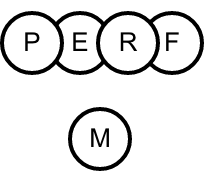}}\\
	\vspace{.5cm}
	\subfloat{\includegraphics[width=.6cm]{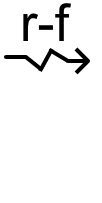}}
	\subfloat{\includegraphics[width=2.5cm]{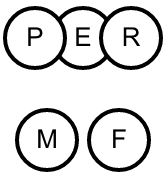}}
	\subfloat{\includegraphics[width=.6cm]{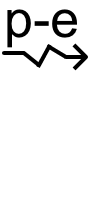}}
	\subfloat{\includegraphics[width=2.5cm]{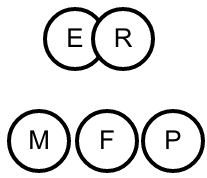}}
	\subfloat{\includegraphics[width=.6cm]{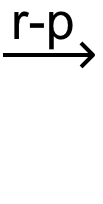}}
	\subfloat{\includegraphics[width=2.5cm]{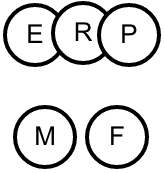}}
	\subfloat{\includegraphics[width=.6cm]{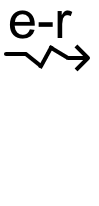}}
	\subfloat{\includegraphics[width=2.5cm]{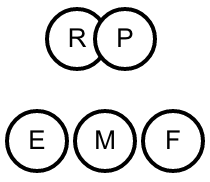}}
	\subfloat{\includegraphics[width=.6cm]{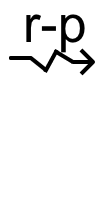}}
	\subfloat{\includegraphics[width=2.5cm]{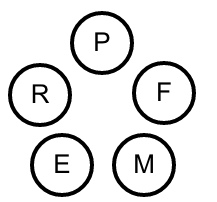}}
	\caption{Reactions in the \emph{ERK}-pathway where \emph{F} denotes \emph{Raf*-1}, 
		\emph{M} denotes \emph{MEK}, \emph{E} denotes \emph{ERK}, \emph{R} denotes \emph{RKIP}, and \emph{P} denotes the phosphorylation of the bonded molecule }\label{ERK}
\end{figure}

In Figure~\ref{ERK} we demonstrate the extracellular-signal-regulated kinase  (\emph{ERK}) pathway, also known as \emph{Ras/Raf-1, MEK, ERK} pathway, which  is one of the major signalling cassettes of the mitogen activated protein kinase (\emph{EMAPK}) signalling pathway. The ERK pathway is a chain of 
proteins in the cell that delivers mitogenic and differentiation signals from the membrane of a cell 
to the \emph{DNA} in the nucleus, and is regulated by the protein \emph{RKIP}. 
The starting point 
of the pathway is when a signalling molecule binds to a receptor on the cell surface and is spatially organised so that, when a signal arrives at the membrane, it can be transmitted to the nucleus via a cascade of biological reactions that involves protein kinases. A kinase is an enzyme that catalyses the transfer of a phosphate group from a donor molecule to an acceptor. The main $\mathit{MAPK/ERK}$  kinase kinase (\emph{MEKK}) component is the kinase component \emph{Raf-1} that phosphorylates the serine residue on the \emph{MAPK/ERK} kinase \emph{MEK}. We denote  \emph{Raf*-1} with  \emph{F}, \emph{MEK} with  \emph{M}, \emph{ERK} with \emph{E}, \emph{RKIP} with \emph{R}, and  the phosphorylation of the bonded molecule is denoted by {P}.

The pathway begins with the activation of the protein kinase of \emph{Raf-1} by the $G$ protein 
\emph{Ras}  that has been activated near a receptor on the cell's membrane. 
\emph{Ras}  activates a kinase \emph{Raf-1} 
to become \emph{Raf*-1}, which is generally known as a mitogen-activated protein 
kinase kinase kinase (\emph{MAP-KKK}) and can be inhibited by \emph{RKIP}. Subsequently, as we may see in Figure~\ref{ERK}, 
\emph{Raf*-1}~($F$) may bind with \emph{MEK}~($F\bond M$) by facilitating 
the next step in the cascade (\emph{MAPKK}), which is the phosphorylation of the \emph{MEK}~($F \bond M \bond P$) 
protein and the release of \emph{Raf*-1}~($M \bond P$).  
The phosphorylated \emph{MEK}~($M\bond P$) activates a mitogen-activated protein kinase, \emph{ERK}~($E \bond M \bond P$), which in turn becomes 
phosphorylated and releases \emph{MEK}~($P \bond E$). Finally, the  phosphorylation of \emph{MAPK}
allows the phosphorylated \emph{ERK}~($P\bond E$)  to function as an enzyme and translocate in order to signal the nucleus. Now the regulation sequence consumes the phosphorylated \emph{ERK}~($P \bond E$) in order to deactivate  \emph{RKIP}~($R$) from regulating \emph{Raf*-1}~($F$). 
Therefore, when \emph{RKIP} binds \emph{Raf*-1}~($R \bond F$), the resulting complex binds to a 
phosphorylated \emph{ERK}~($P\bond E\bond R\bond F$). In the end, the complex breaks releasing \emph{Raf*-1}~($P \bond E \bond R$), 
which can get involved in the cascade and after the phosphorylation of \emph{RKIP}~($E\bond R\bond P$)  the system releases \emph{ERK}~($R \bond P$) and the phosphorylated  \emph{RKIP}.

\begin{figure}
	\centering
	\setlength{\abovecaptionskip}{-2pt}
	\setlength{\belowcaptionskip}{-2pt}
	\subfloat{\includegraphics[width=15cm]{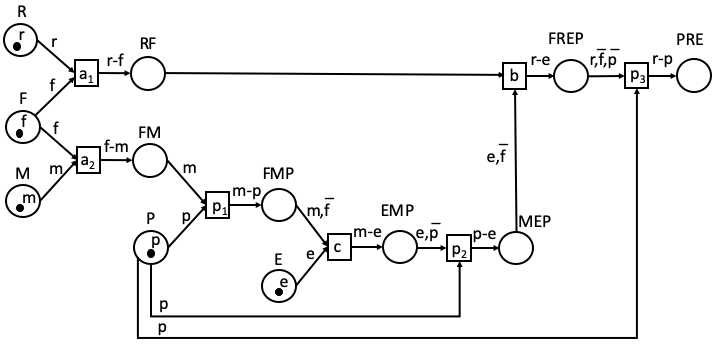}}
	\caption{ERK-pathway example in reversing Petri nets}\label{ERK reversing}
\end{figure}

We now describe the biochemical reactions of the ERK signalling pathway as the RPN demonstrated in Figure~\ref{ERK reversing}. On this RPN we represent molecules as tokens that can bond with each other, thus creating more complex molecules, and these composite molecules can be dissolved back to single tokens. The building blocks of the system are the base tokens representing the associated molecules. 

We begin our execution from the already activated  \emph{Raf-1} 
kinase that has become \emph{Raf*-1}. The molecule of \emph{Raf*-1} is represented by base token $f$ and resides in place  $F$. The token availability of base $m$, which represents \emph{MEK} in place $M$ enables the firing of  transition $a_2$ denoting that $f$ has bonded with $m$ and thus creating molecule $f\bond m$. The firing of transition $a_2$ facilitates 
the next step in the cascade, which is transition $p_1$ representing the phosphorylation of $m$ as the binding between $p$ and $m$. Since transition $a_2$ has enabled the execution of $p_1$ the transition $a_2$ is now reversed and therefore releases $f$ back to place $F$. This reversal in necessary
for the next step of the execution where the absence of $f$
is a condition for transition $c$ to fire. Indeed, in transition
$c$, the phosphorylated $m$ is now able to 
activate the kinase \emph{ERK}  denoted by base $e$ and thus creating a bond between $m\bond e$ along with $p$ that shows that $m$ is already phosphorylated. In the next step, transition $p_1$ reverses in order to release $p$, which can then be used in the firing of transition $p_2$ to phosphorylate $e$  and therefore creating the molecule $p\bond e$.  After transition $p_2$, transition $c$ is reversed in order to release $m$ back to $M$. Finally, after the  phosphorylation of $p\bond e$, transition $a_1$ executes in order to bond $f\bond r$ where $r$ represents molecule $RKIP$. Base $r$ functions as an enzyme and  by enabling transition $b$ represents the passing of the signal to the nucleus which can then  consume $p\bond e$ by creating a connected component between $f \bond r \bond e \bond p$. In the end, the complex breaks by reversing $a_1$ in order to release $f$ and $p_2$ to release $p$ which then in action $p_3$ phosphorylates $r$. Finally, the system reverses $b$ to release $e$ followed by the reversal of $p_3$, which releases both $r$ and $p$ and therefore returns the system back to its initial marking.

We show below an execution of the reversing Petri net that illustrates the process until the signal that arrived at the membrane is transmitted  to the nucleus. The following states of the net (with histories omitted) represent a cascade of reactions that involve protein kinases $F$, $M$, $E$, $P$, $R$, with initial marking 
$M_0$ such that  $M_0(R)=\{r\}$, $M_0(F)=\{f\}$,  $M_0(M)=\{m\}$,  $M_0(P)=\{p\}$,  $ M_0(E)=\{e\}$,
and $M_0(p)= \emptyset$ for all remaining places. (In the following, the markings of places with no
tokens are omitted.) 
\begin{tabbing}
	$M_0\xrightarrow{a_2}M_1$,\; \;\= where\; \;\= 
	$M_1(R)=\{r\}$,  $M_1(P)=\{p\}$,  $M_1(E)=\{e\} $,
	\\ \hspace{1.45in} $M_1(FM)= \{f\bond m\} $
	\\
	$M_1\xrightarrow{p_1}M_2$,\> where\> 
	$M_2(R)=\{r\}$,  $  M_2(E)=\{e\} $,
	\\ \hspace{1.45in}  $M_2(FMP)=\{f\bond m, m\bond p\} $
	\\
	$M_2\rtrans{a_2}M_3$,\>  where\> 
	$M_3(R)=\{r\}$,  $M_3(F)=\{f\}$,  $M_3(E)=\{e\} $,
	\\ \hspace{1.45in} $M_3(FMP)=\{m\bond p\} $
	\\
	$M_3\xrightarrow{c}M_4$,\>  where\> 
	$M_4(R)=\{r\}$,  $M_4(F)=\{f\} $,
	\\ \hspace{1.45in} $M_4(EMP)=\{m\bond e, m\bond p\} $  
	\\
	$M_4\rtrans{p_1}M_5$,\>  where\> 
	$M_5(R)=\{r\}$,  $M_5(F)=\{f\} $,  $M_5(P)=\{p\} $,
	\\ \hspace{1.45in} $M_5(EMP)=\{m \bond e\} $
	\\
	$M_5\xrightarrow{p_2}M_6$,\>  where\> 
	$M_6(R)=\{r\}$,  $M_6(F)=\{f\} $,
	\\ \hspace{1.45in} $M_6(MEP)=\{m\bond e, e \bond p\} $  
	\\
	$M_6\rtrans{c}M_7$,\>  where\> 
	$M_7(R)=\{r\}$,  $M_7(F)= \{f\}$,  $M_7(M)=\{m\} $,
	\\ \hspace{1.45in} $M_7(MEP)=\{e\bond p\} $  
	\\
	$M_7\xrightarrow{a_1}M_8$,\>  where\> 
	$M_8(M)=\{m\} $,
	$M_8(RF)=\{r\bond f\} $,    
	\\ \hspace{1.45in} $M_8(MEP)=\{e\bond p\} $ 
	\\
	$M_8\xrightarrow{b}M_9$,\>  where\> 
	$M_9(M)=\{m\} $,  
	$M_9(FREP)=\{r\bond f, r\bond e, e\bond p\} $        
	\\
	$M_9\rtrans{a_1}M_{10}$,\>  where\> 
	$M_{10}(M)=\{m\}$,  $M_{10}(F)=\{f\} $,  
	\\ \hspace{1.45in} $M_{10}(FREP)=\{r\bond e, e\bond p\} $       
	\\
	$M_{10}\rtrans{p_2}M_{11}$,\>  where\> 
	$M_{11}(M) =\{m\}$,  $M_{11}(F)=\{f\}, M_{11}(P)=\{p\} $,  
	\\ \hspace{1.45in}  $M_{11}(FREP)=\{r\bond e\} $    
	\\
	$M_{11}\xrightarrow{p_3}M_{12}$,\>  where\> 
	$M_{12}(M)=\{m\}$,  $  M_{12}(F)=\{f\} $,  
	\\ \hspace{1.45in} $M_{12}(PRE)=\{r\bond e, p\bond r\} $     
	\\
	$M_{12}\rtrans{b}M_{13}$,\>  where\> 
	$M_{13}(M)=\{m\}$,  $ M_{13}(F)=\{f\} M_{13}(E)\{e\} $,  
	\\ \hspace{1.45in} $M_{13}(PRE)=\{p\bond r\} $   
	\\
	$M_{13}\rtrans{p3}M_{14}$,\>  where\> 
	$M_{14}(R)=\{r\}$,  $ M_{14}(F)= \{f\}$,  $ M_{14}(M)=\{m\} $,  \\ \hspace{1.45in} $M_{14}(P)=\{p\}$,  $ M_{14}(E)=\{e\} $
\end{tabbing}


\subsection{Transactions with Compensations}

\begin{figure}[t]
	\centering
	\hspace{.7cm}
	\subfloat{\includegraphics[width=9cm]{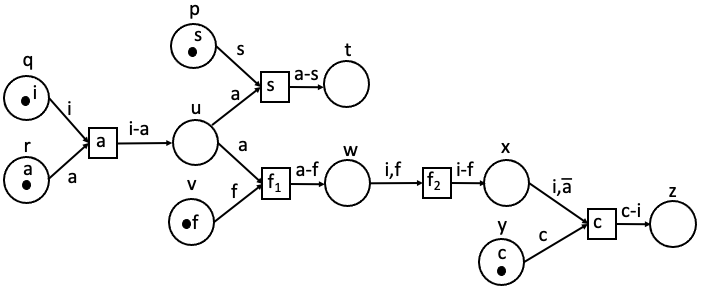}}\\
	\subfloat{\includegraphics[width=.7cm]{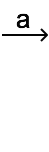}} 
	\subfloat{\includegraphics[width=9cm]{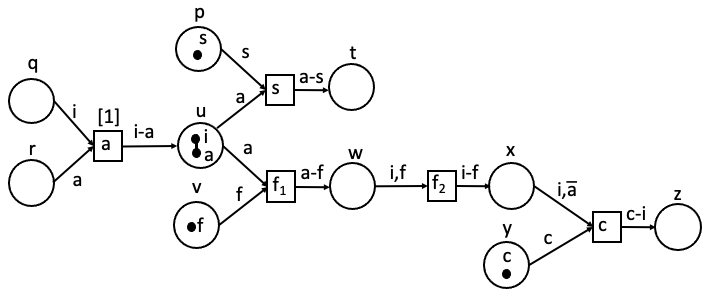}}\\
	\subfloat{\includegraphics[width=.7cm]{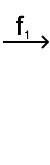}}
	\subfloat{\includegraphics[width=9cm]{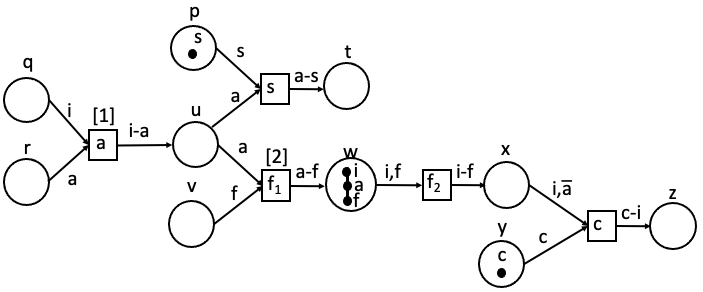}}\\
	\subfloat{\includegraphics[width=.7cm]{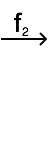}}
	\subfloat{\includegraphics[width=9cm]{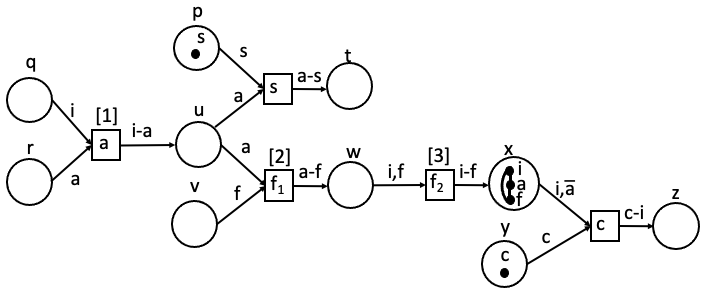}}
	\caption{Transaction processing - forward execution}
	\label{ftrans}
\end{figure}

\begin{figure}[h]
	\centering
	\subfloat{\includegraphics[width=.7cm]{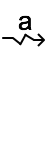}} 
	\subfloat{\includegraphics[width=9cm]{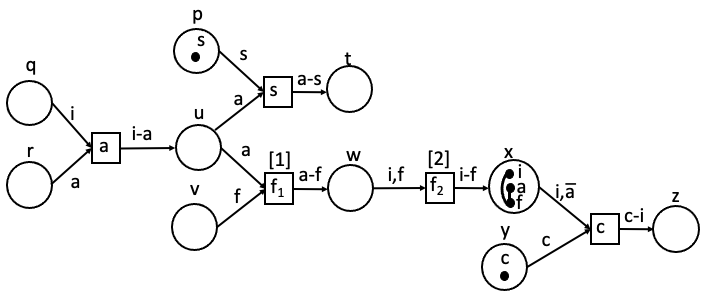}}\\
	\subfloat{\includegraphics[width=.7cm]{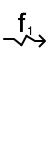}}
	\subfloat{\includegraphics[width=9cm]{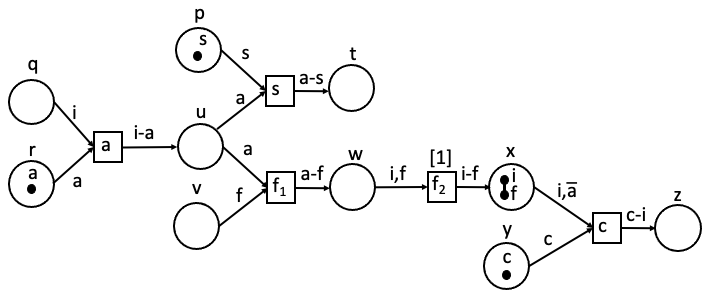}}\\
	\subfloat{\includegraphics[width=.7cm]{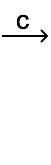}}
	\subfloat{\includegraphics[width=9cm]{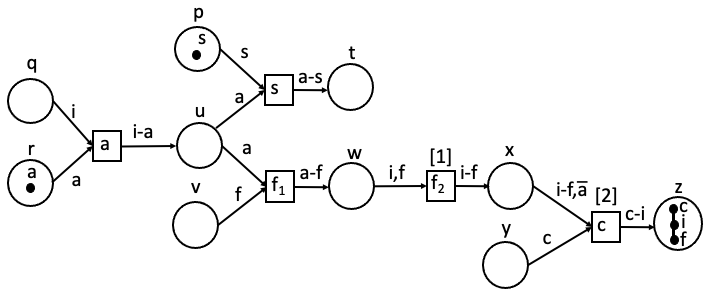}}\\
	\subfloat{\includegraphics[width=.7cm]{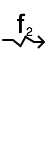}}
	\subfloat{\includegraphics[width=9cm]{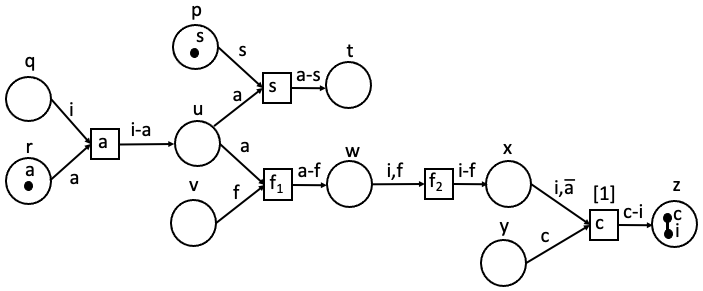}}
	\caption{Transaction processing: out-of-causal-order execution }
	\label{trans}
\end{figure}

Transaction processing manages sequences of operations, also called transactions, that can either succeed 
or fail as a complete unit.  Specifically, a long-running transaction  aggregates smaller atomic transactions, and typically use a coordinator to complete or abort the transaction. An atomic transaction is an indivisible and irreducible series of  operations such that either all occur, or nothing occurs~\cite{Trans}.

Long-running transactions consist of a sequence of steps that avoid locks on non-local resources and use compensation to handle failures. Each of these steps may either succeed, in which case the flow of control moves on to the next atomic 
step in the sequence, or it may
fail, in which case a compensating transaction is often used to undo failed transactions and restore the 
system to a previous state.   In contrast to rollback in atomic  transactions, compensation restores the original state, or an equivalent, and it is business-specific. 
If all steps of the transaction execute successfully then the transaction is considered as successful and
it is committed. 


The definition of causal reversibility has spawned various reversible extensions of concurrent languages that are used for validating formal connections between causal-consistent reversibility and reliability as well as studying its consequences. It enables a new strategy for debugging concurrent systems, where the different speed of processes that are replaying an execution looking for a bug may cause different behaviours. 
There have been proposed reversible process calculi used to build constructs for reliability, and in particular communicating transactions with compensations~\cite{Kutavas} where interacting transactions with compensations have been mapped into a reversible calculus with alternatives in~\cite{LaneseLMSS13}. ~\cite{Kutavas} uses transactions with compensations, which are computations that either succeed, or their effects are reversed and then a compensation is executed by a dedicated ad-hoc piece of code. 
Behavioural equivalences for communicating transactions with compensation have been studied in~\cite{Kutavas2,Kutavas3}.

In Figures~\ref{ftrans} and~\ref{trans} we consider a model of such a transaction.  Specifically in Figure~\ref{ftrans} we demonstrate the forward execution of a failed transaction and in~\ref{trans} we demonstrate the compensation part of the transaction execution which follows the strategy of out-of-causal-order reversing. Due to the size of the 
net we restrict our attention to a transaction
with only one step. 
The computation starts with the initialisation step $a$ which never fails. The intuition is as follows: for the execution of the transaction to commence it is necessary for token $i$ to be available. This token
is bonded with token $a$ in which case transition $a$ can be executed with the effect of creating the bond $i\bond a$ in place $u$.
At this stage there are two possible continuations. The first possibility is that the bond $i\bond a$ will participate in transition $s$ which 
models the successful completion of  step $a$ as well as the transaction, yielding the bond $i\bond a\bond s$. The second possibility that a transaction can fail at any stage after step $a$. In this case, token $f$ comes in place and the failure is implemented via transitions $f_1$ and $f_2$ as follows: 
To begin with in action $f_1$, token $f$ is bonded with token $a$ repressing that the transaction has failed, whereas in action $f_2$ token $i$ is bonded
with token $f$ indicating that the initialisation has failed thus triggering reversal. At this stage the compensation comes in place (token $c$) where the intention is that step $a$ should be undone
by undoing transition $a$. Note that this will have to be done according to  our out-of-causal-order
definition since transition $a$ was followed by $f_1$ and $f_2$ which have not been undone. Only once this is accomplished,
will the precondition of transition $c$, namely $\overline{a}$, be enabled. In this case, transition $c$ can be executed leading
to the creation of bond $i\bond c$ in place $z$.


\section{ Concluding Remarks}

This chapter proposes a reversible approach to Petri nets~\cite{RPNs,RPNscycles} that allows the modelling
of reversibility as realised by backtracking, causal-order reversing and out-of-causal-order
reversing. To the best of our knowledge, this is the first such proposal in the 
context of Petri nets. For instance, the works of~\cite{PetriNets,BoundedPNs} 
introduce reversed transitions in a Petri net and study various decidability
problems in this setting. This approach, however, does not precisely capture
reversible behaviour due to the property of backward conflict in PNs. On the contrary,~\cite{ Unbounded} is concerned with causal order reversal where a subclass of Petri nets can be restructured by adding effect-reversals that do not affect the computational behaviour of the model. Moreover,
the works of~\cite{RPT,RPlaceTrans} propose a causal semantics for P/T nets by identifying
the causalities and conflicts of a P/T net through unfolding it into an
equivalent occurrence net and subsequently introducing appropriate reverse
transitions to create a coloured Petri net that captures a causal-consistent
reversible semantics. The colours in this net capture causal histories.

On the other hand, our proposal consists of a reversible approach to Petri 
nets, where the formalism supports the reversible semantics without explicitly 
introducing reverse transitions. This is achieved with the use of bonds of 
tokens, which can be thought of as
colours and, combined with the history function of the semantics, capture 
the memory of an execution as needed to implement reversibility. Furthermore, 
the approach allows to implement both causal-order and out-of-causal-order reversibility.  

  As in~\cite{RPT,RPlaceTrans}, 
our goal has been to allow a causally-consistent semantics reflecting causal 
dependencies as a partial order, and allowing an event to be reversed only if all 
its consequences have already been undone. To achieve this goal we have defined a causal dependence relation that resorts to the \emph{marking} of a net. 
As illustrated via examples (e.g. see Figures~\ref{cycles1} and~\ref{cycles2}), 
this is central in capturing causal dependencies and the intended causal-consistent semantics. Our dependence relation is strong enough to capture partial order causality even in the absence of bonds. Specifically, the introduction of bonds can be handled by representing tokens as colours similarly to coloured Petri nets. Therefore, a simplification of the model can be proposed without including bonds that will still preserve the causal-consistent semantics of RPNs. The resulting framework would be closely related to coloured Petri nets thus possibly inheriting various theoretical results proven for the traditional model. 

In a related line of work, we are also investigating the expressiveness relationship
between RPNs and Coloured Petri Nets. Specifically, in~\cite{RPNtoCPN} a subclass of RPNs with
trans-acyclic structures has been encoded in coloured PNs. Currently, we are 
extending this work with ultimate objective
to provide and prove the correctness of the translation between the 
two formalisms and analyse the
associated trade-offs in terms of Petri net size.

 Another possible direction for future work would be the extension of RPNs with directed bonds. This would enable the framework to model double bonds as defined in biochemistry, where a covalent bond between two atoms involves four bonding electrons as opposed to two in a single bond.

\chapter{Token Multiplicity in Reversing Petri Nets}
\label{sec:multi}

%

 In the previous chapter we have introduced a form of reversing Petri nets that assumes tokens to be unique and does not allow transitions to break bonds. In this chapter we focus on relaxing these restrictions, to develop reversible semantics in the presence of bond destruction, and to allow multiple tokens of the same base/type to occur in a model.
 
  By allowing the destruction of bonds  in forward transitions 
   we alter our perception of what the effect of a transition is, as additionally to the addition of new bonds the destruction of existing bonds should also be considered as the effect of a transition. We show the associated semantics of all three forms of reversible computing in this setting. 
  
  We also enhance the modelling flexibility of reversing Petri nets by introducing token multiplicity. We study how the partial order of causality in the original RPN model is affected and explain what modifications to the causal semantics are needed in order to provide a satisfactory treatment of reversibility. The causal semantics of such extended nets can often no longer be described solely in terms of a partial order, thus we introduce a new form of reversibility that follows disjunctive causality and we justify our proposal by modelling an example of a biochemical reaction.

\section{Destruction of Bonds in Reversing Petri Nets.}

As our original RPN model is inspired by biochemistry and since concerted chemical reactions involve the breaking and making of bonds in a single step, a consideration of transitions that allow the simultaneous bonding and destructing of molecules is essential to this understanding. We motivate our design decisions by giving an example of a concerted reversible chemical reaction that allows the simultaneous creation and destruction of bonds. 

\begin{figure}[t]
	\begin{center}
		\hspace{1cm}
		\includegraphics[height=4.2cm]{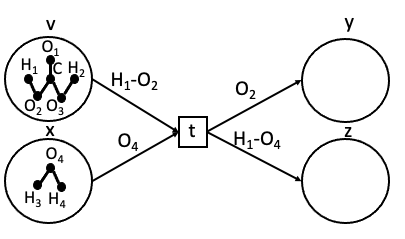}\\
		\includegraphics[width=.7cm]{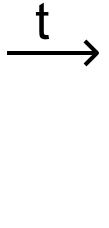}
		\includegraphics[height=4cm]{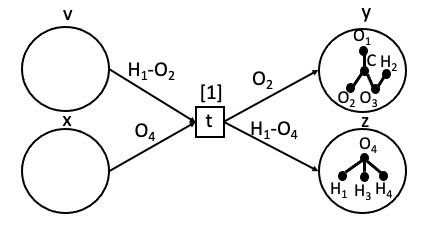}
	\end{center}
	\caption{Reversible chemical reaction}
	\label{carbonic}
\end{figure}

A reversible chemical reaction is a reaction where the reactants and products react together to give the reactants back. Weak acids, such as carbonic acid, and bases, such as water, undertake reversible reactions. Carbonic acid is a chemical compound with the chemical formula $H_2CO_3$. It is also a name sometimes given to solutions of carbon dioxide in water also known as carbonated water, because such solutions contain small amounts of $H_2CO_3$. In the example presented in Figure~\ref{carbonic}, carbonic acid $H_2CO_3$ and water $H_2O$ react to form  bicarbonate  $HCO_3$ and hydronium $H_3O$. 

We may now proceed to define \emph{Single Reversing Petri Nets (SRPNs)} that extend the model of Chapter 3 by allowing transitions to break bonds during forward execution. As with the original RPN model we use the tuple  $(A,P,B,T,F)$ to define SRPN structures that consist of bases, places, bonds, transitions and labelled directed arcs. 
%
%
In this section we only consider the extension of destructing bonds thus we still consider each token to have a unique name. 
Allowing connected tokens to fork in different outgoing places requires close attention to the existing connected components so that we do not clone tokens. To avoid duplicating tokens we still require directed arcs to ensure token preservation as defined by the well-formedness of RPNs. However, compared to the original model, we have eliminated one condition as we no longer require existing bonds to be preserved on the outgoing arcs of a transition. 
%
%
\begin{definition}\label{well-formedSingle}{\rm 
		A SRPN $(A,P,B,T,F)$ is \emph{well-formed} if it satisfies the following conditions for all $t\in T$:
		\begin{enumerate}
			\item $A\cap \guard{t} = A\cap \effects{t}$, 
			\item $ F(t,x)\cap F(t,y)=\emptyset$ for all $x,y\in P$, $x\neq y $. 
		\end{enumerate}
}\end{definition}

According to the above we have that: (1) transitions do not
erase tokens, and  (2) tokens and bonds cannot be cloned into more than one outgoing place.


 We may now define the various types of execution for single reversing Petri nets where forward transitions are able to break bonds. As with the original RPNs in this extension we restrict our attention to  well-formed SRPNs $(A,P,B,T,F)$ with initial  marking $M_0$ such that for all $a\in A$, $|\{ x \mid a\in M_0(x)\} | = 1$.  
 
\subsection{Forward Execution}

In this section we consider the standard, forward execution of SRPNs. As before, for a transition to fire in the forward direction we require the corresponding token and bond availability. As we now allow connected tokens to break their bonds during forward execution we require the bonds that connect them to be a requirement for the transition to fire. Formally: 

\begin{definition}\label{forwardSingle}{\rm
		Consider a SRPN $(A, P,B,T,F)$, a transition $t\in T$, and a state $\state{M}{H}$. We say that
		$t$ is \emph{forward-enabled} in $\state{M}{H}$  if the following hold:
		\begin{enumerate}
			\item  if $a\in F(x,t)$,  for some $x\in\circ t$, then $a\in M(x)$, and  if $\beta\in F(x,t)$,  for some $x\in\circ t$, then $\beta\in M(x)$,

			\item 
			for all $a,b \in F(x,t)$, $x\in \circ t$ where $(a,b) \in M(x)$ then $(a,b) \in F(x,t)$, and
			\item 
			if $a \in F(t,y_1)$, $b \in F(t,y_2)$, $y_1,y_2\in t\circ$, $y_1\neq y_2 $ then $a\not\in \connected( b, (M(x) -\guard{t}) \cup \effects{t}),x\in \circ t$. 
		\end{enumerate}
}\end{definition}

Thus, $t$ is enabled in state $\state{M}{H}$ if (1) all tokens and bonds required for the transition to take place
are available in the incoming places of $t$, (2) if a pre-existing bond  appears in an incoming place of the transition and its tokens are required for the transition to fire then the bond should also appear as a requirement on the incoming arcs (we do not recreate bonds), and (3) 
if two tokens are transferred by a transition to different outgoing places then these tokens should not remain connected when removing the incoming arcs and adding the outgoing arcs. Transferring tokens that are connected either directly or indirectly to different places without breaking their bonds it would result in token duplication. As such, we require for the tokens that fork to different places not to be connected when executing the transitions, thus any bonds that exist between them  in the incoming places of the transition will be destructed by the specifications on the arcs of the  transition.

\begin{figure}[t]
	\begin{center}
		\hspace{1cm}
		\includegraphics[height=3cm]{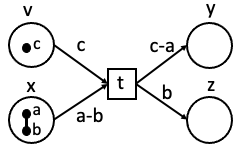}
		\hspace {.5cm}
		\includegraphics[width=.7cm]{figures/t.png}
		\hspace {.5cm}
		\includegraphics[height=3cm]{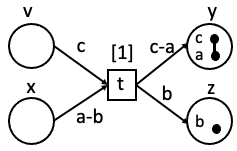}
	\end{center}
	\caption{Forward execution}
	\label{singleforwardFig}
\end{figure}

During forward execution the new bonds created by a transition are
exactly those that occur in the outgoing edges of a transition but not in the incoming edges and the bonds that are broken are those that occur in the incoming edges of a transition but not in the outgoing edges. Thus, firing a transition in the forward direction recreates the marking by removing the bonds that occur on the incoming arcs but adding the bonds that occur in the outgoing arcs. 
Specifically, firing a transition in the forward direction is defined as follows:

\begin{definition}{\rm \label{forwSingle}
		Given a SRPN $(A, P,B,T,F)$, a state $\langle M, H\rangle$, and a transition $t$ enabled in 
		$\state{M}{H}$, we write $\state{M}{H}
		\trans{t} \state{M'}{H'}$
		where $H'$ is updated as in Definition~\ref{forw} and:
		\[
		\begin{array}{rcl}
		M'(x) & = & \left\{
		\begin{array}{ll}
		M(x)-\bigcup_{a\in F(x,t)}\connected(a,M(x))  & \textrm{if } x\in \circ{t} \\
		M(x)\cup \bigcup_{ a\in F(t,x)\cap F(y,t)}\connected(a,(M(y) - F(y,t))\cup F(t,x)) & \textrm{if }  x\in t\circ\\
		M(x), &\textrm{otherwise}
		\end{array}
		\right.
		\end{array}
		\]
}\end{definition}

Thus, when a transition $t$ is executed in the forward direction, all tokens and bonds
occurring in its outgoing arcs are relocated from the input places to the output places along with their connected components. The  SRPN in Figure ~\ref{singleforwardFig}  represents the destruction of bond $a\bond b$ and the creation of  bond $c\bond a$ where the new bond  $c \bond a$ is relocated in place $y$ and token $b$ is relocated in place $z$. The history is updated as usual.

\subsection{Backtracking}

\begin{figure}[t]
	\begin{center}
		\hspace{1cm}
		\includegraphics[height=3cm]{figures/singleforward2.png}
				\hspace {.5cm}
		\includegraphics[width=.7cm]{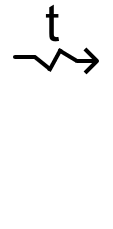}
				\hspace {.5cm}
		\includegraphics[height=3cm]{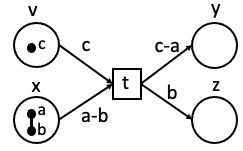}
	\end{center}
	\caption{Backtracking execution }
	\label{singlebacktrackingFig}
\end{figure}

Let us now proceed to backtracking. As with the original model the destruction of bonds does not affect $bt$-enabledness thus we define a transition to be \emph{bt}-enabled if it was the last executed transition:

\begin{definition}\label{bt-enabledSingle}{\rm
		Consider a SRPN $(A, P,B,T,F)$, a state $\state{M}{H}$, and a transition $t\in T$. We say that $t$ is \emph{$bt$-enabled} in
		$\state{M}{H}$ as in Definition~\ref{bt-enabled}.
}\end{definition}

The effect of backtracking a transition in a single reversing Petri net with bond destruction is shown in Figure~\ref{singlebacktrackingFig} which reverses the execution of the reversing Petri net by recreating the bond $a\bond b$ and returning it to its initial place $x$ as well as breaking the bond  $ c\bond a$ and returning $c$ to place $v$. Thus backtracking execution is defined as follows:

\begin{definition}\label{br-defSingle}{\rm
		Given a SRPN $(A, P,B,T,F)$, a state $\langle M, H\rangle$, and a transition $t$ with history $k=max(H(t))$ that is $bt$-enabled in $\state{M}{H}$, we write $ \state{M}{H}
		\btrans{t} \state{M'}{H'}$
		where $H'$ is updated as in Definition~\ref{br-def} and:
		\[
		\begin{array}{rcl}
		M'(x) & = & \left\{
		\begin{array}{ll}
		M(x)\cup\bigcup_{ y \in t\circ, a\in F(x,t)\cap F(t,y)}\connected(a,(M(y)-F(t,y))\cup F(x,t)),  & \textrm{if } x\in \circ{t} \\
		M(x)- \bigcup_{a\in F(t,x)}\connected(a,M(x)) , & \textrm{if }  x\in t\circ\\
		M(x) &\textrm{otherwise}
		\end{array}
		\right.
		\end{array}
		\]
		
}\end{definition}

\subsection{Causal Reversal}


\begin{figure}[t]
	\begin{center}
		\hspace{.5cm}
		\includegraphics[height=4.5cm]{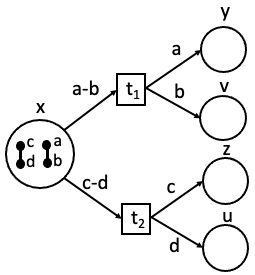}
		\includegraphics[width=.7cm]{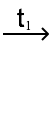}
		\includegraphics[height=4.5cm]{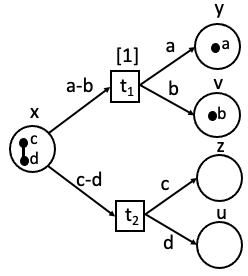}\\
		\includegraphics[width=.7cm]{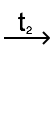}
		\includegraphics[height=4.5cm]{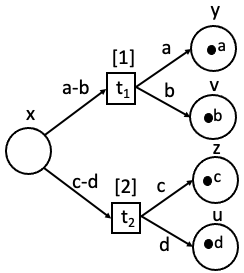}
		\includegraphics[width=.7cm]{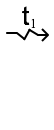}
		\includegraphics[height=4.5cm]{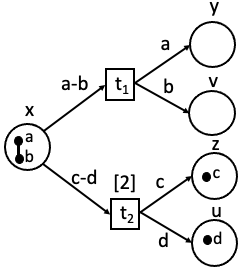}\\
		\hspace{0.5cm}
		\includegraphics[width=.7cm]{figures/t2r.png}
		\includegraphics[height=4.5cm]{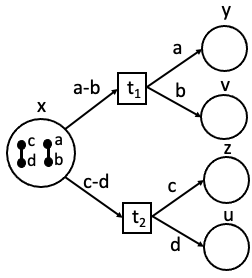}
	\end{center}
	\caption{Causal-order execution}
	\label{singlecausalFig}
\end{figure}

%
%

We introduce destruction of bonds in causal reversibility and show that 
no  modifications are needed to the notions of causal dependence and causal enabledness of the original reversing Petri nets. 
As expected,
we consider a transition $t$ to be enabled for causal-order reversal only if all
transitions that are causally dependent on it have either been reversed or not 
executed. 
%
We may now define that a transition is enabled for causal-order reversal as follows:
\begin{definition}\label{co-enabledSingle}{\rm
		Consider a SRPN $(A, P,B,T,F)$, a state $\cstate{M}{H}{\prec}$, 
		and a transition $t\in T$. Then $t$ is $c$-enabled  in $\cstate{M}{H}{\prec}$  as in Definition~\ref{co-enabled}.
%
}\end{definition}


Reversing a transition in a causally-respecting order is implemented similarly to
backtracking, i.e. the tokens are moved from the outgoing places to the incoming places of the transition,  all bonds
created by the transition are broken and all bonds destructed by the transition are reconstructed. In addition, the history function is updated in the same manner as in backtracking, where we remove the key of the reversed transition, and the
causal dependence relation removes all references to the reversed transition occurrence. The example in Figure~\ref{singlecausalFig} represents two concurrent transitions that have been executed and reversed in different orders. 

\begin{definition}\label{co-defSingle}{\rm
		Given a SRPN $(A, P,B,T,F)$, a state $\langle M, H,\prec\rangle$, and a transition $t$ $c$-enabled in $\state{M}{H}$, we write $\cstate{M}{H}{\prec}
		\ctrans{t} \cstate{M'}{H'}{\prec'}$ for $M'$ and $H'$ as in Definition~\ref{br-defSingle}, and $\prec'$  such that 
		\begin{eqnarray*}
			\prec' &=& \{((t_1,k_1), (t_2,k_2)) \in \prec \mid k_2\neq k\}
		\end{eqnarray*}
		
}\end{definition}

\begin{theorem}\label{singlemain}{\rm Consider executions $\state{M}{H} \fctrans{\sigma_1} \state{M_1}{H_1}$ and $\state{M}{H}\fctrans{\sigma_2} \state{M_2}{H_2}$. Then, $\sigma_1\asymp\sigma_2$ if and only if   $\state{M_1}{H_1}\asymp\state{M_2}{H_2}$.
	}
\end{theorem}

\paragraph{Proof}
The proof of the theorem follows as a corollary of  Theorem~\ref{multimain},  which will be presented in Section 4.2 since SRPNs are a special instance of Multi Reversing Petri Nets.  
\proofend

\subsection{Out-of-Causal Order}

\begin{figure}[t]
	\begin{center}
		\hspace{1cm}
		\includegraphics[height=3.15cm]{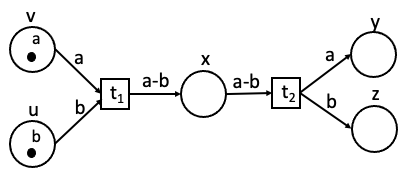}\\
		\includegraphics[width=1cm]{figures/t1.png}
		\includegraphics[height=3.1cm]{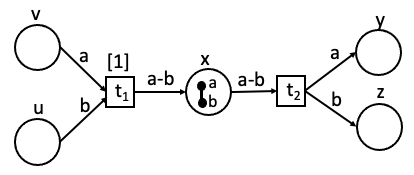}\\
		\includegraphics[width=1cm]{figures/t2.png}
		\includegraphics[height=3.3cm]{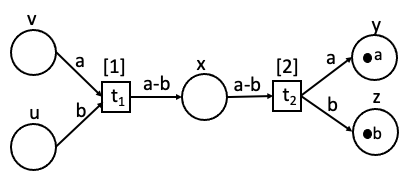}\\
		\includegraphics[width=1cm]{figures/t1r.png}
		\includegraphics[height=3.2cm]{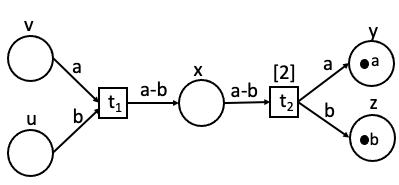}\\
		\includegraphics[width=1cm]{figures/t2r.png}
		\includegraphics[height=2.9cm]{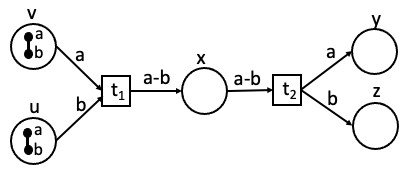}
	\end{center}
	\caption{Out-of-causal order paradox}
	\label{grandfather}
\end{figure}

We may now proceed to out-of-causal order reversibility which in the original model of RPNs allows any transition to reverse as long as it is an executed transition. When allowing the destruction of bonds  during forward execution,  this form of reversing presents a peculiarity since we allow bonds to be broken during forward execution then there is the possibility to execute two forward transitions that have the opposite effect. Consider the example in Figure~\ref{grandfather} where the first transition creates a bond and the second transition destructs the same bond. When reversing the first transition in out-of-causal order, then we try to reverse a bond that has already been broken by the second transition. In this way we negate a bond that was necessary for the effect of the second transition and thus create inconsistencies such as the token duplication presented in the final net of Figure~\ref{grandfather}.  In this case reversing a bond that was required for the already executed following transition leads to inconsistencies in regards to token preservation, an important feature of reversing Petri nets and  reversible computation in general. As such, we assume transitions like these to be irreversible and we only allow out-of-causal reversal in transitions that do not generate these paradoxes. 

 We begin by noting that in out-of-causal-order reversibility any executed transition can be reversed at any time as long as its effect has not been reversed by a forward transition.

\begin{definition}\label{o-enabledSingle}{\rm
		Consider a SRPN $(A,P,B,T,F)$, a state $\state{M}{H}$, and a transition $t\in T$. We say that transition $t$ is \emph{$o$-enabled} in 
		$\state{M}{H}$, if (1) $H(t)\neq \emptyset$ and:
		\begin{enumerate}[a]
			\item for all	$(a,b)\in \guard{t}$,	$(a,b)\not \in \effects{t}$ then $\not\exists t',$ $ k' \in  H(t')$ where   $k'>k$ such that $(a,b)\in \effects{t'}$,	$(a,b)\not \in \guard{t'}$ 
		\item for all $(a,b)\in \effects{t}$,	$(a,b)\not \in \guard{t}$ then $\not\exists t', k' \in H(t')$  where $k'>k$ such that $(a,b)\in \guard{t}$,	$(a,b)\not \in \effects{t}$ 
		\end{enumerate}
}\end{definition}

 The definition states that  for $t$ to be $o-$enabled then (1) the transition should be executed, (2)(a) if the transition breaks a bond during forward execution then it should not be followed by an executed transition that has destructed the same bond, and (2)(b) if the transition creates a bond during forward execution then it should not be followed by a transition that has destructed the same bond. Requirements (2)(a) and (2)(b) are used to avoid token duplication as in the case of~\ref{grandfather}. 
 
As in the original model we define the last transition that a connected component has participated and the last place where it should be relocated. 

\begin{definition}\label{lastSingle}{\rm
		{	Given a SRPN $(A,P,B,T,F)$, an initial marking $M_0$,  
			a history $H$, and 
			a set of bases and bonds $C\subseteq A\cup B$ we write:
			\[
			\begin{array}{rcl}
			\lastt{C,H} &=& \left\{
			\begin{array}{ll}
			t , \;\;\hspace{2cm}\textrm{ if }\exists t, \; \effects{t}\cap C\neq \emptyset, \; H(t)\neq \emptyset, \mbox{ and }\\
			\hspace{0.2in} \hspace{2.9cm} \not\exists t', \;\effects{t'} \cap C\neq \emptyset, \; H(t') \neq \emptyset, 	\\ 	
			\hspace{4.9cm}max(H(t'))\geq max (H(t)) \\
			\bot,  \;\hspace{6.7cm}\textrm{ otherwise }
			\end{array}
			\right.
			\end{array}
			\]
			\[
			\begin{array}{rcl}
			\lastp{C,H} &=& \left\{
			\begin{array}{ll}
			x , \;\;\textrm{ if }t=\lastt{C,H}, \{x\} = \{y\in t\circ 
			\mid F(t,y)\cap C \neq \emptyset\}\\
			\hspace{3.3cm} \textrm{or, if } \bot=\lastt{C,H}, C\subseteq M_0(x)\\
			\bot, \hspace{6.6cm} \;\textrm{ otherwise }
			\end{array}
			\right.
			\end{array}
			\]}
}\end{definition}

Note that similarly to backtracking and causal order we recreate broken bonds and re-break created bonds by removing the bonds in the outgoing arcs and adding the bonds in the incoming arcs. Transition reversal in an out-of-causal order can thus be defined as follows: 

\begin{definition}\label{oco-defSingle}{\rm
		Given a SRPN $(A, P,B,T,F)$, an initial marking $M_0$, a state $\langle M, H\rangle$ and a transition $t$ with history $k$ that is $o$-enabled in $\state{M}{H}$, we write $\state{M}{H}
		\otrans{t} \state{M'}{H'}$
		where $H'$ is defined as in Definition~\ref{co-defSingle} and  we have:
		\begin{eqnarray*}
			M'(x) & = & \Big(M(x)\cup \bigcup_{a\in M(y)\cap \effects{t}, \lastp{C_{a,y},H'} =x}C_{a,y}\Big )  	 
			 -\Big(\effect{t}  \cup \bigcup_{a\in M(x)\cap \effects{t}, \lastp{C_{a,x},H'}\neq x}C_{a,x}\Big) 
		\end{eqnarray*}
		where we use the shorthand $C_{b,z} = \connected(b,\{\connected(c,M(z))|c\in A, z\in P\}-\effects{t}) \cup \guard{t})$ for $b\in A$, $z\in P$.
}\end{definition}

\begin{figure}[t]
	\begin{center}
		\includegraphics[height=3cm]{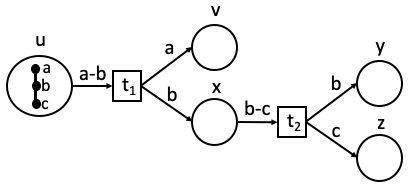}
		\includegraphics[width=1cm]{figures/t1.png}
		\includegraphics[height=3.5cm]{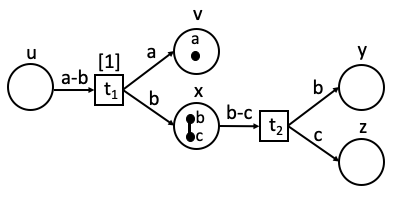}\\
		\includegraphics[width=1cm]{figures/t2.png}
		\includegraphics[height=3.5cm]{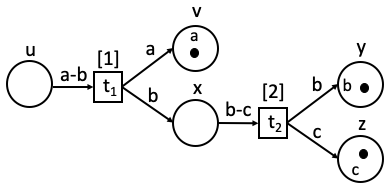}
		\includegraphics[width=1cm]{figures/t1r.png}\\
		\includegraphics[height=3.5cm]{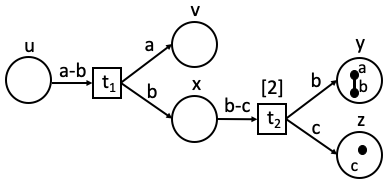}
		\includegraphics[width=1cm]{figures/t2r.png}
		\includegraphics[height=3.1cm]{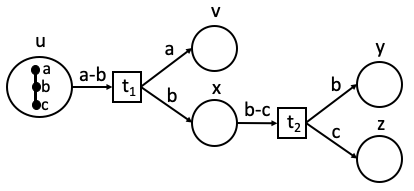}
	\end{center}
	\caption{Out-of-causal order execution}
	\label{singleioutFig}
\end{figure}

Thus, when a transition $t$ is reversed in an out-of-causal-order fashion all bonds that were created by
the transition are undone and all bonds destructed are recreated. If the destruction of a bond divides a component 
into smaller connected sub-components then each of these
sub-components should be relocated (if needed) back to the place where the sub-complex would have existed if transition $t$ never took place, i.e., exactly after the last transition 
that involves tokens from the sub-complex. Otherwise if the recreation of a bond creates a larger component then this component should be moved (if needed) to the place where the complex would have existed if transition $t$ never took place, i.e. exactly after the last transition that involves tokens from the bigger complex. The example in Figure~\ref{singleioutFig} represents the ou-of-causal reversal of two bond breaking transitions $t_1$ and $t_2$. By initially reversing $t_1$ the bond $a\bond b$ is reconstructed and the history of $t_1$ is eliminated. Since reversing a transition is equivalent to skipping the transition in the net, then the bond $a\bond b$ is transferred to place $y$ as if transition $t_1$ has never been executed.

\section{ Token Multiplicity}

 We now proceed to explore token multiplicity in reversing Petri nets. Allowing multiple tokens of the same type to occur in a model entails that tokens of the same type are allowed to execute the same transition. As a transition can be fired by different sets of tokens this introduces possible nondeterminism when going backwards. This nondeterminism is also known as backward conflicts since multiple different tokens are allowed to reverse the same transition resulting in different states. 
%

In order to define reversible semantics for RPNs in the presence of backward conflict we have identified two different approaches. 
The first approach is inspired by the individual token interpretation presented in~\cite{individual} and the second by the collective token interpretation presented in~\cite{ConfStruct,ZeroSafe}.
The two approaches differ on the way they handle backward conflicts, however, both of them remain abstract enough while doing justice to the truly concurrent nature of Petri nets. 
We observe that the individual token interpretation is accompanied by a set of desirable theoretical properties while the collective token interpretation is well suited for modelling a variety of possible applications.

According to the individual token philosophy, tokens are distinguished based on their causal path~\cite{NonSeq,IndividualTokens}. 
The approach distinguishes tokens as individual 
by providing precise correspondence between the token instances and their past. 
Specifically, the model keeps track of where the tokens come from and therefore identifies the causal links between transitions in terms of a partial order.  
In this partial order, causal dependencies  are explicitly defined as an unfolding of an occurrence net which is 
an acyclic net that does not have backward conflicts. 
%
%
%
This approach ensures backward determinism which is a crucial property of reversible systems. 

\begin{figure}[tb]
	\centering
	\includegraphics[height=3.5cm]{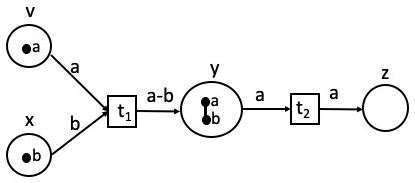}
	\includegraphics[width=1cm]{figures/t1.png}
	\includegraphics[height=3.5cm]{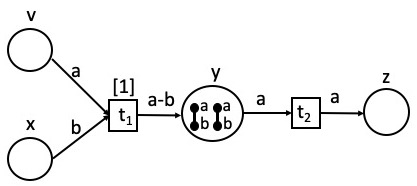}
	\includegraphics[width=1cm]{figures/t2.png}\\
	\hspace{-1.2cm}
	\includegraphics[height=3.5cm]{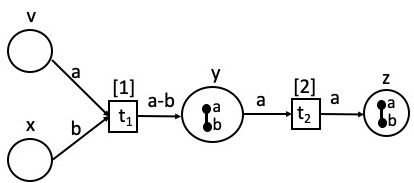}
	\caption{Individual token interpretation}
	\label{iti}
	\vspace{-.2cm}
\end{figure}

Let us consider the example in Figure~\ref{iti}, which illustrates backward determinism as understood by the individual token interpretation. As already discussed  in Chapter 3 the causal relationship between transitions is defined as the manipulation of common tokens. Based on this relationship we are able to uniquely identify the transition that can be reversed by a particular token. If we consider the example in Figure~\ref{iti} after the execution of $t_1$ there are two identical connected components in the middle place. If the component that was already there was used to fire $t_2$, then there is no causal link between the two transitions.  If the component produced by $t_1$ was used to fire $t_2$, then $t_2$ is causally dependent on $t_1$. Depending on how the causal relationships are defined the behaviour of reversible actions changes, as a causal link between $t_1$ and $t_2$ means that transition $t_1$ is unable to reverse until $t_2$ has also reversed. 
%

On the other hand, based on the collective token philosophy, when multiple tokens of the same type reside in the same place then these  tokens are  indistinguishable. The rationality behind this approach is that in the example of Figure~\ref{iti} the preconditions for firing transition $t_2$ do not change and consequently $t_2$ is always independent of $t_1$. This means that all that is known by the model is the amount of token occurrences of a specific type and their location in the marking.  

\section{Multi Reversing Petri Nets}

 In this section we propose an extension of the SRPN model, multi reversing Petri nets, by allowing multiple tokens of the same type as well as the possibility for 
 transitions to break bonds 
 under the individual token interpretation. Thus, we allow tokens of the same type to fire the same transition when going forward, however when going backwards tokens will be able to reverse only the transitions that they have fired. Therefore, the individuality of tokens of the same type is imposed by their causal path. 
 
 We formulate four firing rules for multi reversing Petri nets under the individual token interpretation with multiple tokens, namely forward, backtracking, causal-order reversing, and out-of-causal-order reversing. We then proceed to translate multi reversing Petri nets and single reversing Petri nets into Labelled  Transition Systems (LTSs)~\cite{LTS}. We compare the expressive power offered by multi tokens against that of single tokens, in terms of the associated Labelled  Transition Systems. We conclude that reversing Petri nets with single tokens are equally expressive as reversing Petri nets with multiple tokens.
 

We present multi reversing Petri nets (MRPNs) which are 
Petri net structures with multiple tokens of the same type, which we refer to as multi-tokens that allow transitions to be reversed. Formally, a MRPN
is defined as follows:
\begin{definition}{\rm
		A \emph{multi reversing Petri net} (MRPN) is a tuple $(P,T,  A, A_V, B, F)$ where:
		\begin{enumerate}
			\item $P$ is a finite set of \emph{places} and
			$T$ is a finite set of \emph{transitions}.
			\item $A$ is a finite set of \emph{base} or \emph{token types} ranged over by $a, b,\ldots$
			\item $A_V$ is a finite set of \emph{token variables}. We write $\type(v)$ for the type
			of variable $v$ and assume that $\type(v) \in A$ for all $v\in A_V$.
			\item $B\subseteq A\times A$ is a finite set of undirected \emph{bond} types ranged over
			by $\beta,\gamma,\ldots$ We use the notation $a \bond b$ for a bond $(a,b)\in B$.  
			\item $F : (P\times T  \cup T \times P)\rightarrow {\cal P}( A_V \cup (A_V \times A_V))$ 
			is a set of directed labelled \emph{arcs}.
		\end{enumerate}
}\end{definition}

A multi reversing Petri net is built on the basis of a set of \emph{tokens} or \emph{bases}. These
are organized in a set of token types $A$, where each token type is associated with a set of token instances. $A_I$ defined as follows:

\begin{definition}\label{recursive}{\rm	
		Given a multi reversing Petri net $(P, T, A, A_V,B,F)$ the set of token instances $A_I$ is recursively defined by:
		\begin{itemize}
			\item[--] $(a,*,i)$, $i\in \mathbb{N}$, $a\in A$, and	
			\item[--] $(a_i,k,u)$ where $a_i\in A_I$, $k\in \mathbb{N}$ and $u \in  \{*\} \cup \{A_V \}$.
		\end{itemize} 
		For $a_i,a_j \in A_I$ we  use the notation $a_i \overline{\in}a_j$ if (i) $a_i=a_j$ or (ii) $a_j=(a_j',k,u)$, $a_i \overline{\in}a_j'$. 
		 Moreover, we define 
		 	\[
		 a_j\downarrow a_i = \left\{
		 \begin{array}{ll}
		 a_i \hspace{6.1cm} \textrm{ if } a_j=(a_i,k,u) \; \\
		 (a_j'\downarrow a_i,k_j,u_j) \hspace{2.4cm}   \textrm{ if }  a_j=(a_j',k_j,u_j), a_j'\neq a_i\;
		 \end{array}
		 \right.			\]
		 
}\end{definition}

The set of token instances $A_I$  corresponds to the basic entities that occur in a system. Initially tokens of type $a\in A$ are denoted by $(a,*,i)$ where $type((a,*,i))=a$ and $i\in\mathbb{N}$ is a unique number that distinguishes tokens from each other. As computation proceeds the tokens evolve by extending their memory whenever they fire a transition in the forward direction or decreasing their memory whenever they execute a transition in reverse. Note that in a token of the form $(a_i,k,u)$, $k$ is the key of the last transition the token has engaged in and $u$ the variable to which the token was assigned, with $u=*$ for tokens that participate in the transition but do not correspond to any variables.  Token instances may occur as 
stand-alone elements but they may also merge together to form \emph{bonds}. Bond instances are denoted by the set $B_I$ and are formed similarly to the other variations of RPNs.  For $a_i,a_j \in A_I$ we  use the notation $a_i \overline{\in}a_j$ to denote that token $a_i$ has evolved to $a_j$. As such, the memories of $a_i$ are also part of the memory of $a_j$.  For $a_i,a_j \in A_I$ we  use the notation 	 $a_j\downarrow a_i$ to denote the removal of a specific memory in token $a_j$. Specifically, by replacing $(a_i,k,u)$  with $a_i$ we remove  from $a_j$ the memory of transition occurrence $(t,k)$ where $a_i$, and as a result $a_j$, has participated in. 

As with the original RPN model, places and transitions are connected via  labelled directed arcs. These labels are derived from $A_V \cup (A_V \times A_V)$. They express the requirements and the effects of the transition based solely on the type of tokens consumed. Thus, any token corresponding to the same type as the variable on the labelled arc is able to participate in the transition.  
More precisely, if $F(x,t) = U\cup V$, 
where $U\subseteq A_V$, $V\subseteq 
A_V \times A_V$, 
this implies that for the transition to fire for each $u\in U$ a distinct token instance of type $type(u)$ is 
required. These instances should be bonded together according to $V$. 
Similarly, if 
$F(t,x) = U\cup V$, where $U\subseteq A_V$, $V\subseteq 
A_V \times A_V$,
this implies that during the forward execution of the transition
for each $u\in U$ a token instance of type $type(u)$ will be 
transmitted to place $x$ by the transition, in addition to  the bonds specified by
$V$, some of which will be created as an effect of the transition.
We make the assumption
that if $(u,v)\in V$ then $u,v\in U$.  


We restrict our attention to well-formed MRPNs defined as follows: 
 
\begin{definition}\label{multiwell-formed}{\rm 
		A multi reversing Petri net $(P,T,A,A_V,B,F)$ is \emph{well-formed}, if for all $t\in T$:
		\begin{enumerate}
			\item $A_V\cap \guard{t} = A_V\cap \effects{t}$, and
			\item $ F(t,x)\cap F(t,y)\cap A_V=\emptyset$ for all $x,y\in P$, $x\neq y $. 
		\end{enumerate}
}\end{definition}
Thus, a multi  reversing Petri net is well-formed if (1) whenever a variable exists in
the incoming arcs of a transition then it also exists on the outgoing arcs, which implies that transitions do not
erase tokens, and  (2) 
tokens/bonds cannot be cloned into more than one outgoing places.

As with RPNs the association of token/bond instances to places is called a \emph{marking}  such that 
$M: P\rightarrow 2^{A_I\cup B_I}$, where we assume that if $(u,v)\in M(x)$ then $u, v\in M(x)$. 
In addition, we employ the notion of a \emph{history}, which~assigns a memory to each
transition $H : T\rightarrow 2^\mathbb{N}$. 
Intuitively, a history of $H(t) = \emptyset$ for some $t \in T$ captures that the transition has not taken place, or 
every execution of it has been reversed, and a history
of $ k \in H(t)$, captures that the transition was executed as the  $k^{th}$ transition occurrence.
Note that $|H(t)|>1$ may
arise due to cycles but also due to the consecutive execution of the transition by different token
instances. A pair of a marking and a history, $\state{M}{H}$, describes a \emph{state} of a MRPN 
with $\state{M_0}{H_0}$ the initial state, where $H_0(t) = \emptyset$ for all $t\in T$ and if $a_i \in M_0(x),x\in P$, then $a_i=(a,*,i), i \in \mathbb{N}, a\in A$. Graphically, token variables $u\in F(x,t) \cap A_V$ of type $type(v)=a$ are denoted by $u:a$ over the corresponding arc $F(x,t)$ (respectively for $F(t,x)$). 

Finally, we define $\connected(a_i,C)$, where $a_i\in A_I$ and $C\subseteq 2^{A_I\cup B_I}$,
to be the tokens connected
to $a_i$  as well as the bonds creating these connections according to 
set $C$, in the usual way. 

 SRPNs are a special case of MRPNs as tokens in SRPNs correspond to tokens in MRPNs with the associated memories ignored. Additionaly, tokens are explicitly requested on the directed arcs of SRPN transitions where in MRPNs a variable is used to represent tokens of the same type.  


\section{Semantics Under the Individual Token Interpretation}

We may now define the forward and backward execution within multi reversing Petri nets. Note that as in Section 4.1 we allow transitions to break bonds and we
restrict our attention to  well-formed MRPNs $(P,T,A,A_V,B,F)$ with initial
marking $M_0$ such that for all $a_i\in A_I$, $|\{ x \mid a_i\in M_0(x)\} | = 1$.  

\subsection{Forward Execution}
During the forward execution of a transition in a MRPN, 
a set of tokens and bonds, as specified by the incoming arcs of the transition, are selected and
moved to the outgoing places of the transition, as specified by the transition's outgoing arcs, possibly
forming or destructing bonds, as necessary. Due to the presence of multiple instances of the same token
type, it is possible that different token instances are selected during the transition's execution.

A transition is forward-enabled in a state $\state{M}{H}$ of a MRPN if there exists a selection of token instances 
available at the incoming places of the transition 
matching the requirements  on the transitions incoming arcs. Also the transition should not recreate bonds or clone tokens. Formally:

\begin{definition}\label{multifenabled}{\rm
		Given a MRPN $(P,T, A, A_V, B, F)$, a state $\state{M}{H}$, and a transition $t$, we say that $t$ is 
		\emph{forward-enabled} in $\state{M}{H}$  if there exists an injective function 
		$U_f:\guard{t}\cap A_V\rightarrow A_I$ such that:
		
		\begin{enumerate}
			\item for all $u\in F(x,t)$, $x\in \circ t$, then $U_f(u)\in M(x)$ where $\type(u)=\type(U_f(u))$, and for all $(u,v)\in F(x,t)$, for some $x\in \circ t$,  then $(U_f(u),U_f(v))\in M(x)$,
			\item for all $u,v \in F(x,t), x\in \circ t$ and $(U_f(u),U_f(v)) \in M(x)$, then $(u,v) \in F(x,t)$, and 
			\item if $u \in F(t,y_1)$, $v \in F(t,y_2)$, $y_1,y_2\in t\circ $, $y_1\neq y_2$ then $U_f(u)\not\in \connected( U_f(v), (M(x) -$\\$\pre{t,U_f}) \cup$$ \post{t,U_f}),x\in \circ t$.
		\end{enumerate}
	where 
	$ \pre{t,U}=\{U(u)|u\in F(x,t), x \in \circ t\} \cup \{(U(u),U(v))|(u,v)\in F(x,t), x \in \circ t\} $ and $\post{t,U} =\{U(u)|u\in F(t,y), y \in t\circ \} \cup \{(U(u),U(v))|(u,v)\in F(t,y), y\in t\circ \} $.
}\end{definition}

Thus, $t$ is enabled in state $\state{M}{H}$ if (1)  there is a type-respecting assignment of token instances in the incoming places of the transition to the variables on the incoming edges, with the token instances originating from the appropriate input places and where tokens are connected with bonds as required by the transition's incoming edges,
(2) if the selected token instances are bonded together in an incoming place of the transition then the bond should also exist on the variables labelling the incoming arcs   (thus transitions do not recreate bonds), and  (3) 
if two token instances are transferred by a transition to different outgoing places then these tokens should not remain connected when removing the selected incoming tokens and adding the selected outgoing tokens (we do not clone tokens). We use $\pre{t,U}$ and $\post{t,U}$ to help us identify the effect of the transition $t$ on the particular selection of token instances $U$. 
%
We refer to $U_f$ as a forward enabling assignment.

We now define the incoming token/bond instances as:

\begin{definition}\label{multisetsf1}{\rm
		Given a MRPN $(P,T, A, A_V, B, F)$, a state $\state{M}{H}$, a transition $t$ and an
		enabling assignment $U_f$, we define  $^\bullet\!{U}_f :P\rightarrow 2^{A_I\cup B_I }$ to be a function that assigns to each place a set of incoming token and bond instances that are used for the firing of $t$:
		
		$^\bullet\!{U}_f (x)=\bigcup_{u\in F(x,t)} \connected(U_f(u), M(x))$ 
		
}\end{definition}

We now define the outgoing token/bond instances as:

\begin{definition}\label{multisetsf2}{\rm
		Given a MRPN $(P,T,A,A_V,B,F)$, a state $\state{M}{H}$, a transition $t$, and an enabling assignment $U_f$, we define $U_f^\bullet :P\rightarrow 2^{A_I\cup B_I}$ to be a function that assigns to each place a set of  outgoing token/bond instances of $t$:
		
		$U_f^\bullet (x)=  \bigcup_{u\in F(t,x),U_f(u)\in M(y)} \connected(U_f(u), (M(y)-\pre{t,U_f}) \cup \post{t,U_f})$
		
}\end{definition}

To execute a transition $t$ according to an enabling assignment $U_f$, 
the selected token instances, along with their connected components,
are relocated to the outgoing places of the transition as specified
by the outgoing arcs, with bonds created and destructed accordingly. 
Furthermore,
the history of the executed transition is updated in the standard way. As the same transition can be executed by different tokens of the same type we indicate transition firings by $(t,k)$ in order to be able to identify the set of tokens that have participated in this specific transition occurrence. In Figure~\ref{multiforwardFig} we observe the change in history of transition $t$, as well as, the change in the name of the token instances $(a,*,1)$, and $(b,*,1)$ to $((a,*,1),1,u)$ and $((b,*,1),1,v)$, respectively. Thus, the memory of token instance $(a,*,1)$ is extended to indicate that $(a,*,1)$ has participated in transition $t$ with history identifier $1$ corresponding to variable $u$.  Similarly, for token instance $(b,*,1)$ and variable $v$. Specifically,
we define:

\begin{figure}[t]
	\begin{center}
		\textbf{}
		\includegraphics[height=3cm]{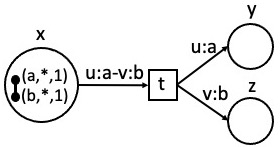}
		\\
		\vspace{.8cm}
		\includegraphics[width=1cm]{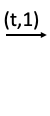}
		\includegraphics[height=3.5cm]{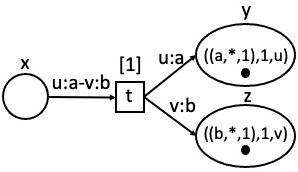}
	\end{center}
	\caption{Forward execution}
		\label{multiforwardFig}
\end{figure}

\begin{definition}{\rm \label{multiforward}
		Given a MRPN $(P,T, A, A_V, B, F)$, a state $\langle M, H\rangle$, a transition $t$ that is enabled in state $\langle M, H\rangle$,  and an enabling assignment $U_f$, 
		 we write $\state{M}{H}
		\trans{(t,k)} \state{M'}{H'}$
		where $k=max(\{0\} \cup \{k'|k'\in H(t'),t'\in T\})+1$ and for all $x\in P$:
		\begin{eqnarray*}
			M'(x) =   (M(x)- ^\bullet\!{U}_f(x)) 
		\cup (\bigcup_{a_i\in U_f^\bullet(x)} (a_i,k,V(a_i)) 
		\cup \bigcup_{(a_i,b_i)\in U_f^\bullet(x)} ((a_i,k,V(a_i)), (b_i,k,V(b_i)))
		\end{eqnarray*}
	where
		\[\hspace{1.9cm}
  V(a_i) = \left\{
	\begin{array}{ll}
u \hspace{3.6cm} \textrm{ if } U_f(u)=a_i \; \\
	* \hspace{4.1cm}  \;\textrm{ otherwise }
	\end{array}
	\right.			\]
		\[
		\hspace{0.2in}\mbox{and}\hspace{0.2in}
		H'(t') = \left\{
		\begin{array}{ll}
		H(t') \cup  \{k\} \hspace{2.8cm} \textrm{ if } t' = t  \; \\
		H(t'), \hspace{3.1cm}  \;\textrm{ otherwise }
		\end{array}
		\right.			\]
}\end{definition} 

The following proposition states that tokens are preserved throughout forward execution such that the amount of tokens of the same type remains the same. 
\begin{proposition} [Token preservation]  \label{multiprop1}    {\rm Consider a MRPN $(P,T,A,A_V,B,F)$, a state $\state{M}{H}$ and a transition 
			$\state{M}{H} \trans{(t,k)} \state{M'}{H'}$. Then for all $a_i\in A_I\cap M_0(z)$, $z\in P$ we have $|\{a_j\mid a_j\in M(x)\cap A_I, x\in P, a_i\overline{\in}a_j\}|=|\{a_j'\mid a_j'\in M'(y)\cap A_I, y\in P, a_i \overline{\in} a_j'\}|=1$.
	}\end{proposition}
	\paragraph{Proof}
	The proof  follows from the definition of forward execution and relies on the well-formedness of
	MRPNs. Consider a MRPN $(P,T,A,A_V,B,F)$, a state $\state{M}{H}$   such that $|\{a_j \mid a_j\in M(x)\cap A_I, x\in P, a_i\overline{\in}a_j\}| = 1$ for
	some $a_i\in A_I\cap M_0(z)$, $z\in P$, and suppose 
	$\state{M}{H} \trans{(t,k)} \state{M'}{H'}$ such that $|\{a_j'\mid a_j'\in M'(y)\cap A_I, y\in P, a_i \overline{\in}a_j'\}| = n$. Let $a_j\in A_I$. Two cases exist:
	\begin{enumerate}
		\item $a_j\in \connected(b_j,M(x))$ for some $b_j=U_f(v), v\in F(x,t)$. According to Definition~\ref{multisetsf1}, 
		we have that $a_j\in ^\bullet\!\!{U_f(x)}$, which by Definition~\ref{multiforward} implies that $a_j\not\in M'(x)$.
		On the other hand,  by  Definition~\ref{multiwell-formed}(1),
		$v\in \effects{t}$. Thus, there exists $y\in t\circ$, such that $v\in F(t,y)$. Note that this $y$
		is unique by Definition~\ref{multiwell-formed}(2).  As a result, by Definition~\ref{multisetsf2}, 
		$a_j\in U_f^\bullet (y)$ which by Definition~\ref{multiforward} yields $a_j\overline{\in}a_j'$, $a_j'\in M'(y)$ such that $a_i\overline{\in}a_j'$.\\
		Now suppose that $a_j\in \connected(c_j,M(x))$ for some  $c_j\neq b_j$, $u\in F(t,y'),U_f(u)=c_j$. Then, by Definition~\ref{multiwell-formed}(2), 
		it must be that $y = y'$. As a result, we have that $n=|\{a_j'\mid a_j'\in M'(y')\cap A_I, y'\in P, a_i \overline{\in} a_j' \}| = |\{a_j \mid a_j \in M(x)\cap A_I, x\in P, a_i\overline{\in}a_j\}|=1$ and 
		the result follows.
		\item $a_j\not\in \connected(b_j,M(x))$ for all $v\in F(x,t),U_f(v)=b_j$, $x\in P$. This implies that 
		$1=|\{a_j \mid a_j \in M(x)\cap A_I, x\in P, a_i\overline{\in}a_j\}|= |\{a_j \mid a_j \in M'(x), x\in P,a_i\overline{\in}a_j\}|=n$ and the result follows.
	\end{enumerate}
	\proofend

\subsection{Backtracking}

Let us now proceed to backtracking. 
A transition can be reversed in a certain state if it was the last executed transition  and there exist token instances in its output places 
that match the requirements on its outgoing arcs. 
To capture this, we define transition occurrence as follows:

\begin{definition}\label{multitransOccurr}{\rm
		Consider a MRPN $(P,T,A,A_V,B,F)$, a state $\state{M}{H}$, and a transition $t$. We refer to $(t,k)$  as a \emph{transition occurrence} in $\state{M}{H}$ if $k\in H(t)$.
}\end{definition}

We now define the notion of backtracking enabledness
as follows.
\begin{definition}\label{multibt-enabled}{\rm
		Consider a MRPN $(P,T, A, A_V, B, F)$, a state $\state{M}{H}$, and a transition occurrence $(t,k)$.
		We say that $(t,k)$ is \emph{bt-enabled} in $\state{M}{H}$ if (1)  $k\in H(t)$ with $k\geq k'$ for all $k' \in H(t')$, $t'\in T$, and (2) there exists an injective function 
		$U_b:\effects{t}\cap A_V\rightarrow  A_I$ such that:
 	\begin{enumerate}[(a)] 
			\item  for all $u\in F(t,x)$, $x\in t\circ$, we have $U_b(u)=(a_i,k,u),$ $ U_b(u) \in M(x)$ where $type(u)=type(U_b(u))$, and
			\item for all $(u,v)\in F(t,x)$, $x\in t\circ$, we have $(U_b(u),U_b(v))=((a_i,k,u),(b_i,k,v))$, $ (U_b(u),$\\$U_b(v)) \in M(x)$.
		\end{enumerate}
}\end{definition}

Thus, a transition $t$ is $bt$-enabled in $\state{M}{H}$ if  
(1) it was the last transition to be executed, and (2) there exists a type-respecting assignment of token instances in the outgoing places of the transition, to the variables on the outgoing
edges of the transition, and where the tokens are connected with bonds as required by
the transition's outgoing edges.
We refer to $U_b$ as a backtracking enabling assignment.

Similarly to forward execution, the following definition selects the incoming connected components and the outgoing connected components. Note that the incoming connected components are selected based on the outgoing arcs of the transition and the outgoing connected components are selected based on the incoming arcs. We now define the  incoming token/bond instances as:

\begin{definition}\label{multisetsb1}{\rm
		Given a MRPN $(P,T, A, A_V, B, F)$, a state $\state{M}{H}$, a transition $t$ and an
		enabling assignment $U_b$, we define  $^\bullet\!{U}_b:P\rightarrow 2^{A_I\cup B_I}$ to be a function that assigns to each place a set of incoming token and bond instances that are used for the backtracking of $t$:
		
		$^\bullet\!{U}_b(x)=\bigcup_{u\in F(t,x)} \connected(U_b(u), M(x)) $
		
}\end{definition}

We now define the outgoing token/bond instances as:

\begin{definition}\label{multisetsb2}{\rm
		Given a MRPN $(P,T, A, A_V, B, F)$, a state $\state{M}{H}$, a transition $t$, and an enabling assignment $U_b$, we define $U_b^\bullet :P\rightarrow 2^{A_I\cup B_I}$ to be a function that assigns to each place a set of outgoing token/bond instances of $t$:
		
		$U_b^\bullet (x)=  \bigcup_{u\in F(x,t),U_b(u)\in M(y)} \connected(U_b(u), (M(y)-\post{t,U_b}) \cup \pre{t,U_b})$ 
		
}\end{definition}

\begin{figure}[t]
	\begin{center}
		\hspace{.8cm}
		\includegraphics[height=3cm]{figures/forward1.jpeg}\\
		\includegraphics[width=1cm]{figures/t,1.png}
		\includegraphics[height=3.3cm]{figures/forward2.jpeg}\\
		\includegraphics[width=1cm]{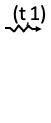}
		\includegraphics[height=3cm]{figures/forward1.jpeg}
	\end{center}
	\caption{Backtracking execution }
	\label{multibacktrackingFig}
\end{figure}

To implement the reversal of a transition $t$ according to a backtracking enabling assignment $U_b$, 
the selected instances are relocated from  the outgoing places of the transition to the incoming places,
as specified
by the incoming arcs of the transition, with bonds created and destructed accordingly. In Figure~\ref{multibacktrackingFig} the backtracking execution of Figure~\ref{multiforwardFig} is illustrated, where we can observe the history of the reversing transition  being eliminated and the token instances returning to their initial place. Specifically we define:

\begin{definition}{\rm \label{multibt}
		Given a MRPN $(P,T, A, A_V, B, F)$, a state $\langle M, H\rangle$,  a transition occurrence $(t,k) $ that is $bt-$enabled and an
		enabling assignment $U_b$,
		 we write $\state{M}{H}
		\btrans{(t,k)} \state{M'}{H'}$
		where for all $x\in P$:
		\begin{eqnarray*}
			M'(x) =   (M(x)- ^\bullet\!{U}_b(x))
			\cup (\bigcup_{(a_i,k,u)\in U_b^\bullet(x)
			} a_i 
		\cup \bigcup_{((a_i,k,u),(b_i,k,v))\in U_b^\bullet(x)
		} (a_i,b_i))
		\end{eqnarray*}
		\[
		\hspace{0.2in}\mbox{and}\hspace{0.2in}
		H'(t') = \left\{
		\begin{array}{ll}
		H(t') - \{k\}, \hspace{.5cm} \textrm{ if } t' = t  \; \\
		H(t'), \hspace{.7cm}  \;\textrm{ otherwise }
		\end{array}
		\right.			\]	
		
}\end{definition} 

The following proposition states that tokens are preserved throughout backtracking execution such that the amount of tokens of the same type remains the same. 
 
\begin{proposition}  [Token preservation] \label{multiprop2}
	{\rm Consider a multi reversing Petri net $(P,T,A,A_V,B,F)$, a state $\langle M, H\rangle$, and a transition 
		$\state{M}{H} \btrans{(t,k)} \state{M'}{H'}$. Then for all $a_i\in A_I\cap M_0(z)$, $z\in P$ we have $|\{ a_j \mid a_j \in M(x)\cap A_I,x\in P, a_i\overline{\in}a_j|=|\{ a_j' \mid a_j' \in M'(y)\cap A_I,y\in P, a_i\overline{\in}a_j'\}|=1$.
}\end{proposition}
\paragraph{Proof}
The proof of the result follows the definition of backward 
execution and relies on the well-formedness of
multi reversing Petri nets. 
Consider a MRPN $(P,T,A,A_V,B,F)$, a state $\langle M, H\rangle$ such that for all $a_i\in A_I\cap M_0(z)$, $z\in P$ we have $|\{ a_j \mid a_j \in M(x)\cap A_I,x\in P, a_i\overline{\in}a_j\}|=1$, and suppose  $\state{M}{H} \btrans{(t,k)} \state{M'}{H'}$ such that $|\{ a_j' \mid a_j' \in M'(y)\cap A_I,y\in P, a_i \overline{\in} a_j'\}|=n$.  Two cases exist:
\begin{enumerate}
	
	\item $a_j\in \connected(b_j,M(x))$ for some $b_j=U_b(v),v\in F(t,x)$. 
	Let us choose $b_j$ such that $a_j\in \connected(b_j,(M(x) - \post{t,U_b}) \cup \pre{t,U_b}  )$. Note
	that such a $b_j$ must exist, otherwise 
	the forward execution of $t$ would not have transferred $a_j$ along with $b_j$ to place $x$.
	
	According to Definition~\ref{multisetsb1}, 
	we have that $a_j \in ^\bullet\!{U_b(x)}$, 
	which implies that $a_j\not\in M'(x)$.
	On the other hand, note that by the definition of well-formedness, Definition~\ref{multiwell-formed}(1),
	$v\in \guard{t}$. Thus, there exists $y\in \circ t$, such that $v\in F(y,t)$. Note that this $y$
	is unique by Definition~\ref{multiwell-formed}(2).   As a result, by Definition~\ref{multisetsb2}, 
	$a_j \in U_b ^\bullet(y)$. Since
	$v\in F(y,t)\cap F(t,x)$, $a_j\in \connected(b_j,(M(x)- \post{t,U_b}) \cup \pre{t,U_b}) $, this implies that $a_j' \overline{\in} a_j, a_j'\in M'(y)$ where $a_i \overline{\in} a_j'$. 
	
	Now suppose that $a_j\in \connected(c_j,(M(x)- \post{t,U_b}) \cup \pre{t,U_b})$, $c_j\neq b_j$, and $c_j\in F(y',t)$. Since
	 $a_j\in \connected(b_j,(M(x)- \post{t,U_b}) \cup \pre{t,U_b})$, it must be that
	$ \connected(b_j,(M(x)- \post{t,U_b}) \cup \pre{t,U_b})=\connected(c_j,(M(x)- \post{t,U_b}) \cup \pre{t,U_b})$. Since $b_j$ and $c_j$ are
	connected to each other but the connection was not created by transition $(t,k)$ (the connection is
	present in $(M(x)- \post{t,U_b}) \cup \pre{t,U_b}$), it must be that the connection was already present before
	the forward execution of $t$ and, by token uniqueness, we conclude that  $y=y'$ and therefore  $1=|\{ a_j \mid a_j \in M(x)\cap A_I,x\in P, a_i\overline{\in}a_j\}|=|\{ a_j' \mid a_j' \in M'(y')\cap A_I,y'\in P, a_i\overline{\in}a_j'\}|=n$.
	\item $a_j\not\in \connected(b_j,M(x))$ for all $b_j=U_b(v),v\in F(t,x)$. This implies that 
	 $|\{ a_j\mid a_j \in M(x)\cap A_I,x\in P,a_i\overline{\in}a_j\}|=|\{ a_j' \mid a_j' \in M'(x)\cap A_I,x\in P,a_i\overline{\in}a_j'\}|=1$ and the result follows.
\end{enumerate}
	\proofend

We may establish a loop lemma:
	\begin{lemma}[Loop]\label{multiloopb}{\rm 
			For any forward transition $\state{M}{H}\trans{(t,k)}\state{M'}{H'}$ there exists a backward transition
			$\state{M'}{H'} \btrans{(t,k)} \state{M}{H}$ and vice versa. 
	}\end{lemma}
	\paragraph{Proof}
	{Suppose $\state{M}{H}\trans{(t,k)}\state{M'}{H'}$. Then $t$ is clearly $bt$-enabled in $H'$. Furthermore,
		$\state{M'}{H'} \btrans{(t,k)} \state{M''}{H''}$ where $H'' = H$. In addition, all tokens and bonds
		involved in transition $t$ (except those that have been created but including those that have been broken by $t$) will be returned from the outgoing places
		of transition $t$ back to its incoming places. Specifically, for all $a_i\in A_I$, it
		is easy to see by the definition of $\btrans{}$  that $a_i\in M''(x)$ if and only if $a_i\in M(x)$.
		Similarly,  for all $\beta_i\in B_I$,  $\beta_i\in M''(x)$ if and only if
		$\beta_i\in M(x)$. The opposite direction can be argued similarly, only this
		time tokens and bonds involved in transition $t$ will be moved from the incoming places to
		the outgoing places of transition $t$.}
	\proofend

\subsection{Causal Order Reversing}

We now move on to  reversing transitions in causal order. Causal dependence is 
determined by the path that tokens follow: two transition occurrences are causally 
dependent, if a token produced by the one occurrence was subsequently used to
fire the other.
To capture this type of dependencies, we adopt the following 
definition of causal dependence.

\begin{definition}\label{multico-dep}{\rm
		Consider a MRPN $(P,T,A,A_V,B,F)$, 
		a state  $\state{M}{H}$ and suppose $(t,k)$ and $(t',k')$ are transition occurrences in $\state{M}{H}$.
		We say that $(t',k')$ \emph{causally depends} on $(t,k)$ denoted by $(t,k)\prec 
		(t',k')$, if $k<k'$ and  there exists $a_i\in M(x),x\in P,$ such that $(a_j,k,u)\overline{\in} a_i$ and $(a_j',k',u')\overline{\in}a_i$.
}\end{definition}

As tokens in multi reversing Petri nets are associated with their causal path, we are able to identify the transitions that each token has participated in by observing the memory of the token. When the keys of two transitions belong to the memory of the same token then it means that this token has participated in both transitions. Thus, a transition occurrence $(t',k')$ causally depends on a preceding 
transition occurrence $(t,k)$ if one or more tokens used during the 
firing of $(t',k')$ was also used for the firing of $(t,k)$.

A transition can be reversed in a certain state if there are no transitions causally following it and there exist token instances in its output places 
that match the requirements on its outgoing arcs. Specifically, we define the notion of reverse enabledness
as follows.
\begin{definition}\label{multico-enabled}{\rm
		Consider a MRPN $(P,T, A, A_V, B, F)$, a state $\state{M}{H}$, and a transition occurrence $(t,k)$.
		We say that transition occurrence $(t,k)$ is \emph{c-enabled} in $\state{M}{H}$ if (1) there is no transition occurrence $(t', k' ) \in \state{M}{H}$ with $(t,k) \prec (t',k')$, and (2)  there exists an injective function 
		$U_c:\effects{t}\cap A_V\rightarrow  A_I$ such that:
\begin{enumerate}[(a)] 
			\item 
			for all $u\in F(t,x)$, $x\in t \circ$, we have $U_c(u)=(a_i,k,u), U_c(u) \in M(x) $ where $type(u)=type(U_c(u))$, and 
			\item for all $(u,v)\in F(t,x)$, $x\in t \circ$, we have $(U_c(u),U_c(v))=((a_i,k,u),(b_i,k,v)), $\ $(U_c(u),$\\$U_c(v)) \in M(x)$. 
		\end{enumerate}
}\end{definition}

Thus, a transition occurrence $(t,k)$ is c-enabled in $\state{M}{H}$ if  
(1) there are no transitions causally dependent on it, and (2) there exists a type-respecting assignment of token instances in the outgoing places of the transition, to the variables on the outgoing
edges of the transition, and where the tokens are connected with bonds as required by
the transition's outgoing edges.  
We refer to $U_c$ as a causal enabling assignment.

We now define the incoming token/bond instances as:

\begin{definition}\label{multisetsc1}{\rm
		Given a MRPN $(P,T, A, A_V, B, F)$, a state $\state{M}{H}$, a transition $t$ and an
		enabling assignment $U_c$, we define  $^\bullet\!{U}_c:P\rightarrow 2^{A_I\cup B_I}$ to  be a function that assigns to each place a set of  incoming token and bond instances that are used for the reversing of $t$ where for all $x\in P$, $^\bullet\!{U}_c(x)$ is defined as $^\bullet\!{U}_b(x)$ in Definition~\ref{multisetsb1} with $U_b$ replaced by $U_c$. 
		
}\end{definition}

We now define the outgoing token/bond instances as:

\begin{definition}\label{multisetsc2}{\rm
		Given a MRPN $(P,T, A, A_V, B, F)$, a state $\state{M}{H}$, a transition $t$, and an enabling assignment $U_c$, we define $U_c^\bullet :P\rightarrow 2^{A_I\cup B_I}$ to be a function that assigns to each place a set of   outgoing token/bond instances of $t$ where for all $x\in P$, $U_c^\bullet(x)$ is defined as $ U_b ^\bullet(x)$ in Definition~\ref{multisetsb2} with $U_b$ replaced by $U_c$. 
		
}\end{definition}

\begin{figure}[t]
	\begin{center}
		\includegraphics[height=4cm]{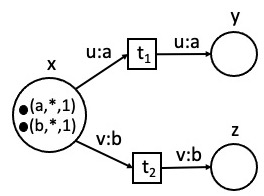}
		\includegraphics[width=1cm]{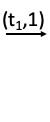}
		\includegraphics[height=4cm]{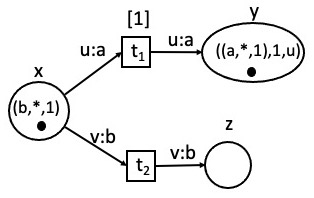}
		\includegraphics[width=1cm]{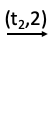}
		\includegraphics[height=4cm]{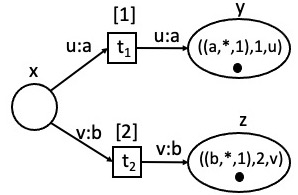}
		\includegraphics[width=1cm]{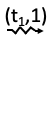}
		\includegraphics[height=4cm]{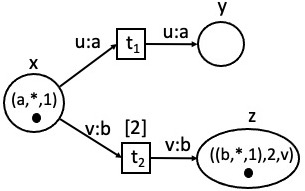}
		\includegraphics[width=1cm]{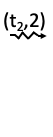}
		\includegraphics[height=4cm]{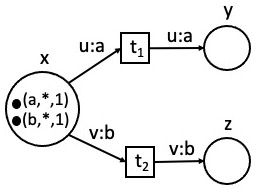}
	\end{center}
	\caption{Causal-order execution}
	\label{multicausalFig}
\end{figure}

To implement the reversal of a transition $t$ according to a causal enabling assignment $U_c$, 
the selected instances are relocated from  the outgoing places of the transition to the incoming places,
as specified
by the incoming arcs of the transition, with bonds created and destructed accordingly. In Figure~\ref{multicausalFig} we can observe causal order reversal of transitions $t_1$ and $t_2$ where the history of transitions and the memories of token/bond instances are updated as defined by the definition bellow:

\begin{definition}{\rm \label{multicausal}
		Given a MRPN $(P,T, A, A_V, B, F)$, a state $\langle M, H\rangle$,  a transition occurrence $(t,k)$ that is $c-$enabled and an
		enabling assignment $U_c$, 
		 we write $\state{M}{H}
		\ctrans{(t,k)} \state{M'}{H'}$
		where $H$ is updated as in Definition~\ref{multibt} and for all $x\in P$:
		\begin{eqnarray*}	
			M'(x) =   (M(x)- ^\bullet\!{U}_c(x))
			\cup (\bigcup_{(a_i,k,u)\in U_c^\bullet(x )
			} a_i 
			\cup \bigcup_{((a_i,k,u),(b_i,k,v))\in U_c^\bullet(x)
		} (a_i,b_i))
		\end{eqnarray*}
		
		
		
		
}\end{definition}

The following proposition states that tokens are preserved throughout backtracking execution such that the amount of tokens of the same type remains the same. 

\begin{proposition}  [Token preservation] \label{multiprop2C}
	{\rm Consider a multi reversing Petri net $(P,T,A,A_V,B,F)$, a state $\langle M, H\rangle$, and a transition 
		$\state{M}{H} \ctrans{(t,k)} \state{M'}{H'}$. Then for all $a_i\in A_I\cap M_0(z)$, $z\in P$ we have $|\{ a_j \mid a_j \in M(x)\cap A_I,x\in P, a_i\overline{\in}a_j|=|\{ a_j' \mid a_j' \in M'(y)\cap A_I,y\in P, a_i \overline{\in}a_j'\}|=1$.
}\end{proposition}
\paragraph{Proof}
The proof follows along the same lines as that of Proposition~\ref{multiprop2} with $\btrans{}$ replaced by $\ctrans{}$. 
\proofend

	We may now establish the causal consistency of our semantics. 
	First, we define 
	some auxiliary notions. Given a transition $\state{M}{H}\fctrans{(t,k)}\state{M'}{H'}$,
	we say that the \emph{action} of the transition is
	$(t,k)$ if $\state{M}{H}\trans{(t,k)}\state{M'}{H'}$ and $(\underline{t,k})$ 
	if $\state{M}{H}\ctrans{(t,k)}\state{M'}{H'}$
	and we may write $\state{M}{H}\fctrans{(\underline{t,k})}\state{M'}{H'}$. 
	We use $\alpha$ to
	range over $\{(t,k), \mid t\in T\}$, $\underline{\alpha}$ to range over $\{(\underline{t,k}), \mid t\in T\}$. 
Given an execution 
	$\state{M_0}{H_0}\fctrans{\alpha_1}\ldots 
	\fctrans{\alpha_n}\state{M_n}{H_n}$, we say that the \emph{trace} 
	of the execution is
	$\sigma=\langle \alpha_1,\alpha_2,\ldots,\alpha_n\rangle$, 
	 and write
	$\state{M}{H}\fctrans{\sigma}\state{M_n}{H_n}$. Given $\sigma_1 = 
	\langle \alpha_1,\ldots,\alpha_k\rangle$, $\sigma_2 = 
	\langle \alpha_{k+1},\ldots,\alpha_n\rangle$,
	we write $\sigma_1;\sigma_2$ for $\langle \alpha_1,\ldots,\alpha_n\rangle$. 
	We may also use the notation $\sigma_1;\sigma_2$ when 
	$\sigma_1$ or $\sigma_2$ is a single transition.\\
	As in RPNs, the execution of a MRPN can be partitioned
	as a set of independent flows of executions
	running through the net. We capture these flows by the notion of
	causal paths:
	\begin{definition}\label{multico-path}{\rm
			Given a MRPN $(P,T, A, A_V, B, F)$, a state $\state{M}{H}$ and transition occurrences
			$(t_i,k_i)$ in $\state{M}{H}$, $1\leq i \leq n$, we say that
			$(t_1,k_1),\ldots,(t_n,k_n)$ is a \emph{causal path} in $\state{M}{H}$,
			if $(t_i,k_i)\prec (t_{i+1}, k_{i+1})$, for all $0\leq i < n$.
	}\end{definition}
	Based on this concept, we define the notion of causal equivalence for histories by 
	requiring that two histories $H$ and $H'$ are causally equivalent if and only if they contain the same causal paths: 
	\begin{definition}\label{multieq}{\rm Consider a MRPN $(P,T, A, A_V, B, F)$ and two executions
			$\state{M}{H} \fctrans{\sigma} \state{M'}{H'}$ and
			$\state{M}{H}\fctrans{\sigma'} \state{M''}{H''}$. Then the histories
			$H'$ and $H''$ are \emph{causally equivalent}, denoted by $H'\asymp H''$, if
			for each causal path   $(t_1,k_1),\ldots,(t_n,k_n)$ 
			in $\state{M'}{H'}$, there is a causal path $(t_1,k_1'),\ldots,
			(t_n,k_n')$ in $\state{M''}{H''}$, and vice versa.
	}\end{definition}
\begin{figure}[t]
	\begin{center}
		\includegraphics[height=2.2cm]{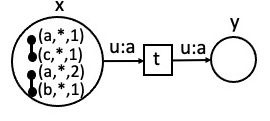}
		\includegraphics[width=.7cm]{figures/t,1.png}
		\includegraphics[height=2.2cm]{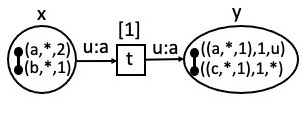}
		\includegraphics[width=.7cm]{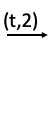}
		\includegraphics[height=2.4cm]{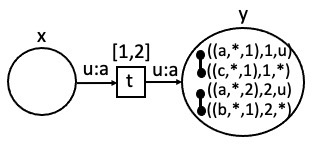}\\
		\includegraphics[height=2.2cm]{figures/equivalent.jpeg}
		\includegraphics[width=.7cm]{figures/t,1.png}
		\includegraphics[height=2.2cm]{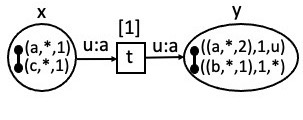}
		\includegraphics[width=.7cm]{figures/t,2.png}
		\includegraphics[height=2.4cm]{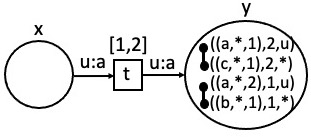}
	\end{center}
	\caption{Equivalent markings}
	\label{equivalentmarkings}
\end{figure}
	Now we define causal equivalence of markings as the equivalence where markings consist of identical  token instances participating in the same transitions that have been assigned different keys. Two equivalent markings can be observed in Figure~\ref{equivalentmarkings}, where in the first execution we fire $t$ with $((a,*,1),(c,*,1))$ first and then with $((a,*,2),(b,*,1))$, and in the second execution we fire with  $((a,*,2),(b,*,1))$ first and then  $((a,*,1),(c,*,1))$. This results in equivalent markings, i.e. markings consisting of connected components that have tokens of the same type used to fire the same transitions.

	\begin{definition}\label{multieqmarkings}{\rm Consider a MRPN $(P,T, A, A_V, B, F)$ and two executions
			$\state{M}{H} \fctrans{\sigma} \state{M'}{H'}$ and
			$\state{M}{H}\fctrans{\sigma'} \state{M''}{H''}$. Then the markings
			$M'$ and $M''$ are \emph{causally equivalent}, denoted by $M'\asymp M''$, if
			for each $a_i\in M'(x)$ where $(a_j,k,u) \overline{\in}a_i,$ $k\in H'(t),$  $u\in \guard{t},t\in T$ there exists  $a_i'\in M''(x)$ where$ (a_j',k',u) \overline{\in}a_i'$ such that $k' \in H''(t)$  and vice versa.\\
			We extend this notion and write $\state{M}{H}\asymp\state{M'}{H'}$ 
			if and only if $M\asymp M'$ and $H\asymp H'$.
	}\end{definition}
	We may now establish the Loop lemma. 
	\begin{lemma}[Loop]\label{multiloopc}{\rm 
			For any forward transition $\state{M}{H}\trans{(t,k)}\state{M'}{H'}$ there exists a backward transition
			$\state{M'}{H'} \ctrans{(t,k)} \state{M}{H}$ and for any backward transition $\state{M}{H} \ctrans{(t,k)} \state{M'}{H'}$ there exists a forward transition $\state{M'}{H'}\trans{(t,k')}\state{M''}{H''}$ where $\state{M}{H}\asymp \state{M''}{H''}$.
	}\end{lemma}
	\paragraph{Proof} The proof of the first direction follows 
	along the same lines as that of Lemma~\ref{multiloopb} with $\btrans{}$
	replaced by $\ctrans{}$. For the other direction, suppose $\state{M}{H}\ctrans{(t,k)}\state{M'}{H'}\trans{(t,k')}\state{M''}{H''}$. 
	To begin with, we may observe that, as with Lemma~\ref{multiloopb}, by Definitions~\ref{multiforward} and~\ref{multicausal}, the tokens involved in transition $t$ will be transferred to the incoming places of $t$ and then back to the outgoing places leading to  $M\asymp M''$.
	To show that $H\asymp H''$, we observe that $H=H''$ with the exception of
	$t$, where, if $k\in H(t)$,
	and $k'=\max(\{0\}\cup\{k''|(t',k'')\in H'(t'), t'\in T\})+1$, then
	$H''(t) = (H(t)-\{k\})\cup\{k'\}) $.
	Furthermore, since $t$ is $c$-enabled in $\state{M}{H}$, $(t,k)$ must
	be the last transition occurrence in all the causal paths it occurs in,
	and we may observe that $H''$ contains the same causal paths with 
	$(t,k)$ replaced by $(t,k')$.
	As a result it must be that $H\asymp H''$ and the result follows.
	\proofend
	\begin{definition}{\rm
			Consider a MRPN $(P,T,A,A_V,B,F)$, two actions $\alpha_1$ and $\alpha_2$, and a state $\state{M}{H}$. Then  $\alpha_1$ and $\alpha_2$ are said to be \emph{concurrent} in state $\state{M}{H}$, if whenever
			$\state{M}{H}\fctrans{\alpha_1}\state{M_1}{H_1}$ and $\state{M}{H}\fctrans{\alpha_2}\state{M_2}{H_2}$
			then $\state{M_1}{H_1}\fctrans{\alpha_2}\state{M'}{H'}$ and $\state{M_2}{H_2}\fctrans{\alpha_1}\state{M''}{H''}$,
			and $\state{M'}{H'} \asymp \state{M''}{H''}$.
	}\end{definition}
	As in the original RPNs two actions are concurrent when they can be executed in any order while preserving path equivalence.
	
	\begin{figure}[t]
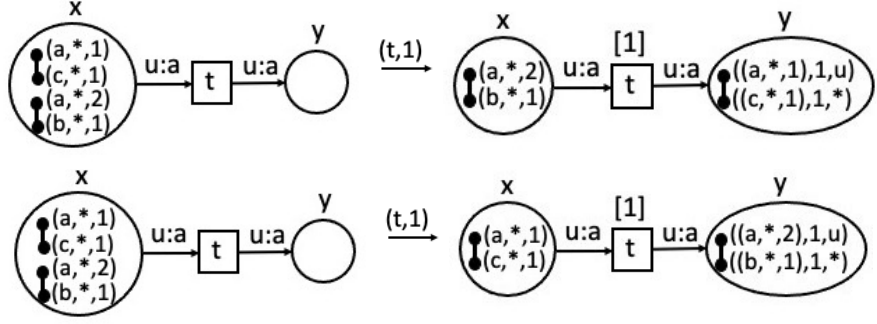

		\begin{center}
			\includegraphics[height=2.2cm]{figures/equivalent.jpeg}
			\includegraphics[width=.8cm]{figures/t,1.png}
			\includegraphics[height=2.2cm]{figures/equivalent1.jpeg}
			\includegraphics[height=2.2cm]{figures/equivalent.jpeg}
			\includegraphics[width=.8cm]{figures/t,1.png}
			\includegraphics[height=2.2cm]{figures/equivalent2.jpeg}
		\end{center}
		\caption{Non-equivalent transition firings}
		\label{equivalent}
		\end{figure}
				 
	\begin{definition}{\rm
			Consider a MRPN $(P,T,A,A_V,B,F)$, and two actions $\state{M_1}{H_1}
			\trans{(t,k_1)} \state{M_1'}{H_1'}$ and $\state{M_2}{H_2}
			\trans{(t,k_2)} \state{M_2'}{H_2'}$. Then  $(t,k_1)$ and $(t,k_2)$ are said to be \emph{equivalent} if for all $ (a_i,k_1,u)\in M_1'(x)$ there exists $(a_i,k_2,u)\in M_2'(x)$ for some $u \in \guard{t}$.
	}\end{definition}
Since a transition can be executed in the forward direction by different token instances, as long as they respect the arc requirements, then it is possible for the same transition to fire using different connected components. As these connected components might consist of different token instances then it is possible to fire the same transition resulting in markings that are not equivalent. Consider the example in Figure~\ref{equivalent}, where firing transition $t$ with bond $((a,*,1),(c,*,1))$ will result in a different marking than firing the transition with bond $((a,*,2),(b,*,1))$. Thus, two transition occurrences are said to be equivalent when they execute the same transition by manipulating the same token instances. 
	\begin{definition}\label{multico-executions}{\rm Consider a multi reversing Petri nets $(P,T, A, A_V, B, F)$ and two executions $\state{M_1}{H_1}
			\trans{\sigma_1} \state{M_1'}{H_1'}$  and $\state{M_2}{H_2}
			\trans{\sigma_2} \state{M_2'}{H_2'}$. 
			 \emph{Causal equivalence on executions}, is the least equivalence relation closed under composition of traces such that 
			if
			(i) $\sigma_1=(t_1,k_1);(t_2,k_2)$ and $\sigma_2= (t_2,k_1);(t_1,k_2)$ where $(t_1,k_1)$ and $(t_2,k_2)$ are concurrent actions in state $\state{M_1}{H_1}=\state{M_2}{H_2}$, 
			(ii)  $\sigma_1=(t,k) ; (\underline{t,k})$ and $\sigma_2=\epsilon$, and (iii) $\sigma_1=(\underline{t,k_1});(t,k_2)$ and $\sigma_2=\epsilon$ where  $(t,k_1)$ and $(t,k_2)$ are equivalent actions according to states $\state{M_1'}{H_1'}$ and $\state{M_2'}{H_2'}$. If the executions $\state{M_1}{H_1}
			\trans{\sigma_1} \state{M_1'}{H_1'}$  and $\state{M_2}{H_2}
			\trans{\sigma_2} \state{M_2'}{H_2'}$ are causally equivalent then we say that traces $\sigma_1$ and $\sigma_2$ are also causally equivalent denoted by $\sigma_1 \asymp \sigma_2$. 
	}\end{definition}
	The first clause states that in two causally-equivalent executions concurrent actions
	may occur in any order and the second clause states that it is possible to ignore transitions that have occurred in both
	the forward and the reverse direction. The third clause states that it is possible to ignore equivalent transitions that have occurred in both
the reverse and forward direction. Note that unlike $(t,k) ; (\underline{t,k}) \asymp \epsilon$, we require these transitions to be equivalent as with token multiplicity it is possible to fire again a reversed transition by manipulating different connected tokens of the same type. These two transitions should be equivalent in order to be ignored so that they will produce the same marking, as explained for Figure~\ref{equivalent}.  

	The following proposition establishes that two transition instances belonging
	to distinct causal paths are in fact concurrent transitions, and thus can be executed in any order. 
	\begin{proposition}\label{multiconc-transitions}{\rm
			Consider a MRPN $(P,T, A, A_V, B, F)$ and suppose $\state{M}{H}
			\trans{(t_1,k_1)} \state{M_1}{H_1}\trans{(t_2,k_2)} \state{M_2}{H_2}.$
			If
			there is no causal path $\pi$ in $\state{M_2}{H_2}$ with $(t_1,k_1)\in\pi$ and $(t_2,k_2)\in\pi$,
			then
			$(t_1,k_1)$ and $(t_2,k_2)$ are concurrent transition occurrences in $\state{M}{H}$. 
	}\end{proposition}
	\paragraph{Proof}
	Since there is no causal path containing both $(t_1,k_1)$ and $(t_2,k_2)$ in $\state{M_2}{H_2}$,
	we conclude that $(t_1,k_1)\not\prec (t_2,k_2)$.
	This implies that there is no token that has participated in both  transition occurrences and they can
	be executed in any order, leading to the same marking. Thus, they are 
	concurrent in $\state{M}{H}$.
	\proofend 
	We note that causally-equivalent states can execute the same transitions.
	\begin{proposition}{\rm
			Consider a MRPN $(P,T,A,A_V,B,F)$ and states $\langle M_1, H_1\rangle\asymp\langle M_2, H_2\rangle$. 
			Then $\state{M_1}{H_1}\fctrans{(t,k_1)}\state{M_1'}{H_1'}$ if and only if  
			$\state{M_2}{H_2}\fctrans{(t,k_2)}\state{M_2'}{H_2'}$, where $\langle M_1', H_1'\rangle\asymp\langle M_2', H_2'\rangle$. 
			\label{multiextendStates}
	}\end{proposition}
	\paragraph{Proof}
	It is easy to see that if a transition $(t,k_1)$ is enabled in $\state{M_1}{H_1}$
	it is also enabled in $\state{M_2}{H_2}$. Specifically, there exists an enabling assignment $U_1$ for $(t,k_1)$ and $U_2$ for $(t,k_2)$ such that they manipulate the same components that have been assigned different keys. 
	 Therefore  if  $\state{M_1}{H_1}\fctrans{(t,k_1)}\state{M_1'}{H_1'}$
	then $\state{M_2}{H_2}\fctrans{(t,k_2)}\state{M_2'}{H_2'}$ where
	$M_1'\asymp M_2'$, and vice versa. In order
	to show that $H_1'\asymp H_2'$ two cases exist:
	\begin{itemize}
		\item Suppose $t$ is a forward transition corresponding to transition
		occurrences $(t,k_1)$ and  $(t,k_2)$ in each state respectively. 
		Suppose that $(t',k_1')\prec (t,k_1)$.
		Then, $\exists a_i, (a_1',k_1',u) \overline{\in}a_i$ and $(a_1,k_1,v) \overline{\in}a_i$, $a_i\in M_1'(x)$. Since $H_1 \asymp H_2$ this implies that $(t',k_2') \prec (t,k_2)$. 
		Therefore, for all causal paths $\pi$ in $\state{M_1}{H_1}$,
		if the last transition occurrence of $\pi$ causes $(t,k_1)$ then
		$\pi;(t,k_1)$ is a causal path of $\state{M_1'}{H_1'}$ and, if not, then $\pi$ is
		a causal path in $\state{M_1'}{H_1'}$. The same holds for causal paths
		in $\state{M_2}{H_2'}$ and $(t,k_2)$. Consequently, we deduce that $H_1'\asymp H_2'$, as required.
		\item Suppose that $t$ is a reverse transition and consider the causal paths of
		$H_1'$ and $H_2'$. Since $t$ is a reverse transition, there exists no transition occurrence
		in $\state{M_1}{H_1}$  caused by $(t,k_1)$
		and no transition occurrence in $\state{M_2}{H_2}$ caused by $(t,k_2)$. 
		As such, $(t,k_1)$ and  $(t,k_2)$ are the last
		transition occurrences in all paths in $\state{M_1}{H_1}$ and $\state{M_2}{H_2}$,
		respectively, in which they belong.
		Reversing the transition occurrences results in their elimination from these causal paths. 
		Therefore, we observe that for each causal path in $\state{M_1'}{H_1'}$ there is
		an equivalent causal path in $\state{M_2'}{H_2'}$, and vice versa. 
		Thus $H_1'\asymp H_2'$ as required.
		\proofend
	\end{itemize}
Note that the above result can be extended to sequences of transitions:
\begin{corollary}\label{multiconc-paths}{\rm
		Consider a MRPN $(P, T,A,A_V,B,F)$ and states $\langle M_1, $ $H_1\rangle\asymp\langle M_2, H_2\rangle$. 
		Then $\state{M_1}{H_1}\fctrans{\sigma}\state{M_1'}{H_1'}$ if and only if  
		$\state{M_2}{H_2}\fctrans{\sigma}\state{M_2'}{H_2'}$, where $\langle M_1', H_1'\rangle\asymp\langle M_2', H_2'\rangle$.  
}\end{corollary}

	The main result, Theorem~\ref{multimain} below, states  that  two computations beginning in the same initial state lead to equivalent
	states if and only if the two computations are causally equivalent. Specifically,
	if two executions from the same state reach causally-equivalent states by executing transitions $\sigma_1$
	and $\sigma_2$, then the two executions are causally
	equivalent and vice versa. This guarantees the consistency of the approach since reversing transitions in causal order is in a sense equivalent
	to not executing the transitions in the first place. Reversal does not give rise
	to previously unreachable states, on the contrary, it gives rise to causally-equivalent markings and
	histories due to the different keys being possibly assigned because of the different ordering of transitions.
	\begin{theorem}\label{multimain}{\rm Consider executions $\state{M}{H} \fctrans{\sigma_1} \state{M_1}{H_1}$ and $\state{M}{H}\fctrans{\sigma_2} \state{M_2}{H_2}$. Then, $\state{M}{H} \fctrans{\sigma_1} \state{M_1}{H_1}$ and $\state{M}{H}\fctrans{\sigma_2} \state{M_2}{H_2}$  are causally equivalent executions if and only if   $\state{M_1}{H_1}\asymp\state{M_2}{H_2}$.
		}
	\end{theorem}
	For the proof of Theorem~\ref{multimain} we employ some intermediate results. To begin, the lemma below
	states that causal equivalence allows the permutation of reverse and forward transitions that have no causal relations between them. 
	Therefore, computations are allowed to reach for the maximum freedom of choice going backward and then continue forward.
	\begin{lemma}\label{multiperm}{\rm \
			Let $\state{M}{H} \frtrans{\sigma}\state{M'}{H'}$ be an execution. Then there exist traces $r,r'$ both forward such that $\state{M}{H} \frtrans{\sigma}\state{M'}{H'}$ and  $\state{M}{H} \frtrans{\underline{r};r'}\state{M''}{H''}$ are causally equivalent executions
			where $\state{M'}{H'}\asymp \state{M''}{H''}$.
	}\end{lemma}
	\paragraph{Proof}
	We prove this by induction on the length of  $\sigma$ and the
	distance from the beginning of $\sigma$ to the earliest pair of transitions
	that contradicts the property $\underline{r};r'$. If there is no such
	contradicting pair, then the property is trivially satisfied.  If not, we distinguish the following cases:
	\begin{enumerate}
		\item If the first contradicting pair is of the form $(t,k);(\underline{t,k})$ then we have  $\state{M}{H}\fctrans{\sigma_1}\state{M_1}{H_1}\fctrans{(t,k)}\state{M_2}{H_2}\fctrans{(\underline{t,k})}\state{M_3}{H_3}\fctrans{\sigma_2}\state{M'}{H'}$ where $\sigma=\sigma_1;(t,k);(\underline{t,k});\sigma_2$. By the Loop Lemma~\ref{loopc} $\state{M_1}{H_1}=\state{M_3}{H_3}$, which yields $\state{M}{H} \fctrans{\sigma_1} \state{M_1}{H_1} \fctrans{\sigma_2} \state{M'}{H'}$. Thus we may remove the two transitions from
		the sequence, the length of $\sigma$ decreases, and the proof follows
		by induction.
		\item If the first contradicting pair is of the form $(t,k);(\underline{t',k'})$, then
		we observe that the specific occurrences of $(t,k)$ and $(\underline{t',k'})$ must be
		concurrent. Specifically, we have $ \state{M}{H} 
		\fctrans{\sigma_1}\state{M_1}{H_1}\fctrans{(t,k)}\state{M_2}{H_2}\fctrans{(\underline{t',k'})}
		\state{M_3}{H_3}\fctrans{\sigma_2}\state{M'}{H'}$ where $\sigma=\sigma_1 ; (t,k)
		;(\underline{t',k'}); \sigma_2$. Since
		action $(t',k')$ is being reversed it implies that all 
		transition occurrences that are causally dependent on it have either
		not been executed up to this point or they have already been reversed. This implies
		that in $\state{M_2}{H_2}$ it was not the case that $(t,k)$ was causally dependent on $(t',k')$.
		As such, by Proposition~\ref{multiconc-transitions} $(\underline{t',k'})$ and $(t,k)$ are concurrent transitions
		and $(t',k')$ can be reversed before
		the execution of $t$
		to yield $ \state{M}{H} \fctrans{\sigma_1}\state{M_1}{H_1}\fctrans{(\underline{t',k'})}\state{M_2'}{H_2'}
		\fctrans{(t,k'')}\state{M_3'}{H_3'}\fctrans{\sigma_2}\state{M''}{H''}$ where $\state{M_3}{H_3} \asymp \state{M_3'}{H_3'}$ and $(t,k'')$ is an equivalent transition to $(t,k)$ as in Definition~\ref{equivalent}. Note that it is possible for $k''=k$ if $(t',k')$ was not last the transition to be executed in the forward direction before $(t,k)$, otherwise $k'' \neq k$. This results in a
		later earliest contradicting pair and by induction the result follows.
		\item If the first contradicting pair is of the form $(t,k);(\underline{t,k'})$, then we have  $\state{M}{H}\fctrans{\sigma_1}\state{M_1}{H_1}\fctrans{(t,k)}\state{M_2}{H_2}\fctrans{(\underline{t,k'})}\state{M_3}{H_3}\fctrans{\sigma_2}\state{M'}{H'}$, where $\sigma=\sigma_1;(t,k);(\underline{t,k'});\sigma_2$. Then  $(t,k)$ and $(\underline{t,k'})$ are not the same transition occurrence as transition $(t,k')$ reverses with a different key value than the forward execution $(t,k)$ thus they do not cancel each other out. As $(t,k')$ reverses before $(t,k)$ this means that $(t,k)$ and $(\underline{t,k'})$ must be  concurrent and by applying similar arguments as those in (2) 
		we observe that the specific occurrences of $(t,k)$ and $(\underline{t,k'})$ can be swapped 
		to yield $ \state{M}{H} \fctrans{\sigma_1}\state{M_1}{H_1}\fctrans{(\underline{t,k'})}\state{M_2'}{H_2'}
		\fctrans{(t,k'')}\state{M_3'}{H_3'}\fctrans{\sigma_2}\state{M''}{H''}$ where $\state{M_3}{H_3} \asymp \state{M_3'}{H_3'}$ and $(t,k'')$ is an equivalent transition to $(t,k)$ as in Definition~\ref{equivalent}. Note that it is possible for $k''=k$ if $(t,k')$ was not the last transition to be executed in the forward direction before $(t,k)$, otherwise $k'' \neq k$. 
		This results in a
		later earliest contradicting pair and by induction the result follows.
		\proofend
	\end{enumerate}
	From the above lemma we may conclude the following corollary. The result establishes that causal-order
	reversibility is consistent with standard forward execution in the sense that causal execution will not generate
	states that are unreachable in forward execution:
	\begin{corollary}\label{multiequivalent-executions}{\rm\ \
			Suppose that  $H_0$ is the initial history. If $\state{M_0}{H_0} \fctrans{\sigma} \state{M}{H}$, and $\sigma$ is a
			trace with both forward and backward transitions then
			there exists a transition $\state{M_0}{H_0}\fctrans{\sigma'}\state{M'}{H'}$ where $\state{M}{H}\asymp \state{M'}{H'}$, 
			and $\sigma'$ a trace of forward transitions.
	}\end{corollary}
	\paragraph{Proof} According to
	Lemma~\ref{multiperm}, $\sigma\asymp \underline{r};r'$ where both $r$ and $r'$ are forward
	traces. Since, however, $H_0$ is the initial history it must be that $r$ is empty. This
	implies that $\state{M_0}{H_0}\fctrans{r'}\state{M'}{H'}$, $\state{M}{H} \asymp \state{M'}{H'} $ 
	and $r'$ is a 
	forward trace. Consequently, writing $\sigma'$ for $r'$, the result follows.
	\proofend
	\begin{lemma}\label{multishort}{\rm\ \
			Suppose $\state{M}{H}\fctrans{\sigma_1}\state{M_1}{H_1}$ and
			$\state{M}{H}\fctrans{\sigma_2}\state{M_2}{H_2}$, where $\state{M_1}{H_1}\asymp \state{M_2}{H_2}$ and
			$\sigma_2$ is a forward trace. Then, there exists a forward trace
			$\sigma_1'$ such that  $\state{M}{H}\fctrans{\sigma_1'}\state{M_1'}{H_1'}$  and $\state{M}{H}\fctrans{\sigma_1}\state{M_1}{H_1}$ are causally equivalent executions.
	}\end{lemma}
	\paragraph{Proof}
	If  $\sigma_1$ is forward, then $\sigma_1 = \sigma_1'$ and the result follows
	trivially. Otherwise, we may prove the lemma by induction on the length of
	$\sigma_1$.
	We begin by noting that, by Lemma~\ref{multiperm},
 $\sigma_1 \asymp\underline{r};r'$ and $\state{M}{H}\fctrans{\underline{r};r'}
	\state{M_1}{H_1}$. 
	Let  $(\underline{t,k})$ be
	the last action in $\underline{r}$. 
	Given that $\sigma_2$ is a forward execution that 
	simulates $\sigma_1$, it must be that $r'$ contains a forward execution 
	of transition $t$ manipulating the same tokens since  $\state{M_1}{H_1}$ and $\state{M_2}{H_2}$ contain the same causal 
	paths involving transition $t$ (if not we would have 
	$\state{M_1}{H_1}\not \asymp \state{M_2}{H_2}$ leading to a contradiction). 
	Consider the earliest such  occurrence in $r'$ to be $(t,k')$ an equivalent transition to $(t,k)$. If $(t,k')$ is the first
	transition in $r'$ and as it is equivalent to $(t,k)$ the Loop Lemma~\ref{loopc} can be applied to remove the pair of opposite transitions
	and the result follows by induction. Otherwise, suppose 
	$\state{M}{H}\fctrans{\underline{r_1}}\fctrans{(\underline{t,k})}
	\fctrans{{r_1'}} \state{M_3}{H_3} \fctrans{(t*,k*)}\fctrans{(t,k')}\state{M_4}{H_4}
	\fctrans{r_2'}\state{M_1}{H_1}$, where $r = r_1;(t,k)$ and $r'=r_1';(t*,k*);(t,k');r_2$.
	Two cases exist:
	\begin{enumerate}
		\item Suppose $(t*,k*)\in\sigma_2$. Let us denote by $num(\alpha,\sigma)$, 
		the number of executions of action $\alpha$ in a sequence of
		transitions $\sigma$ where $\alpha$ represents all transition occurrences of transition $t$ manipulating the same connected components. We observe that since $\sigma_2$ contains no
		reverse executions of $t$, it must be that $num(\alpha,r') = num(\alpha,\sigma_2) + num(\underline{\alpha},r)$.
		Suppose that the transition occurrences of $(t*,k*)$ and $(t,k')$ as shown in the execution
		belong to a common causal path. We may extend this path with the succeeding 
		occurrences of $\alpha$ and obtain a causal path such that $(t*,k*)$ is succeeded by
		$num(\alpha,\sigma_2) + num(\alpha,r)$ occurrences of $\alpha$. We observe that it is impossible to
		obtain such a causal path in $\state{M_2}{H_2}$, since $(t*,k*)$ is followed by fewer occurrences of $\alpha$ in $\sigma_2$.
		This contradicts the assumption that $H_1\asymp H_2$. We conclude that the transition
		occurrences of $(t,k')$ and $(t*,k*)$ above do not belong to any common causal path and,
		therefore, by Proposition~\ref{multiconc-transitions}, the two transition occurrences are
		concurrent in $\state{M_3}{H_3}$.
		\item
		Now suppose that $(t*,k*)\not\in \sigma_2$. Since $k*\in H_1(t*)$ it must
		be that $H_2(t*)\neq \emptyset$ and $|H(t*)| = |H_1(t*)|= |H_2(t*)|$. As such, it
		must be that $(t*,k'*)\in r$ and that its reversal has preceded the reversal of $(t,k)$.
		Let us suppose that the transition occurrences of $(t*,k*)$ and $(t,k')$ as shown in the execution
		belong to a common causal path. This implies that a causal path
		with $(t*,k'*)$ preceding $(t,k)$ also occurs in $H_2$ as well as in $H$. If we
		observe that $(t*,k'*)$ has reversed before $(t,k)$ we conclude that  $(t*,k'*)$ does not cause the preceding occurrence of $(t,k)$. As such
		there is no causal path within $\state{M}{H}$ or $\state{M_2}{H_2}$ containing both $(t,k)$ and $(t*,k*)$, which results in a contradiction. We
		conclude that the forward occurrences of $(t,k')$ and $(t*,k*)$ are,  by Proposition~\ref{multiconc-transitions}, concurrent in $\state{M_3}{H_3}$.
	\end{enumerate}
	Given the above, we conclude that we may swap the occurrences of $(t,k')$ and $(t*,k*)$
	to obtain $\state{M}{H}\fctrans{\underline{r_1}}\fctrans{(\underline{t,k})}
	\fctrans{{r_1'}}  
	\state{M_3}{H_3} \fctrans{(t,k'')}\fctrans{(t*,k''*)}\state{M_4'}{H_4'}\fctrans{r_2'}\state{M_1''}{H_1''}$ where $\state{M_4}{H_4}\asymp \state{M_4'}{H_4'}$ and, by Corollary~\ref{multiconc-paths},  $\state{M_1}{H_1}\asymp \state{M_1''}{H_1''}$.
	By repeating the process for the remaining transition occurrences in $r_1'$, this implies
	that we may permute $(t,k')$ with transitions in $r_1'$ to yield the sequence $(\underline{t,k});(t,k')$. By the Loop Lemma~\ref{loopc} we may remove the pair of opposite
	transitions and obtain a shorter equivalent trace, also
	equivalent to $\sigma_2$ and conclude by induction.
	\proofend \\
	We may now proceed with the proof of Theorem~\ref{multimain}:
	\paragraph{Proof of Theorem~\ref{multimain}}
	Suppose that we have $\state{M}{H} \fctrans{\sigma_1} \state{M_1}{H_1}$, $\state{M}{H} \fctrans{\sigma_2} \state{M_2}{H_2}$
	with  $\state{M_1}{H_1}\asymp\state{M_2}{H_2}$. 
	We prove that  $\state{M}{H} \fctrans{\sigma_1} \state{M_1}{H_1}$ and $\state{M}{H} \fctrans{\sigma_2} \state{M_2}{H_2}$ are causally equivalent executions thus giving  $\sigma_1\asymp \sigma_2$ by using a lexicographic induction on the pair consisting 
	of the sum of the lengths of $\sigma_1$ and $\sigma_2$ and the depth of the earliest disagreement
	between them. By Lemma~\ref{multiperm} we may suppose that $\sigma_1$ and $\sigma_2$
	satisfy the property
	$\underline{r};r'$. Call $(t_1,k_1)$ and $(t_2,k_2)$ the earliest actions where they disagree. There are three
	cases in the argument depending on whether these are forward or backward.
	\begin{enumerate}	
		\item If $(\underline{t_1,k_1})$ is backward and $(t_2,k_2)$ is  forward, we have $\sigma_1=\underline{r};(\underline{t_1,k_1});u$ 
		and $\sigma_2=\underline{r};(t_2,k_2);v$ for some $r,u,v$. Lemma~\ref{multishort} applies to $(t_2,k_2);v$, 
		which is forward, and $(\underline{t_1,k_1});u$, which contains both forward and backward actions and thus,
		by the lemma, it  has a shorter forward equivalent. Thus, $\sigma_1$ has a shorter forward 
		equivalent and the result follows by induction.	
		\item If $(t_1,k)$ and $(t_2,k)$ are both forward  then it must be the
		case that $\sigma_1 = {\underline{r};r'};(t_1,k); u$
		and $\sigma_2 =  {\underline{r};r'}; (t_2,k); v$, for some $r$, $u$, $v$. Note that
		it must be that an equivalent transition to $t_1$ appears in $ v$ and an equivalent transition to $t_2$ appears in $u$. 
		If not, we would have $H_1\not \asymp H_2$, which contradicts the assumption that ${H_1}\asymp {H_2}$.
		As such, we may write $\sigma_1 =  {\underline{r};r'};(t_1,k);u_1;(t_2,k_2);u_2$, 
		where $u=u_1;(t_2,k_2);u_2$
		and $(t_2,k_2)$ is the first occurrence of $t_2$ in $u$ manipulating the same tokens as $(t_2,k)$. Consider $(t*,k*)$ the 
		action immediately preceding $(t_2,k_2)$. We may observe that $(t*,k*)$ and $(t_2,k_2)$ 
		cannot belong to
		a common causal path in $\state{M_1}{H_1}$, since an equivalent causal path is 
		impossible to
		exist in $\state{M_2}{H_2}$. This is due to the assumption that $\sigma_1$ 
		and $\sigma_2$ coincide up to transition sequence  {$\underline{r};r'$}. 
		Thus, we may conclude by  Proposition~\ref{multiconc-transitions} that $(t*,k*)$ and 
		$(t_2,k_2)$ are in fact concurrent and can be swapped.
		The same reasoning may be used
		for all transitions preceding $(t_2,k_2)$ up to and including
		$(t_1,k)$, which leads to the conclusion that
		$\sigma_1\asymp  {\underline{r};r'};(t_2,k);(t_1,k_1); u_1;u_2$. This results in an 
		equivalent execution of the same length with a later earliest divergence with 
		$\sigma_2$ 
		and the result follows by the induction hypothesis.	
		\item If $(t_1,k_1)$ and $(t_2,k_2)$ are both backward, we have $\sigma_1=\underline{r};(\underline{t_1,k_1});u$ 
		and $\sigma_2=\underline{r};(\underline{t_2,k_2});v$ for some $r,u,v$. Two cases exist:
		\begin{enumerate}
			\item If $(\underline{t_1,k_1'})$ occurs in $v$, then we have that $\sigma_2=\underline{r};(\underline{t_2,k_2});\underline{v_1};(\underline{t_1,k_1'});v_2$.
			Given that $t_1$ reverses right after $\underline{r}$ in
			$\sigma_1$, we may conclude that there is no transition occurrence
			at this point that causally depends on $(t_1,k_1')$. As such it
			cannot have caused the transition occurrences of $(t_2,k_2)$ and
			${v_1}$ whose reversal precedes it in $\sigma_2$.  
			This implies that the reversal of $(t_1,k_1')$
			may be swapped in $\sigma_2$ with each of the preceding
			transitions, to give
			$\sigma_2\asymp\underline{r};(\underline{t_1,k_1'});(\underline{t_2,k_2});\underline{v_1};v_2$.
			This results in an equivalent execution of the same length with a later earliest divergence 
			with $\sigma_1$ and the result follows by the induction hypothesis.
			\item If $(\underline{t_1,k_1'})$ does not occur in $v$, this implies that $(t_1,k_1')$, an equivalent transition of $(t_1,k_1)$ occurs
			in the forward direction in $u$, i.e. $\sigma_1=\underline{r};(\underline{t_1,k_1});u_1;(t_1,k_1');u_2$, where $u = u_1;(t_1,k_1');u_2$ with
			the specific occurrence of $(t_1,k_1')$ being the first such occurrence in $u$. 
			Using similar arguments as those in Lemma~\ref{multishort},
			we conclude that $\sigma_1\asymp\underline{r};(\underline{t_1,k_1});(t_1,k_1');u_1;u_2
			\asymp \underline{r};u_1;u_2$, an equivalent execution of shorter length for $\sigma_1$ and the result follows by the induction hypothesis.
		\end{enumerate}	
		We may now prove the opposite direction. Suppose that 
		$\state{M}{H}\fctrans{\sigma_1}\state{M_1}{H_1}$ and $\state{M}{H}\fctrans{\sigma_2}\state{M_2}{H_2}$ are causally equivalent executions thus $\sigma_1 \asymp \sigma_2$. 
		We will show that $\state{M_1}{H_1}\asymp \state{M_2}{H_2}$.
		The proof is by induction on the number of rules, $k$, applied to establish 
		the equivalence $\sigma_1 \asymp \sigma_2$.
		For the base case we have $k=0$, which implies that $\sigma_1=\sigma_2$
		and the result trivially follows. For the inductive step, let us assume that 	$\state{M}{H}\fctrans{\sigma_1}\state{M_1}{H_1}$,  $\state{M}{H}\fctrans{\sigma_2}\state{M_2}{H_2}$, and $\state{M}{H}\fctrans{\sigma_1'}\state{M_1'}{H_1'}$ are causally equivalent executions thus
		$\sigma_1\asymp \sigma_1'\asymp \sigma_2$, where $\sigma_1$ can be transformed
		to $\sigma_1'$ with the use of $k=n-1$ rules and $\sigma_1'$ can be transformed
		to $\sigma_2$ with the use of a single rule. By the induction hypothesis,
		we conclude that $\state{M}{H}\fctrans{\sigma_1'}\state{M_1'}{H_1'}$, where 
		$\state{M_1}{H_1}\asymp \state{M_1'}{H_1'}$. We need to show that $\state{M_1'}{H_1'}\asymp 
		\state{M_2}{H_2}$. Let us write
		$\sigma_1' = u;w;v$ and $\sigma_2 = u;w';v$, where $w$, $w'$ refer
		to the parts of the two executions where the equivalence rule has been applied.
		Furthermore, suppose that
		$\state{M}{H}\fctrans{u}\state{M_u}{H_u}\fctrans{w}\state{M_w}{H_w}\fctrans{v}\state{M_1'}{H_1'}$ and
		$\state{M}{H}\fctrans{u}\state{M_u}{H_u}\fctrans{w'}\state{M_w'}{H_w'}\fctrans{v}\state{M_2}{H_2}$.
		Three cases exist:
		\begin{enumerate}
			\item $w= (t_1,k_1);(t_2,k_2)$ and $w'=(t_2,k_1);(t_1,k_2)$ with $(t_1,k_1)$ and $(t_2,k_2)$ concurrent
			\item $w=(t,k);(\underline{t,k})$ and $w'=\epsilon$
			\item $w=(\underline{t,k});(t,k')$ and $w'=\epsilon$ with $(t,k)$ and $(t,k')$ equivalent. 
		\end{enumerate}
		In all the cases above, we have that $\state{M_w}{H_w}\asymp \state{M_w'}{H_w'}$:
		for (a) this follows by the definition of concurrent transitions, whereas
		for (b) and (c) by the Loop Lemma. Given the equivalence of these two
		states, by Corollary~\ref{multiequivalent-executions}, we have that $\state{M_w}{H_w}\fctrans{v}\state{M_1'}{H_1'}$  and
		$\state{M_w'}{H_w'}\fctrans{v}\state{M_2}{H_2}$, where $\state{M_1'}{H_1'}
		\asymp \state{M_2}{H_2}$, as required. This completes the proof.
		\proofend
	\end{enumerate}

\subsection{Out-of-Causal Order}

In this form of reversibility we allow  events to reverse  without the need to respect causality as long as the transition is executed and its effect (creation/destruction of a bond) has not been undone. 

\begin{definition}\label{multioco-enabled}{\rm
		Consider a MRPN $(P,T, A, A_V, B, F)$, a state $\state{M}{H}$, and a transition occurrence  $(t,k)$ in $\state{M}{H}$.
		We say that $(t,k)$ is \emph{o-enabled} in $\state{M}{H}$ 
		if
there exists an injective function 
		$U_o:\effects{t}\cap A_V\rightarrow  A_I$ such that:
\begin{enumerate} 
			\item for all $u\in F(t,x)$, $x \in t\circ$,we have $U_o(u)=a_j$, $(a_i,k,u) \overline{\in} a_j$, $U_o(u)\in M(y) $ 	for some $y$ where $type(u)=type(U_o(u))$,
			\item  for all $(u,v)\in F(t,x)$, $x\in t\circ$ we have $(U_o(u),U_o(v))=(a_j,b_j),(a_i,k,u)\overline{\in} a_j, $$(b_i,k,v)$\\$\overline{\in} b_j, (U_o(u),U_o(v))\in M(y)$, 
			\item for all
			$(u,v)\in \guard{t}$, $(u,v)\not \in \effects{t}$ and $U_o(u)=a_j$, $U_o(v)=b_j$, $(a_i',k',u')\overline{\in} a_j$, $ (b_i',k',v')\overline{\in}b_j$ then $\not\exists t',$ $k' \in H(t')$ where $k'>k$ such that $(u',v')\in \effects{t'}$,	$(u',v')\not \in \guard{t'}$, and
			\item for all
			$(u,v)\in \effects{t}$,	$(u,v)\not \in \guard{t}$ and $U_o(u)=a_j,$  $U_o(v)=b_j,$ $(a_i',k',u')\overline{\in} a_j,$ $(b_i',k',v')\overline{\in}b_j$ then $\not\exists t',$ $k' \in H(t')$ where $k'>k$  such that $(u',v')\in \guard{t}$,	$(u',v')\not \in \effects{t}$. 
			
		\end{enumerate}
}\end{definition}

Thus, a transition  occurrence $(t,k)$ is o-enabled in $\state{M}{H}$ if (1) and (2) there exists a type-respecting assignment of token instances in the outgoing places of the transition, to the variables on the outgoing edges of the transition, and where the instances are connected with bonds as required by
the transition's outgoing edges.  Finally, (3) and (4) require the effect of the transition, i.e. breaking or creating a bond, not to have been undone by a following forward transition.
We refer to $U_o$ as an out-of-causal-order enabling assignment.

Summing up, the effect of reversing a transition in out-of-causal order is that all bonds created by the transition
are undone and all bonds broken by the transition are redone. This may result in tokens backtracking in the net, in the case where the reversal of a transition causes a coalition of bonds 
to be broken down into
a set of subcomponents and moving forward in the net, in the case where the reversal of a transition recreates a coalition into a larger component. 
In both cases the component should be relocated (if needed) after the last transition in which this sub-coalition participated. 
To capture this 
we introduce the following: 

\begin{definition}\label{multilast}{\rm
		{	Given a MRPN $(P,T, A, A_V, B, F)$, an initial marking $M_0$,  
			a history $H$, and 
			a set of bases and bonds $C\subseteq A_I\cup B_I$ we write:
			\[
			\begin{array}{rcl}
			\lastt{C,H} &=& \left\{
			\begin{array}{ll}
			(t,k) , \;\; \hspace{.3cm}\textrm{ if }\exists t, \; (a_j,k,u)\overline{\in}a_i, a_i \in C, \; k \in H(t), u \neq * \mbox{ and }\\
			\hspace{0.25in} \not\exists t', \; (b_j,k',u')\overline{\in}b_i,b_i \in C, \;  k'\in H(t'), u'\neq * \; , 	
			k'>k \\
			\bot,  \;\textrm{ otherwise }
			\end{array}
			\right.
			\end{array}
			\]
			\[
			\begin{array}{rcl}
			\lastp{C,H} &=& \left\{
			\begin{array}{ll}
			x , \;\;\textrm{ if }(t,k)=\lastt{C,H}, \{x\} = \{y\in t\circ 
			\mid (a_j,k,u)\overline{\in} a_i,a_i \in C, u \in F(t,y) \}\\
			\hspace{0.25in} \textrm{or, if } \bot=\lastt{C,H},  C\in M_0(x)\\
			\bot,  \;\textrm{ otherwise }
			\end{array}
			\right.
			\end{array}
			\]}
}\end{definition}
Thus, if the tokens from component $C$ have been manipulated by some previously-executed
transition,  then $\lastt{C,H}$ is the last executed such transition.
Otherwise, if no such transition exists (e.g., because all transitions
involving $C$ have been reversed), then $\lastt{C,H}$ is undefined 
($\bot$). Similarly, $\lastp{C,H}$ is the outgoing place connected to $t$  
with common tokens with $C$, if $\lastt{C,H}\neq \bot$  assuming that such
a place is unique, or the place in the initial marking in which $C$ 
existed if $\lastt{C,H}= \bot$, and undefined otherwise.

The following definition defines all tokens to be removed from a place because their last transition has changed and their current place is not the outgoing place of last. 

\begin{definition}\label{multisetso1}{\rm
		Given a MRPN $(P,T, A, A_V, B, F)$, a state $\state{M}{H}$, an o-enabled transition occurrence $(t,k)$, a history $H'$ as in Definition~\ref{multicausal}, and an
		enabling assignment $U_o$, we define  $^\bullet\!{U}_o:P\rightarrow 2^{A_I\cup B_I}$ to be a function that assigns to each place a set of  incoming token and bond instances:
		\begin{eqnarray*}
			^\bullet\!{U}_o(x)  &= &  \post{t,U_o}
			  \cup \;\{C_{a_i,x}\mid \exists 
			a_i\in M(x),
			x \neq{\lastp{C_{a_i,x},  H'}}\}\hspace{6cm}
		\end{eqnarray*}
		where we use the shorthand $C_{b_i,z} = \connected(b_i,(\{\connected(c_i,M(z))|c_i \in A_I,z\in P\}-\post{t,U_o})\cup$\\$ \pre{t,U_o})$ for $b_i \in A_I$. 
		
}\end{definition}

We now define the outgoing tokens as the tokens that remain or move to these places because it is an outgoing place of their last transition. 

\begin{definition}\label{multisetso2}{\rm
		Given a MRPN $(P,T, A, A_V, B, F)$, a state $\state{M}{H}$, an o-enabled transition occurrence $(t,k)$, a history $H'$ updated as in Definition~\ref{multicausal} and an enabling assignment $U_o$ we define $U_o^\bullet :P\rightarrow 2^{A_I\cup B_I}$ to be a function that assigns to each place a set of outgoing token/bond instances:
		\begin{eqnarray*}
			U_o^\bullet (x)  =  
			\;\{\lastA{a_i,L,k}|\exists a_i\in M(y),L=\lastt{C_{a_i,y},H'}, x=\lastp{C_{a_i,y},H'}\} \cup \hspace{1.7cm}\\
				\;\{(\lastA{a_i,L,k},\lastA{b_i,L,k})|\exists (a_i,b_i)\in M(y),L=\lastt{C_{a_i,y},H'}, x=\lastp{C_{a_i,y},H'}\}
			\end{eqnarray*}
				where \\
					\[
				\begin{array}{rcl}
				\lastA{a_i,L,k} = \left\{
				\begin{array}{ll}
				(a_m,k_m,u_m)\downarrow a_k,  \hspace{1.6cm} 
				 \;\textrm{ if } 
				\not\exists (a_m',k_m',u_m')\overline{\in}a_i,  
				 k'\geq k_m' >k_m  	\textrm{ where}  \\ \hspace{2.5cm}
		(a_m,k_m,u_m)\overline{\in}a_i, L=(t',k')  \textrm{ and } (a_k,k,u)\overline{\in} (a_m,k_m,u_m)   \\
				(a,*,i), \hspace{6.6cm} \;\textrm{ if } (a,*,i) \overline{\in} a_i, L=\bot
				\end{array}
				\right.
				\end{array}
				\]

				

\hspace{-.7cm} and we use the shorthand $C_{b_i,z}=\connected(b_i,(\{\connected(c_i,M(z))|c_i \in A_I,z\in P\}-\post{t,U_o})\cup$\\$ \pre{t,U_o})$ for $b_i \in A_I$.
		
}\end{definition}

The above definition reconstructs connected components by undoing the effect of the transition and by removing from tokens the memory of the transition along with the memories that where recorded later than their last transition. The definition uses $C_{a_i,y}$ to reconstruct the component as a result of breaking or creating bonds during reversal. By $\lastA{a_i,L,k}$ we indicate the updated memory of token $a_i$ by removing transition $(t,k)$ and all memories executed later than its last transition $L=(t',k')$. Specifically, $(a_m,k_m,u_m)$ is the latest memory taken before and including its last transition $L=(t',k')$. $(a_m,k_m,u_m)\downarrow a_k$ indicates that the memory of transition $(t,k)$ has been removed from  $(a_m,k_m,u_m)$. In the case that there is no last transition the initial token $(a,*,i)$ is returned. In this way we remove the implicit memories where the transition has not actively participated in. As demonstrated in the example of Figure~\ref{memories} after the reversal of $(t_1,1)$ the implicit memory of transition $(t_2,2)$ is removed from the token $(((a,*,1),1,u),2,*)$ along with the memory of transition $(t_1,1)$ as defined below:

\begin{figure}[t]
	\begin{center}
		\hspace{-1cm}
		\includegraphics[height=4.3cm]{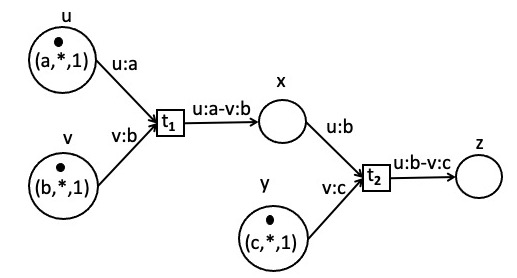}\\
	\hspace{-.6cm}	\includegraphics[width=.8cm]{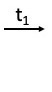,1.png}
		\includegraphics[height=4cm]{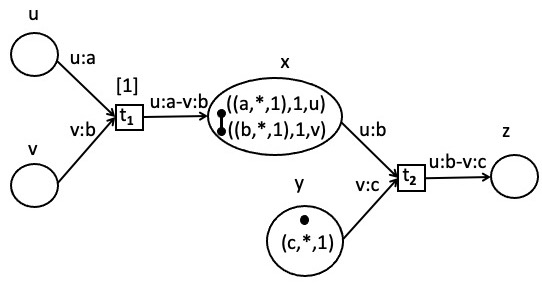}\\
		\includegraphics[width=.8cm]{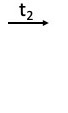,2.png}
		\includegraphics[height=3.8cm]{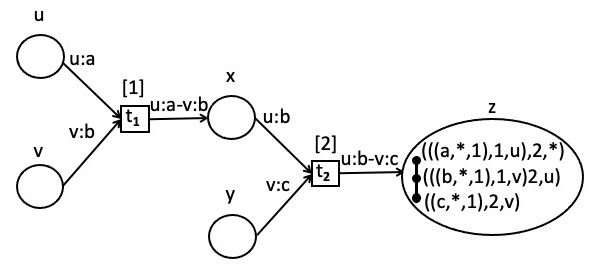}\\
		\includegraphics[width=.8cm]{figures/t1,1r.png}
	\includegraphics[height=3.8cm]{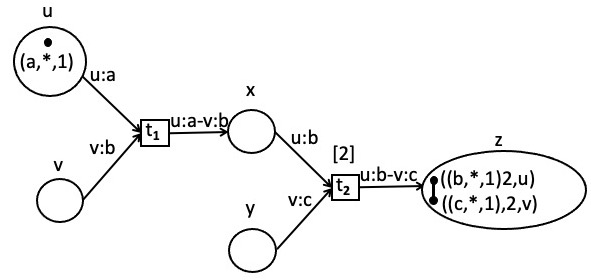}\\
	\end{center}
	\caption{Updating the memories of tokens in out-of-causal-order reversibility}
	\label{memories}
\end{figure}

\begin{definition}{\rm \label{multioco}
		Given a MRPN $(P,T, A, A_V, B, F)$, a state $\langle M, H\rangle$,  a transition occurrence $(t,k) $ that is o-enabled and an
		enabling assignment $U_o$, 
		we write $\state{M}{H}
		\otrans{(t,k)} \state{M'}{H'}$
		where $H' $ is updated as in Definition~\ref{multicausal} and for all $x\in P$:
		\begin{eqnarray*}
			M'(x) =   (M(x)- ^\bullet\!{U}_o(x))
			\cup {U}_o^\bullet(x)
		\end{eqnarray*}
}\end{definition}

Thus, when a transition $t$ is reversed in an out-of-order fashion all bonds that were created by
the transition are undone and all bonds broken by the transition are reconstructed. If the destruction of a bond divides a component 
into smaller connected components then each of these
components should again be relocated (if needed) back to the place 
where the complex would have existed if transition $t$ never took place, i.e., exactly after the last transition 
that involves tokens from the sub-complex. Otherwise when a recreation of a bond creates a larger connected component then this component should be relocated (if needed) to the place where the complex would have existed if transition $t$, never took place, i.e., exactly after the last transition that involves tokens from the bigger complex. Token memories are updated by removing the memory of the reversed transition $(t,k)$ and removing all implicit memories of transitions executed later than their last transition. 
 Also the history is update as defined in Definition \ref{multicausal}.

\begin{figure}[t]
	\begin{center}
		\includegraphics[height=2.8cm]{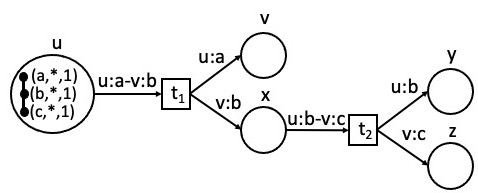}
		\includegraphics[width=.8cm]{figures/t1,1.png}
		\includegraphics[height=3cm]{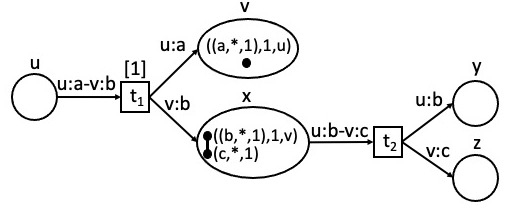}
		\includegraphics[width=.8cm]{figures/t2,2.png}
		\includegraphics[height=3cm]{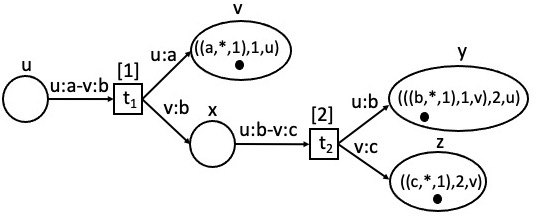}
		\includegraphics[width=.8cm]{figures/t1,1r.png}
		\includegraphics[height=3cm]{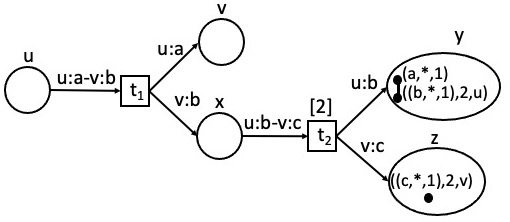}
		\includegraphics[width=.8cm]{figures/t2,2r.png}
		\includegraphics[height=2.8cm]{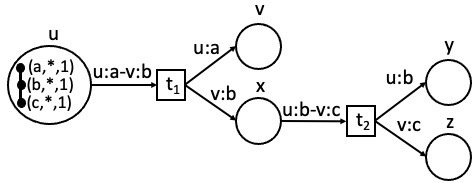}
	\end{center}
	\caption{Out-of-causal order execution}
	\label{multioutFig}
\end{figure}

From the example in Figure~\ref{multioutFig} we observe that after the execution of transitions $t_1$ and $t_2$,  the component $a\bond b\bond c$ has been broken to three parts $a,b$, and $c$ located in different places. The reversal of $t_1$ recreates the bond between $a\bond b$ and since $b$ last participated in $t_2$ then $a\bond b$ is moved to the outgoing place of $t_2$ as it would have happened if we had skipped the execution of $t_1$. However when $t_1$ and $t_2$  are reversed then our system resets and there are no transitions holding the tokens further down the execution of the RPN and therefore the component $a\bond b \bond c$ returns to its initial place. Note that the history and the token instances are updated accordingly.

	The following results describe how tokens and bonds are manipulated
	during out-of-causal-order reversibility, where we write $\fotrans{}$ for $\trans{}\cup\otrans{}$.
	\begin{proposition}\label{multimarkings}{\rm
			Suppose $\state{M}{H} \fotrans{(t,k)}\state{M'}{H'}$ and let $a_i,a_i'\in A_I$ where
			$a_i\in M(x)$ and $a_i'\in M'(y)$. If $(t,k)$ is a forward occurrence with $U_f$ then  $C=\connected(a_i,(\{\connected(b_i,M(z))|b_i \in A_I,z\in P\}-\pre{t,U_f})\cup \post{t,U_f})$ and $C'= \connected(a_i',M'(y))$  such that for all $a_i \in C$ and $a_i' \in C'$, $a_i \overline{\in}a_i'$ and if  $(t,k)$ is a reverse transition with $U_o$ then $C'=\connected(a_i',M'(y))$ and $C=\connected(a_i,(\{\connected(b_i,M(z))|b_i\in A_I,z\in P\})-\post{t,U_o})\cup \pre{t,U_o})$ such that for all $a_i \in C$ and $a_i'\in C'$, $a_i' \overline{\in}a_i$.
		}
	\end{proposition}
	\paragraph{Proof}
	The proof is straightforward by the definition of the firing rules.
	\proofend
	\begin{proposition}\label{multiprop5}
		\rm Given a MRPN $(P,T,A,A_V,B,F)$, an initial state 
		$\langle M_0, H_0\rangle$, and an execution
		$\state{M_0}{H_0} \fotrans{(t_1,k_1)}\state{M_1}{H_1} \fotrans{(t_2,k_2)}\ldots 
		\fotrans{(t_n,k_n)}\state{M_n}{H_n}$ the following hold for all $0\leq i \leq n$ where $a_i\in A_I\cap M_0(z)$, $z\in P$,
		 $|\{ a_i  \mid a_i\in M_i (x)\cap A_I, x\in P, a_0\overline{\in} a_i\}| = |\{ a_{i+1} \mid a_{i+1}\in M_{i+1} (y)\cap A_I, y\in P, a_0\overline{\in} a_{i+1}\}|=1$, 
		and  $a_i\in M_i(x)$ where 
		$x=\lastp{\connected(a_i,M_{i}(x)),H_i}$. 
		%
	\end{proposition}
	\paragraph{Proof}
	Consider a MRPN $(P,T,A,A_V,B,F)$, an initial state 
	$\langle M_0, H_0\rangle$, and an execution
	$\state{M_0}{H_0} \fotrans{(t_1,k_1)}\state{M_1}{H_1} \fotrans{(t_2,k_2)}\ldots \fotrans{t_n,k_n}\state{M_n}{H_n}$. The 
	proof is by induction on $n$.
	\paragraph{Base Case} For  $n=0$, by our assumption of token 
	uniqueness and the definitions of $\mathsf{last}_P$ and
	$\mathsf{last}_T$ the claim follows trivially.
	\paragraph{Induction Step} Suppose the claim holds for all but the last transition
	and consider
	transition $(t_n,k_n)$. Two cases exist, depending on whether $t_n$ is a forward or a reverse transition:
	\begin{itemize}
		\item Suppose that $(t_n,k_n)$ is a forward transition. Then by Proposition \ref{multiprop1},  for all $a_0\in A_I\cap {M_0(z)}$, $z\in P$,
		$|\{a_n \mid a_n\in M_n (x)\cap A_I, x\in P, a_o\overline{\in}a_n\}| = 1$. 
		Additionally, we may see that if $a_n\in M_n(x)$ two cases exists.
		If $a_{n-1}\in \connected(b_{n-1},M_{n-1}(y))$, for some $b_{n-1}$ where $U_f(v)=b_{n-1}, v \in F(t_n,z)$
		then $x=z=\lastp{\connected(a_{n-1},M_{n-1}(x)), H_n}$ and $a_{n-1}\overline{\in}a_n$ then $\lastp{\connected(a_n,M_n(x)),H_n}=x=z$.  Otherwise, it must be that $a_{n-1}\in M_{n-1}(x)$
		where, by the induction hypothesis, $x = \lastp{\connected(a_{n-1},M_{n-1}(x)), H_{n-1}}$. Since $a_{n-1}\not\in\connected(b_{n-1},M_{n-1}(y))$  we may deduce that    $\connected(a_{n-1},M_{n-1}(x))= \connected(a_n,M_{n}(x))$, leading to $x
		= \lastp{\connected(a_n,M_{n}(x)), H_{n}}=\lastp{\connected(a_{n-1},M_{n-1}(x)), H_n}$. Thus, the result follows.
		%
		\item {Suppose that $(t_n,k_n)$ is a reverse transition. 
			Consider $a_{n-1}\in A_I$ with $a_{n-1}\in M_{n-1}(x)$ for some $x\in P$. Two cases exist:
			\begin{itemize}
				\item Suppose $C=\connected(a_{n-1},(\{\connected(b_{n-1},M(z))|b_{n-1}\in A_I,z\in P\}-\post{t,U_o})\cup \pre{t,U_o})$ where $\lastt{C,H_n}=\bot$. Then, it  must be that for $C'=\connected(a_n,M_n(y))$ where $a_n\overline{\in} a_{n-1}$ by Proposition~\ref{multimarkings} where $C' \subseteq M_0(y)$.  Suppose
				that this is not the case. By the induction hypothesis, there exists some $t_i$
				in the execution such that $\exists \beta_i\in C'$ and $\beta_i \not \in M_0(y)$, if $\beta_i$ is produced by $t_i$,  or $\exists \beta_i \in M_0(y)$ and $\beta_i \not \in C'$, if $\beta_i$ is destructed by $t_i$. This however implies
				that $t_i$ is a transition that has manipulated the connected
				component $C$, which contradicts
				our assumption of $\lastt{C,H_n}=\bot$. 
				Therefore, $a_n\in M_n(y)$, where $a_n\in M_0(y)$ and by Proposition~\ref{multimarkings} $a_{n-1} \overline{\in} a_n$ which gives $y = \lastp{C', H_n}$
				and the result follows.
				\item Suppose $C=\connected(a_{n-1},(\{\connected(b_{n-1},M(z))|b_{n-1}\in A_I,z\in P\}-\post{t,U_o})\cup \pre{t,U_o})$ where $\lastt{C,H_n}=(t_k,k)$. 
				Then, it  must be that there exists a unique $y\in t_k
				\circ$ 
				such that $c_{n-1} \in C$ where $ (c_k,k,v)\overline{\in}c_{n-1},v \in  F(t_k,z)$. Suppose
				that this is not the case. Then for $C'=\connected(a_n,M_n(x))$ there must exist some $\beta_{n}=(a_{n},c_{n})\in C'$ with $(a_k,k,u)\overline{\in}a_{n}$, $u\in F(t_k,y_1)$, $v\in F(t_k,y_2)$, and $y_1\neq y_2$. By the induction hypothesis, there exists some $t_i$
				in the execution such that $(a_i,k_i,u_i)\overline{\in} a_{n}$ and $(c_i,k_i,v_i)\overline{\in} c_{n}$, where $(u_i,v_i)\in F(t_i,y_i)$ and $k_i>k$ which was not
				reversed. This however implies
				that $t_i$ is a transition that has manipulated the connected
				component $C$ later than 
				$(t_k,k)$, which contradicts
				our assumption of $\lastt{C,H_n} = t_k$. Therefore, there exists a unique $y\in t_k
				\circ$ 
				such that 
				$a_n\in M_n(y)$. Furthermore, by Proposition~\ref{multimarkings} $a_n \overline{\in}a_{n-1}$ which gives
				$y = \lastp{C', H_n}$
				and the result follows.
			\end{itemize}
		}
		\proofend
	\end{itemize}

	\begin{lemma}[Loop]\label{multiloopo}{\rm 
			For any forward transition $\state{M}{H}\trans{(t,k)}\state{M'}{H'}$ there exists a reverse
			transition $\state{M'}{H'} \otrans{(t,k)} \state{M}{H}$. 
	}\end{lemma}
	\paragraph{Proof}
	Suppose $\state{M}{H}\trans{(t,k)}\state{M'}{H'}$. Then $t$ is clearly $o$-enabled 
	in $H'$. Furthermore, $\state{M'}{H'} \otrans{(t,k)} \state{M''}{H''}$ where $H''=H$
	by the definition of $\otrans{}$. 
	In addition, for all $a_i\in A_I$, we distinguish two cases. 
	If  for some $a_i \in M(x)$,
	$\not\exists (a_i,k,u) = a_i',a_i'\in M'(y)$,  
 then we may see that $a_i\in M'(y)$ and 
	$a_i\in M''(y)$, and the result follows. Otherwise, if  $\exists (a_i,k,u) = a_i'$ then $a_i'\in M'(z)$ where $u\in F(t,z)$. Furthermore suppose $a_i''\in M''(w)$. By proposition~\ref{multimarkings} for $C=\connected(a_i,\{\connected(b_i,M(z))|b_i\in A_I,z\in P\}-\pre{t,U_f} \cup \post{t,U_f})$ and $C_1'=\connected(a_i',M'(y))$ we have $a_i \overline{\in} a_i'$ and for  $C_2'=\connected(a_i',\{\connected(b_i',M'(z))|b_i'\in A_I,z\in P\}-\post{t,U_o} \cup \pre{t,U_o})$ and $C''=\connected(a_i'',M''(w))$ we have $a_i'' \overline{\in} a_i'$. Since $H=H''$ we have $w=\lastp{C'',H'')}=\lastp{C,H)}=y$ and the result follows. 
	\proofend 
	
	As in the original RPN model, the opposite direction of the 
	lemma does not hold. 
The following result
	establishes that the placement of a connected component is
	uniquely determined by the last transition to have manipulated it.
	\begin{proposition}\label{multisecond}{\rm Consider   executions $\state{M_0}{H_0} 
			\fotrans{\sigma_1} \state{M_1}{H_1}$ and $\state{M_0}{H_0} 
			\fotrans{\sigma_2} \state{M_2}{H_2}$,
			and a token $a_i \in A_I, a_i \in M_1(x) \cap  M_2(y)$
			for some $x, y\in P$. Then, 
			$\lastt{\connected(a_i,M_1(x)),H_1}=\lastt{\connected(a_i,M_2(y)),H_2}$ implies
			$x=y$.
	}\end{proposition}
	\paragraph{Proof} 
	Consider executions $\state{M_0}{H_0} \fotrans{\sigma_1} 
	\state{M_1}{H_1}$, $\state{M_0}{H_0} \fotrans{\sigma_2} 
	\state{M_2}{H_2}$  and a token $a_i$ as 
	specified by the lemma. Further, let us assume that $\lastt{\connected(a_i,M_1(x)),H_1} = \lastt{\connected(a_i,M_2(y)),H_2}$. Two cases exist:
	\begin{itemize}
		\item $\lastt{\connected(a_i,M_1(x)),H_1} = \lastt{\connected(a_i,M_2(y)),H_2}=\bot$.
		This implies that no transition has manipulated any of the tokens
		and bonds in the two connected components. As such, by Proposition~\ref{multiprop5}, $\connected(a_i,M_1(x))\subseteq M_0(x)$ and $\connected(a_i,M_2(y))\subseteq M_0(y)$,
		and we conclude that $x=y$ as required.
		\item   $\lastt{\connected(a_i,M_1(x)),H_1} = \lastt{\connected(a_i,M_2(y)),H_2} = (t,k)$. 
		This implies that there exists $b_i \in \ \connected(a_i,M_1(x)) \cap \connected(a_i,M_2(y))$ such that $b_i=(b_j,k,u),u\in\effects{t}$. By
		Proposition~\ref{multiprop5}, $x=\lastp{\connected(b_i,M_1(x)),H_1}$,
		$y=\lastp{\connected(b_i,M_2(y)),H_2}$. Since we have that  $\lastt{\connected(a_i,M_1(x)),H_1} = \lastt{\connected(a_i,M_2(y)),H_2}$
		we conclude that $ \lastp{\connected(b_i,M_1(x)),H_1}=\lastp{\connected(b_i,M_2(y)),H_2}$, thus, $x=y$ as required.
		\proofend
	\end{itemize}

As in the original RPN model we confirm
the relationship between the enabledness conditions for each of backtracking, causal-order, and out-of-causal-order reversibility.
\begin{proposition}\label{multienable}{\rm\ \
		Consider  a state $\langle M, H\rangle$, and a transition occurrence $(t,k)$. Then, if $(t,k)$ is $bt$-enabled in  $\langle M, H\rangle$
		it is also $c$-enabled. Furthermore, if  $(t,k)$ is $c$-enabled in  $\langle M, H\rangle$
		then it is also $o$-enabled.
	}
\end{proposition}
\paragraph{Proof}
The proof is immediate by the respective definitions.
\proofend

The following result establishing
that during causal-order reversibility a component is
returned to the place following the last transition
that has manipulated it or, if no such transition exists,
in the place where it occurred in the initial marking. 
\begin{proposition}\label{multilast-as-co}
	\rm Given a multi reversing Petri net $(P,T,A,A_V,B,F)$, an initial state 
	$\langle M_0, H_0\rangle$, and an execution
	$\state{M_0}{H_0} \fctrans{(t_1,k_1)}\state{M_1}{H_1} \fctrans{(t_2,k_2)}\ldots 
	\fctrans{(t_n,k_n)}\state{M_n}{H_n}$. Then for all $a_i\in A_I$, 
	$a_i\in M_n(x)$ where $x=\lastp{\connected(a_i,M_{n}(x)),H_n}$.
\end{proposition}
\paragraph{Proof}
The 
proof is by induction on $n$ and it follows along similar lines to the proof of Proposition~\ref{multiprop5}.
\proofend

We may now verify that the causal-order and out-of-causal-order reversibility have the same effect in MRPNs when reversing a $c$-enabled transition.
\begin{proposition}\label{multic-to-o}{\rm\ \
		Consider a state $\langle M, H\rangle$ and a transition occurrence $(t,k)$ $c$-enabled in  $\langle M, H\rangle$.
		Then, $\langle M, H\rangle\ctrans{(t,k)}\langle M', H'\rangle$ if and only if $\langle M, H\rangle\otrans{(t,k)}\langle M', H'\rangle$.
	}
\end{proposition}
\paragraph{Proof}
Let us suppose that $(t,k)$ is $c$-enabled and $\langle M, H\rangle\ctrans{(t,k)}
\langle M_1, H_1\rangle$. 
By Proposition~\ref{multienable}, $(t,k)$ is also $o$-enabled. Suppose $\langle M, H\rangle\otrans{(t,k)}\langle M_2, H_2\rangle$.
It is easy to see that in fact $H_1 = H_2$ (the two histories are 
as $H$ with the exception that $H_1(t) = H_2(t) = H(t)-\{k\}$). 

To show that $M_1=M_2$ first we observe that for all
$a_i\in A_I$, by Proposition~\ref{multilast-as-co} we have $a_i\in M_1(x)$ where
$x=\lastp{\connected(a_i,M_1(x)),H_1}$ and by
Proposition~\ref{multiprop5} we have $a_i\in M_1(y)$ where
$y=\lastp{\connected(a_i,M_2(y)),H_2}$. We may also see
that $\connected(a_i,M_1(x)) =  \connected(a_i,M_2(y))$. Since in addition we have $H_1=H_2$
the result follows.
\proofend

An equivalent result can be obtained for backtracking.

\begin{proposition}\label{multib-to-c}{\rm\ \
		Consider  a state $\langle M, H\rangle$, and a transition occurrence $(t,k)$, $bt$-enabled in  $\langle M, H\rangle$.
		Then, $\langle M, H\rangle\btrans{(t,k)}\langle M', H'\rangle$ if and only if $\langle M, H\rangle\otrans{(t,k)}\langle M', H'\rangle$.
	}
\end{proposition}
\paragraph{Proof}
Consider a state $\langle M, H\rangle$ and suppose
that transition occurrence $(t,k)$ is $bt$-enabled and $\langle M, H\rangle\btrans{(t,k)} \langle M', H'\rangle$. Then,
by Proposition~\ref{multienable}{, there exists $k\in H(t)$, such that {for all $t'\in T$, $k'\in H(t')$, it holds that
		$k\geq k'$}. This implies
	that} $(t,k)$ is also $c$-enabled, and by the definition of $\ctrans{}$, we conclude that 
$\langle M, H\rangle{\ctrans{(t,k)}} \langle M', H'\rangle$. Furthermore, by Proposition~\ref{multic-to-o}
$\langle M, H\rangle{\otrans{(t,k)}} \langle M', H'\rangle$, and the result follows.
\proofend

As in RPNS we obtain the following corollary confirming the "universality" of $\otrans{}$.
\begin{corollary}\label{multiconnect}{\rm\ \
		$\btrans{} \subset\ctrans{}\subset \otrans{}$. 
	}
\end{corollary}


\subsection{Example of MRPNs under the Individual Token Interpretation}

\begin{figure}[t]
	\centering
	\subfloat{\includegraphics[width=.7cm]{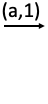}}
	\subfloat{\includegraphics[width=.85cm]{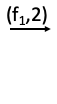}}
	\subfloat{\includegraphics[width=.78cm]{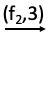}}
	\subfloat{\includegraphics[width=.81cm]{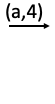}}
	\subfloat{\includegraphics[width=.8cm]{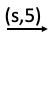}}
	\subfloat{\includegraphics[width=12cm]{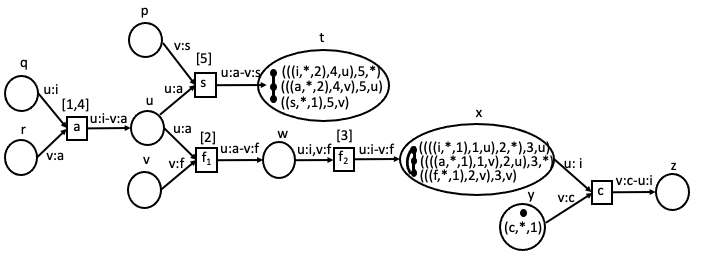}}\\
	\subfloat{\includegraphics[width=.7cm]{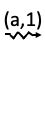}}
	\subfloat{\includegraphics[width=12cm]{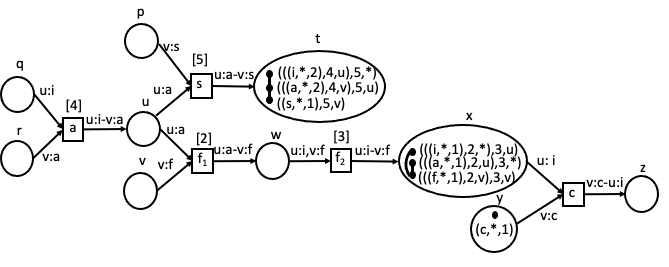}}
	\subfloat{\includegraphics[width=.8cm]{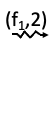}}\\
	\subfloat{\includegraphics[width=12cm]{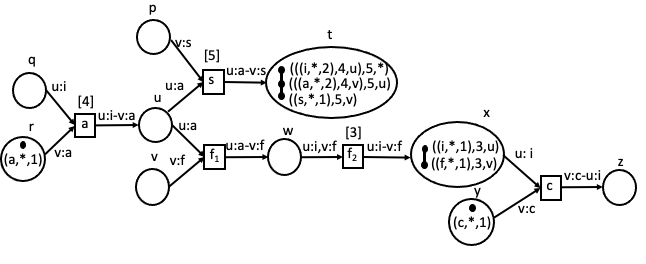}}
	\caption{Transaction processing with multitokens}
	\label{multitrans}
\end{figure}
The individual token interpretation can be used to model systems that require an association between the modelled system and its processes. Specifically, in~\cite{causal}, the classical notion of a process is given as a run of the modelled system, obtained by choosing one of the alternatives in case of a conflict. As such, the individual token interpretation is able to represent such processes as token memories. These memories record all occurrences of the transitions and places visited during a run, together with the causal dependencies between them, which are given by the flow relation of the net. Causal semantics of the system are thus obtained by associating with tokens the processes running in a net. According to the individual token interpretation, causal dependencies are a central aspect in the dynamic evolution of a net. 
 In this case, the actual order of execution of concurrent transitions in the net is invisible when reversing, but all the causal dependencies are preserved.

Let us consider the multicasting system of transaction processing in Figure~\ref{multitrans}. In this example we demonstrate a multi reversing Petri net that corresponds to the example in  Figure~\ref{trans} of the original RPN model. An agent can simultaneously execute multiple transactions in the same system, thus, several processes are running in parallel. In case one transaction fails whereas the rest of the transactions have been successfully completed, the system should be able to correspond transaction initialisations  to failed transactions so that only failed initialisations can be reversed. This can be done by associating each transaction token with its process indicating which transitions the token has traversed and whether one of these transitions represents failure. In this way the individual token interpretation can be used to  coordinate and synchronize  a system consisting of multiple  transactions in case out-of-causal reversal is necessary due to failures.  Multi reversing Petri nets rely on the memories of tokens as a mechanism which is used to express their behaviours. 

Specifically, in this example we have two transaction tokens of type $a$, $(a,*,1)$ and $(a,*,2)$, which can participate in the same transitions. We randomly select one of the two transaction tokens, token $(a,*,1)$, to be involved in a sequence of failed transitions, whereas token $(a,*,2)$ will  be executed successfully. 
As we have a failed transaction in the model we should reverse in out-of-causal order transition $a$ to be able to proceed with the compensation transition $c$. In this example the approach of individual token interpretation plays an important role as it is essential to reverse the occurrence of transition $a$ that is associated with the failed transaction rather than the transaction that has been completed successfully. Thus, by observing the memory of the tokens we are able to identify that the failed transaction corresponds to transition occurrence $(a,1)$ and we may proceed by reversing $(a,1)$ in order to release the failed transaction token.

	\section{ Multi Reversing Petri Nets vs Reversing Petri Nets}
	In this section we present two translations from reversing Petri nets to Labelled Transition Systems (LTS), one from RPNs with single tokens and one from RPNs with multi tokens. This serves to establish the equivalence between the two models by showing that for every MRPN there is a SRPN which is equivalent in terms of the underlining LTS. 
 Labelled Transition Systems (LTS) are defined as follows:
\begin{definition}\label{LTS}{\rm  A labelled transition system is a tuple $(Q, E, \rightarrow, I)$ where:
		\begin{itemize}
			\item $Q$ is a countable set of states,
			\item $E$ is a countable set of actions,
			\item $\rightarrow \subseteq Q \times E \times Q$ is  the step transition relation, and 
			\item $I \in Q$ is the initial state.
		\end{itemize}
}\end{definition}
Henceforth, we write $p \trans{u} q$ for $(p, u, q) \in \rightarrow$.

Here $p \trans{u} q$ means that the represented system can transition from state $p$ to state $q$ by performing action $u$. 

When used for comparing systems, LTSs are considered modulo a suitable semantic equivalence. For our purposes, we employ the following notion of isomorphism of reachable parts, 	$\cong_{\cal{R}}$:

\begin{definition}\label{iso}{\rm 
		Two LTSs $A=(Q^A,E^A,\rightarrow, I^A)$ and $B=(Q^B,E^B,\rightarrow, I^B)$ are isomorphic, written $A \cong B$, if they
		differ only in the names of their states and events, i.e. if there are bijections
		$\beta :Q^A \rightarrow Q^B$ and $\eta :E^A \rightarrow E^B$ such that $\beta (I^A)=I^B$, and, for $p,q\in Q^A$, 
		$u \in E^A: \beta (p)\trans{\eta (u)}_B\beta (q)$ iff $p\trans{u}_A q$.
}\end{definition}

The set $\cal{R}(Q)$ of reachable states in $A = (Q, E, \rightarrow, I)$ is the smallest set such that $I$ is reachable and whenever $p$ is reachable and $p \trans{u} q$ then $q$ is reachable.
We write $A \cong_{\cal{R}} B$ if ${\cal{R}}(A)$ and ${\cal{R}}(B)$ are isomorphic.
To check $A \cong_{\cal{R}} B$ it suffices to restrict to subsets of $Q^A$ and $Q^B$ that contain all reachable states, and construct an isomorphism between the resulting LTSs.

We now give the translation from reversing Petri nets with multi and single tokens into labelled transition systems.  In what follows we write $\fstrans{}$ for $\trans{} \cup \rtrans{}$ where $\rtrans{}$ could be any of $\btrans{}, \ctrans{}$, and $ \otrans{}$ with single tokens and $\fmtrans{}$ the equivalent for mutli tokens. 

\begin{definition}{\rm  Let $N = (P,T,A,A_v,B,F)$ be a net with multi tokens and initial marking $M_0$ and $N' = (A', P, B', T', F')$ be a net with single tokens and initial marking $M'_0$ . Then ${\cal{H}}_m(N,M_0) = (2^{{A_I \cup B_I}},(T \times \mathbb{N}),\fmtrans{}, M_0)$
		is the LTS associated with $N$ under the multi  token interpretation, and ${\cal{H}}_s(N',M_0') = (2^{{A' \cup B'}^P}, T', \fstrans{}, M'_0)$ is the LTS associated with $N'$ under the single token interpretation.
		\label{LTSs}
}\end{definition}

 The following theorem says that reversing Petri nets under the single token interpretation are at least as expressive as reversing Petri nets under the multi token interpretation, in the sense that any LTS that can be denoted by a net under the latter interpretation can also be a denoted by a net under the former interpretation. 

\begin{theorem}{\rm For every multi reversing Petri net $N = (P,T,A,A_v,B,F)$ with initial markings $M_0$ there is a single reversing Petri net $N' = (A', P, B', T', F')$  with initial marking $M_0'$ such that ${\cal{H}}_s(N',M_0') \cong_{\cal{R}}  {\cal{H}}_m(N,M_0)$.
}\end{theorem}

\paragraph{Proof} Let $N = (P,T,A,A_v,B,F)$ be a MRPN with initial marking $M_0$.  We construct a SRPN as $N' = (A', P, B', T', F')$ with initial marking $M_0'$ as follows:
\begin{itemize}
	\item $A'=\{a_i | (a,*,i)\in A_I\cap M_0(x),x\in P\}$
	\item $B'=\{(a_i,b_i)| (a,*,i),(b,*,i) \in A_I\cap M_0(x),x\in P\}$
	\item $T'=\{t_s|s\in S, S=\{(a_1,...,a_n)\in (A')^n|type(a_i)=type(v_i), (v_1,...,v_n)=\guard{t_m}\cap A_V,t_m\in T\}\}$
	\item	\[
	F'(x,y) = \left\{
	\begin{array}{ll}
	a_i, \hspace{.4cm} \textrm{ if } x\in P, y\in T', t_m\in T, v\in F(x,t_m)\cap A_V, type(v)=type(a_i),a_i\in A' \; \\
	a_i, \hspace{.4cm}  \textrm{ if } x\in T', y\in P, t_m\in T, v\in F(t_m,y)\cap A_V, type(v)=type(a_i),a_i\in A' \; \\
	(a_i,b_i), \hspace{.8cm} \textrm{ if } x\in P, y\in T', t_m\in T,  (u,v)\in F(x,t_m), u,v\in A_V, type(u)=\\ \hspace{6.1cm}type(a_i), type(v)=type(b_i),a_i,b_i\in A' \; \\
	(a_i,b_i), \hspace{.8cm} \textrm{ if } x\in T', y\in P, t_m\in T,  (u,v)\in F(t_m,y),u,v\in A_V,  type(u)=\\ \hspace{6.1cm}type(a_i), type(v)=type(b_i),a_i,b_i\in A'\; \\
	\end{array}
	\right.			\]
	\item  
	\[
	M_0'(x) = \left\{
	\begin{array}{ll}
	a_i, \hspace{3.7cm} \textrm{ if } (a,*,i)\in M_0(x)\cap A_I,x\in P \hspace{2cm} \; \\
	\emptyset, \hspace{7.1cm}  \;\textrm{ otherwise \hspace{2.3cm}} 
	\end{array}
	\right.			\]
\end{itemize} 
 We denote $M_s \in
2^{{A' \cup B'}}$ and $M_m \in 2^{{A_I \cup B_I}}$ for any possible
marking in each respective RPN type, and $t_s\in T'$, $t_m\in T$ for transitions. For  ${\cal{H}}_s(N',M_0') \cong_{\cal{R}}  {\cal{H}}_m(N,M_0)$ to hold it
must be that $\beta(M_m)=M_s$ and $\eta(t_m,k_m) =t_s$. We define $\beta(M_m)=M_s$ for all $x\in P$ if there exists $a_i,b_i \in A_I \cap M_m(x)$ and $(a_i,b_i) \in  B_I \cap M_m(x)$ then  there exists $ a,b \in A' \cap M_s(x)$ and $(a,b) \in B'\cap  M_s(x)$  such that  $type(a_i)=type(a)$ and $type(b_i)=type(b)$. We also define $\eta(t_m,k_m) =t_s$ if for all $v \in A_V$ where $v \in F(x,t_m) \cap F(t_m,y)$ for some $x\in \circ t_m, y\in t_m \circ$ then there exists $ a\in A'$ where $a\in F'(x,t_s)\cap F'(t_s,y)$ such that $type(v)=type(a)$.  Similarly for bonds, if $(u,v)\in F(x,t_m)$ for some $x\in \circ t_m$ then there exists$ (a,b)\in B'$ where $(a,b) \in F'(x,t_s)$ such that $type(u)=type(a)$ and $type(v)=type(b)$ (respectively for $F(t_m,y)$). 

Now for the mapping of transition firings from $N'$ to $N$ we have two cases depending on the form of execution:
\begin{itemize}
	\item $\state{M_m}{H_m}\trans{(t_m,k_m)}\state{M_m'}{H_m'}$ and $\state{M_s}{H_s}\trans{t_s}\state{M_s'}{H_s'}$ are both forward executions. During forward execution in both models tokens  along with their connected components are transferred from the incoming to the outgoing places breaking or creating bonds according to the specifications on the incoming and outgoing arcs. Let us assume that $\beta(M_m)=M_s$.  For $\beta(M'_m)=M'_s$ to hold by the respective definitions of $\trans{}$ of each RPN model it must be that $\eta(t_m,k_m)=t_s$. The bijections $\beta$ and $\eta$ constitute an isomorphism
	between the reachable parts of ${\cal{H}}_s(N',M_0')$ and ${\cal{H}}_m(N,M_0)$, and the result follows. 
	\item $\state{M_m}{H_m}\rtrans{(t_m,k_m)}\state{M_m'}{H_m'}$ and $\state{M_s}{H_s}\rtrans{t_s}\state{M_s'}{H_s'}$ are both reverse  executions where $\rtrans{}$ represents either $\btrans{}$, $\ctrans{}$, or $\otrans{}$.  In all three cases the transition is reversed by undoing the effect of the transition, i.e. breaking or creating a bond, and transferring the resulting components in the incoming places of the transition, for backtracking and causal order, or to the outgoing place of the last participating transition in out-of-causal order reversibility. We make again the same arguments as in forward execution and by assuming that $\beta(M_m)=M_s$ we show that $\beta(M'_m)=M'_s$ holds by the respective definitions of $ \btrans{}, \ctrans{},\otrans{}$ only if $\eta(t_m,k_m)=t_s$. Again the bijections $\beta$ and $\eta$ constitute an isomorphism
	between the reachable parts of ${\cal{H}}_s(N',M_0')$ and ${\cal{H}}_m(N,M_0)$, and the result follows. 
	
	 \proofend
\end{itemize}

 \begin{theorem}{\rm For every single reversing Petri net $N = (A, P, B, T, F) $ with initial marking $M_0$ there is a multi reversing Petri net $N' = (P',T',A',A_v,B',F')$ with initial marking $M_0'$ such that ${\cal{H}}_m(N',M_0') \cong_{\cal{R}}  {\cal{H}}_s(N,M_0)$.
 }\end{theorem}

\paragraph{Proof}

The proof follows trivially as SRPNs are a special instance of MRPNs with single tokens. 
\proofend

\begin{figure}[t]
	\centering
	\subfloat[Reversing Petri net with multi tokens]{\includegraphics[height=4cm]{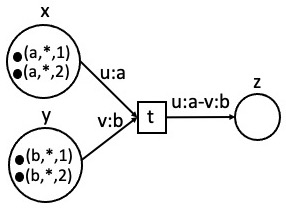}}
	\hspace{2cm}
	\subfloat[Reversing Petri net with single tokens]{\includegraphics[height=5cm]{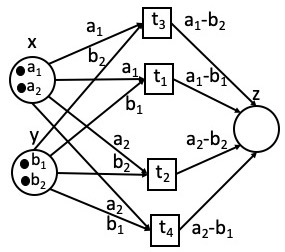}} 
	\caption{Equivalent RPNs with multi and single tokens}
	\label{multiFig}
\end{figure}

%
%

\begin{figure}[t]
	\centering
	\subfloat[LTS for reversing Petri net with multi tokens]{\includegraphics[height=3cm]{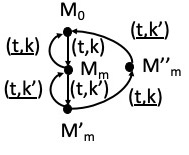}}
	\hspace{2cm}
	\subfloat[LTS for reversing Petri net with single tokens]{\includegraphics[height=7cm]{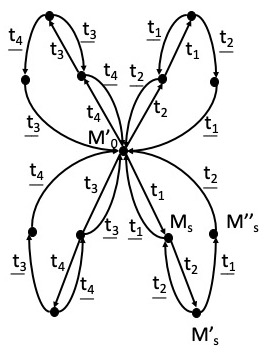}} 
	\caption{Labelled transition systems for the reversing Petri nets in Figure \ref{multiFig}}
	\label{LTSFig}
\end{figure}

%

In Figure~\ref{multiFig} we present a MRPN $N$ and its respective SRPN $N'$. From $N$ we are able to obtain the $SRPN$ $N'$ by constructing the unique tokens $a_1$, $a_2$, $b_1$ and $b_2$ each of them representing one of the tokens $(a,*,1)$, $(a,*,2)$, $(b,*,1)$, and $(b,*,2)$ respectively. The places are the same in both RPN models. The amount of transitions  constructed for the SRPN is dependent on the type of variables required for each MRPN transition and the amount of tokens representing that type.  Specifically for each token of type $a$ associated with the variable $v, type(v)=a$ a respective transition is constructed in the SRPN. Thus, in this example two tokens of type $a$ represent the token variable $u$ and two tokens of type $b$ represent the token variable $v$. As both variables $u$ and $v$ are required for the transition to fire four combinations of tokens of type $a$ and $b$ exist resulting in four different transitions. On that note, the arcs between the places and the constructed transitions follow the  token/bond variable specifications in the MRPN expressed by the combinations of tokens in the SRPN. 

 Let the LTSs in Figure~\ref{LTSFig} capture the complete state space  of the respective RPNs in Figure~\ref{multiFig}. The equivalence of $N$ and $N'$ manifests itself as an isomorphism of reachable parts of the associated LTSs. 
  Letters like $t$ and $t_1$ stand for different events labelled $t$. In fact the first step of ${\cal{H}}_s(N',M_0')$ is $(M_0',t_1,M_s)$ where the first step of ${\cal{H}}_m(N,M_0)$ is $(M_0,(t,k),M_m)$. As we can see $\beta(M_0)=M_0'$ since $type(a_1)=type((a,*,1))$, $type(a_2)=type((a,*,2))$, $type(b_1)=type((b,*,1))$, and $type(b_2)=type((b,*,2))$. The same can be observed for all reachable markings in both RPNs. We also know that $\eta(t,k) =t_1$ since $type(u)=type(a_1)$ and $type(v)=type(b_1)$. The same equivalence applies between $t$ and the rest of the transitions in the SRPN. 
Therefore, ${\cal{H}}_s(N',M_0') \cong_{\cal{R}} {\cal{H}}_m(N,M_0)$.

\section{ Semantics Under the Collective Token Interpretation }

In this section we describe a new form of  reversibility based on the collective token interpretation philosophy according to which tokens of the same type are not distinguished. We note that this philosophy is maintained by various application domains e.g. recourse aware systems or systems from biology. This approach is implemented as another firing rule for  multi reversing Petri nets. Unlike forms of reversing under the individual token interpretation this approach focuses on the local nature of reversing Petri nets and introduces a new approach on reversing systems where the interest is on the location of tokens rather than the relations between transitions.

\begin{figure}[tb]
	\centering{
		\includegraphics[height=3cm]{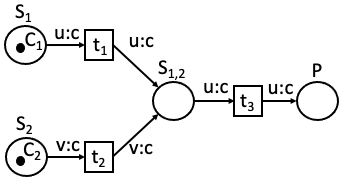}\hspace{0.3in}
	\includegraphics[height=2cm]{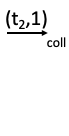}\hspace{0.3in}
\includegraphics[height=3cm]{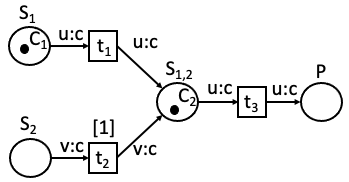}\hspace{0.3in}
\includegraphics[height=2cm]{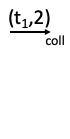}\hspace{0.3in}
\includegraphics[height=3cm]{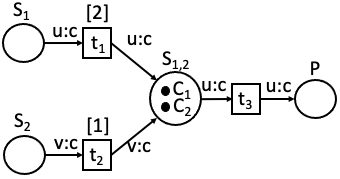}\hspace{0.3in}
	\includegraphics[height=2cm]{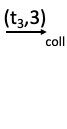}\hspace{0.3in}
\includegraphics[height=3cm]{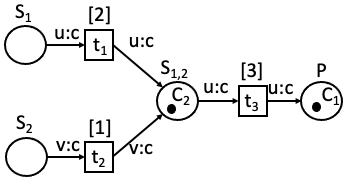}\hspace{0.3in}
\includegraphics[height=2cm]{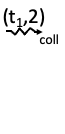}\hspace{0.3in}
\includegraphics[height=3cm]{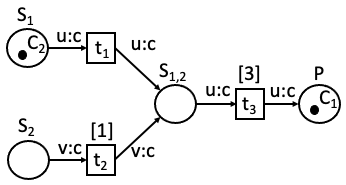}
}
\caption{Students buying present for their teacher}
\label{coll1}
\end{figure}

\begin{figure}[tb]
	\centering
	{\includegraphics[height=3cm]{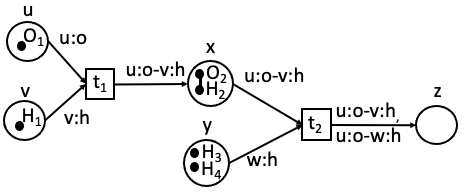}}
		{\includegraphics[height=2cm]{figures/t1,1coll.PNG}}	{\includegraphics[height=3cm]{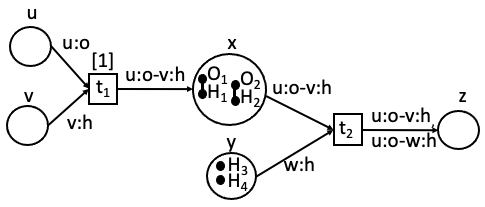}}	{\includegraphics[height=2cm]{figures/t2,2coll.PNG}}	{\includegraphics[height=3cm]{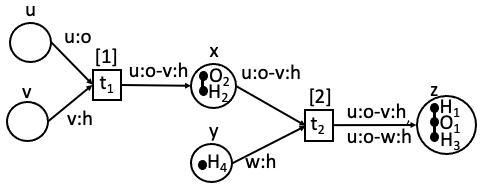}}	{\includegraphics[height=2cm]{figures/t1,1rcoll.PNG}}	{\includegraphics[height=3cm]{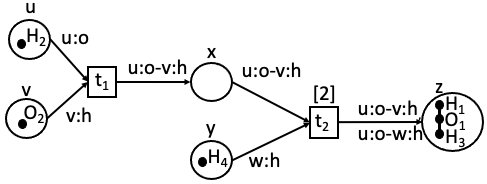}}
	\caption{Chemical reaction for the creation of water molecules}
	\label{coll2}
\end{figure}

To better understand the purpose of the collective token interpretation consider the example in Figure~\ref{coll1}, originally presented in~\cite{individual}. This example illustrates  the situation where two students want to buy a present for their teacher. In places $S_1$ and $S_2$ we are able to find their coins, indicated by tokens $C_1,C_2$. The actions $t_1$ and $t_2$ indicate the contribution of the coins where action $t_3$ indicates the act of buying the present that only costs one coin. After the contributions are made and the present has been bought one coin remains in place $S_{1,2}$ that needs to be returned. Based on the collective token interpretation this coin can be returned to any of the students since the purchase has been caused by the contribution of both of them. Whereas in the individual token interpretation we only have one option when going backwards that is predefined during forward execution. Reversal can therefore be identified by keeping track of the student whose coin has been used for the purchase. 

In this case the collective token philosophy is a fairer description, in which the buying of the present is caused by a disjunction of the two contributions, whereas the individual philosophy suggests that the present is bought from the contribution of either one student or the other. Thus, we relax the requirement of backward determinism of reversible computation and propose a variation where we can reverse any of the transitions where students contribute their coins and return the coin to a randomly selected student. 
Therefore, the collective token philosophy gives rise to more subtle causal relationships between transitions that cannot be expressed by partial order. This relation where a transition could be causally dependent on either of two transitions is called disjunctive causality. In this case the system admits only one execution where the disjunctive causality is realised as  $t_1\lor t_2$ causing $t_3$. 
Whereas in the individual token interpretation causality is given as a partial order and we have two separate executions, one where $t_1$ causes $t_3$ and another execution where $t_2$ causes $t_3$.

Under the collective token interpretation, two transitions are considered to be causally independent when the preconditions for the execution of one do not change by the execution of the other. To understand the effect of this let us consider another example in Figure~\ref{coll2} which describes the process of creating two water molecules. In this example, transitions $t_1$ and $t_2$ create a water molecule $H \bond O \bond H$. Both $t_1$ and $t_2$ are enabled in the initial marking since there already exists a hydroxide element $O_2\bond H_2$ in $y$ and another one can be created by the execution of $t_1$. The execution of $t_1$ will place another hydroxide molecule $O_1\bond H_1$ in place $y$ which results in two bonds of an identical type in the same place. According to the collective token interpretation, these transitions are always considered to be concurrent because the execution of the one does not preclude the execution of the other. Since the enabling condition of $t_2$ does not depend on the execution of $t_1$ but only on the existence of a token of type $o\bond h$, then there is no reason in distinguishing whether the molecule was a result of $t_1$ or not. 

This frame of mind when it comes to causal relations also reflects on how we perceive reversibility. Lets us assume that during the execution of $t_2$ the $O_1\bond H_1$ molecule that has been produced by $t_1$ bonds with $H_3$ to create the water molecule $H_1\bond O_1 \bond H_3$. In the collective token interpretation we are allowed to proceed with the reversal of $t_1$ since we already have the required $o\bond h$ tokens in the outgoing place $y$. When reversing, we undo the effect of $t_1$ by breaking the bond $O_2\bond H_2$ without distinguishing whether this was the pre-existing molecule or the exact one that has been produced during forward execution. However, in the individual token interpretation, we keep track of the tokens that have executed each transition, which means that $t_1$ cannot be reversed by the pre-existing $O_2\bond H_2$ since it can only be reversed after the reversal of $t_2$.


In the following subsections we present an additional firing rule for  multi reversing Petri nets under the collective interpretation. As we are no longer interested in explicitly distinguishing tokens that participate in specific transitions we no longer update the memories of token instances as triples of the form $(a,k,u)$. We assume that
for any token  of type $a$ there may exist a finite number of \emph{token instances}. 
We denote initial token $(a,*,i)$ with $a_i$ and we use $A_I$ for the set of all token instances. Tokens are distinguished by their index $i$ in order to avoid introducing multisets to the model. Tokens of the same type have identical capabilities on firing transitions and can participate only in transitions with variables of the same type. 

The collective token interpretation is proposed as an additional form of reversibility because, due to its local nature, it allows reversing to states that cannot be reached through forward execution. For this reason it does not follow backtracking, causal, and non-causal semantics because it involves and proposes reversal based on a distinct notion of causality closely related to disjunctive causality (rather than the usual partial-order causality). Tokens of the same type are considered to be indistinguishable and when a transition involving a certain set of tokens is to be reversed, any set of tokens of the needed types can be employed to reverse the transition. As a result, different components may be involved in the forward and backward execution of a transition and reversing a transition may lead to states not reachable by forward-only execution. 
Furthermore, the approach is light in terms of memory and preserves the local nature of classical Petri nets. 
We note that an alternative approach could be followed to enable the definition of causal-order and out-causal-order reversibility by considering two tokens to be indistinguishable only if they belong to equivalent connected components. However, we have opted for the present approach, motivated by the applications at hand as well as the philosophy described above.

	\subsection{Forward Execution}
	
	We may now redefine the forward firing rule by ignoring the memories of token instances in order to allow tokens to reverse transitions that they have not participated in. Forward enabledness is defined in the same manner as in the individual token interpretation where token instances are selected non-deterministically as long as they respect the variable types required by the transition's incoming arcs. Based on the selected  forward enabling assignment $U_f$ we are able to identify the tokens that are removed from places as defined by $^\bullet\!{U_f}$ and the tokens that are added to places as defined by $U_f^\bullet$. Thus a transition firing executed in the forward direction will transfer a set of token and bond instances, as specified by the incoming arcs of the transition, to the outgoing places of the transition, as specified by the transition's outgoing arcs, possibly forming or destructing bonds, as necessary. In the collective token interpretation token instances are relocated without being updated with memories of their past transitions. Furthermore, the history of the executed transition is updated accordingly.

\begin{definition}{\rm \label{collforward}
		Given a MRPN $(P,T, A, A_V, B, F)$, a state $\langle M, H\rangle$, a transition $t$ that is enabled in state $\langle M, H\rangle$,  and an
		enabling assignment $U_f$,
we write $\state{M}{H}
		\transcoll{(t,k)} \state{M'}{H'}$
		where $k=max(\{0\} \cup \{k'|k'\in H(t''),t''\in T\})+1$ and for all $x\in P$:
		\begin{eqnarray*}
			M'(x) =   (M(x)- ^\bullet\!{U}_f(x))
			\cup U_f^\bullet (x) 
		\end{eqnarray*}
		\[
		\hspace{0.2in}\mbox{and}\hspace{0.2in}
		H'(t') = \left\{
		\begin{array}{ll}
		H(t') \cup  \{k\} \hspace{2cm} \textrm{ if } t' = t  \; \\
		H(t'), \hspace{2.3cm}  \;\textrm{ otherwise }
		\end{array}
		\right.			\]
}\end{definition} 

%

\begin{figure}[t]
	\begin{center}
		\hspace{.6cm}
		\includegraphics[height=3cm]{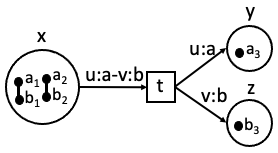}\\
		\includegraphics[height=2cm]{figures/t,1.png}
		\includegraphics[height=3cm]{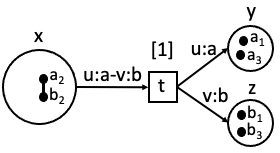}
	\end{center}
	\caption{Forward execution under the collective token interpretation} \label{collf}
	
\end{figure}

As demonstrated in Figure~\ref{collf} a bond of type $a\bond b$ is selected from the incoming places of the transition as required by the variables labelled on the incoming arcs. The result of firing the transition is transferring the tokens into their respective outgoing places by breaking the bond between them. Unlike the individual token interpretation we do not update the memories of the token instances as we are only interested in the location of tokens, i.e. the effect of the transitions, as well as, the fact that a transition has been executed. 

	\subsection{Reversing Execution}
	We now move on to  reversing transitions. A transition can be reversed in a certain state if it has been previously executed and 
there exist token instances in its output places 
that match the requirements on its outgoing arcs. Note that compared to the individual token interpretation, in the collective approach we ignore the causal paths assigned to the tokens during forward execution. As such, tokens are allowed to reverse any transition as long as they respect the variable types, independently on whether the tokens were explicitly used for firing this particular transition occurrence. Specifically, we define the notion of collective reverse 
enabledness as follows:

\begin{definition}\label{renabled}{\rm
		Consider a MRPN $(P,T, A, A_V, B, F)$, a state $\state{M}{H}$, and a transition occurrence $(t,k)$.
		We say that $(t,k)$ is \emph{coll-enabled} in $\state{M}{H}$ if 
		there exists an injective function 
		$U_{coll}:\effects{t}\cap A_V\rightarrow  A_I$ such that:
		\begin{enumerate}
			\item  for all
			$u\in F(t,x)$, $x\in t\circ$, we have $U_{coll}(u)\in M(x)$ where $\type(u)= \type(U_{coll}(u))$, and for all $(u,v)\in F(t,x)$, then  $(U_{coll}(u),U_{coll}(v))\in M(x)$,
			
			\item  If $u,v \in F(t,x), x\in t \circ $ and  $(U_{coll} (u), U_{coll}(v)) \in M(x)$ then $(u,v) \in F(t,x)$, and 
				\item if $u \in F(y_1,t), v \in F(y_2,t), y_1,y_2\in \circ t, y_1\neq y_2 $ then $U_f(u)\not\in \connected( U_f(v), (M(x) -\post{t,U_{coll}}) \cup \pre{t,U_{coll}}),x\in \circ t$.
		\end{enumerate}
}\end{definition}

Thus, a transition occurrence  $(t,k)$ is reverse-enabled based on the collective token interpretation in $\state{M}{H}$ if  
(1)  there exists a type-respecting assignment of token instances,
from the instances in the out-places of the transition, to the variables on the outgoing
edges of the transition, and where the instances are connected with bonds as required by the transition's outgoing edges.  Furthermore, (2) if the selected token instances are bonded together in an outgoing place of the transition then the bond should also exist on the variables labelling the outgoing arcs  (thus we do not recreate existing bonds), and  (3)
if two tokens are transferred by a transition to different incoming places then these tokens should not remain connected when removing the selected outgoing tokens and adding the selected incoming tokens (we do not clone tokens). 
 We refer to $U_{coll}$ as a reversal enabling assignment.

We now define the incoming token/bond instances as:

\begin{definition}\label{multisetscoll1}{\rm
		Given a MRPN $(P,T, A, A_V, B, F)$, a state $\state{M}{H}$, a transition occurrence  $(t,k)$ and an
		enabling assignment $U_{coll}$, we define  $^\bullet\!{U}_{coll}:P\rightarrow 2^{A_I\cup B_I}$ to be a function that assigns to each place a set of  incoming token and bond instances that are used for the firing of $t$:
		
		$^\bullet\!{U}_{coll}(x)= \bigcup_{u\in f(t,x)} \connected(U_{coll}(u), M(x))$
		
}\end{definition}

We now define the outgoing token/bond instances as:

\begin{definition}\label{multisetscoll2}{\rm
		Given a MRPN $(P,T, A, A_V, B, F)$, a transition $t$,  and a state $\state{M}{H}$ and an enabling assignment $U_{coll}$, we define $U_{coll}^\bullet :P\rightarrow 2^{A_I\cup B_I}$ to be a function that assigns to each place a set of  outgoing token/bond instances of $t$:\\
		
		$U_{coll}^\bullet (x)=  \bigcup_{u\in f(x,t),U_{coll}(u)\in M(y)} \connected(U_{coll}(u),(M(y)-\post{t,U_{coll}})\cup \pre{t,U_{coll}}$
		
}\end{definition}

To implement the reversal of a transition $t$ according to a reversal enabling assignment $U_{coll}$, the selected instances are relocated  from  the outgoing places of the transition to the incoming places, as specified by the incoming arcs of the transition, with bonds created and destructed accordingly. Note that compared to the individual token interpretation in the collective approach we do not update the causal paths assigned to the tokens during forward execution.

\begin{definition}\label{collective}{\rm
		Given a MRPN $(P,T, A, A_V, B, F)$, 
		a state $\langle M, H\rangle$, and a transition occurrence $(t,k)$ that is reverse-enabled in $\state{M}{H}$  with 
		$U_{coll}$ a reversal enabling assignment, we write $\state{M}{H}
		\colltrans{(t,k)} \state{M'}{H'}$   where for all $x$:
		\[
		M'(x) =  ( M(x)- ^\bullet\!{U}_{coll}(x))
		\cup  U_{coll}^\bullet (x)\]
		\[
		\hspace{0.3in}\mbox{and}\hspace{0.3in}
		H'(t') = \left\{
		\begin{array}{ll}
		H(t') - \{k\}, \hspace{0.3cm} \;\;\textrm{ if } t' = t  \; \\
		H(t'),\hspace{1cm}  \;\textrm{ otherwise }
		\end{array}
		\right.
		\]
}\end{definition}

\begin{figure}[t]
	\begin{center}
		\hspace{1cm}
		\includegraphics[height=3cm]{figures/collf.png} \\
		\includegraphics[height=2.5cm]{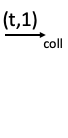}
		\includegraphics[height=3cm]{figures/collf1.png}\\
		\includegraphics[height=2.5cm]{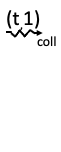}
		\includegraphics[height=3cm]{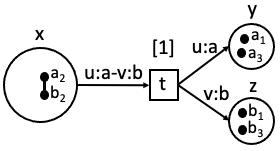}
	\end{center}
	\caption{Forward execution under the collective token interpretation.} \label{collr}
	
\end{figure}

In Figure~\ref{collr} we may observe the reverse execution of the net presented in Figure~\ref{collf}. As two tokens of type $a$ and $b$ are located in the outgoing places of the transition matching the requirements of the variables in the labelled outgoing arcs, we are able to non-deterministically select a pair to reverse transition $t$. As such it is not necessary to reverse the transition with the exact token instances that have contributed in its execution thus we select $a_3 \bond b_3$ to reverse. 

%

\subsection{Case Study for the Collective Token Interpretation }

Biological reactions, pathways, and reaction networks have been extensively studied in the literature 
using various  techniques,  including  process  calculi and Petri nets.  Initial research was mainly 
focused  on reaction rates by modelling and simulating networks of reactions, in order to
analyse or predict the common paths through the network.  Reversibility
was not considered explicitly.  Later on
reversibility  started  to  be  taken  into  account,  since  it  plays  a  crucial  role  in
many processes, typically by going back to a previous state in the system.

\subsubsection{Autoprotolysis of Water}

\begin{figure}
	\centering
	\subfloat{\includegraphics[width=14cm]{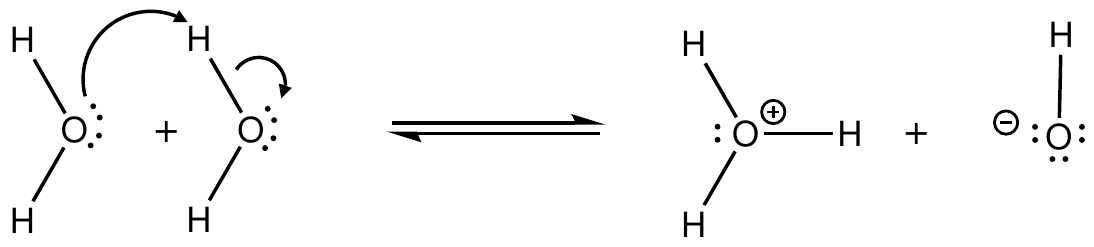}}
	\caption{Autoprotolysis of water}\label{autoprotolysis}
\end{figure}

In this case study we consider a chemical reaction that transfers a hydrogen atom between two water molecules. This
reaction is known as the \emph{autoprotolysis of water} and is shown 
in Figure~\ref{autoprotolysis}. There, $O$ indicates an oxygen atom and $H$ a hydrogen atom. The lines indicate bonds. Positive and negative charges on atoms are shown by $\oplus$ and $\ominus$
respectively. The meaning of the curved arrows and the dots will be explained in the next paragraphs.
The reaction is reversible and it takes place at a relatively 
low rate, making pure water slightly conductive. We have chosen this reaction 
as our example reaction, since it is non-trivial but manageable, and has some
interesting aspects to be represented.

To model the reaction we need to understand why it takes place and what causes it.
The main reason is that the oxygen in the water molecule is \emph{nucleophilic}, meaning it has the tendency to bond to another atomic nucleus, which would serve as an \emph{electrophile}. This is because oxygen has a high 
electro-negativity, therefore it attracts electrons and has an abundance of electrons around it. The electrons around the atomic nucleus are arranged on electron shells, where only those in the outer shell participate in bonding. Oxygen has four electrons in its outer shell, which are not involved in the initial bonding with hydrogen atoms. These electrons form two \emph{lone pairs} of two electrons each, which can form new bonds (lone pairs are shown in Figure~\ref{autoprotolysis} by pairs of dots). All this makes oxygen nucleophilic: it tends to connect 
to other atomic nuclei by forming bonds from its lone pairs. Since oxygen attracts electrons, the hydrogen atoms in water
have a positive partial charge and oxygen has a negative partial 
charge. 

The reaction starts when an oxygen in one water molecule is attracted by a hydrogen in another 
water molecule due to their opposite charges. This results in a \emph{hydrogen bond}. 
This bond is formed out of 
the electrons of one of the lone pairs of the oxygen. The large curved arrow in Figure~\ref{autoprotolysis} indicates the movements of the electrons. Since a hydrogen atom cannot have more 
than one bond, 
the creation of a new bond is compensated by breaking the existing hydrogen-oxygen bond (indicated by the small curved arrow). When this happens, the two electrons, which formed the original hydrogen-oxygen bond, remain with the oxygen. Since a hydrogen contains 
one electron and one proton, it is only the proton that is transferred, so the process can be called a proton transfer as well as a hydrogen transfer. The forming of the new bond and the breaking of the old bond are \emph{concerted}, meaning that 
they happen together without a stable 
intermediate configuration. As a result we have reached the state where one oxygen atom
has three bonds to hydrogen atoms and is positively charged, represented on the right side of the reaction in 
 Figure~\ref{autoprotolysis}. This molecule is called \emph{hydronium} and is written as $\mathrm{H_3O^+}$. The other oxygen atom bonds to only one hydrogen and is negatively charged, having an 
electron in surplus. This molecule is called a \emph{hydroxide} and is written as $\mathrm{OH^-}$. 

Note that the reaction is reversible: the oxygen that lost a hydrogen can 
pull back one of the hydrogens from the other molecule, the  $\mathrm{H_3O^+}$ molecule. This is the case since the negatively 
charged oxygen is a strong nucleophile and the hydrogens in the $\mathrm{H_3O^+}$ molecule are 
all positively charged. Thus, any of~the hydrogens 
can be removed, making both oxygens formally uncharged, and restoring the two water 
molecules. In Figure~\ref{autoprotolysis} the curved arrows are given for the reaction going from left to right. Since the reaction is reversible (indicated by the double arrow) there are corresponding
electron movements when going from right to left. These are not given in line with usual conventions, but can be~inferred.

In this simple reaction, the forward and the reverse step consist of two steps each. The breaking of the old and the forming of the new bond occur simultaneously. This means that there is no strict causality of actions, since none of them can be called the cause of the overall reaction. Furthermore, the reverse step can be done with a different atom to the one used during the forward step because each of the molecules are in a sense identical and in practice there does not exist a single ``reverse'' path corresponding to a forward one.

It should be noted that there are two types of bonding modelled here. Firstly, we have the initial bonds where two atoms contribute an electron each. Secondly, the \emph{dative} or \emph{coordinate bonds} are formed where both electrons come from one atom (an oxygen in this case). Both are \emph{covalent bonds}, and once formed they cannot be distinguished. Specifically, in the oxygen with three bonds all bonds are the same and no distinction can be made. If one of the bonds is broken by a deprotonation (as in the autoprotolysis of water) the two electrons are left behind and they form a lone pair. If the broken bond was not previously formed as a dative bond, the electrons changed their ``role''. This explains why any proton can be transferred in the reverse reaction and not just the one that was 
involved in the forward path.

\begin{figure}
	\centering
	\subfloat{\includegraphics[width=12cm]{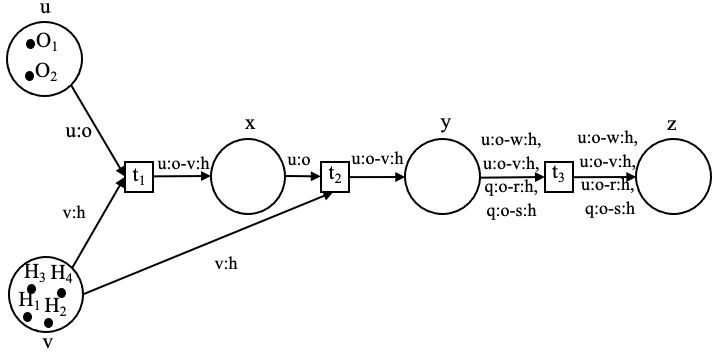}}
	\subfloat{\includegraphics[width=1.2cm]{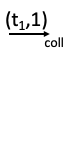}}\\
	\subfloat{\includegraphics[width=12cm]{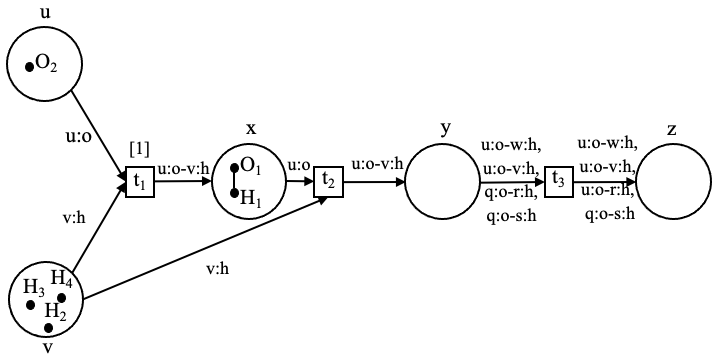}}
	\subfloat{\includegraphics[width=1.2cm]{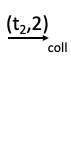}}\\
	\hspace{-1cm}
	\subfloat{\includegraphics[width=12cm]{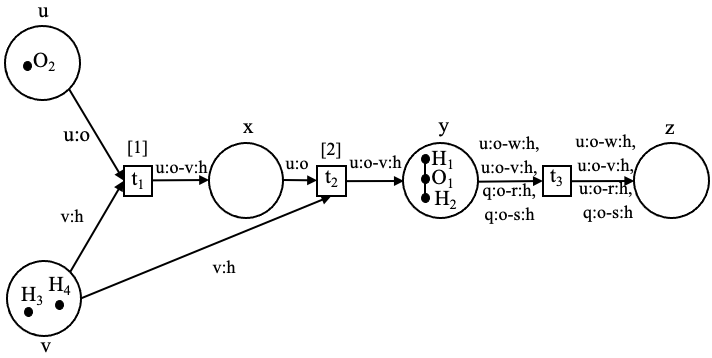}}	\caption{RPN model of the formation of a water molecule}
	\label{RPNmodel}
\end{figure}

\begin{figure}
        \centering
        \subfloat{\includegraphics[width=1.2cm]{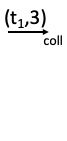}}
        \subfloat{\includegraphics[width=1.2cm]{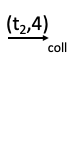}}
        \subfloat{\includegraphics[width=12cm]{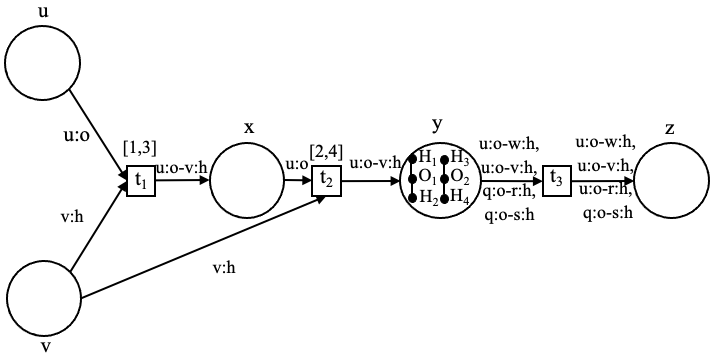}}\\
        \hspace{1.5cm}
        \subfloat{\includegraphics[width=1.2cm]{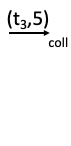}}
        \subfloat{\includegraphics[width=12cm]{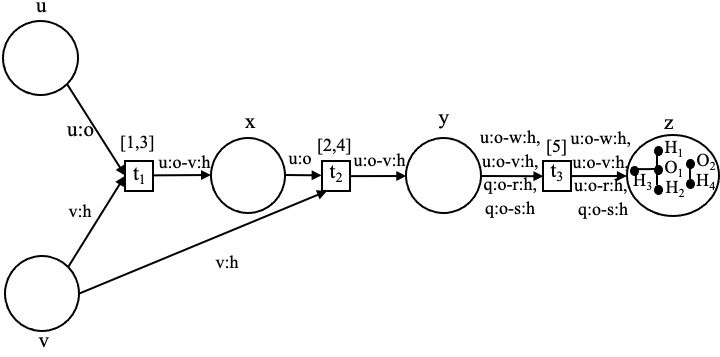}}\\
        \hspace{1.5cm}
        \subfloat{\includegraphics[width=1.1cm]{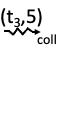}}
        \subfloat{\includegraphics[width=12cm]{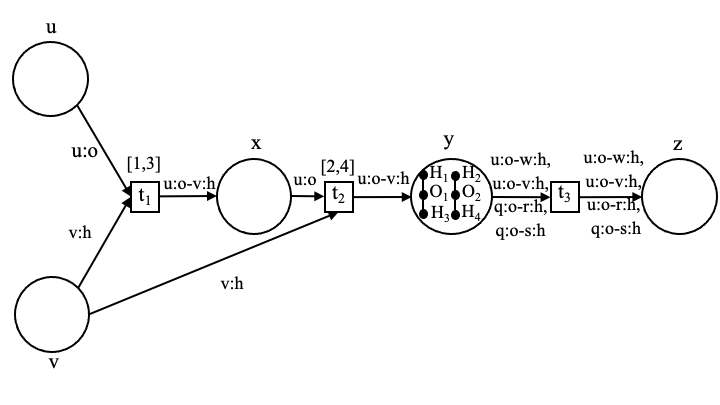}}
        \caption{RPN model of the execution of the autoprotolysis of water}
        \label{execution}
\end{figure}

\noindent
\subsubsection{ Reversing Petri Net representation }
Figure~\ref{RPNmodel} shows the graphical representation of the forming of a water molecule as a RPN.
In this model, we assume two token types, $h$ for hydrogen
and $o$ for oxygen. They are instantiated via four token instances of $h$ ($H_1$, $H_2$, $H_3$, and $H_4$) and two token instances of
$o$, ($O_1$ and $O_2$). The net consists of five places and three transitions and the edges between
them are associated with token variables and bonds, where we assume that $\type(u)=\type(q) = o$
and $\type(v)=\type(w)=\type(r)=\type(s)=h$. 
Looking at the transitions, transition $t_1$ models the formation of a bond between a hydrogen
token and an oxygen token. Precisely, the transition stipulates a selection of two such
molecules with the use
of variables $u$ and $v$ on the incoming arcs of the transition which are bonded
together, as described in the outgoing arc of the transition.
Subsequently, transition $t_2$ completes
the formation of a water molecule by selecting an oxygen token from place $x$ and
a hydrogen token from place $v$ and forming a bond between them, placing the resulting component at place $y$. Note that the selected oxygen instance in this transition will be connected to a hydrogen token via a bond 
created by transition $t_1$; this bond is preserved and the component resulting from the creation of
the new $o-h$ bond will be transferred to place $y$.
Finally, transition $t_3$ models the autoprotolysis reaction: assuming the existence of two distinct oxygen
instances, as required by the variables of type $o$ and $h$ on the incoming arc of the transition, connected
with hydrogen instances as specified in $F(y,t_3)$, the transition breaks the bond $q-r$ and forms the
bond  $u-r$.  As such, assuming the existence of two water molecules at place $y$, the transition
will form  a hydronium  ($H_3^+O$) and a hydroxin ($OH^-$) molecule in place $z$ of the net.
The reversibility semantics of RPNs ensures that  reversing the transition $t_3$ will result in the re-creation
of  two water molecules placed at $y$, while the use of variables allows the formation of water molecules consisting
of different bonds between the hydrogen and oxygen instances. 

The first net in Figure~\ref{execution}  shows the system after the execution of transition $t_1$ with enabling assignment $U_f(v)=H_1$, $U_f(u)=O_1$. 
Subsequently,
we have the model after execution of transition $t_2$ with enabling assignment $U_f(v)=H_2$,
$U_f(u)=O_1$, creating the bond $O_1-H_2$, thus forming the first water molecule. A second
execution of transitions $t_1$ and $t_2$ results in the second molecule of water in the system,
placed again at place $y$, as shown in the third net in the figure. At this state, transition
$t_3$ is forward-enabled and,  with enabling assignment $U_f(u)=O_1, U_f(q)=O_2, U_f(w)=H_1, 
U_f(v)=H_2,U_f(r)=H_3,U_f(s)=H_4$, we have the creation of the hydronium and hydroxide
depicted at place $z$ in the fourth net of the figure. At this stage, transition $t_3$ is
now reverse-enabled and the last net in the figure illustrates the state resulting after
reversing $t_3$ with reversal enabling assignment $U_{coll}(u)=O_1, U_{coll}(v)=O_2, U_{coll}(w)=H_1, 
U_{coll}(v)=H_2,U_{coll}(r)=H_3,U_{coll}(s)=H_4$.

\remove{
Reversing Petri Nets are Petri net structures that assume tokens to be distinct and persistent. During the execution of transitions individual tokens can be bonded/unbonded with each other, and the creation/destruction of these bonds is considered to be the effect of a transition, whereas their destruction/creation is the effect of the transition's reversal. }

\section{Concluding Remarks}

This chapter has focused on relaxing the restrictions of the RPN model presented in the previous chapter,  
by allowing multiple tokens of the same base/type to occur in a model and developing reversible semantics in the presence of bond destruction. We have extended our formalism with multi tokens by following the individual token philosophy, which defines the notion of causality in reversible systems as a partial order. The individuality of identical tokens can be imposed by their causal path, which allows identical tokens to fire the same transition when going forward, however, when going backwards tokens will be able to reverse only the transitions that they have fired. 
 Additionally, our work provides the reversible semantics for out-of-causal-order reversibility and shows how the presence of bond destruction affects transition enabledness. Finally, we have shown that the expressive power of RPNs with multi tokens is equivalent to the expressive power of RPNs with single tokens.

 Another approach on extending our formalism with multiple tokens is that of the collective token philosophy. Our experience strongly suggests that resource management systems can be studied and understood in terms of the collective token interpretation of RPNs~\cite{RCbook}. Reversing Petri Nets are a natural choice to model and analyse biochemical reaction systems, such as the autoprotolysis of water, which by nature has multi-party interactions, is inherently concurrent, and features reversible behaviour.

  The autoprotolysis of water has also been modelled by the Calculus of Covalent Bonding (CCB) as well as the Bonding Calculus~\cite{RCbook}. All three models can perform the forward reaction using any of the hydrogens involved. In RPNs the feature of token multiplicity  and
the use of variables allows to non-deterministically
select different combinations of atoms of a particular element when creating molecules.  Unlike Bonding calculus, CCB and RPNs are able to express concerted actions, since a transition simultaneously destroys  a water molecule and creates a hydronium whose  reversal results in the opposite effect. CCB and RPNs can perform the reverse reaction by transferring arbitrary hydrogens, whereas the Bonding Calculus permits only the transfer of exactly those hydrogens that were used in the forward reaction. 
%

	The other criterion for comparing the formalisms for the modelling of chemical reactions is to ask if they enable the same actions as they appear in reality.  Each of the three formalisms does not permit a $H_3O$ molecule to be formed directly. Furthermore, CCB and the bonding calculus allow one reaction which is not realistic: If there are many water molecules and therefore several hydroxide and water molecules at the same time, it is possible that the remaining hydrogen is transferred from the hydroxide to a water. In reality, this is not possible since the hydroxide is strongly negatively charged and no hydrogen bond can form. However, this is not the case for RPNs since, on the one hand, a transition’s conditions make restrictions on the types of molecules that will participate in a transition firing or its reversal and, on the other hand, places impose a form of locality for molecules. For instance, in the autoprotolysis example, each place is the location of specific types of molecules, e.g., transition $t_3$ modelling the autoprotolysis reaction is only applied on water molecules and its reversal only on pairs of a hydronium and a hydroxide molecule, as required.

\chapter{Controlling Reversibility in Reversing Petri Nets}
\label{sec:control}

In this chapter we extend the framework of reversing Petri nets  with a mechanism for controlling reversibility~\cite{RC19,ieee,coll}. This control is enforced with
the aid of conditions associated with transitions, whose satisfaction acts as 
a guard for executing the transition
in the forward/backward direction.
The conditions are 
enunciated within a simple logical language expressing 
properties relating to available tokens. The mechanism may capture environmental conditions, e.g., changes in temperature, or 
the presence of faults. 
Note that conditional transitions can also be found in existing Petri net models, e.g., in~\cite{CPN}, a 
Petri-net model that associates transitions and arcs with expressions. 
The resulting model is general enough to capture a wide range of systems, in this context we give an overview of   several properties  of reversing Petri nets that could be used to analyse the behaviour of these systems. We conclude this section with the model of a novel antenna selection (AS) algorithm which inspired the development of our framework.

	    
\section{Controlled Reversing Petri Nets}

In this section we extend the multi
reversing Petri nets of Chapter 4, by associating transitions with conditions that control their execution and reversal, enunciated on  data values associated with tokens. We introduce controlled reversible semantics under the collective token interpretation where transitions are controlled and can break bonds, and tokens are indistinguishable and can curry data values. Specifically, we define:

\begin{definition}{\rm
		A \emph{Controlled Reversing Petri Net} (CRPN) is a tuple $(N, \Sigma,D, C_F,C_R, I)$ where:
		\begin{enumerate}
			\item $N$ is a multi reversing Petri net.
			\item $\Sigma$ forms a finite set of data types with $D$ the associated
			set of data values where $type_{\Sigma}(d)\in \Sigma,$ $d\in D$.
			\item $C_F:T\rightarrow COND_{A_V}$ is a function that assigns a forward condition to each 
			transition $t\in T$.
			\item $C_R:T\rightarrow COND_{A_V}$ is a function that assigns a reverse condition to each 
			transition $t\in T$.
			\item $I : A_I \rightarrow D$ is a function that associates a 
			data value from $D$ to each token instance $a_i\in A_I$ such that $type_{\Sigma}(I(a_i))=type_{\Sigma}(a_i)$.
		\end{enumerate}
}\end{definition}

As in multi reversing Petri nets a controlled reversing Petri net is built on the basis of a set of \emph{tokens} or \emph{bases}. These
are organized in a set of token types $A$, where each token type is associated with a set of token instances $A_I$. Variable tokens are associated with a data type such that for all $u\in A_V$, $type_{\Sigma}(u) \in \Sigma$. Places and transitions have the standard meaning and are connected
via directed arcs which are labelled by a set of elements from $A_V \cup (A_V \times A_V) $. 
Finally, we define  $C\subseteq A_I\cup B_I$ in the expected way according to MRPNs. We also assume that the CRPN model is well formed with distinct tokens.

In addidtion, in CRPNS token instances are associated with data values via function $I$. These data values have a type from the set $\Sigma$, and we write $type_{\Sigma}(d)$ to denote the type of a data value. 
Transitions are associated with conditions $COND_{A_V}$ which constitute additional preconditions for a transition to fire. Conditions are boolean expressions over a set of variables $A_V$ that evaluate to either "TRUE" or "FALSE" determining the behaviour of the net. The function $C_F$ assigns a forward condition to each transition that needs to be satisfied during forward execution, whereas, $C_R$ assigns a reverse condition that needs to be satisfied during reverse execution.  Graphically we indicate conditions below their respective transitions as $C_F(t)/C_R(t)$ and in case where  $C_F(t)=!C_R(t)$ only $C_F(t)$ is presented. 

Conditions are built via a simple propositional language whose basic building blocks are relations on the data types $\Sigma$  applied on expressions involving token values of a CRPN model. An instantiation of such a language for arithmetic expressions follows, though this can be generalised for more complex types. 
Therefore, the grammar of the expression $COND_{A_V}$ is defined as follows: 

$\phi := 
\;\;\neg \phi \;\;|\;\;\phi_1 \lor \phi_2\;\;|e_1 > e_2\;\;$


$e:= \;\;a_i.x\;\;|\;\; u\;\;|\;\; d\;\; 
|\;\;(e) \;\;| \;\; if\;\; \phi\;\; e_1\;\; else\;\; e_2\;\; | \;\;e_1 + e_2 \;\;| \;\;e_1 - e_2 \;\;|\;\; e_1 \times e_2\;\; | \;\;e_1 / e_2 \;\;$ 

%
%
 
The free variables in a  condition $\phi$ are denoted by $Free(\phi) \subseteq A_V$.  Variable assignments $V: A_V \cap Free(\phi) \rightarrow A_I$  are the mappings of a token instance to a free variable. We require the variable assignments  to respect the types of data values associated with the respective variable token instances such that for $V(u)=a_i$, $a_i\in A_I\cap M(x)$, $x\in P$ we have $type_{\Sigma}(u)= type_\Sigma(a_i)$. The variable assignment $V$ of a transition covers (at least) all variables from $\guard{t}$ such that $u\in \guard{t} \cap Free(\phi)) = \emptyset$ ($\effects{t}\cap Free(\phi) = \emptyset$ for reverse transitions).
 

 Conditions are evaluated based on data values associated with the token instances of the model and functions/predicates over the associated data types.
 Given a transition $t$ with condition $\phi$, the corresponding variable  assignment $V$, a marking $M$, and an assignment function $I$, we evaluate the condition $\phi$ as follows:

	\[
E(\phi,V,M,I)= \left\{
\begin{array}{ll}
\neg E(\phi', V, M,I),   \hspace{7.5cm}  \;\textrm{ if } \phi =\neg \phi'\\
E(\phi_1, V, M, I)  \; \lor \;  E(\phi_2, V, M, I),\hspace{3.8cm}  \;\textrm{ if }   \phi=\phi_1 \; \lor \; \phi_2 \\
Eval(e_1, V,M,I)>Eval(e_2,V,M,I) \hspace{3.6cm}  \;\textrm{ if }   \phi=e_1 \; > \; e_2
\end{array}
\right.
\]

	\[
 Eval(e,V,M,I)= \left\{
\begin{array}{ll}
Eval(e_1, V, M, I), \hspace{1.4cm} \;\textrm{ if } e=if \; \phi \;  then \; e_1 \; else \; e_2, \;E(\phi, V,M,I)=T  \\
Eval(e_2, V, M, I), \hspace{1.4cm}  \;\textrm{ if } e=if \; \phi \;then \; e_1 \; else \; e_2,\;E(\phi , V,M,I)=F  \\
Eval(e_1, V, M, I)  \; \diamond \;  Eval(e_2, V, M, I),  \hspace{.4cm}
  \;\textrm{ if }  e=e_1  \; \diamond \;e_2,
   \diamond\in \{ +, -, \times, / \} \\
 I(V(u)), \hspace{7.5cm} \;\textrm{ if } e=u, u\in A_V\\
 I(a_i), \hspace{7.2cm} \;\;\textrm{ if } e=a_i.x, a_i\in M(x)  \;\\
 0,\hspace{7.8cm}  \;\textrm{ if }e=a_i.x,a_i\not \in M(x) \; \\ 
 d, \hspace{8.7cm} \;\textrm{ if } e=d, d\in D\\
\end{array}
\right.
\]

The function $E: (COND_{A_V} \times A_I \times (2^{A_I \cup B_I}) \times D) \rightarrow BOOL$ evaluates the condition of a transition into a boolean value and the function $Eval: (COND_{A_V} \times A_I \times (2^{A_I \cup B_I}) \times D) \rightarrow D$ evaluates the data value of an arithmetic expression.  The truth value of conditions depends on the interaction of the logical operators and their component conditions. Arithmetic conditions have the value "TRUE" if the relation exists between the two expressions and "FALSE" otherwise. We resolve nested conditions with recursive evaluation  where variable assignments $V(v)=a_i$ are substituted by the data value of the selected token instance $I(a_i)$. Elements of the form $a_i.x$ are substituted  by the data value of the token instance  $I(a_i)$ if the token instance exists in place $x$, and $0$ if not ($0$ is the identity element dependent on the application). 

\subsection{Controlled Forward Execution}
A transition is forward-enabled in a MRPNs, if there exists a selection of token instances 
available at the incoming places of the transition 
matching the requirements  on the transitions incoming arcs. Also the transition should not recreate bonds and duplicate tokens. 
 The addition of conditions in CRPNs requires additionally for the forward condition of the transition to evaluate to TRUE according to the variable  assignment $V_f$. Formally:

\begin{definition}\label{confenabled}{\rm
		Given a CRPN $(N, \Sigma, D, C_F,C_R, I)$, a state $\state{M}{H}$, and a transition $t$, we say that $t$ is 
		\emph{controlled-forward-enabled} in $\state{M}{H}$  if there exist two injective functions $U_f:\guard{t}\cap A_V\rightarrow A_I$ and 
		 $V_f:A_V \cap Free(COND_{A_V}) \rightarrow A_I$ such that:
		\begin{enumerate}
			\item transition $t$ is forward-enabled in $N$ (Definition~\ref{multifenabled}),
			\item for all $u \in Free(C_F(t))$ then $V_f(u)=a_i$, $a_i\in A_I\cap M(x),$ $x\in P$ such that $\type_{\Sigma}(u)=\type_{\Sigma}(a_i)$,
			\item if $u\in F(x,t),$ $x\in P$, and $u\in Free(C_F(t))$ then $V_f(u)=U_f(u)$, and  
			\item   $E(C_F(t),V_f,M,I)$= TRUE. 
		\end{enumerate}
}\end{definition}

Thus, $t$ is enabled in state $\state{M}{H}$ if  (1) it is also forward-enabled in MRPNs,
(2) there is a type respecting assignment of data instances to variables, (3) variables that appear on both the arcs and the condition should have the same variable assignment, (4) if the transition bears a forward condition $C_F(t)$, then by substituting the variables with the selected data values $E(C_F(t),V_f,M,I)$ evaluates to TRUE. Note that different selections of 
token and bond instances may yield different evaluations to a transition's condition.
Thus, for some selections the transition may be enabled whereas for others not.
Note that if $E(C_F(t),V_f,M,I)$ or $C_F(t)$ is not defined, then the validity of the input is only dependent on  the first two conditions.
We refer to $V_f$ as a  variable assignment.

\begin{figure}[tb]
	\centering
	\hspace{.7cm}
	\subfloat{\includegraphics[height=3.3cm]{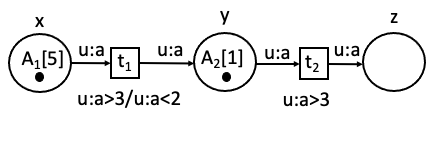}}\\
	\subfloat{\includegraphics[height=3cm]{figures/t1,1coll.png}}
	\subfloat{\includegraphics[height=3.6cm]{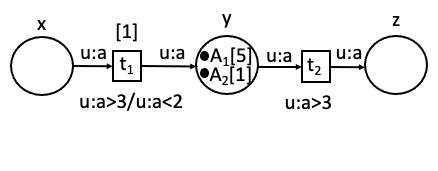}} \\
	\subfloat{\includegraphics[height=3cm]{figures/t2,2coll.png}}
	\subfloat{\includegraphics[height=3.6cm]{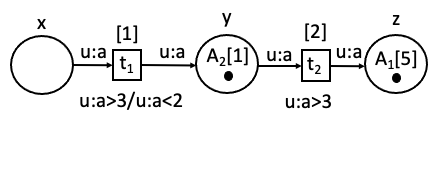}}\\
	\caption{Forward execution in CRPNs}
	\label{CF}
	
\end{figure}

As in MRPNs, when a transition $t$ is executed in the forward direction, all tokens and bonds occurring in its outgoing arcs are relocated from the input to the output places along with their connected components. The history of $t$ is extended accordingly:

\begin{definition}{\rm \label{con-forw}
		Given a CRPN $(N, \Sigma,D, C_F,C_R, I)$, a state $\langle M, H\rangle$, and a transition $t$ controlled-forward enabled in $\state{M}{H}$ with $U_f$ an enabling assignment and $V_f$ a variable assignment, we write $\state{M}{H}
			\transcoll{(t,k)}\state{M'}{H'}$
		where $M'$ and $H'$ are updated as in $N$ (Definition~\ref{collforward}). 
}\end{definition}

Figure~\ref{CF} demonstrates the forward execution of a CRPN with two transitions. The forward conditions of both transitions require a variable assignment for variable $u$ of type $type(u)=a$ to be greater than three ($u>3$). Only token $A_1[5]$ is able to satisfy these conditions and thus fire both transitions. As such, token $A_1[5]$ has been selected as a variable assignment for variable $u$ which is then used to fire transition $t_1$ followed by $t_2$. 

\subsection{Controlled Reversibility}
We now move on to controlled reversibility. The following definition enunciates that a transition $t$ is controlled-reverse enabled if it is also reverse-enabled in multi reversing Petri nets under the collective token interpretation. Furthermore, we require that 
the reverse condition of  the transition is satisfied according to the variable assignment $V_{coll}$. 

\begin{definition}\label{con-rev-enabled}{\rm
	Consider a	CRPN $(N, \Sigma,D, C_F,C_R, I)$, a state $\state{M}{H}$, and a transition  occurrence $(t,k)$. Then $(t,k)$ is controlled-reverse-enabled in  $\state{M}{H}$ if there exist two injective functions $U_{coll}:\effects{t}\cap A_V\rightarrow A_I$ and 
	$V_{coll}:A_V \cap Free(COND_{A_V})  \rightarrow A_I $ such that:
		\begin{enumerate}
			\item $t$ is $coll$-enabled in in $N$ (Definition~\ref{renabled}),
			\item for all $u \in Free(C_R(t))$ then $V_{coll}(u)=a_i$,  $a_i\in A_I\cap M(x),$ $x\in P$ such that $\type_{\Sigma}(u)=\type_{\Sigma}(a_i)$,
			\item if $u\in F(t,x),$ $x\in P$  and $u\in Free(C_R(t))$ then $V_{coll}(u)=U_{coll}(u) $, and 
			\item $E(C_R(t),V_{coll},M,I)$= TRUE. 
		\end{enumerate}	
}\end{definition}

Thus, a transition occurrence $(t,k)$ is reverse-enabled in $\state{M}{H}$ if  
(1) it is $coll-$enabled in MRPNs, (2) (3) the definition makes similar requirements to Definition~\ref{confenabled} only this time in (4) it requires for the reverse condition of the transition to evaluate to TRUE. 
Note that while a selection of tokens may yield the reversal of the transition impossible by setting $E(C_R(t),V_{coll},M,I)=$ FALSE, another selection, and its associated variable assignment $V_{coll}$ may be such that $E(C_R(t),V_{coll},M,I)=$ TRUE. This is determined by the data values associated to the token instances by the assignment $I$. We refer to $V_{coll}$ as a reversal variable assignment.

When a transition occurrence $(t,k)$ is reversed  the marking and history of the controlled reversing Petri net are updated according to the respective definition of MRPNs under the collective token interpretation. 

\begin{definition}\label{con-rev}{\rm
		Given a CRPN $(N, \Sigma,D, C_F,C_R, I)$, 
		a state $\langle M, H\rangle$, and a transition occurrence $(t,k)$ controlled-reversed-enabled in $\state{M}{H}$ with $U_{coll}$ a reversal enabling assignment and $V_{coll}$ a reversal variable assignment, we write $ \state{M}{H}
		\colltrans{(t,k)} \state{M'}{H'}$
		where $M'$ and $H'$ are updated as in  $N$ (Definition~\ref{collective}). 
%
		
}\end{definition}

\begin{figure}[tb]
	\centering
	\subfloat{\includegraphics[height=2.9cm]{figures/t1,1coll.png}}
	\subfloat{\includegraphics[height=3cm]{figures/t2,2coll.png}}
	\subfloat{\includegraphics[height=3.6cm]{figures/cf3.png}}\\
	\hspace{1.5cm}
	\subfloat{\includegraphics[height=3cm]{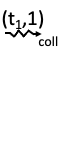}}
	\subfloat{\includegraphics[height=3.6cm]{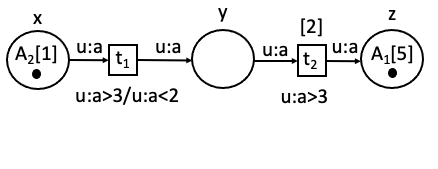}} \\
	\caption{Reverse execution in CRPNs}
	\label{CR}
	
\end{figure}

Figure~\ref{CR} is a continuation of the forward execution of Figure~\ref{CF}. The reverse condition of transition $t_1$ requires for the variable assignment of token variable $u$ of type $type(u)=a$ to be smaller than two ($u<2$), whereas, the reverse condition of transition $t_2$ requires for  a token variable $u$ of type $type(u)=a$ to be smaller than three ($u<3$). Note that in the case of transition $t_2$ we have  $C_F(t_2)=!C_R(t_2)$, thus, the reverse condition is omitted from the figure. In this example, only transition $t_1$ is able to reverse as token $A_2[1]$ can be used for the variable assignment of $u$ enabling the reversal of transition $t_1$. 

		\begin{figure}[tb]
		\centering
		\hspace{.5cm}
		\subfloat{\includegraphics[height=4cm]{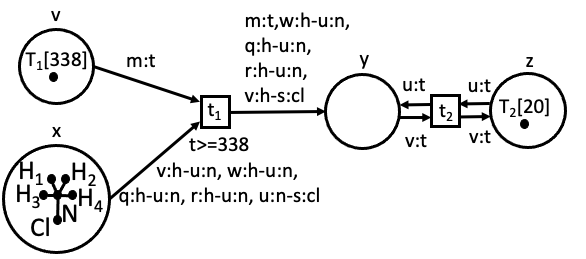}}\\
		\subfloat{\includegraphics[height=2cm]{figures/t1,1coll.png}}
		\subfloat{\includegraphics[height=3.6cm]{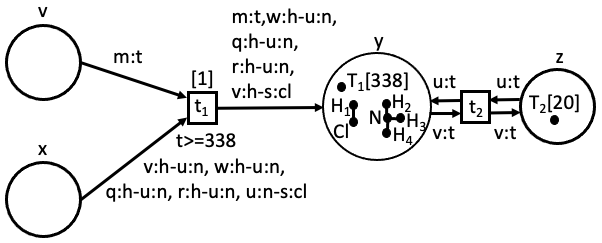}} \\
		\subfloat{\includegraphics[height=2cm]{figures/t2,2coll.png}}
		\subfloat{\includegraphics[height=3.6cm]{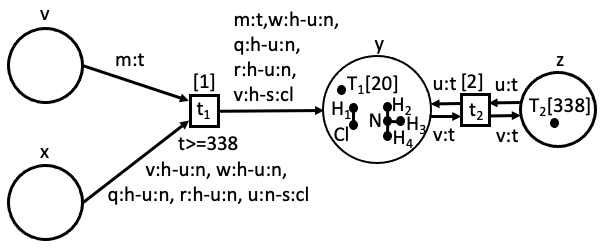}}\\
		\subfloat{\includegraphics[height=2cm]{figures/t1,1rcoll.png}}
		\subfloat{\includegraphics[height=4cm]{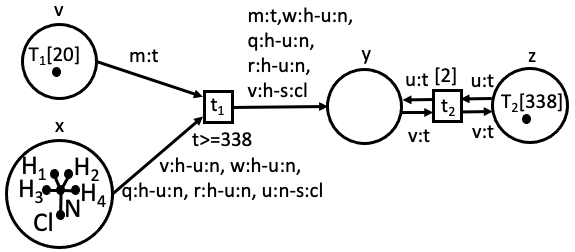}} 
		\caption{Ammonium chloride chemical reaction}
		\label{chloride}
		
	\end{figure}

Let us consider a more complicated example of a reversible chemical reaction that depends on environmental 
	conditions. Ammonium chloride ($NH_4Cl$)  is an inorganic compound that  decomposes into ammonia ($NH_3$) 
	and hydrogen chloride gas ($HCl$). This decomposition is a reversible reaction that occurs when ammonium chloride 
	is heated to over $338$ degrees Celsius. The two gases ammonia and hydrogen chloride can then 
	react together in cooler temperatures to reform the solid ammonium chloride and therefore reverse the decomposition. 
	The recommended storing temperature of ammonium chloride is $15^\circ$C to $25^\circ$C.
	
	The model of this reaction is shown in the initial marking of Figure~\ref{chloride}. Here we assume
	the token types $A=\{H,N, Cl, T\}$, with the first three bearing the expected meaning and type $T$
	capturing different temperatures. In particular $T$ has instances $T_1$ and $T_2$, bearing
	values $I(T_1) = 338$ and $I(T_2)=20$. These are placed in places $v$ and $z$, respectively.
	In place $x$, the initial marking contains the component $NH_4Cl$. In transition $t_1$, with
	condition $I(t)\geq 338$, the ammonium chloride decomposes into  $NH_3$ and $HCl$, assuming
	that a $T$ token with value at least $338$ is present. This is the case, thus, the transition
	takes place as shown in the second marking of the figure. If the temperature decreases, as implemented
	in transition $t_2$ where token instance $T_1$ exchanges places with token instance $T_2$, $I(T_2) = 20$, 
	then the reversal of the transition is enabled leading to the reversal of the decomposition, as shown
	in the last marking of the figure.

\section{Behavioural Properties for Controlled Reversing Petri Nets}	    
	 A major strength of Petri nets is their support for analysis of various properties and problems associated with concurrent systems~\cite{murata}. Two types of properties can be studied within reversing Petri net models based  on whether they are dependent on the initial marking, or are independent of the initial marking. The former type of properties is referred to as marking-dependent or behavioural properties, whereas the latter type of properties is called structural properties. In this section, we discuss only basic behavioural properties and their analysis.

	\paragraph{Reachability} Reachability is a fundamental basis for studying the dynamic properties of systems. The firing of an enabled transition will change the token distribution in a net according to the firing rules. A sequence of firings will result in a sequence of states. A state $\state{M_n}{H_n}$, is said to be reachable from a state $\state{M_0}{H_0}$ if there exists a sequence of firings that transforms $\state{M_0}{H_0}$ to $\state{M_n}{H_n}$. A firing or transition sequence is denoted by $\sigma = t_1;t_2;
	...;t_n$. If $\state{M_n}{H_n}$ is reachable from $\state{M_0}{H_0}$ by $\sigma$ we write $\state{M_0}{H_0}\frtrans{\sigma} \state{M_n}{H_n}$.  The set of all possible sates reachable
	from $\state{M_0}{H_0}$ in a net $N$ is denoted by $R(N, \state{M_0}{H_0})$ or simply $R(\state{M_0}{H_0})$. The set of all possible firing sequences from $\state{M_0}{H_0}$ in a net $N$ with initial state $\state{M_0}{H_0}$ is denoted by $L(N,\state{M_0}{H_0})$ or simply $L(\state{M_0}{H_0})$.
	Now, the reachability problem for controlled reversing Petri nets is the problem of finding if $\state{M_n}{H_n} \in R(\state{M_0}{H_0})$ for a given state $\state{M_n}{H_n}$ in a net $N$ with initial marking $M_0$. In some applications, one may be interested in the markings of a subset of places and not care about the rest of places in a net. This leads to a submarking reachability problem which is the problem of finding if $\state{M_j}{H_j} \in R(\state{M_0}{H_0})$,where $\state{M_j}{H_j}$ is any marking whose restriction to a given subset of places agrees with that of a given marking $\state{M_n}{H_n}$. 
	
	\begin{definition}{\rm
			Given a CRPN $N$ an initial state $\state{M}{H}$ and an execution $\state{M}{H}$ $\frtrans{\sigma}\state{M'}{H'}$  then $\state{M'}{H'}$ is reachable from $\state{M}{H}$ in the net $N$ and the set of all states reachable from $\state{M}{H}$ is denoted by  $R(N,\state{M}{H})$ or simply $R(\state{M}{H})$. The set of all possible firing sequences from $\state{M}{H}$ in $N$ is denoted by $L(N,\state{M}{H})$ or simply $L(\state{M}{H})$.
			
	}\end{definition}

As a state in CRPNs constitutes a combination of both a marking and a history most of the properties are defined based on both parameters. However, depending on the needs of the modelled system the reachability properties  can be redefined to ignore the status of the history. The reachability property is redefined as follows:

	\begin{definition}{\rm
		Given a CRPN $N$ an initial state $\state{M}{H}$ and an execution $\state{M}{H}$ $\frtrans{\sigma}\state{M'}{H'}$  then marking  $M'$ is reachable from $\state{M}{H}$ in the net $N$ denoted by $M' \in R(\state{M}{H})$.
		
}\end{definition}

	\begin{figure}[h]
	\centering
	\subfloat{\includegraphics[height=4cm]{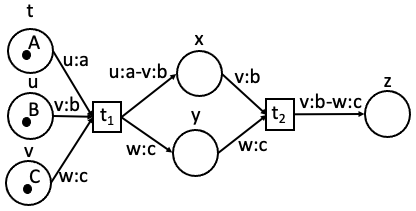}}
		\\
	\subfloat{\includegraphics[height=4cm]{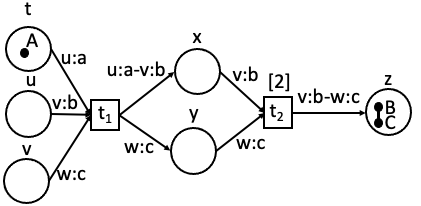}} 
	\caption{Reachability property}
	\label{reachability}
\end{figure}

Let us consider the example in Figure~\ref{reachability}. The figure on the top represents the initial marking $\state{M_0}{H_0}$. The figure on the bottom represents a desired marking $\state{M}{H}$ of the same controlled reversing Petri net. The reachability property questions whether $\state{M}{H}$ is reachable from the initial marking $\state{M_0}{H_0}$, such that $\state{M}{H} \in R(\state{M_0}{H_0})$. We can see that indeed $\state{M}{H}$ is reachable from $\state{M_0}{H_0}$ through the firing sequence $\sigma= (t_1,1); (t_2,2) ;\underline{(t_1,1)}$ such that $\state{M_0}{H_0}\frtrans{\sigma} \state{M}{H}$.

	\paragraph{Home state}
In many applications, it is not necessary to get back to the initial state as long as one can get back to some (home) state. For example in various electronic devices, home states may be reached automatically after periods of inactivity, or may be forced to be reached by resetting the device. Also in self-stabilising systems, reaching a failed state can be recovered from automatically reaching a non-erroneous home state. A state $\state{M’}{H'}$ is said to be a home state if, for each state $\state{M}{H}$ in $R(\state{M_0}{H_0})$, $\state{M’}{H'}$ is reachable from $\state{M}{H}$.
	
	\begin{definition}{\rm
			Given a CRPN $N$, a state $\state{M}{H}$ is a home state if $\state{M}{H}\in R(N,\state{M'}{H'})$ from every state $\state{M'}{H'}\in R(N,\state{M_0}{H_0})$ and $N$ is reversible if $\state{M_0}{H_0}$ is a home state.
		}
	\end{definition}
	
	As with the reachability property, home state can be redefined to ignore the history. In this way we have a more flexible notion of a home state where only the location of tokens constitutes a home state. 
	
	\begin{definition}{\rm
			Given a CRPN $N$, a marking $M$ is a home state if $M\in R(N,\state{M'}{H'})$ from every marking $M'\in R(N,\state{M_0}{H_0})$ and $N$ is reversible if $M_0$ is a home state.
		}
	\end{definition}

			\begin{figure}[h]
		\centering
		\subfloat{\includegraphics[height=4.5cm]{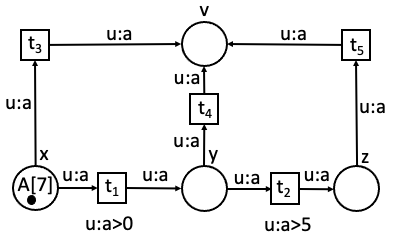}}
		\\
		\subfloat{\includegraphics[height=4.5cm]{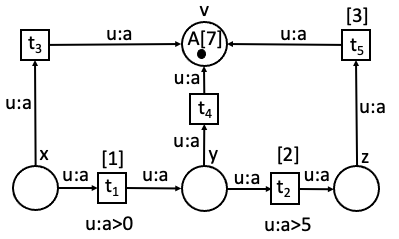}} 
		\caption{Home state property}
		\label{home}
	\end{figure}

Now consider the controlled reversing Petri net in Figure~\ref{home}. Given as initial state $\state{M_0}{H_0}$ the controlled reversing Petri net on the top we can observe that the state on the bottom $\state{M}{H}$ is a home state for that controlled reversing Petri net. Since only transitions $t_1$ and $t_2$ are irreversible then when executing any other transition we can reverse their execution and proceed with the firing sequence $(t_1,1);(t_2,2);(t_5,3)$ leading to the home state $\state{M}{H}$. Also note that this controlled  reversing Petri net is not reversible since $t_1$ and $t_2$ are not reversible and therefore after the are execution we cannot return to the initial marking $\state{M_0}{H_0}$.

	\paragraph{Liveness}
	The concept of liveness is closely related to the complete absence of deadlocks in operating systems. A controlled reversing Petri net $N$ with initial marking $M_0$ is said to be live (or equivalently $\state{M_0}{H_0}$ is said to be a live marking for $N$) if, no matter what state has been reached from $\state{M_0}{H_0}$, it is possible to ultimately fire any transition of the net by progressing through some further firing sequence. This means that a live controlled reversing Petri net guarantees deadlock-free operation, no matter what firing sequence is chosen.

	\begin{definition}{\rm
		A CRPN $N$ it is said to be live (or equivalently $\state{M_0}{H_0}$ is said to be a live state for $N$) if for all $\state{M}{H} \in R(N,\state{M_0}{H_0})$ then  there exists $  t \in T$ such that $t$ enabled in  $\state{M}{H}$. 
	}\end{definition}

Liveness is an ideal property for many systems. However, it is impractical and too costly to verify this strong property for some systems such as the operating system of a large computer. Thus, we relax the liveness condition and define different levels of liveness.

	\begin{definition}{\rm
			Given a CRPN $N$ and a transition $t \in T$ then $t$ is said to be:
			\begin{enumerate}
				\item dead($L0-$live) if $t \notin L(\state{M_0}{H_0}$),
				\item $L1-$live (potentially fire-able) if $\mid \{ t\mid t \in L(\state{M_0}{H_0})\}\mid=1$,
				\item $L2-$live if given any positive integer $k \mid \{t \mid t \in L(\state{M_0}{H_0})\}\mid=k $,
				\item $L3-$live if $\mid \{  t \in L(\state{M_0}{H_0})\} \mid =\infty$,
				\item $L4-$live or live if $t$ is $L1-$live for all$\state{M}{H},\state{M}{H} \in R(\state{M_0}{H_0})$, and
				\item $Lk$-live if every transition in the net is $Lk$-live, $k=0,1,2,3,4$.    	
			\end{enumerate}
		}
	\end{definition}

	$L4-$liveness is the strongest and corresponds to the liveness defined earlier.  It is easy to see the following implications: $L4$-liveness $\implies$ $L3-$liveness  $\implies$ $L2-$liveness  $\implies$ $L1-$liveness. We say that a transition is strictly $Lk-$live if it is $Lk-$live but not $L(k +1)-$live, $k = 1, 2, 3$.
	
	In the case of liveness the history parameter of a state cannot be ignored as history in controlled reversing Petri nets plays an important role when deciding if a transition is reversed enabled in a specific state. For example non executed transitions indicated by $H(t)=\emptyset$ cannot be reversed.  
	
	\begin{figure}[h]
	\centering
	\subfloat{\includegraphics[height=5.5cm]{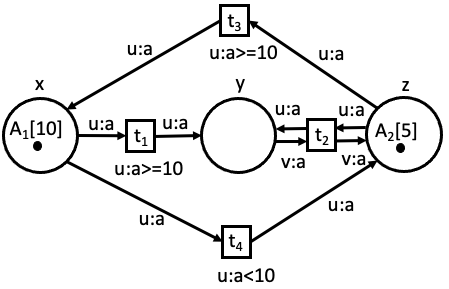}}
	\caption{Liveness property}
	\label{liveness}
\end{figure}

Consider the controlled reversing Petri net in Figure~\ref{liveness}. We observe that the execution $(t_1,1); (t_2,2); (t_3,3);(\underline{t_1,1});(t_4,4)$ can be repeated infinitely. 


	\paragraph{Deadlock} The concept of deadlock in controlled reversing Petri nets is the inability to proceed with the execution of a transition. Therefore, a controlled reversing Petri net has a deadlock when there are no other transitions that can be executed in the forward or reverse transition. 
	
	\begin{definition}{\rm
			Given a CRPN $N$  a deadlock is as state $ \state{M}{H}$, $\state{M}{H} \in R(N,\state{M_0}{H_0})$ such that there exists no  $ t\in T$ such that $t$ (controlled-forward/controlled-reverse) enabled in $\state{M}{H}$.
	}\end{definition}

As deadlock is equivalently defined as L0-liveness then the history as part of a state is an important element when deciding if a transition can fire. 

	\begin{figure}[h]
	\centering	\subfloat{\includegraphics[height=1.8cm]{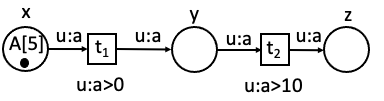}}
\\
	\subfloat{\includegraphics[height=2cm]{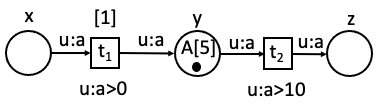}} 
	\caption{Deadlock property}
	\label{deadlock}
\end{figure}

Consider the CRPN in Figure~\ref{deadlock}. The controlled reversing Petri net on the top figure represents the initial marking  $\state{M_0}{H_0}$ and the controlled reversing Petri net in the bottom figure represents the resulting marking $\state{M}{H}$ after the execution of $t_1$, such that $\state{M_0}{H_0}\trans{(t_1,1)}\state{M}{H}$. Since transition $t_2$ requires $u>10=TRUE$, $type(u)=a$ in order to be executed in forward direction then it is not fireable by $A[5]$. Also $t_1$ requires $u<0 =$TRUE in order to reverse then it is irreversible by $A[5]$. Therefore in state $\state{M}{H}$ there are no transitions fireable by the only token $A[5]$ and therefore the reversing Petri net has reached a deadlock. 

		\paragraph{Siphon} 
		A non-empty subset of places $P_S$ in a CRPN is called a siphon if for all $\state{M}{H}$ where $M(x)= \emptyset$, $x\in P_S$ and for all $\state{M'}{H'}$ then $\state{M'}{H'} \in R(\state{M}{H})$ we have  $M'(y) = \emptyset$, $y \in P_S$. 
		A siphon has a behavioural property that if it is token-free under some marking, then it remains token-free under each successor marking. 
		 It is easy to verify that the union of two siphons is again a siphon. A siphon is called a basic siphon if it cannot be represented as a union of other siphons. All siphons in a CRPN can be generated by the union of some basis siphons. A siphon is said to be minimal if it does not contain any other siphon. A minimal siphon is a basis siphon, but not all basis siphons are minimal.
		 
		  Exiting a siphon highly depends on whether a transition is executable in either the forward or reverse  direction. The execution of transitions in controlled reversing Petri nets depends both on the satisfaction or violation of conditions but also on the form of execution i.e. whether we are firing in forward or reverse. As reversibility allows the execution of transitions in both forward and reverse execution this means that a fully reversible MRPN cannot have siphons as when exiting a siphon we always have the possibility of reversing in order to enter the siphon again. As such the use of conditions disables transitions from reversing when necessary and therefore irreversible transitions do not constitute entrance in a siphon area. 

		\paragraph{Trap} A non-empty subset of places $P_T$ in a CRPN is called a trap if for all $\state{M}{H}$ where $M(x)\neq \emptyset$, $x\in P_T$ then for all $\state{M'}{H'}$ where $\state{M'}{H'} \in R(\state{M}{H})$ we have $M'(y) \neq \emptyset$, $y \in P_T$. 
		A trap has a behavioural property that if it is marked (i.e., it has at least one token) under some marking, then it remains marked under each successor marking. It is easy to verify that the union of two traps is again a trap. A trap is called a basic trap if it cannot be represented as a union of other traps. All traps in a CRPN can be generated by the union of some basis traps. A trap is said to be minimal if it does not contain any other trap. A minimal trap is a basis trap, but not all basis traps are minimal.
		
		Similarly to siphons, traps are a behavioural property in which when entering the trap region we are unable to execute transitions outside that region. As MRPNs can be fully reversible the introduction of conditions disables transitions from reversing and thus being able to exit the trap. 
		
				\begin{figure}[h]
			\centering
			\subfloat{\includegraphics[height=3cm]{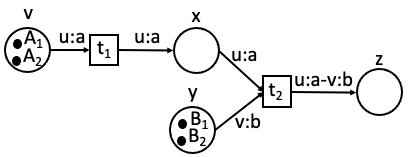}}
			\\
			\subfloat{\includegraphics[height=3.2cm]{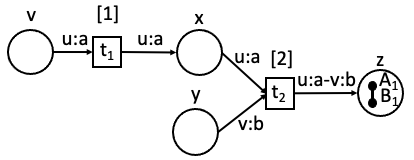}} \\
			\subfloat{\includegraphics[height=3cm]{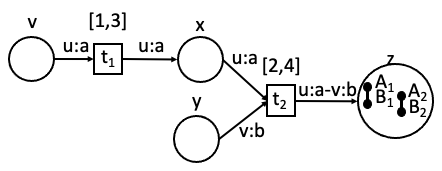}} 
			\caption{Coverability property}
			\label{coverability}
		\end{figure}
	\paragraph{Coverability} A state $\state{M}{H}$ in a controlled reversing Petri net $N$ with $\state{M_0}{H_0}$ is said to be coverable if there exists a marking $\state{M’}{H'}$ in $R(\state{M_0}{H_0})$ such that $M’(x) \subseteq  M(x)$ for each $x\in P$ in the net and $H’(t) \subseteq  H(t)$ for each $t\in T$ in the net. Coverability is closely related to $L1$- liveness (potential firability). Let $\state{M}{H}$ be the minimum marking needed to enable a transition $t$. Then $t$ is dead (not $L1$- live) if and only if $\state{M}{H}$ is not coverable. That is, $t$ is $L1$-live if and only if $\state{M}{H}$ is coverable.
	
	\begin{definition}{\rm 
			Given a CRPN $N$ and a state $\state{M}{H}$ is said to be coverable if there exists $\state{M'}{H'} \in R(\state{M_0}{H_0})$ such that $\mid\{a_i|a_i\in M(x), type(a_i)=a\}\mid \leqslant \mid\{a_i|a_i \in M(x), type(a_i)=a\}\mid,$ for all $a \in A$ and if $k\in H(t)$ then $k\in H'(t)$, $t\in T$. 
		}
	\end{definition}

Coverable states can be similarly defined as: 

	\begin{proposition}{\rm
			Given a CRPN $N$ and a state $\state{M}{H}$ is said to be coverable if there exists $\state{M'}{H'} \in R(\state{M_0}{H_0})$ such that $\state{M'}{H'}\geq\state{M}{H}$.
		}
	\end{proposition}
	\begin{proposition}{\rm
			Given a CRPN $N$ and a state $\state{M}{H}$ is said to be coverable iff $\state{M}{H} \in R(\state{M_0}{H_0})$.
	}\end{proposition}

	Similarly, to the reachability property, coverability can be defined in terms of a marking and thus ignoring the history part of state. 
	
	\begin{definition}{\rm 
			Given a CRPN $N$ and a marking $M$ is said to be coverable if there exists $M' \in R(\state{M_0}{H_0})$ such that $\mid\{a_i|a_i\in M(x), type(a_i)=a\}\mid \leqslant \mid\{a_i|a_i \in M(x), type(a_i)=a\}\mid,$  for all $a \in A$. 
		}
	\end{definition}

	Consider the CRPN in Figure~\ref{coverability}. The first  controlled reversing Petri net on the top corner is the initial marking $\state{M_0}{H_0}$. The second controlled reversing Petri net is the desired state $\state{M}{H}$ that we want to check if its coverable by some marking  $\state{M'}{H'} $ reachable from the initial marking such that $\state{M'}{H'} \in R(\state{M_0}{H_0})$. The final controlled reversing Petri net is the marking $\state{M'}{H'}$ which is reachable from $\state{M_0}{H_0}$ by the execution $\sigma=(t_1,1);(t_2,2);(t_1,3);(t_2,4)$,  $\state{M_0}{H_0} \frtrans{\sigma}\state{M'}{H'}$ which covers $\state{M}{H}$ such that  $\state{M'}{H'}\geq\state{M}{H}$. Note that the marking derived from the execution $\sigma'={(t_1,1);(t_2,2)}$, $\state{M_0}{H_0} \frtrans{\sigma'}\state{M''}{H''}$ also covers $\state{M}{H}$.

	\paragraph{Persistence}
	A controlled reversing Petri net $N$  with initial marking $\state{M_0}{H_0}$ is said to be persistent if, for any two enabled transitions, the firing of one transition will not disable the other. A transition in a persistent net, once it is enabled, will stay enabled until it fires.  Persistence is closely related to conflict-free nets, and a safe persistent net can be transformed into a marked graph by duplicating some transitions and places.

	\begin{definition}{\rm
			Given a CRPN $N$ and a state $\state{M}{H} \in R(\state{M_0}{H_0})$ then $N$ is said to be persistent if for all $t_1,t_2  \in T$, $t_1,t_2$ enabled in $\state{M}{H}$ and $\state{M}{H} \trans{t_1} \state{M'}{H'}$ then $t_2$ enabled in $\state{M'}{H'}$ and vice versa. 
			
	}\end{definition}

As persistence is dependent on the enabledness of transitions, it cannot be defined solely by the marking of a state. 

%
%

\begin{figure}[h]
	\centering
	\hspace{2.2cm}
	\subfloat{\includegraphics[height=3.5cm]{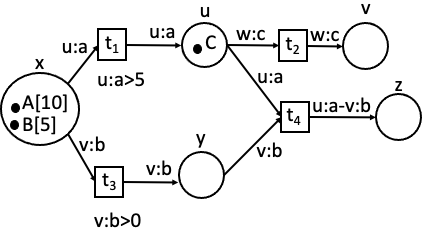}}\\
	\subfloat{\includegraphics[height=2.5cm]{figures/t1,1coll.png}}
	\subfloat{\includegraphics[height=2.5cm]{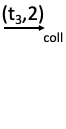}}
	\subfloat{\includegraphics[height=4cm]{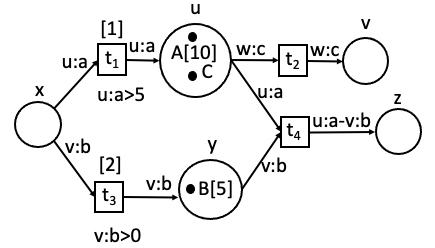}}
	\caption{Persistence property}
	\label{persistence}
\end{figure}

Consider the controlled reversing Petri net in Figure~\ref{persistence}. In the first controlled reversing Petri net we observed the initial marking $\state{M_0}{H_0}$ where transitions $t_1$ and $t_3$ are simultaneously forward enabled and  the execution of one does not preclude the execution of the other. On the second controlled reversing Petri net we observe the marking $\state{M}{H}$ after the execution of both $t_1$ and $t_3$ where transitions $t_2$ and $t_4$ are simultaneously forward enabled. Since transitions $t_1$ and $t_3$ are irreversible by $A[10]$ and $B[5]$ respectively then only transitions $t_2$ and $t_4$ can be executed and the execution of one of them does not preclude the execution of the other. Hence the controlled reversing Petri net is indeed persistent. Note that if transitions $t_1$ and $t_3$ were not irreversible then the reversal of transition $t_1$ would have precluded the execution of transition $t_2$ and $t_4$. 
\remove{		
	\paragraph{Initial Marking} A reversing Petri net is said to be sourcable when there exists a sequence from a current state to the initial marking. 
	
	\begin{definition}{\rm 
			Given a CRPN $N$ a current state $\state{M}{H}$ then $\state{M_0}{H_0}$ is sourceable if there exists a sequence $\state{M}{H} \frtrans{\sigma} \state{M_0}{H_0}$. 	
			
	}\end{definition}

	\begin{figure}[h]
	\centering
	\subfloat{\includegraphics[height=3cm]{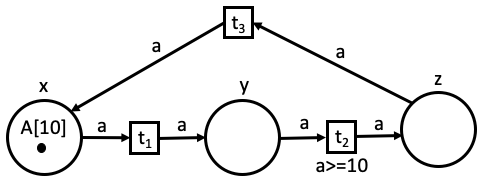}}
	\caption{Initial marking property}
	\label{initial }
\end{figure}
}

\section{Case Study}

%
%
 Antenna selection in distributed Massive MIMO (Multiple Input Multiple Output)~\cite{gaomassive2015} antenna arrays is an important optimisation
	problem on a complex system comprised of a large number of simple,
	similar-behaving components. While a large number of antennas offers
	diversity, spatial multiplexing opportunities, interference suppression
	and redundance \cite{ozgur2013spatial}, not all antennas contribute
	the same, and powering all of them is not optimal \cite{hoydismassive2013}.
	Optimal transmit antenna selection for large antenna arrays is computationally
	demanding \cite{gao2018massive}, so suboptimal approaches are pursued
	for real time use. 
	
	Petri nets are a convenient tool for modelling and control of networks,
	and have been applied in higher layers of ISO OSI model for wireless
	networks \cite{heindl2001performance}.
	Centralized AS in DM MIMO (distributed, massive, multiple input, multiple output) systems is computationally complex, demands a large information exchange, and the communication channel between antennas and users changes rapidly. The reliability of distributed multiple-antenna systems depends on fault tolerance and recovery which align naturally with reversibility. 
	 We therefore introduce a CRPN-based, distributed, time-evolving solution with reversibility, asynchronous execution and local condition tracking for reliable performance and fault tolerance. In this setting, we use our expressive controlling mechanism in order to manage the pattern and the direction of computation in order to deal with error recovery or to provide the main focus of the computation. The internal control mechanism validates the conditions when the addition of an antenna improves the sum capacity and violates the condition in case an antenna is consider to be no longer useful triggering reversal which removes the antenna from the selected set.


Reversible models conserve quantities, both in the sense of energy and matter. In the case of our controlled reversing Petri net model, it preserves the number of tokens: in the particular implementation it means that a constant number of antennas will be used at all times, which is advantageous in terms of planning and hardware resource deployment and represents an improvement compared to the previous localised antenna selection in which the number of antennas was an emergent property. Conservation of the token count can also represent a fixed power budget,
when we use the controlled reversing Petri net for power control, constant user count if we perform user selection, etc. 

Reversibility allows the resource management algorithm to go back to a previous state and take a different execution route in case of an antenna becomes faulty in one or more places (in our presented case, antennas). Since there has been a fault and no forward transition is possible we should reverse the last transition and thus see the token returning to a properly-functioning antenna.

Reversing the evolution of a controlled reversing Petri net is a logical
behaviour in some use cases. Without movement of users in the grid, the selected set of antennas is concentrated in a predefined state. With users moving, the antennas coordinate their tracking. Once there is no more need for their activity, the tokens return to the initial positions with simple reversal of their trajectories. At the same time, the whole network does not have to be reversed, as parts of it could still be engaged with serving users.

\subsection{Reversing Petri Net Representation}

\begin{figure}[t]
	\centering
	\subfloat{\includegraphics[width=15.5cm]{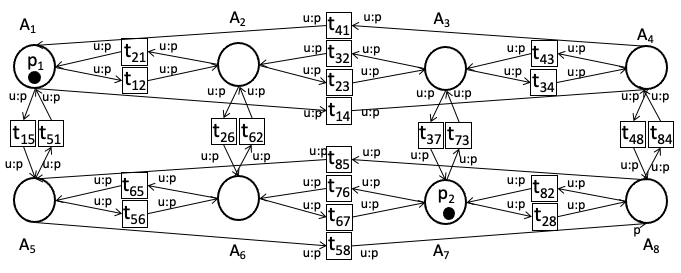}}
	\caption{Antenna selection on massive-MIMO}
	\label{grid}
\end{figure}

 In this section, the semantics for collective reversibility 
as realised in the framework of controlled reversing Petri nets (CRPNs), 
  dynamically illustrate antenna selection in massive-MIMO based on how the proposed algorithm is implemented and how it changes based on different operating scenarios. 
This section presents a new method for antenna selection which divides antennas into virtual sub-neighbourhoods whose dynamic behaviour can be observed by being simulated in controlled reversing Petri nets. 

For real-life systems such as massive-MIMO the state space is too large to illustrate and therefore we decompose the whole system into smaller subsystems. This decomposition divides the whole model into two sub-models the first one being the power distribution among antennas and the second one being the memory mechanism that controls the execution of transitions. We call the power distribution model the high-level layer illustrating the exchanges of power between antennas as well as the neighbouring connections between them.  The low-level layer consists of the memory mechanism that collects information about the powered antennas in a neighbourhood and thus executing transitions in forward or reverse direction depending on whether the required conditions are satisfied. Ideally, the two sub-models can be merged together resulting in a complete model, which then includes both the power allocation of the system and the controlled decision steps.

\paragraph{\textbf{High-level layer}} The model in Figure~\ref{grid} illustrates the higher-level net of the antenna selection algorithm. We demonstrate a sample neighbourhood of eight base station antennas with random distributed topology. Every eight antennas are considered to belong in the same neighbourhood by allowing each antenna to be bidirectionally linked to four other antennas we create overlapping neighbourhoods. The maximum number of enabled antennas in this example is two and the power token is transferred from one antenna to the other through directly connected links. Note that resource allocation systems like antenna selection can greatly benefit from the collective token interpretation since the existence of any power token $p$ in the corresponding place should be able to execute a transition in either the forward or reverse direction.

\begin{figure}[tb]
	\centering
	\subfloat[antennas and users]{\includegraphics[height=6.5cm]{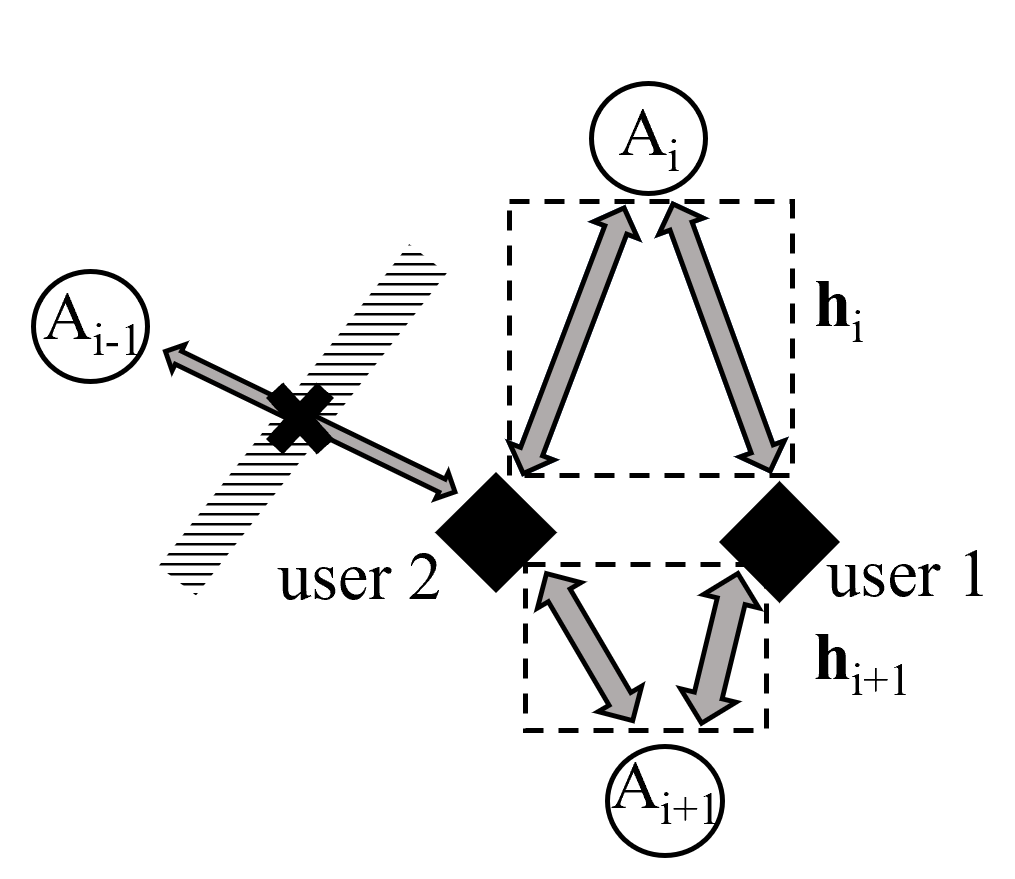}}
	\hspace{2cm}
	\subfloat[a part of the CRPN model]{\includegraphics[height=5.5cm]{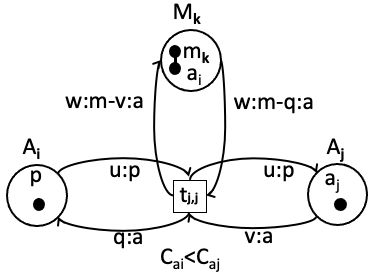}} 
	\caption{CRPN for antenna selection in DM MIMO (large antenna array)}
	\label{mechanism}
	\vspace{-.2cm}
\end{figure}

\paragraph{\textbf{Low-level layer}} The search for a suitable set of antennas is a sum capacity maximization problem:
\begin{equation}
\mathcal{C}=\max_{\mathbf{P},\mathbf{H_{c}}}\log_{2}
\det\left(\mathbf{I}+\rho\frac{N_R}{N_{TS}} \mathbf{H_{c}}\mathbf{P}\mathbf{H_{c}}^{H}\right)\label{capac}
\end{equation} 
where $\rho$ is the signal to noise ratio, $N_{TS}$ the number of antennas 
selected from  a total of $N_T$ antennas, $N_{R}$ the number of users, 
$\mathbf{I}$ the $N_{TS}\times N_{TS}$ identity matrix, $\mathbf{P}$ a diagonal
$N_{R}\times N_{R}$ power matrix; $\mathbf{H_{c}}$ is the $N_{TS}\times N_{R}$
submatrix of $N_{T}\times N_{R}$ channel matrix $\mathbf{H}$. Instead of
centralized AS, in our approach sum capacity is calculated locally for small 
sets of antennas (neighbourhoods), switching on only antennas which improve the
capacity as for example in Figure~\ref{mechanism}(a), we demonstrate the case where antenna $A_{i-1}$ decreases sum capacity and therefore it will not be selected. 

In the CRPN interpretation, we present the antennas by places $A_1,\ldots,A_n$, where $n=N_T$, and the  overlapping neighbourhoods
by places $M_1,\ldots,M_h$. These places are connected together via transitions 
$t_{i,j}$, connecting $A_i$, $A_j$ and $M_k$,
whenever there is a connection link between antennas $A_i$ and $A_j$. The transition captures that based on the neighbourhood knowledge in place $M_k$, antenna $A_i$ may be preferred
over $A_j$ or vice versa (the transition may be reversed). 

To implement the intended mechanism, we employ three types of tokens. 
First, we have the power tokens $p_i$, which are of the same type and therefore by the collective token interpretation are able to execute/reverse any transition that requires them. 
If token $p_i$ is located on place $A_i$, antenna $A_i$ is considered to be on.
The transfer of these tokens results into new antenna selections, ideally converging
to a locally optimal solution.
Second, tokens $m_1,\ldots,m_h$, each represent one neighbourhood. 
Finally, $a_1,\ldots,a_n$, represent the
antennas. The tokens are used as follows:
Given transition $t_{i,j}$ between antenna places $A_i$ and $A_j$ in
neighbourhood $M_k$, transition $t_{i,j}$ is enabled if token $p$ is
available on $A_i$, token $a_j$ on  $A_j$, and bond $(a_i,m_k)$
on $M_k$, i.e.,
$F(A_i,t_{i,j}) = \{u\}$, $type(u)=p$, $F(A_j,t_{i,j})= \{v\}$, $type(v)=a$, and
$F(M_k,t_{i,j})=\{(q,w),q,w\}$, $type(q)=a$ and $type(w)=m$. This configuration
captures that antennas $A_i$ and $A_j$ are on and off, respectively.
(Note that the bonds between token $m_k$ and tokens of type $a$
in $M_k$ capture the active antennas in the neighbourhood.)
Then, the effect of the transition
is to break the bond $(a_i,m_k)$, and release token $a_i$ to place
$A_i$, transferring the power token to $A_j$, and creating the bond
$(a_j,m_k)$ on $M_k$, i.e.,
$F(t_{i,j}, A_i) = \{q\}$, $F(t_{i,j},A_j)= \{u\}$, and $F(t_{i,j},M_k)
= \{(v,w),v,w\}$.
The mechanism achieving this for two antennas can be seen in Figure~\ref{mechanism}(b).

Finally, to capture the transition's condition, an antenna token $a_i$ is associated with data vector $I(a_i) =
\mathbf{h}_i$, $type_{\Sigma}(\mathbf{h}_i)= \mathbb{R}^2$ ($=\mathbb{C}$), i.e., the 
corresponding row of $\mathbf{H}$. 
The condition constructs the matrix  $\mathbf{H}_c$ of
(\ref{capac}) by collecting the
data vectors $\mathbf{h}_i$ associated with the antenna tokens $a_i$
in place $M_k$: 
$\mathbf{H}_c=(\mathbf{h}_1,...,\mathbf{h}_n)^T$ where
$\mathbf{h}_i=I(a_i)$ if $a_i\in M_k$, otherwise $\mathbf{h}_i=(0\;\ldots\;0)$. 
The transition $t_{i,j}$ will occur if the
sum capacity calculated for all currently active antennas 
(including $a_i$), $\mathcal{C}_{a_i}$, is less than the sum capacity calculated for the same
neighbourhood with the antenna $A_i$ replaced by $A_j$,
$\mathcal{C}_{a_j}$, i.e., $\mathcal{C}_{a_i}<\mathcal{C}_{a_{j}}$. Note that if
the condition is violated, the transition may be executed in the reverse direction.

\subsection{Property Analysis in Massive  MIMO Systems}


The behaviour of a modern wireless communications system (in our example, a massive MIMO system) can be thought as an aggregation of multiple networks employing varying levels of coordination and communication. Various resources (electromagnetic spectrum, power, physical infrastructure) are continuously managed, the network of users interacts with the network of base stations, computation and communication intertwine. Hardware faults happen, working modes change, and the modern networks are supposed to handle these unexpected events seamlessly.
The general idea behind the aggregation method is to substitute complex Petri net structures by simple ones observing some important properties of the model like e.g. deadlock, siphons, traps, reachability etc. Intelligent monitoring of a massive distributed network requires careful specification and modelling in order to analyse its constraints and deliverables, as well as to avoid hazards, waste of resources and security threats and therefore be used by engineers and designers for what-if analysis and experimentation.


On that note, the modelled MIMO system can be used for formal verification of properties such as the  deadlock property, siphons and traps. Optimisation processes in wireless communications in general should be converging fast to a steady state with minimal computational burden in order to enable real time application in high mobility scenarios. 
 Therefore, we could use the deadlock property to define the final antenna selection where our algorithm converges.
%
%
Similarly, in the case of siphons we know that once a token escapes the siphon region the token will never return to that region. As such once a power token escapes to an antenna outside a siphon region it will never consider that region as computationally better than the antenna that the token is already in. In this way we are able to define non critical antennas given a predefined set of specifications of our MIMO system. In the opposite manner once a token enters a trap region it will never escape that region, i.e. it will never consider antennas outside that region as better antennas than the already selected ones. As such we are able to identify critical antennas that improve sum capacity given a predefined set of specifications for our MIMO system.
%
%
%
Finally, the ability of our model to allow tokens to carry data yields in customized performance properties which can be quantified by specific metrics that provide the average measure of the probability with which an error is encountered. 

\section{Concluding Remarks}

In this chapter we have extended MRPNs with conditions that control reversibility~\cite{ieee,RC19}, and we have applied our framework in the context of wireless communications. 
Our formalism introduces conditional transitions that permit the system to manage the  pattern and direction of computation.  It allows systems to reverse under specified conditions leading to previously visited states or even new ones without the need of additional forward actions.  A possibility to extend the model exists by introducing arc expressions that will perform operations on the data values associated with the manipulating tokens. 

We have shown how the reversible structure of CRPNs is amenable to  implementations from wireless communications in terms of distributed antenna selection and is expressive enough to encode reversible processes. 
This experience has illustrated that resource management can be studied and understood in terms of CRPNs as, along with their visual
nature, they offer a number of features, such as token persistence, that is especially relevant in these contexts. 
\chapter{Conclusions}\label{sec:Conclusions}
\section{Summary}


 This thesis proposes a reversible approach to Petri Nets~\cite{RPNs,RPNscycles} that allows the modelling of reversibility as realised by backtracking, causal reversing, and out-of-causal-order reversing. Our proposal allows transitions to reverse at any time leading to previously visited states or even to new ones without the need of additional forward actions. Moreover, this interpretation of Petri Nets has the capability of reversing without the need of an extensive memory. To enable this, additional machinery has been necessary
 to capture causal dependencies in the presence of cycles.  This machinery identifies  a causal dependence relation that resorts to the marking of a net and is partnered  along with stack histories for each transition that record all previous occurrences. To the best of our knowledge, this is the first such proposal, since the related previous work~\cite{BoundedPNs,PetriNets}, having a different aim, implemented a very liberal way of reversing computation in Petri nets by introducing additional reversed transitions. On the contrary, in~\cite{Unbounded}, reversibility is achieved by adding new places to non-reversible Petri nets while preserving their computation, which however is only possible in a subclass of Petri nets and it is only focused on causal reversal. The works of~\cite{RPT,RPlaceTrans,RON} identify the causal memory of a Petri net by unfolding them into occurrence nets and coloured Petri nets. All of these approaches, including ours, are concerned with reversing single steps of transitions, unlike~\cite{ReversingSteps}, which examines the possibility of reversing the effect of groups of actions.

Other than the technical scope of this work we are also concerned with the theoretical foundations of reversible computation, specifically the different strategies of reversing and their relationships. 
 Through the aid of Petri nets we were able to examine the different strategies of reversing and focus mostly on causal reversing and out-of-causal-order reversing. Causality is one of the most interesting topics within models of concurrency where various interpretations have been proposed throughout time which can be justified either by theoretical properties, or by the implementation of possible applications. We 
 focus on the approach  where dependencies between transitions are determined by the token manipulation performed during an execution.  We prove that the amount of flexibility allowed in causal reversibility indeed yields a causally consistent semantics. 
 
 On a similar note, research on out-of-causal reversibility is very limited since the only related work is that in~\cite{ERK,Bonding}. Therefore, from various examples and theoretical results we have examined the theoretical properties of out-of-causal reversibility and demonstrated that out-of-causal-order reversibility is able to create new states unreachable by forward-only execution.
 
   Most works in the literature discuss out-of-causal reversibility when creating of bonds rather than destructing bonds. In our model, where states are more elaborate since they preserve token evolution, we were able to observe that it is not possible to reverse a transition in out of causal order whose effect no longer exists in the system. This shows that out-of-causal order is not as flexible as one might initially believe.  This applies and should be considered in other formal models, such as process calculi and event structures, independently of how abstract their states are.
   
    Additionally, we establish the relationship between the three forms of reversing and define a transition relation that can capture each of the three strategies modulo the enabledness condition for each strategy. This allows us to provide a uniform treatment of the basic theoretical results.

We continue exploring these reversible strategies in extensions of reversing Petri nets, Multi Reversing Petri Nets (MRPNs), by allowing multiple tokens of the same type to exist in a model and developing reversible semantics in the presence of bond destruction. Our aim was to generalize reversing Petri nets in a setting where multiple tokens that have identical behavioural capabilities can occur in a system. However, allowing multiple instances of identical tokens results in ambiguities when it comes to causal dependencies. In fact, we have distinguished the different ways of introducing reversible behaviour into causal systems with multiple tokens and we explore two directions, namely, the individual token interpretation defined based on partial order~\cite{individual,ZeroSafe} and the collective token interpretation defined based on disjunctive causality. 

We have proposed reversible semantics that  follow the individual token philosophy and therefore achieve precise correspondence between the token instances and their past. The individuality of identical tokens can be imposed by their causal path which allows identical tokens to fire the same transition when going forward, however when going backwards tokens will be able to reverse only the transitions that they have fired. We have also provided the reversible semantics for out-of-causal-order reversibility in the presence of bond destruction. Finally, we show that the expressive power of multi reversing Petri nets  is equivalent to the expressive power of single reversing Petri nets (SRPNs).  However, there is blow-up on the size of SRPNs as multiple transitions in SRPNs can be represented by the same transition in MRPNs as long as the type equivalence between the required tokens and variables is the same. 
 We have also presented a more relaxed form of reversibility following the collective token philosophy and we have given the associated semantics for the respective firing rule. This approach considers all tokens of a certain type to be identical, disregarding their history during execution, and is particularly applicable in the context of resource-aware systems. In the collective token interpretation when multiple tokens of the same type reside in the same place then these tokens are not distinguished. This means that all that is known by the model is the amount of token occurrences of a specific type and their location in the marking.  We have shown how this firing rule relates to the firing rules of the individual token interpretation and how the robustness of this mechanism can be applied in an application from biochemistry known as the  autoprotolysis of water. 

 A subsequent extension of our formalism, called Controlled Reversing Petri Nets (CRPNs), considers approaches for controlling reversibility as for instance in \cite{ERK,LaneseMSS11,compensations}. While various
frameworks make no restriction as to when a transition can be reversed (uncontrolled
reversibility), it can be argued that some means of controlling the conditions
of transition reversal is often useful in practice. 
For instance, in biological phenomena where environmental conditions change or when dealing with fault recovery  where reversal is triggered when a fault is encountered. We therefore have extended our research by proposing conditional executions that indicate the pattern and direction of computation as well as irreversible actions or less likely executable actions.  In fact, we have extended MRPNs with conditions that control reversibility by determining the direction of transition execution. We then provide the main behavioural properties of our controlled model as the specification of models according to these properties can be useful towards their analysis and verification.


Finally, we show the robustness of our control mechanism and the associated behavioural properties by modelling an example from telecommunications of a recently-proposed distributed algorithm for antenna selection. Our application illustrates the ability of CRPNs to not only formalise complex distributed systems, but also to naturally capture controlled execution and conservation of information in a system. The ability of the CRPN solution to act asynchronously and converge fast with minimal computational burden enables real time application of the algorithm even in high mobility scenarios. We have shown how the reversible structure of CRPNs is amenable to implementations from wireless communications in terms of distributed antenna selection and is expressive enough to encode reversible processes. 

%

\section{Current and Future Work}\label{sec:RemainingWork}
The simplicity of the basic user interface of Petri nets has easily enabled extensive tool support over the years, particularly in the areas of model checking, graphically oriented simulation, and software verification. Recently, Petri nets have been associated with a novel paradigm, known as Answer Set Programming (ASP)~\cite{TorstenBook,LifschitzBook}, which is a declarative programming language with competitive solvers that solve a problem by devising a logic program such that models of the program provide the answers to the problem. ASP  applies declarative logic programming techniques that run multiple simulations and parallel evolutions in order to analyse the properties of various modelling domains. Various subclasses of Petri Nets have been translated to ASP, such as regular Petri Nets~\cite{BoundedLTL}, Simple Logic Petri Nets~\cite{ModelBased}, 1-safe Petri nets~\cite{1safe}, general Petri Nets~\cite{Encoding}, timed Petri nets~\cite{ResourceAllocation}, as well as high-level Petri nets~\cite{highLevelASP}.

Given that RPNs are able to model discrete event systems with well-formed semantics, they can also be used for specifying and manipulating the states of a system. Based on that we are currently exploring how ASP can be used to encode reversing Petri nets in an intuitive manner while preserving the modelling power and analyzability of decision problems in Petri net theory~\cite{ASPtoRPNs}. 
 Our implementation allows the enumeration of all possible evolutions of a reversing Petri net simulation as well as the ability to carry out additional reasoning about these simulations. Our long term goal is the development of an ASP-based framework for reasoning about RPN models. The visual nature of Petri nets in combination with reversible computation can help in understanding reversibility through various case studies and explore how reversibility can help in specification, verification, and testing.

In order to further understand how reversibility affects computation, we need to investigate the expressiveness relationship between models that are equipped with reversibility and the forms of traditional models. The trade-offs between traditional models and reversible models, the relations between the several reversible approaches as well as a clarification and classification of the different notions are of particular interest, and need to be studied in depth. This will show whether the expressive power of the traditional model improves from the added feature of reversible computation and whether it affects the decidability problems discussed in~\cite{PetriNets}, regarding reachability and coverability. 

\begin{figure}[h]
	\centering
	\subfloat{\includegraphics[width=11cm]{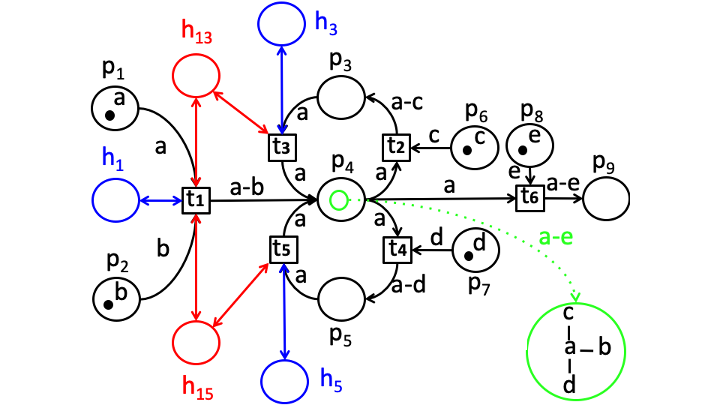}}
	\caption{Translation from reversing Petri nets to coloured Petri nets}
	\label{CPN}
\end{figure}

As such, another translation of RPNs investigates the expressiveness relationship between RPNs and coloured Petri nets where a subclass of RPNs with trans-acyclic structures has been translated into coloured Petri Nets (CPNs) by encoding the structure of the net along with the execution~\cite{RPNtoCPN, RPNtoCPNcycles}. 
The more typical challenges are related with the complexity and the cost of increasing (exponentially) the size of the net. Specifically, we have proposed a structural translation from RPNs to CPNs,
where for each transition we consider both forward and backward instances. Furthermore, 
the translation relies on storing histories and causal dependencies of executed transition sequences in additional
places (Figure ~\ref{CPN}). We have tested the translation on a number of examples, where the CPN-tools~\cite{CPN} was employed to illustrate that the translations conform to the semantics of reversible computation. As a result, we conclude that the principles of reversible computation in the presence of cyclic behaviour can be encoded in the traditional model. 

 As an extension of our work with coloured Petri nets we aim to provide and prove the correctness of our translation and analyse the associated trade-offs in terms of Petri net size. We intend to investigate the expressiveness relationship between reversing Petri nets and coloured Petri nets and explore how reversible computation affects the expressive power of various subclasses of Petri nets.  
 As a general aim, we plan on implementing an algorithmic translation that transforms RPNs to CPNs in an automated manner using the proposed transformation techniques.

From a more practical point of view, we believe that our framework can be applied in fields outside Computer Science, since the expressive power and visual nature offered by Petri nets coupled with reversible computation has the potential of providing an attractive setting for analysing systems (for instance in biology, chemistry or electrical engineering). Our application of RPNs in the antenna selection problem is a pioneering one, and we believe that the RPN approach can be expanded to other resource management problems in electrical engineering, drawing benefits from both the conservation properties of RPNs as well as the ability to run the networks, or their parts, in reverse direction to recover from faults and handle inherently reversible communication phenomena (e.g. receiver/transmitter duality). 
%
  
 Specifically, intelligent monitoring of the electric power system requires careful specification and modelling in order to analyse its constraints and deliverables, as well as to avoid hazards, wastage of resources and security threats.
 In order to observe the global behaviour of the power system, a global model representing the different smart grid components and the different modes of communication between these components is needed. The selected modelling formalism should be intuitive and should support the specification of the appropriate abstraction level where in a higher level we should be  able to represent the various actors in a smart grid and in a lower level we should be able to include consumers/prosumers as well as their electrical appliances. The simulation of the model should be able to support all the features of the smart grid and incorporate all the technologies used in the smart grid; the model can therefore be used by the smart grid engineers and designers for what-if analysis and experimentation.  
We believe that reversing Petri nets satisfy the above requirements and that they can be used in order to represent the dynamic and complex behaviour associated with a smart grid covering all its functionalities ranging from the generation of power to intelligent billing mechanisms. 

Other than electrical engineering, reversibility attracts much interest for its potential in many other application areas ranging from cellular automata, programming languages, circuit design to quantum computing. Of a particular interest is quantum computing which is a form of computing that performs based on quantum mechanical phenomena such as superposition and entanglement. Many of the components in quantum computers, such as databases or modular exponentiation, obey the fundamental laws of physics which are inherently reversible making quantum computations also reversible~\cite{Structural}.
Our aim is to transition from quantum theory to quantum engineering by formally presenting the fundamental rules governing quantum systems, along with methodologies for verification of correctness, safety and reliability of these systems. Due to some essential differences between classical and quantum systems, classical model-checking techniques cannot be directly applied to quantum systems. Therefore, an interesting direction would be to extend RPNs to be able to model the behaviour of systems that combine classical and quantum communication and computation. The aim of this study  could be to extend the application area of reversing Petri nets, which have been very successfully used to model classical engineering systems, by modelling the use and operation of Feynman's quantum computer~\cite{feynman}. Future research could develop model-checking techniques that can be used not only for quantum communication protocols but also for general quantum systems, including physical systems and quantum programs. We envisage that quantum model-checking techniques can be applied for checking physical systems, verification of quantum circuits, analysis and verification of quantum programs, and verification of security of quantum communication protocols. 
%
 Similar work was done in process calculi, the most prominent being qCCS,  a natural quantum extension of CCS~\cite{qccs} and CQP~\cite{cqp}, a combination of the communication primitives of pi-calculus with primitives for measurement and transformation of quantum states. 

%
%
%
%
%


\singlespace

\bibliographystyle{abbrv}

\bibliography{References.bib} 

\appendix

\end{document}